\documentclass[twoside,fleqn,11pt,a4paper]{report}
\usepackage{titlesec}
\usepackage{gensymb}
\usepackage{amsthm,amsmath,amssymb,epsfig,bm,mathrsfs,color}
\usepackage{slashed}
\usepackage[normalem]{ulem}
\usepackage{isotope}
\usepackage[german,english]{babel}
\usepackage{caption}
\usepackage{pdfpages}
\usepackage[hidelinks]{hyperref}

\hypersetup{ pdftitle = {Nuclear systems under extreme conditions:
    isospin asymmetry and strong B-fields}, pdfsubject = {Study of
    superfluid isospin-asymmetric nuclear matter, effects of magnetic
    fields on nuclei and superfluid polarized neutron matter.},
  pdfauthor = {Martin Stein}, pdfkeywords = {Asymmetric nuclear
    matter, Polarized neutron matter, Forces in hadronic systems and
    effective interactions, Theories and models of superconducting
    state, Equations of state of nuclear matter, Hartree-Fock, Strong
    magnetic fields, Sky3D} }

\usepackage[twoside,top=3cm, bottom=4cm, outer=5cm, inner=3.5cm,
headsep=14pt]{geometry}

\usepackage{fancyhdr}

\renewcommand{\headrulewidth}{.4pt}
\usepackage{setspace}

\usepackage{booktabs}
\usepackage{nicefrac}
\usepackage[activate]{pdfcprot}
\usepackage{pdfpages}

\newcommand{\vecp}{{\bm p}}
\newcommand{\veck}{{\bm k}}
\newcommand{\vecB}{{\bm B}}
\newcommand{\vecQ}{{\bm Q}}
\newcommand{\vecR}{{\bm R}}
\newcommand{\vecl}{{\bm l}}
\newcommand{\vecL}{{\bm L}}
\newcommand{\vecJ}{{\bm J}}
\newcommand{\vecS}{{\bm S}}
\newcommand{\vecs}{{\bm s}}
\newcommand{\vecj}{{\bm j}}
\newcommand{\vecsigma}{{\bm \sigma}}
\newcommand{\vecr}{\bm r}
\newcommand{\vecW}{\bm W}
\newcommand{\vecA}{\bm A}
\newcommand{\SD}{^3S_1$-$^3D_1}

\begin{document}

\setcounter{chapter}{0}

\pagenumbering{roman}
\thispagestyle{empty}

\vspace*{3\baselineskip}

\begin{center}

  \noindent
  \begin{Huge}
    \textbf{ Nuclear systems under extreme conditions: isospin
      asymmetry and strong B-fields }
  \end{Huge}

  \vspace*{4\baselineskip}

  \noindent
  \begin{Large} Dissertation\\ zur Erlangung des Doktorgrades\\ der Naturwissenschaften\\
  \end{Large}
  \vspace*{3\baselineskip}

  \noindent
  \begin{Large} vorgelegt beim Fachbereich Physik\\ der Johann Wolfgang Goethe--Universit\"at\\ in Frankfurt am Main\\
  \end{Large}
  \vspace*{3\baselineskip}

  \noindent
  \begin{Large} von\\ Martin Stein\\ aus Frankfurt am Main\\
  \end{Large}
  \vspace*{3\baselineskip}

  \noindent
  \begin{Large} Frankfurt am Main 2015\\ (D30)
  \end{Large}

\end{center}
\clearpage

\thispagestyle{empty} \mbox{} \vspace*{24\baselineskip} \vfill

\begin{large}
  \noindent vom Fachbereich Physik der Johann Wolfgang Goethe--Universit\"at\\ in Frankfurt am Main als Dissertation angenommen\\

\end{large}
\vspace*{3\baselineskip}

\begin{large}
  \noindent
  \begin{tabular}{rl}
    \parbox[t]{0.37\linewidth}{\mbox{} \hfill Dekan} &
                                                       \parbox[t]{0.80\linewidth}{Prof.\ Dr.\ Rene Reifarth} \\[1.5em]
    \parbox[t]{0.37\linewidth}{\mbox{} \hfill Gutachter} & 
                                                           \parbox[t]{0.80\linewidth}{Prof.\ Dr.\ Joachim A.\ Maruhn,\\PD\ Dr.\ Armen Sedrakian} \\[2.5em] \parbox[t]{0.37\linewidth}{\mbox{} \hfill Datum der Disputation } & 
                                                                                                                                                                                                                               \parbox[t]{0.80\linewidth}{16. Dezember 2015} \\[1.5em]
  \end{tabular}\\
\end{large}
\vspace*{3\baselineskip}
\newpage

\onehalfspacing

\setlength{\marginparwidth}{0.6\marginparwidth}

\newlength{\totalmarginwidth}
\setlength{\totalmarginwidth}{\marginparsep}
\addtolength{\totalmarginwidth}{\marginparwidth}

\newlength{\totallinewidth}
\setlength{\totallinewidth}{\linewidth}
\addtolength{\totallinewidth}{\totalmarginwidth}

\fancypagestyle{plain}{\fancyhf{}\fancyfoot[C]{\thepage}
\renewcommand{\headrulewidth}{0pt}}
\setlength{\headwidth}{\totallinewidth}

\setcounter{page}{1}
\renewcommand{\figurename}{Abbildung}
\chapter*{\"Ubersicht}
Diese Doktorarbeit besch\"aftigt sich mit Kernmaterie und Atomkernen
unter extremen Bedingungen, wie sie z.B. in kompakten Sternen
vorkommen k\"onnen. Kapitel~\ref{chap_1} untersucht suprafluide
Neutron-Proton Paarung in Iso\-spin-asymmetrischer Kernmaterie im
$\SD$ Kanal. In Kapitel~\ref{chap_2} untersuchen wir den Einflu\ss{}
starker Magnetfelder auf \isotope[12]{C}, \isotope[16]{O} und
\isotope[20]{Ne}. Abschlie\ss{}end untersuchen wir in
Kapitel~\ref{chap_3} suprafluide Neutron-Neutron Paa\-rung in
Spin-asymmetrischer (polarisierter) Neutronenmaterie im $^1S_0$
Ka\-nal; eine Polarisation kann z.B. durch ein magnetisches Feld
verursacht werden.

In Kapitel~\ref{chap_1} erhalten wir ein reichhaltiges Phasendiagramm
f\"ur Isospin-asym\-metrische Kernmaterie. Ein besseres Verst\"andnis
dieser Materie kann z.B. f\"ur niederenergetische
Schwerionenkollisionen, Supernovaexplosionen oder Atomkerne wichtig
sein. Im \"au\ss{}eren Bereich von Atomen ist die Dichte gering. Dies
f\"uhrt dazu, dass eine Isospin-Asymmetrie die Neutron-Proton-Paarung
kaum unterdr\"uckt. Wir untersuchen die ungepaarte Phase und
verschiedene suprafluide Phasen. Wir untersuchen den Crossover von der
schwach gebundenen Bardeen Cooper Schrieffer (BCS) Phase bei hohen
Dichten hin zum Bose-Einstein-Kondensat (englisch: Bose-Einstein
condensate) (BEC) im Grenzfall starker Kopplung bei niedrigen
Dichten. Au\ss{}erdem untersuchen wir zwei exotische Phasen: Die
Larkin-Ovchinnikov-Fulde-Ferrell (LOFF) Phase, bei der die
Cooper-Paare einen endlichen Schwerpunktsimpuls erhalten. Diese Phase
taucht nur bei hohen Dichten auf. Au\ss{}erdem untersuchen wir eine
Phasenseparation (PS), bei der die Materie in einen
Isospin-symmetrischen Teil in der BCS oder BEC Phase und einen
ungepaarten Teil mit Neutronen\"uberschuss aufgeteilt wird. Die
Phasenseparation kann sowohl bei hohen als auch bei niedrigen Dichten
auftauchen, weshalb wir in der Phasenseparation einen Crossover
erhalten. Der Phasen\"ubergang zwi\-schen LOFF und PS ist erster
Ordnung, alle anderen sind Phasen\"uberg\"ange zweiter
Ordnung. Au\ss{}erdem untersuchen wir den Gap, den Kernel der
Gap-Gleichung, die Wellenfunktionen der Cooper-Paare, die
Besetzungszahlen und die Einteilchenenergien. Im BCS Grenzfall
erhalten wir ein fer\-mio\-ni\-sches und im BEC Grenzfall erhalten wir
ein bosonisches Verhalten. Im Fall der LOFF Phase n\"ahern sich die
oben aufgef\"uhrten Funktionen denen der BCS Phase mit verschwindender
Isospin-Asymmetrie an.

In Kapitel~\ref{chap_2} untersuchen wir den Einfluss eines starken
Magnetfeldes auf die Elemente \isotope[16]{O}, \isotope[12]{C} und
\isotope[20]{Ne}. Diese Elemente k\"onnen z.B. in Wei\ss{}en Zwer\-gen
vorkommen, welche starke Magnetfelder aufweisen k\"onnen. Des Weiteren
k\"onnen diese Elemente bei akkretierenden Neutronensternen eine Rolle
spielen. Bei \isotope[16]{O} und \isotope[12]{C} werden die
Einteilchenenergien mit zunehmendem Magnetfeld
aufgespalten. Au\ss{}erdem werden Bahndrehimpuls und Spin bei starken
Magnetfeldern am Magnetfeld ausgerichtet, diese Ausrichtung wird bei
schwachen Magnetfeldern durch die Spin-Bahn-Kopp\-lung
unterdr\"uckt. Bei starken Magnetfeldern werden bei \isotope[16]{O}
die Energie\-niveaus umbesetzt. Die kol\-lek\-ti\-ve
Flie\ss{}\-ge\-schwin\-dig\-keit in den Atomkernen beschreibt
kreis\-f\"or\-mi\-ge oder nahezu kreisf\"ormige Bahnen um die
Magnetfeldachse. Die Spindichte richtet sich bei starkem
Ma\-gnet\-feld aus. \isotope[20]{Ne} ist bei ver\-schwin\-den\-dem
Ma\-gnet\-feld stark verformt, diese Verformung nimmt mit zunehmendem
Magnetfeld ab.

Das in Kapitel~\ref{chap_3} untersuchte Phasendiagramm f\"ur
polarisierte Neutronenmaterie kann f\"ur Studien in Neutronenmaterie
von gro\ss{}er Bedeutung sein; besonders f\"ur die innere Kruste von
Neutronensternen. Auch f\"ur Untersuchungen an Atomkernen kann es von
Bedeutung sein. Es gibt ph\"anomenologische Hinweise auf
Neutronen-Suprafluidit\"at in Neutronensternen. Das erhaltene
Phasendiagramm besteht nur aus der ungepaarten Phase und der BCS
Phase. Da es keine gebundenen Neutron-Neutron-Paare gibt, kann kein
BEC entstehen. Da die Kopplungsst\"arke im $^1S_0$ Kanal schw\"acher
ist als im $\SD$ Kanal, ist die kritische Temperatur geringer als bei
dem in Kapitel~\ref{chap_1} analysierten Phasendiagramm. F\"ur die
mikroskopischen Funktionen erhalten wir \"ahnliche Resultate wie f\"ur
die BCS Phase in Kapitel~\ref{chap_1}. Au\ss{}erdem haben wir das
f\"ur eine bestimmte Polarisation ben\"otigte Magnetfeld berechnet und
dessen Energie mit der Temperatur des Systems verglichen. Hierbei ist
die magnetische Energie in dem von uns analysiertem Bereich in der
Regel gr\"o\ss{}er als die Temperatur.

\clearpage{\pagestyle{empty}\cleardoublepage}
\thispagestyle{empty}
\chapter*{Zusammenfassung}
\section*{Einleitung}
In dieser Arbeit untersuchen wir Kernmaterie und Atomkerne unter
extremen Bedingungen. In Kapitel~\ref{chap_1} und \ref{chap_3}
untersuchen wir suprafluide Pha\-sen von Kernmaterie
bzw. Neutronenmaterie. In Kapitel~\ref{chap_2} und \ref{chap_3}
untersuchen wir den Einfluss starker Magnetfelder auf Atomkerne
bzw. auf Neutronenmaterie.

Die Untersuchungen der Crossovers mit Einbeziehung von
unkonventionellen Phasen, wie sie in Kapitel~\ref{chap_1} er\"ortert
werden, k\"onnte hilfreich sein bei Untersuchungen von fermionischen
Systemen mit unaus\-ge\-gli\-che\-nem Spin/Flavor in ultrakalten
atomischen Gasen, siehe
z.B.~\cite{2007AnPhy.322.1790S,2008RvMP...80.1215G,2011JPhCS.321a2028S},
farbsupraleitender dichter Quarkmaterie, siehe
z.B.~\cite{2009PhRvD..80g4022S,2010PhRvD..82e6006M,2014PhRvD..89c6009M,2010PhRvD..82g6002F,2012NuPhA.875...94K},
oder anderen verwandten Quantensystemen. Bei niederenergetischen
Schwer\-ionenkollisionen erh\"alt man im Endzustand viele Deuteronen,
welche $\SD$ Kondensation
nahelegen~\cite{1995PhRvC..52..975B}. Gro\ss{}e Atomkerne wie
z.B. $^{92}$Pd k\"onnten Neutron-Proton-Paare
aufweisen~\cite{2011Natur.469...68C}.

Neutron-Neutron-Paarung wird
z.B. in~\cite{2009PhRvC..79c4304M,2013PhRvC..88c4314P,PhysRevC.73.044309,2007PhRvC..76f4316M,2008PhRvC..78a4306I,2009PhRvC..80d5802G,2006pfsb.book..135S}
untersucht. Neutron-Neutron-Paarung kann wichtig f\"ur Studien von
Neutronensternmaterie sein, besonders f\"ur die innere Kruste, und
f\"ur Atomkerne, besonders f\"ur neutronenreiche wie z.B. $^{11}$Li,
welches einen Neutron-Halo besitzt~\cite{PhysRevC.73.044309}. Die
Rotation von Neutronensternen und Anomalien in der Rotation sprechen
f\"ur suprafluide Phasen~\cite{2006pfsb.book..135S}.

Neutron-Proton-Paarung kann eine wichtige Rolle in Supernova-Materie
spielen, in der die Isospin-Asymmetrie gering ist. Sie kann auch im
\"au\ss{}eren Bereich von Atomkernen auftreten; aufgrund der geringen
Dichte un\-ter\-dr\"uckt die dort vorherrschende Isospin-Asymmetrie
die Neutron-Proton-Paarung kaum.

\section*{Suprafluide Materie}
Bei hohen Dichten von $\rho\lesssim\frac12\rho_0$ mit
$\rho_0=0{,}16\,$fm$^{-3}$, was $2{,}8\cdot10^{14}\,$g\,cm$^{-3}$
entspricht, k\"onnen verschiedene suprafluide Phasen auftreten;
hierbei steht $\rho_0$ f\"ur die Kerns\"attigungsdichte. Diese Phasen
sind mathematisch der Phase supraleitender Elektronen sehr
\"ahnlich. \"Ahnlich wie bei Supraleitung muss auch bei
Suprafluidit\"at die Temperatur gering sein, wobei die Temperatur
hierbei gering bez\"uglich der anderen relevanten Energien sein
muss. Tiefe Temperatur bedeutet in diesem Zusammenhang bis zu mehrere
MeV, wobei ein MeV einer Temperatur von $11{,}6\cdot10^9$ K
entspricht.

Bei ausreichend hohen Dichten erreichen die chemischen Potenziale der
Nukleonen Werte, die mit der Ruhemasse von Hyperonen vergleichbar
sind. In diesem Fall kann die Materie mit Hyperonen angereichert
werden, dies kann bei doppelter Kerns\"attigungsdichte geschehen. Wenn
die Dichten sehr gro\ss{} werden, wird der Teilchenabstand kleiner als
der Nukleonenradius und das Confinement kann aufgehoben werden.

Supraleitende bzw. suprafluide Paarung kann zwischen \"ahnlichen
Fer\-mio\-nen auftreten, die sich aufgrund des Pauli Prinzips in
mindestens einer Quantenzahl unterscheiden m\"ussen. So kann z.B. eine
supraleitende bzw. suprafluide Paa\-rung von zwei Elektronen,
Neutronen oder Protonen unterschiedlichen Spins entstehen. Zwei
Nukleonen unterschiedlichen Isospins, also ein Proton und ein Neutron,
k\"onnen auch mit gleichem Spin eine suprafluide Paarung eingehen. Da
der Massenunterschied zwischen Neutronen und Protonen weniger als
0,14\% der Nukleonenmasse betr\"agt, k\"onnen die Effekte, die
aufgrund des Massenunterschiedes auftreten, vernachl\"assigt
werden. Die Ruhemasse eines Neutrons betr\"agt 939,6 MeV und die eines
Protons betr\"agt 938,3 MeV.

Aus der Streutheorie von Nukleonen kann man die kritische Temperatur
verschiedener Paarungskan\"ale
berechnen~\cite{2006pfsb.book..135S}. Bei den f\"ur uns interessanten
Dichten ist der Spin-Triplett $\SD$ Kanal dominant, wobei er f\"ur
Isospin-Triplett-Paarung (Neutron-Neutron-Paarung) aufgrund des
Pauli-Prinzips verboten ist. Der f\"ur Isospin-Triplett-Paarung
dominante Kanal ist der deutlich schw\"achere Isospin-Singulett
$^1S_0$ Kanal.

In Kapitel~\ref{chap_1} untersuchen wir Isospin-Singulett
Spin-Triplett Paarung (Neutron-Proton-Paarung mit gleichem Spin) im
$\SD$ Kanal in Kern\-ma\-te\-rie und in Kapitel~\ref{chap_3}
untersuchen wir Isospin-Triplett Spin-Singulett Paarung
(Neutron-Neutron-Paarung mit unterschiedlichem Spin) im $^1S_0$ Kanal
in Neutronenmaterie. In Isospin-symmetrischer Kernmaterie
bzw. Spin-symmetrischer Neutronenmaterie haben wir f\"ur tiefe
Temperaturen Paarung in der Bardeen Cooper Schrieffer (BCS)
Phase. Hierbei findet eine Paarung zwischen zwei Nukleonen statt,
deren Impuls betragsm\"a\ss{}ig gleich ist, deren Richtung aber
entgegengesetzt ist ($\veck_1=-\veck_2$). G\"unstig f\"ur eine Paarung
sind sowohl eine hohe Zustandsdichte als auch eine hohe
Kopp\-lungs\-st\"ar\-ke. Die Zustandsdichte nimmt mit steigender
Dichte zu, die Kopp\-lungs\-st\"ar\-ke dagegen nimmt ab. Die
Zustandsdichte dominiert f\"ur geringe und die Kopplungsst\"arke f\"ur
hohe Dichten. F\"ur den Gap bei verschwindender Temperatur und
Asymmetrie $\Delta_{00}$ finden wir folgende Relation:
\begin{eqnarray}
  \Delta_{00}&=&2\varepsilon_F\cdot e^{-\frac1{NV}}\,.
\end{eqnarray}
Folglich steigt der Gap f\"ur geringe Dichten mit zunehmender Dichte,
wohingegen er bei hohen Dichten abf\"allt. Die kritische Temperatur
$T_C$ ist proportional zu diesem Gap:
\begin{eqnarray}
  \Delta_{00}=1{,}76\,T_C\,.
\end{eqnarray}
Die Asymmetrie $\alpha$ bezieht sich im Isospin-Singulett
Spin-Triplett Zustand auf eine Isospin Asymmetrie $\alpha_\tau$ und im
Isospin-Triplett Spin-Singulett Zustand auf eine Spin Asymmetrie --
oder auch Polarisation -- $\alpha_\sigma$ mit
\begin{subequations}
  \begin{eqnarray}
    \alpha_\tau&=&\frac{\rho_n-\rho_p}{\rho_n+\rho_p}\,,\\
    \alpha_\sigma&=&\frac{\rho_{n\uparrow}-\rho_{n\downarrow}}{\rho_{n\uparrow}+\rho_{n\downarrow}}\,,
  \end{eqnarray}
\end{subequations}
wobei $\rho_i$ sich auf die jeweilige Anzahldichte der Teilchensorte
$i$ bezieht. $\rho_n$ und $\rho_p$ sind als Summe der jeweiligen Spin
up und Spin down Teilchen zu verstehen;
$\rho_\tau=\rho_{\tau\uparrow}+\rho_{\tau\downarrow},\,\tau\in\{n,p\}$.

\section*{Das Phasendiagramm suprafluider Kern- und Neutronenmaterie}
In Abbildung~\ref{fig_0_01} sehen wir das Phasendiagramm f\"ur
Isospin-Singulett Spin-Triplett Paarung (Neutron-Proton-Paarung mit
gleichem Spin) im $\SD$ Kanal in Kernmaterie und in
Abbildung~\ref{fig_0_02} das Phasendiagramm f\"ur Isospin-Triplett
Spin-Singulett Paarung (Neutron-Neutron-Paa\-rung mit
un\-ter\-schied\-lichem Spin) im $^1S_0$ Kanal. Die Paarung im
$\SD$-Kanal ist n\"aher beschrieben in den Kapitel~\ref{chap_1}
zugrundeliegenden
Publikationen~\cite{2012PhRvC..86f2801S,2014PhRvC..90f5804S}. Um das
Phasendiagramm zu bestimmen, haben wir ein gekoppeltes
Glei\-chungssystem f\"ur den Gap und die Dichten gel\"ost
(Glei\-chungen~\eqref{eq_1_38} und \eqref{eq_1_40}
bzw.~\eqref{eq_3_01} und \eqref{eq_3_20}).

In der Natur wird der Zustand mit niedrigster Energie realisiert. Wir
untersuchen die normale, ungepaarte Phase und verschiedene suprafluide
Pha\-sen. Neben dem BCS untersuchen wir zwei exoti\-sche suprafluide
Pha\-sen, auf die weiter unten genauer eingegangen wird. Welche Phase
die niedrigste freie Energie hat, haben wir mit
Gleichungen~\eqref{eq_1_45} und \eqref{eq_1_47} bzw.~\eqref{eq_3_19}
bestimmt, wobei wir bei Neutron-Neutron-Paarung die M\"oglichkeit
einer Phasenseparation nicht ber\"ucksichtigt haben. F\"ur
Neutron-Neutron-Paa\-rung erhalten wir keinen Bereich, in dem die
Larkin-Ovchinnikov-Fulde-Ferrell (LOFF) Phase am energetisch
g\"unstigsten ist. Neben den verschiedene Phasen untersuchen wir einen
Crossover, der unten genauer erkl\"art wird.

Wir haben separable Paris Potenziale aus~\cite{1984PhRvC..30.1822H}
verwendet. F\"ur die Neutron-Proton-Paarung im $\SD$ Kanal haben wir
das PEST~1 und f\"ur die Neutron-Neutron-Paarung im $^1S_0$ Kanal das
PEST~3 Potenzial verwendet. Die effektive Masse haben wir mit der
Skyrmekraft SkIII aus~\cite{1987PhRvC..35.1539S} berechnet.

\subsection*{Allgemeiner Verlauf}
Wie oben beschrieben, steigt $T_C$ bei geringen Dichten mit
zunehmender Dichte, wohingegen es bei hohen Dichten abf\"allt. Ein
Vergleich der beiden untersuchten Paarungskan\"ale zeigt, dass die
kritische Temperatur des $^1S_0$ Kanals deutlich geringer ist als die
des $\SD$ Kanals.

\begin{figure}[!]
  \begin{center}
    \includegraphics[width=0.8\textwidth]{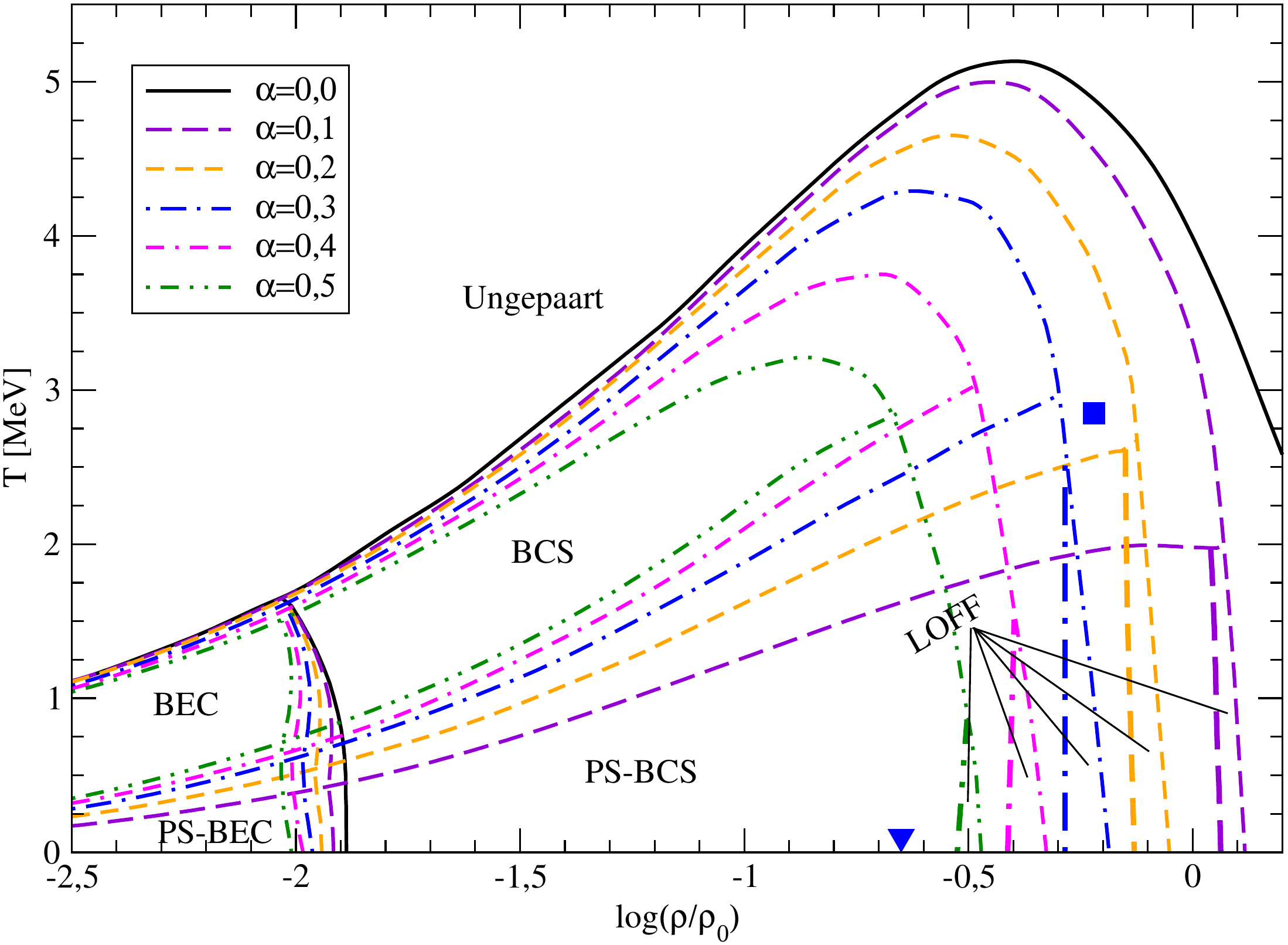}
    \caption[]{Das Phasendiagramm von Kernmaterie im Isospin-Singulett
      Spin-Triplett $\SD$ Kanal in der Temperatur-Dichte Ebene f\"ur
      verschiedene Isospin-Asymmetrien $\alpha$
      aus~\cite{2012PhRvC..86f2801S}. Wir sehen vier Phasen: die
      ungepaarte Phase, die BCS (BEC) Phase, die LOFF Phase und die PS
      (PS-BCS und PS-BEC) Phase. F\"ur gen\"ugend kleine Asymmetrien
      sehen wir zwei trikritische Punkte. F\"ur einen be\-stimm\-ten
      Wert der Asymmetrie fallen diese beiden Werte zusammen und wir
      erhalten einen tetrakri\-tischen Punkt, dargestellt durch ein
      blaues Quadrat. Das blaue Dreieck zeigt den Punkt der
      niedrigsten Dichte und gleichzeitig h\"ochsten Asymmetrie, an
      dem die LOFF Phase existiert.}
    \label{fig_0_01}
  \end{center}
\end{figure}

\begin{figure}[!]
  \begin{center}
    \includegraphics[width=0.8\textwidth]{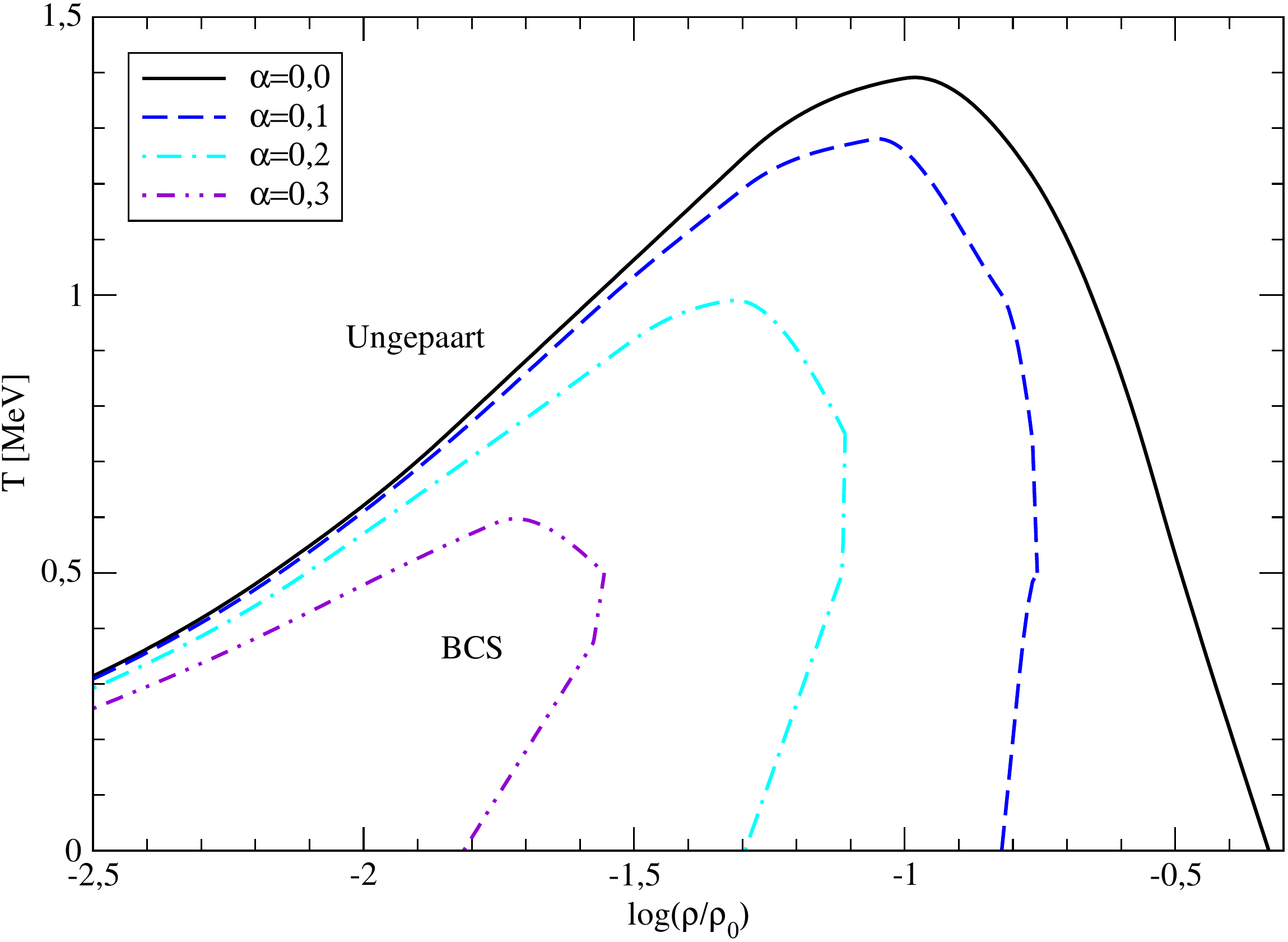}
    \caption[]{Das Pha\-sen\-dia\-gramm von Neutronenmaterie im
      Isospin-Triplett Spin-Singulett $^1S_0$ Kanal in der
      Temperatur-Dichte Ebene f\"ur verschiedene Spin-Asymmetrien
      $\alpha$. Wir sehen die ungepaarte Phase und die BCS Phase.}
    \label{fig_0_02}
  \end{center}
\end{figure}

Als N\"achstes wollen wir auf den Effekt der Asymmetrie eingehen. Wie
oben beschrieben, findet die Paarung in der BCS-Phase zwischen zwei
Teilchen mit betragsm\"a\ss{}ig gleichem aber entgegengerichtetem
Impuls statt. Die Paarung findet hierbei bei tiefen Temperaturen in
der N\"ahe der Fermikante statt. Eine Asymmetrie ver\"andert die
Dichten und somit auch die Fermiimpulse der Paarungspartner. In
asymmetrischer Kernmaterie haben wir mehr Neutronen als Protonen
($\rho_n>\rho_p$), in asymmetrischer Neutronenmaterie gehen wir -- wie
oben -- von einem Spin up \"Uberschuss aus
($\rho_{\uparrow}>\rho_{\downarrow}$). (F\"ur die Rechnungen spielt es
keine Rolle, ob mit einem Spin up oder Spin down \"Uberschuss
gerechnet wird.) Somit gilt auch: $k_{F_n}>k_{F_p}$
bzw. $k_{F_\uparrow}>k_{F_\downarrow}$. Folglich wird die Paarung
durch die Asymmetrie unterdr\"uckt. Die St\"arke der Unterdr\"uckung
h\"angt von der Dichte ab: Im Grenzfall hoher Dichten erhalten wir
Stufenfunktionen f\"ur die Besetzungszahlen. Hierdurch wird der
Bereich um die Fermikante, in dem Paarung stattfinden kann,
gering. F\"ur niedrige Dichten werden die Besetzungszahlen
aufgeweicht, wodurch der Bereich um die Fermikante, in dem Paarung
stattfinden kann, vergr\"o\ss{}ert wird. Somit hat die Unterdr\"uckung
der Paarung durch die Asymmetrie nur bei hohen Dichten starke
Auswirkungen, was auch gut in den Abbildungen~\ref{fig_0_01} und
\ref{fig_0_02} zu sehen ist. Die Pauli-Absto\ss{}ung ist f\"ur geringe
Dichten weniger effektiv.

Im Phasendiagramm f\"ur Neutronenmaterie in Abbildung~\ref{fig_0_02}
erhalten wir bei bestimmten Werten f\"ur Asymmetrie und Dichte eine
untere kritische Temperatur~\cite{2000PhRvL..84..602S}. Bei $T=0$
befindet sich die Neutronenmaterie in der ungepaarten Phase. Eine
Erh\"ohung der Temperatur f\"uhrt bei der unteren kritischen
Temperatur zu einem Phasen\"ubergang in die BCS Phase, eine weitere
Erh\"ohung f\"uhrt in die ungepaarte Phase. Diese untere kritische
Temperatur hat folgenden Grund: F\"ur eine Paarung werden
\"uberlappende Fermikanten ben\"otigt. Eine endliche Asymmetrie
f\"uhrt zu einer Aufspaltung der Fermikanten, folg\-lich wird f\"ur
eine Paarung ein Effekt ben\"otigt, der die Fermikanten auf\-weicht;
z.B. eine entsprechend hohe Temperatur. Im Phasendiagramm f\"ur
Kernmaterie in Abbildung~\ref{fig_0_01} erhalten wir aufgrund der
exotischen Phasen keine untere kritische Temperatur.

\subsection*{Crossover von BCS nach BEC}
Fermionische Suprafl\"ussigkeiten, die im Grenzfall schwacher Kopplung
ein BCS aus schwach gebundenen Cooper-Paaren bilden, gehen \"uber in
ein Bose-Einstein-Kondensat (englisch: Bose-Einstein condensate) (BEC)
aus stark gebundenen bosonischen Dimeren, wenn die St\"arke der
Paarung aus\-rei\-chend gro\ss{}
wird~\cite{1985JLTP...59..195N,1969PhRv..186..456E}. F\"ur Paarung im
$\SD$ Kanal erhalten wir ein BEC aus Deuteronen im Grenzfall starker
Kopplung~\cite{1993NuPhA.551...45A,1995PhRvC..52..975B,1995ZPhyA.351..295S,2001PhRvC..63c8201L,
  2006PhRvC..73c5803S,2009PhRvC..79c4304M,2010PhRvC..81c4007H,2010PhRvC..82b4911J,2013JPhCS.413a2024S,2014JPhCS.496a2008S,2013PhRvC..88c4314P,2013PhRvC..88b5806S,2013arXiv1308.0364S,2013NuPhA.909....8S}.

Im Grenzfall hoher Dichten erhalten wir Paarung ungebundener Teilchen
im BCS. Die Paarung erfolgt hierbei an der Fermikante, die
Kopplungsst\"arke ist schwach. Das mittlere chemische Potenzial der
beiden Paarungspartner $\bar\mu$ ist gr\"o\ss{}er als null. Wenn wir
die Dichte verringern, verringert sich auch das mittlere chemische
Potenzial. Bei geringen Dichten erhalten wir ein BEC aus gebundenen
Teilchen mit negativem mittlerem chemischen Potenzial. Hierbei ist die
Kopplungsst\"arke gro\ss{}.

\begin{figure}[!]
  \begin{center}
    \includegraphics[width=0.7\textwidth]{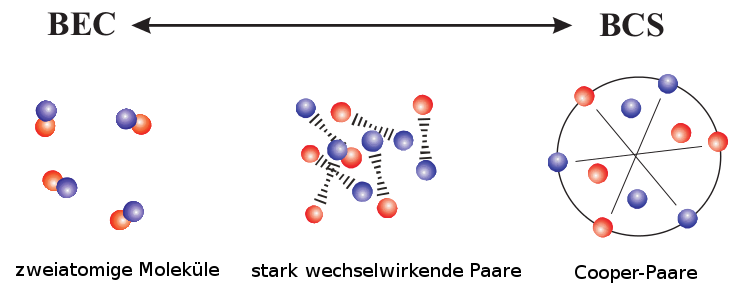}
    \caption[]{Eine Darstellung des Crossovers \"ubernommen
      aus~\cite{2006PhDT.......107R}. Links sehen wir gebundene
      Deuteronen, rechts sehen wir ungebundene Cooper-Paare.}
    \label{fig_0_03}
  \end{center}
\end{figure}

Ein weiteres Kriterium f\"ur den Crossover ist das Verh\"altnis des
mitt\-le\-ren Teilchenabstandes $d$ und dem Abstand von zwei gepaarten
Teilchen $\xi$. Dies ist in Abbildung \ref{fig_0_03}
dargestellt. Rechts sehen wir die Situation, in der ein BCS Bereich
vorkommt, hierbei gibt es Paarung an der Fermifl\"ache von zwei
Teilchen, die eine gro\ss{}e r\"aumliche Distanz haben; $\xi\gg d$.
Links sehen wir stark gebundene Paare, die r\"aumlich von den anderen
Paaren isoliert sind; $\xi\ll d$.

Ein BEC kann nicht in jedem Paarungskanal entstehen. Im $\SD$ Kanal
k\"onnen gebundene Paare entstehen; im Grenzfall verschwindender
Dichte erhalten wir gebundene Deuteronen. Im $^1S_0$ Kanal k\"onnen
keine gebundenen Paare entstehen; im Grenzfall verschwindender Dichte
erhalten wir freie Neutronen. Einen \"Ubergangsbereich kann man
trotzdem
nachweisen~\cite{PhysRevC.73.044309,2007PhRvC..76f4316M,2009PhRvC..79c4304M}.

Der \"Ubergang von einem BCS zu einem BEC ist kein Phasen\"ubergang,
weil keine Symmetrie gebrochen wird. Es handelt sich vielmehr um
Grenz\-f\"al\-le des gleichen Ph\"anomens.

\subsection*{Exotische Phasen}
Neben der ungepaarten Phase und der BCS Phase haben wir exotische
Phasen untersucht. Zum einen eine Phase, bei der die Cooper-Paare
einen endlichen Schwerpunktsimpuls
haben~\cite{2001PhRvC..63b5801S,2003PhRvC..67a5802M,2009PhRvC..79c4304M}. Diese
Phase ist analog zur Larkin-Ovchinnikov-Fulde-Ferrell (LOFF) Phase in
elekt\-ri\-schen
Su\-pra\-lei\-tern~\cite{1965LO,1964PhRv..135..550F}. Wir hatten
gesehen, dass man f\"ur eine Paarung \"uberlappende Fermikanten
ben\"otigt. Diese Fermikanten n\"ahern sich Stufenfunktionen f\"ur
hohe Dichten und tiefe Temperaturen und werden f\"ur hohe Asymmetrien
getrennt, was eine Paarung in der BCS Phase unm\"oglich macht.

\begin{figure}[!]
  \begin{center}
    \includegraphics[width=0.6\textwidth]{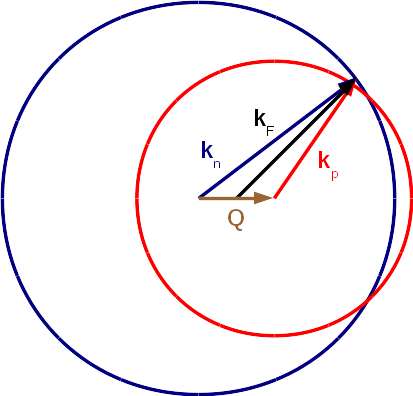}
    \caption[]{Die Fermifl\"ache f\"ur die beiden Komponenten in der
      LOFF Phase. Durch die Verschiebung der Fermifl\"achen erh\"oht
      sich einerseits die kinetische Energie, andererseits kommt es
      durch den \"Uberlapp der Fermifl\"achen zu einer Kondensation
      und somit zu einer Absenkung der Ener\-gie.}
    \label{fig_0_04}
  \end{center}
\end{figure}

Eine M\"oglichkeit trotzdem \"uberlappende Fermikanten zu bekommen,
ist, die Fermikanten gegeneinander zu verschieben. Dies ist in
Abbildung~\ref{fig_0_04} dargestellt. Wir erhalten einen endlichen
Cooper-Paar-Impuls $\vecQ$, der in braun dargestellt ist. Die Kreise
geben die Fer\-mi\-fl\"a\-chen von Neutronen (blau) und Protonen (rot)
an. Wir sehen, dass sich die Fermifl\"achen durch die Verschiebung um
$\vecQ$ kreuzen und es einen Bereich gibt, in dem die Fermifl\"achen
nahe beieinander sind. Au\ss{}erdem sehen wir in Schwarz den Vektor
$\veck_F$, der mit $\vecQ$ einen Winkel von $45\degree$
einschlie\ss{}t und die dazugeh\"origen Vektoren der Neutronen
($\veck_n$ blau) und Protonen ($\veck_p$ rot). Bei diesem Winkel
kompensiert der Cooper-Paar-Impuls die Verschiebung der
Fer\-mi\-fl\"a\-chen sehr gut. Durch den Cooper-Paar-Impuls erh\"oht
sich die ki\-ne\-tische Ener\-gie des Systems. Andererseits wird durch
die Kondensation die Ener\-gie vermindert. Die LOFF Phase bricht die
Translationssymmetrie und ist somit -- im Gegensatz zum Crossover von
BCS nach BEC -- ein Phasen\"ubergang. Nach unseren Rechnungen tritt
die LOFF Phase in Spin-asymmetrischer Neutronenmaterie nicht auf.

Eine weitere Phase, die wir im $\SD$ Kondensat untersucht haben, ist
die Phasenseparation (PS); hier trennt sich die Materie in zwei
Bereiche auf: in einen Isospin-symmetrischen Teil, in dem symmetrische
BCS/BEC Paarung stattfindet und in einen ungepaarten Teil, der einen
starken Neutronen\"uberschuss besitzt. Die Phasenseparation wurde in
kalten atomaren Gasen vorgeschlagen~\cite{2003PhRvL..91x7002B}. Diese
Phase gibt es -- im Gegensatz zum LOFF -- auch bei geringen
Dichten. Den Crossover von BCS zu BEC gibt es auch in der
Phasenseparation. Beim \"Ubergang zur Phasenseparation wird auch eine
Symmetrie gebrochen: Das System ist anschlie\ss{}end nicht mehr
homogen.

Sehr interessant ist auch der Verlauf der Phasen\"uberg\"ange. Wir
erhalten zwei trikritische Punkte, die je nach Asymmetrie an
verschiedenen Phasen angrenzen. F\"ur bestimmte Werte von Dichte,
Temperatur und Asymmetrie fallen diese beiden Punkte zusammen und wir
erhalten einen tetrakritischen Punkt, an dem vier Phasen koexistieren:
LOFF, PS, BCS und die ungepaarte Phase. F\"ur die Ordnung der
Phasen\"uberg\"ange erhalten wir Folgendes: Fast alle
Phasen\"uberg\"ange sind zweiter Ordnung, weil die \"Anderung der
Parameter glatt verl\"auft. Die einzige Ausnahme ist der \"Ubergang
von LOFF nach PS, dort macht der Gap einen Sprung, was einem
Phasen\"ubergang erster Ordnung entspricht.

\subsection*{Mikroskopische Funktionen}
Neben dem Verlauf des Phasendiagramms besch\"aftigt sich diese Arbeit
auch mit mi\-kro\-sko\-pi\-schen Funktionen. Wir haben den Gap, den
Kernel der Gap-Gleichung, die Wellenfunktionen der Cooper-Paare, die
Besetzungszahlen und die Einteilchenenergien berechnet. Dies haben wir
sowohl f\"ur $\SD$ Paarung als auch f\"ur $^1S_0$ Paarung
durchgef\"uhrt. Im $\SD$ Kanal konnten wir auch den Crossover und die
LOFF Phase betrachten, im $^1S_0$ Kanal waren wir auf die BCS Phase
beschr\"ankt. Wir haben keine mi\-kro\-sko\-pi\-schen Funktionen in
der Phasenseparation dargestellt; da sich ein Teil der Materie in
einem symmetrischen BCS/BEC befindet, kann hierbei keine neue
physikalische Erkenntnis gewonnen werden.

\begin{figure}[!]
  \begin{center}
    \begin{minipage}[t]{\textwidth}
      \begin{center}
        \includegraphics[width=0.7\textwidth]{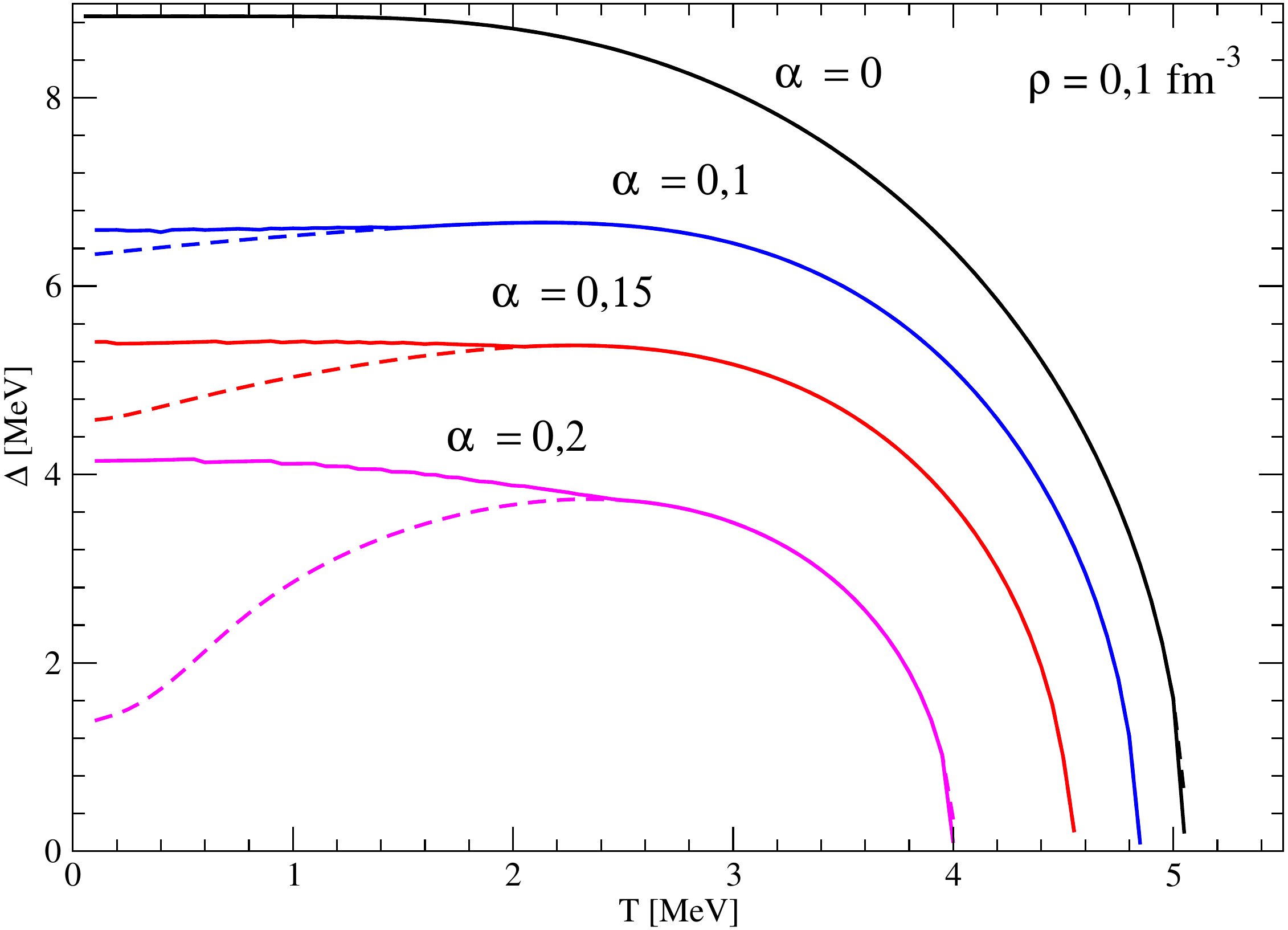}
      \end{center}
    \end{minipage}
    \newline\newline
    \begin{minipage}[t]{\textwidth}
      \begin{center}
        \includegraphics[width=0.7\textwidth]{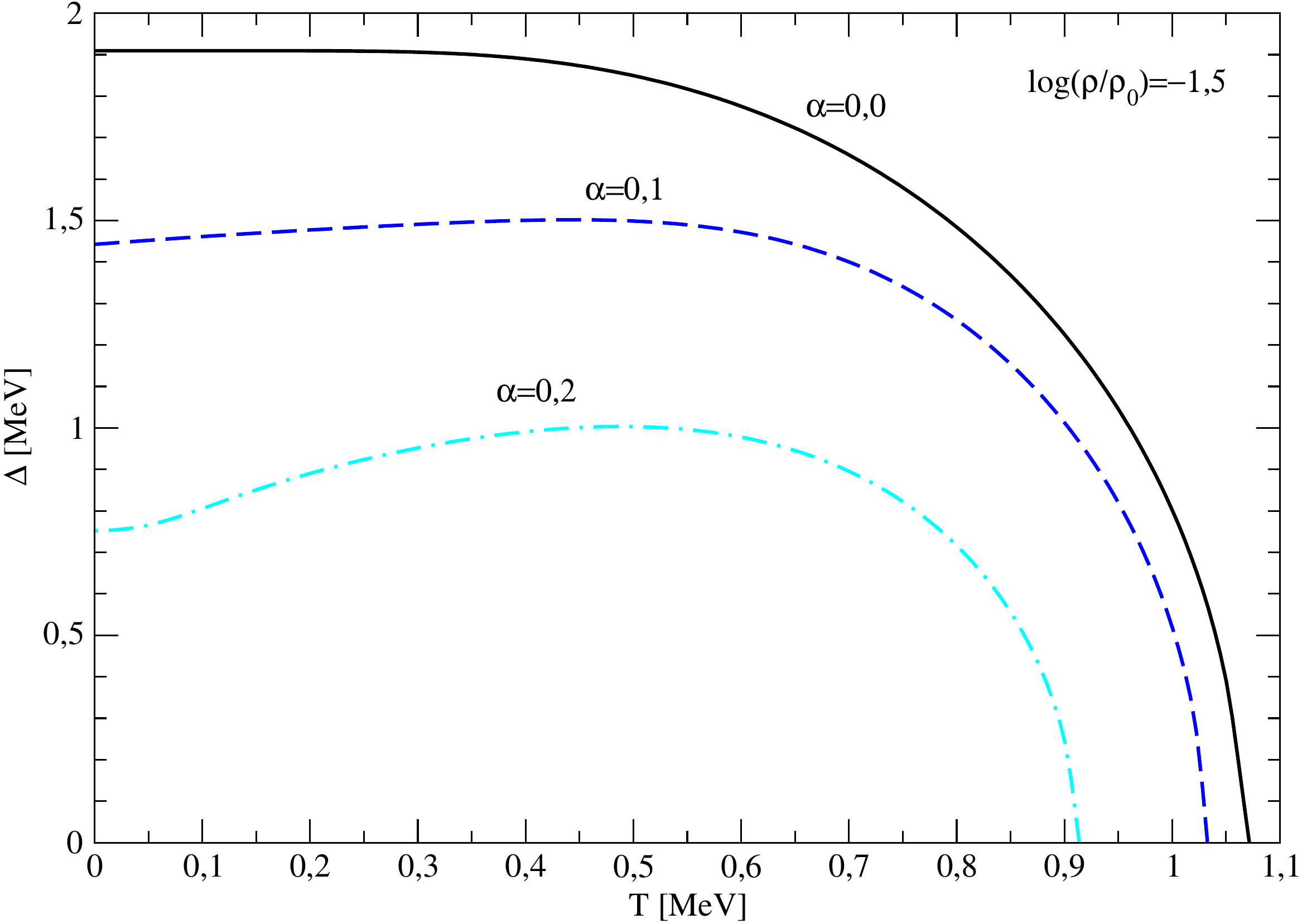}
      \end{center}
    \end{minipage}
    \caption[]{Der Gap als Funktion der Temperatur bei konstanter
      Dichte f\"ur verschiedene Asymmetrien. Oben f\"ur den $\SD$ und
      unten f\"ur den $^1S_0$ Kanal. Im $\SD$ Kanal beziehen sich die
      gestrichelten Linien auf die BCS Phase und die durchgezogenen
      Linien auf die resultierende Phase, BCS oder LOFF.}
    \label{fig_0_05}
  \end{center}
\end{figure}

In Abbildung~\ref{fig_0_05} sehen wir den Gap als Funktion der
Temperatur bei konstanter Dichte f\"ur verschiedene Asymmetrien. Oben
sehen wir den $\SD$ und unten den $^1S_0$ Kanal. Im $\SD$ Kanal
beziehen sich die gestrichelten Linien auf die BCS Phase und die
durchgezogenen auf die resultierende Phase, BCS oder LOFF. Wir sehen,
dass der Gap im $\SD$ Kanal deutlich gr\"o\ss{}er ist als im $^1S_0$
Kanal. Insgesamt sehen wir, dass der Gap f\"ur h\"ohere Asymmetrien
unterdr\"uckt wird. Bei endlicher Asymmetrie und geringer Temperatur
steigt der Gap f\"ur die BCS Phase mit steigender Temperatur,
ansonsten f\"allt er. Dies liegt an dem oben beschriebenen
Zusammenhang, dass die Fermikanten f\"ur hohe Asymmetrien separiert
und f\"ur hohe Temperaturen aufgeweicht werden. In der LOFF Phase
erhalten wir diese Anomalie nicht, weil die Fermikanten verschoben
werden.

\begin{figure}[!]
  \begin{center}
    \begin{minipage}[t]{\textwidth}
      \begin{center}
        \includegraphics[width=0.8\textwidth]{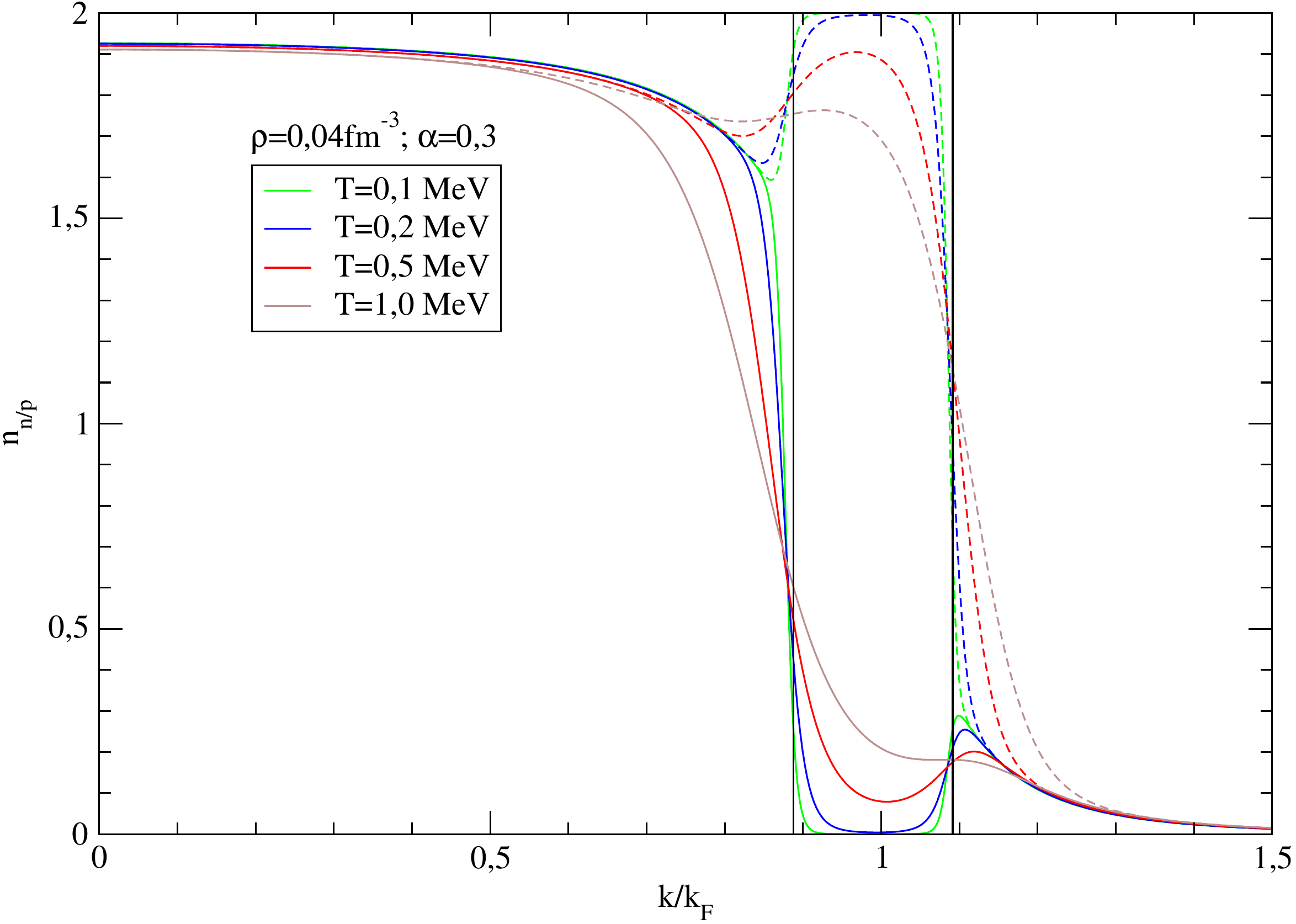}
      \end{center}
    \end{minipage}
    \newline\newline
    \begin{minipage}[t]{\textwidth}
      \begin{center}
        \includegraphics[width=0.8\textwidth]{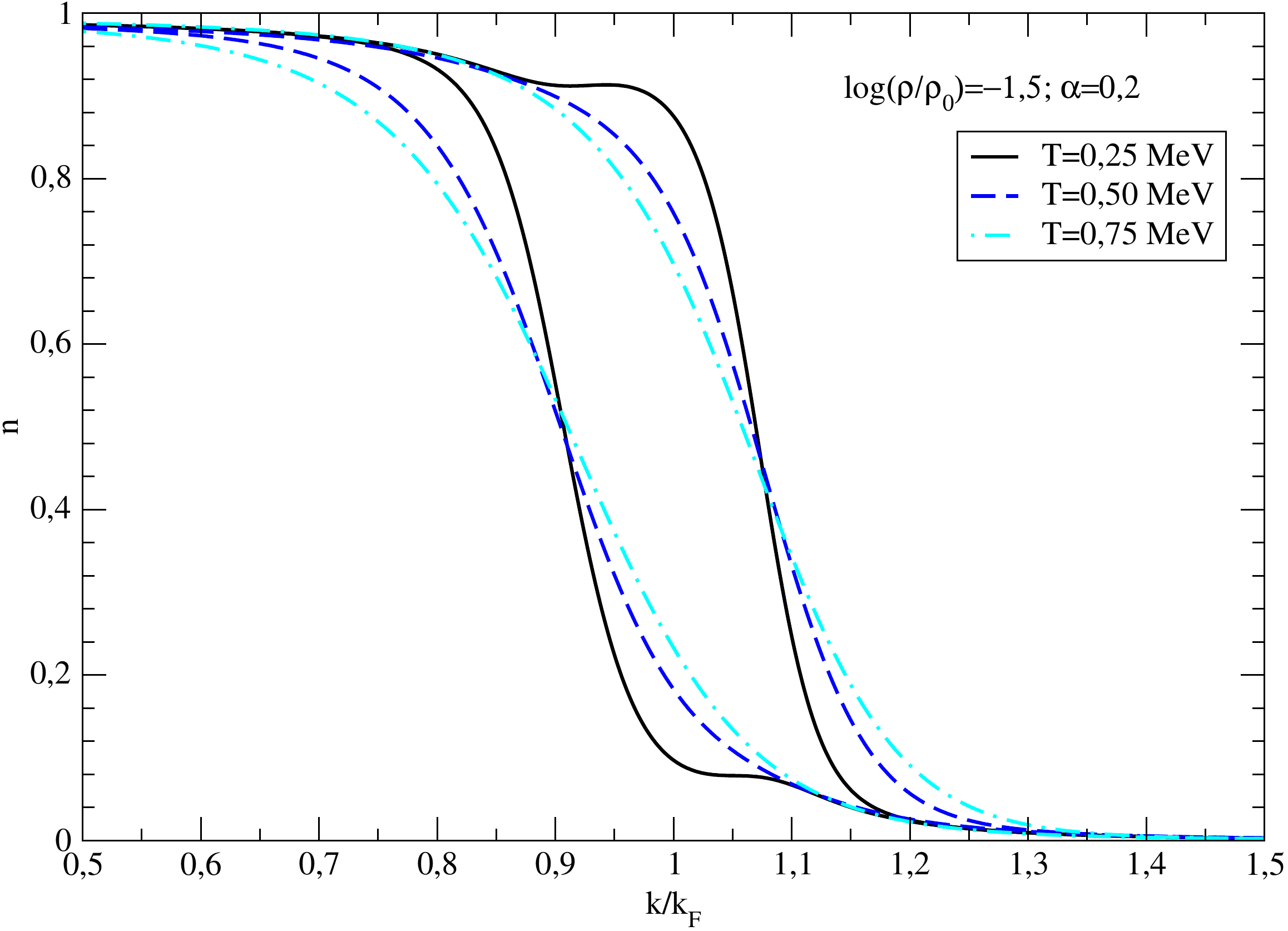}
      \end{center}
    \end{minipage}
    \caption[]{Die Besetzungszahlen der beiden Komponenten als
      Funktion des Impulses bei konstanter Dichte und Asymmetrie f\"ur
      verschiedene Temperaturen. Oben f\"ur den $\SD$ und unten f\"ur
      den $^1S_0$ Kanal. Im $\SD$ Kanal zeigen wir die
      Besetzungszahlen von Neutronen und Protonen und im $^1S_0$ Kanal
      die von Spin up und Spin down Neutronen.}
    \label{fig_0_06}
  \end{center}
\end{figure}

Als N\"achstes wollen wir uns mit den Besetzungszahlen
besch\"aftigen. Diese sind in Abbildung~\ref{fig_0_06} f\"ur beide
Paarungskan\"ale im BCS Limit dargestellt. Wir sehen, dass wir als
grobe Struktur zwei Fermifunktionen haben, die bei
$k_{F_{n/p}}/k_F=(1\pm\alpha)^{1/3}$
bzw. $k_{F_{\uparrow/\downarrow}}/k_F=(1\pm\alpha)^{1/3}$ abfallen; in
der Abbildung f\"ur den $\SD$ Kanal sind die Fermiimpulse der
Neutronen und Protonen durch waagerechte schwarze Linien
dargestellt. Im Fall der $\SD$ Paarung ist das Maximum bei $2$, weil
wir \"uber den Spin summiert haben. Durch die endliche Temperatur
werden die Fermifunktionen aufgeweicht. Neben der normalen Aufweichung
durch die Temperatur kommt ein wei\-te\-rer Effekt durch die Paarung
hinzu: An der Fermikante der Minderheits\-komponente f\"allt auch die
Mehrheitskomponente ab, dann bildet sich eine L\"ucke aus, bis
schlie\ss{}lich an der Fermikante der Mehrheitskomponente die
Minderheitskomponente ansteigt. Wir haben sozusagen einen Abfall
beider Komponenten an der Fermikante mit einer L\"ucke f\"ur
$k_{F_{p}}\lesssim k\lesssim k_{F_{n}}$
bzw. $k_{F_{\downarrow}}\lesssim k\lesssim k_{F_{\uparrow}}$.

Diese L\"ucke ist auch in anderer Hinsicht von Bedeutung. Der Kernel
der Gap-Gleichung liefert in diesem Bereich, in dem Paarung durch
Asymmetrie unterdr\"uckt wird, keinen Beitrag. Die
Ein\-teilchen\-ener\-gien der Minderheitskomponente werden hier
negativ, was zur sogenannten {\it gapless superconductivity} f\"uhrt.

Beim Verringern der Dichte geht die BCS-Phase im Fall von $\SD$
Paarung in ein BEC \"uber. Dieser \"Ubergang von fermionischen
Eigenschaften hin zu bosonischen l\"asst sich bei verschiedenen
untersuchten mi\-kro\-sko\-pi\-schen Funktionen beobachten. In der
LOFF Phase erhalten wir, dass der Cooper-Paar-Impuls die Aufspaltung
der beiden Komponenten stark verringern kann. Insgesamt erhalten wir
bei den mi\-kro\-sko\-pi\-schen Funktionen, dass die LOFF Phase eine
Ann\"aherung an den BCS Fall mit verschwindender Asymmetrie bedeutet.

\section*{Materie in starken magnetischen Feldern}
Starke magnetische Felder k\"onnen in kompakten Sternen
auftreten~\cite{2000ApJ...537..351B,2013PhRvD..88b5008S,2010arXiv1005.4995S,2009PhRvD..79l3001S,2012PhRvC..86e5804C}. Nach
dem Wasserstoffbrennen entwickelt sich ein Stern, je nach Masse, zu
einem Roten Riesen oder Roten \"Uberriesen. Nach der Entwicklung
\"uber einen planetarischer Nebel bzw. eine Supernova entsteht ein
kompakter Stern: ein Wei\ss{}er Zwerg, ein Neutronenstern oder ein
Schwarzes Loch. Das Oberfl\"achenmagnetfeld von Wei\ss{}en Zwergen
betr\"agt $B\approx10^{6}-10^{8}\,$G, das von Neutronensternen
betr\"agt $B\approx10^{12}\,$G~\cite{1991ApJ...383..745L}. Es wurden
Neutronensterne mit Oberfl\"achenmagnetfeldern von
$B\approx 10^{14}-10^{15}\,$G entdeckt; diese Neutronensterne werden
als Magnetare bezeichnet. Es wird vermutet, dass sie in ihrem Inneren
Magnetfelder mit $B \approx 10^{18}\,$G haben
k\"onnen~\cite{2000ApJ...537..351B,2013PhRvD..88b5008S,2010arXiv1005.4995S,2009PhRvD..79l3001S,1991ApJ...383..745L}. Aufgrund
des Virialsatzes kann das Magnetfeld im Inneren eines Neutronensterns
einen Wert von $B\approx10^{18}\,$G und im Inneren von Wei\ss{}en
Zwergen einen Wert von $B\approx10^{12}\,$G nicht
\"uber\-schrei\-ten~\cite{1991ApJ...383..745L}. Starke Magnetfelder
($B\approx10^{16}-10^{17}\,$G) werden in neugeborenen Neutronensternen
in Betracht gezogen~\cite{2012PhRvC..86e5804C}. Die Zusammensetzung
der Elemente in der Kruste von Neutronensternen kann durch starke
Magnetfelder mit $B\gtrsim10^{17}\,$G stark beeinflusst werden; eine
aktuelle Studie \"uber die Ele\-mentenh\"aufigkeit in Neutronensternen
in Abh\"angigkeit des Magnetfeldes kann in~\cite{2012PhRvC..86e5804C}
gefunden werden. Aufgrund der geringen Masse der Sterne, aus denen
sich Wei\ss{}e Zwer\-ge entwickeln, bestehen Wei\ss{}e Zwerge aus
leichteren Elementen als die Kruste eines Neutronensterns; schwere
Wei\ss{}e Zwerge bestehen vermutlich zu gro\ss{}en Teilen aus
Kohlenstoff und Wasserstoff~\cite{1986bhwd.book.....S}. Neon, das
dritte Element, das wir in Kapitel~\ref{chap_2} untersuchen, kommt
auch in Wei\ss{}en Zwergen vor. Diese relativ leichten Elemente
k\"onnen auch bei akkretierenden Neutronensternen vorkommen.

In Kapitel~\ref{chap_2} und \ref{chap_3} untersuchen wir Materie in
starken magnetischen Feldern. In Kapitel~\ref{chap_2} untersuchen wir
Kerne mit der Hartree-Fock-Theorie. Eine n\"ahere Beschreibung findet
sich z.B. in~\cite{2011JPhG...38c3101E}
und~\cite{2014CoPhC.185.2195M}; die folgende Beschreibung st\"utzt
sich auf diese Arbeiten. Der den Rechnungen aus Kapitel~\ref{chap_2}
zugrundeliegende Code ist der Sky3D
Code~\cite{2014CoPhC.185.2195M}. Das ultimative Ziel bei der
Beschreibung von Kernmaterie und Atomkernen ist eine Theorie, die
aufgrund von grundlegenden mi\-kro\-sko\-pi\-schen
Wech\-sel\-wir\-kungen die Eigenschaften gro\ss{}er Systeme
voraussagen kann; die sogenannten ab inito Methoden. F\"ur Kernmaterie
und Atomkerne w\"aren das Nukleon-Nukleon-Wechselwirkungen, bzw. die
noch fundamentaleren Quantenchromodynamik (QCD)
Wechselwirkungen. Derartige Modelle sind zwar f\"ur Coulomb
Wechselwirkungen realisiert, aber nicht f\"ur Kernmaterie, weswegen
man N\"aherungen machen muss. Das andere Extrem ist das
Fl\"us\-sig\-keits\-trop\-fen-Modell (englisch: liquid-drop model)
(LDM). Hierbei werden makroskopische Daten gefittet. Zwischen diesen
beiden Extremen gibt es verschiedene Ans\"atze, z.B. Rechnungen mit
einem selbstkonsistenten mitt\-le\-ren Feld (englisch: self-consistent
mean-field) (SCMF). Diese Modelle arbeiten auf einem
mi\-kro\-sko\-pi\-schem Level, verwenden aber auch effektive
Wechselwirkungen, z.B. wie in unserem Fall Skyrme-Kr\"afte, die eine
verschwindende Reichweite haben.

In Kapitel~\ref{chap_3} untersuchen wir suprafluide Neutronenmaterie
in starken magnetischen Feldern. Vieles deckt sich mit den
Untersuchungen suprafluider Kernmaterie aus Kapitel~\ref{chap_1},
diese Effekte sind weiter oben bereits erkl\"art. Studien zu
Neutron-Neutron-Paarung finden sich
z.B. in~\cite{2009PhRvC..79c4304M,2013PhRvC..88c4314P,PhysRevC.73.044309,2007PhRvC..76f4316M,2008PhRvC..78a4306I,2009PhRvC..80d5802G,2006pfsb.book..135S}. Neutron-Neutron-Paarung
tritt auf, wenn die Isospin-Asymmetrie gro\ss{} genug ist, um die
dominante $\SD$ Paarung von Neutron-Proton-Paaren zu
unterdr\"ucken. Neutron-Neutron-Paarung im $\SD$ Kanal ist aufgrund
des Pauli-Prinzips verboten. Der dominante Kanal f\"ur
Neutron-Neutron-Paarung bei niedrigen Dichten ist der $^1S_0$
Kanal. Dieser Spin-Singulett Kanal wird durch eine Spin-Asymmetrie,
die z.B. durch ein Ma\-gnet\-feld verursacht wird, unterdr\"uckt.

\subsection*{Kerne in starken magnetischen Feldern}
Der in Sky3D gegebene Hamiltonian, der in Anhang~\ref{sec_b_4} n\"aher
beschrieben wird, hat die folgende Form:
\begin{eqnarray}
  \hat h_q&=&U_q(\vecr)-\nabla\cdot\left[B_q(\vecr)\nabla\right]+\mathrm i\vecW_q\cdot\left(\vecsigma\times\nabla\right)+\vecS_q\cdot\vecsigma\nonumber\\
          && -\tfrac{\mathrm i}2\left[\left(\nabla\cdot\vecA_q\right)+2\vecA_q\cdot\nabla\right]\,,
\end{eqnarray}
wobei $q$ den Isospin bezeichnet.

Um Kerne in starken magnetischen Feldern zu untersuchen, haben wir
diesen Hamiltonian wie folgt modifiziert:
\begin{subequations}
  \begin{eqnarray}
    \hat h_{\mathrm{mod},\,q} &=&\hat h_q+ \hat h_{\mathrm{mag},\,q}\,,\\
    \hat h_{\mathrm{mag},\,q} &=&- \left(\vecl\cdot\delta_{q,p} + g_q\frac{\vecsigma}2\right) \cdot \tilde\vecB_q\,,
                                  \label{eq_0_05b}
  \end{eqnarray}
\end{subequations}
hierbei bezeichnet $g_q$ den Land\'e $g$-Faktor und
$\tilde\vecB_q=e\hbar/(2m_qc)\vecB$. Der erste Term ber\"ucksichtigt
die Kopplung des Bahndrehimpulses an das Ma\-gnet\-feld und der zweite
die des Spins. Eine n\"ahere Erkl\"arung von
Gleichung~\eqref{eq_0_05b} befindet sich in
Unterabschnitt~\ref{subsec_2_3_1}.

\begin{figure}[!]
  \begin{center}
    \hspace{0.1\textwidth}
    \includegraphics[width=0.8\textwidth]{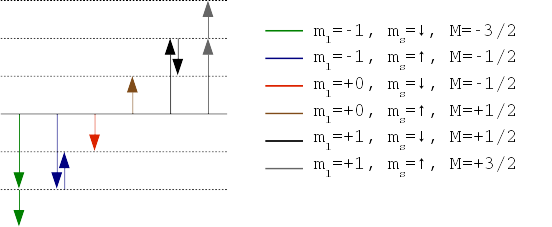}
    \caption[]{Die Quantenzahlen f\"ur verschiedene Zust\"ande. $m_l$
      bezeichnet die $z$-Komponente des Bahndrehimpulses, $m_s$ die
      des Spins und $M$ ist deren Summe.}
    \label{fig_0_07}
  \end{center}
\end{figure}

In Abbildung~\ref{fig_0_07} sind verschiedene Quantenzahlen f\"ur Spin
und Bahndrehimpuls dargestellt. Hierbei haben wir $s$- und
$p$-Zust\"ande ber\"ucksichtigt. Da wir Spin-$1/2$-Teilchen haben, ist
der Spin $s$ immer $1/2$ und somit gilt f\"ur die $z$-Komponente
$m_s=\uparrow,\,\downarrow$, mit $\uparrow=+1/2$ und
$\downarrow=-1/2$. Bei $s$-Zust\"anden erhalten wir f\"ur den
Bahndrehimpuls $l=0$, somit gilt f\"ur die $z$-Komponente
$m_l=0$. Folglich haben wir zwei Zust\"ande, n\"amlich die in
Abbildung~\ref{fig_0_07} rot bzw. braun dargestellten Pfeile. Bei
$p$-Zust\"anden haben wir $l=1$ und somit drei M\"oglichkeiten f\"ur
$m_l$: $m_l=-1,\,0,\,1$. F\"ur Zust\"ande mit $M=\pm3/2$ haben wir je
eine M\"oglichkeit, f\"ur Zust\"ande mit $M=\pm1/2$ haben wir je zwei
M\"oglichkeiten; die entsprechenden Einteilchenzust\"ande berechnen
sich aus Superpositionen dieser Zust\"ande.

Im Folgenden wollen wir unsere Resultate kurz zusammenfassen; eine
ausf\"uhrliche Darstellung findet sich in
Unterabschnitt~\ref{subsec_2_3_3}.

\subsubsection*{Energieniveaus, Bahndrehimpuls und Spin}
\begin{figure}[!]
  \begin{center}
    \includegraphics[width=\textwidth]{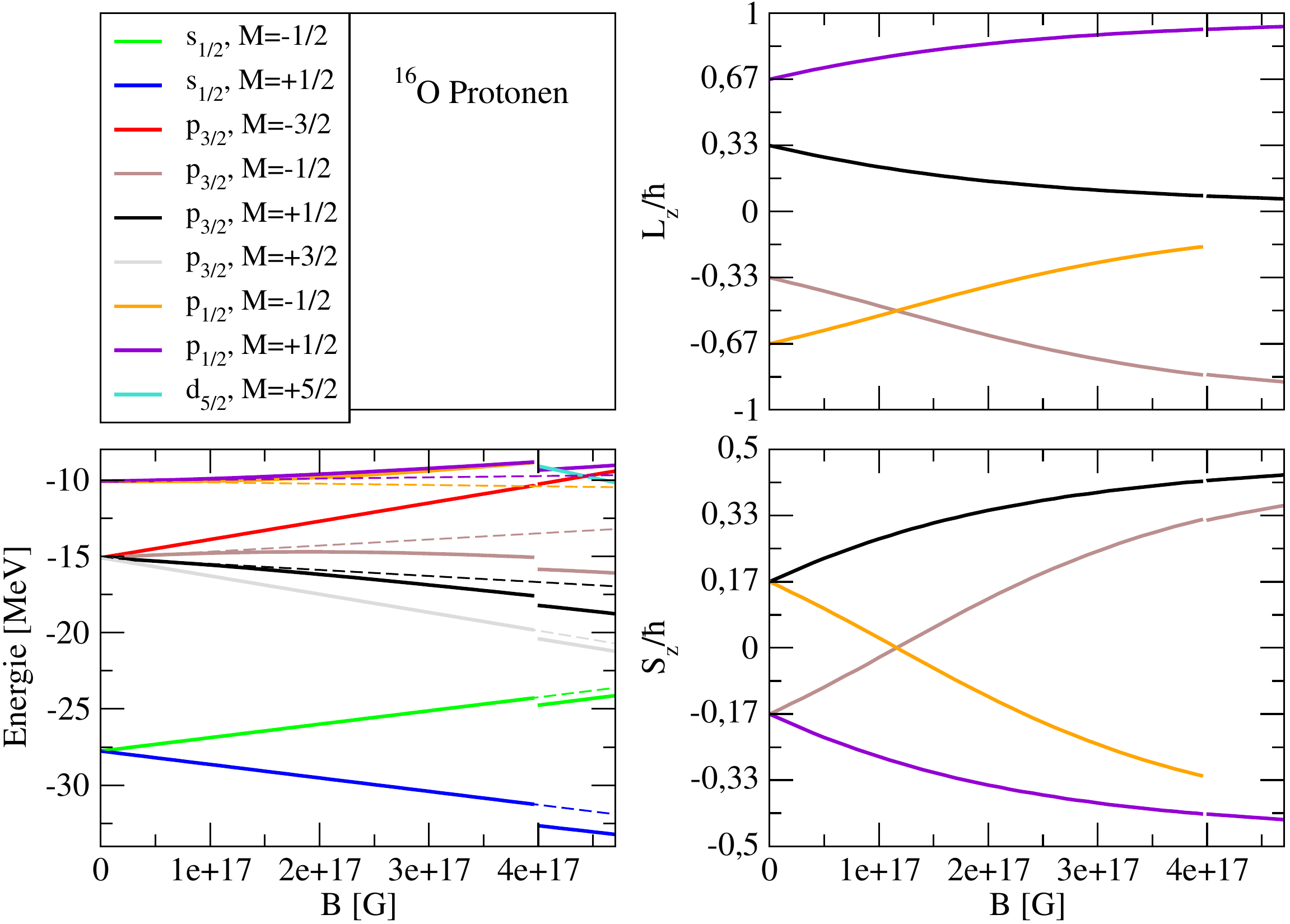}
    \caption[]{Die Energieniveaus (gestrichelte Linie: analytisch,
      durchgezogene Linie: numerisch), $\left<L_z\right>$ und
      $\left<S_z\right>$ als Funktionen des Magnetfeldes f\"ur
      Protonen in \isotope[16]{O}.}
    \label{fig_0_08}
  \end{center}
\end{figure}

\begin{figure}[!]
  \begin{center}
    \includegraphics[width=\textwidth]{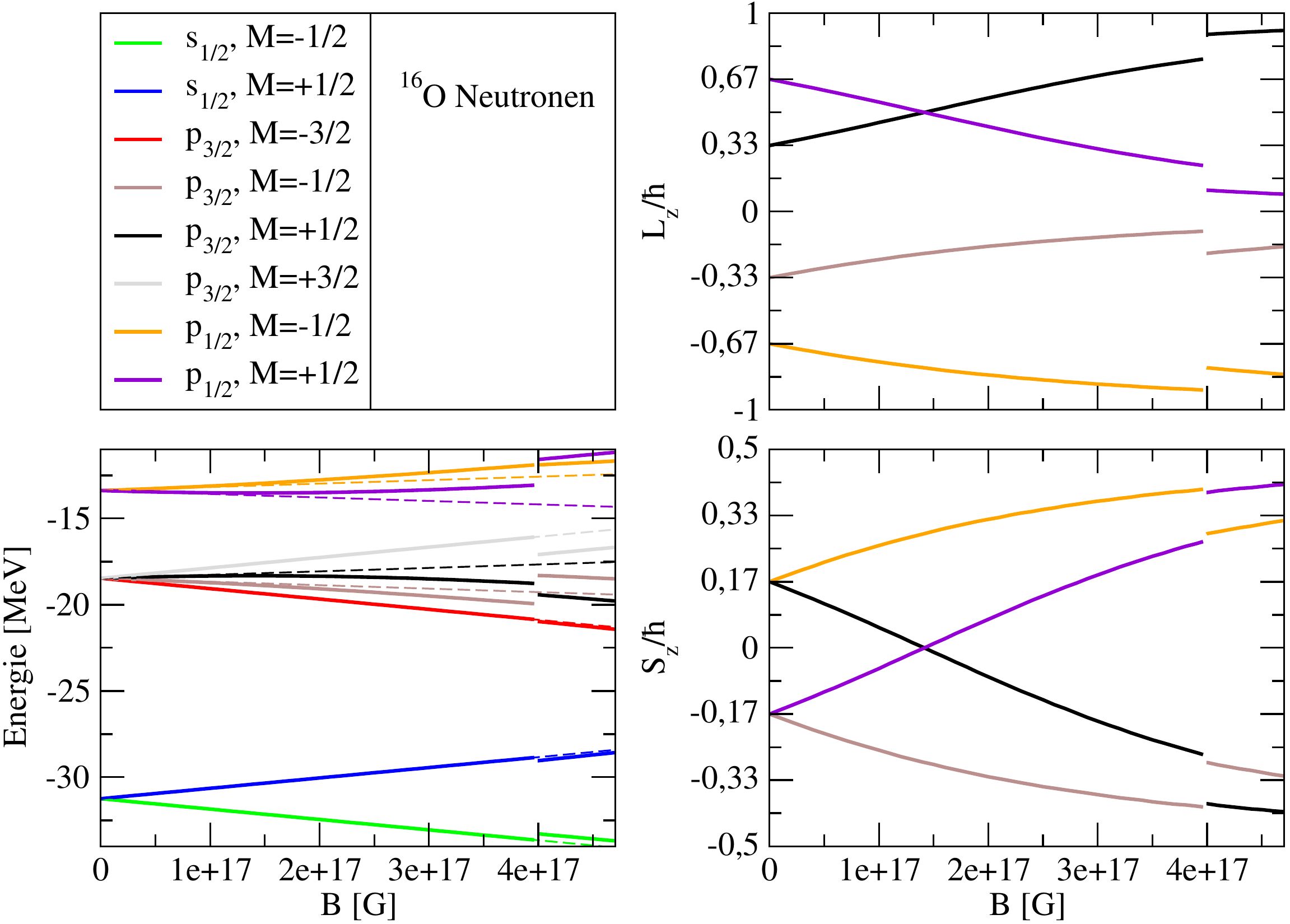}
    \caption[]{Die Energieniveaus (gestrichelte Linie: analytisch,
      durchgezogene Linie: numerisch), $\left<L_z\right>$ und
      $\left<S_z\right>$ als Funktionen des Magnetfeldes f\"ur
      Neutronen in \isotope[16]{O}.}
    \label{fig_0_09}
  \end{center}
\end{figure}

In den Abbildungen~\ref{fig_0_08} und \ref{fig_0_09} sind verschiedene
Gr\"o\ss{}en f\"ur Protonen und Neutronen in \isotope[16]{O}
dargestellt. Links unten sehen wir jeweils die Ener\-gie\-ni\-veaus
der einzelnen Einteilchenzust\"ande. Die $s_{1/2}$, $p_{3/2}$ und
$p_{1/2}$ Zust\"ande sind bei verschwindendem Magnetfeld jeweils
entartet und spalten sich bei nicht verschwindendem Magnetfeld
auf. Bei $B=4.0\cdot10^{17}\,$G erhalten wir eine Umbesetzung der
Energieniveaus. F\"ur $\left<L_z\right>$ und $\left<S_z\right>$
erhalten wir halb- bzw. ganzzahlige Werte f\"ur alle $s$-Zust\"ande
und $p$-Zust\"ande mit $M=\pm3/2$; f\"ur $p$-Zust\"ande mit $M=\pm1/2$
erhalten wir Superpositionen, wie oben erkl\"art. Bei Letzteren haben
wir zwei Effekte: die Spin-Bahn-Kopplung und die Kopplung des
Bahndrehimpulses und des Spins einzeln an das Magnetfeld. Ersteres
dominiert bei schwachen, Letzteres bei starken
Ma\-gnet\-fel\-dern. Deswegen erhalten wir halb- bzw. ganzzahlige
Werte nur im Grenzfall starker Ma\-gnet\-fel\-der. Der Grenzfall
schwacher Magnetfelder wird durch den Zeeman-Effekt und der Grenzfall
starker Ma\-gnet\-fel\-der durch den Paschen-Back-Effekt beschrieben.

\subsubsection*{Spin- und Stromdichte}
\begin{figure}[!]
  \begin{center}
    \includegraphics[width=0.4\textwidth]{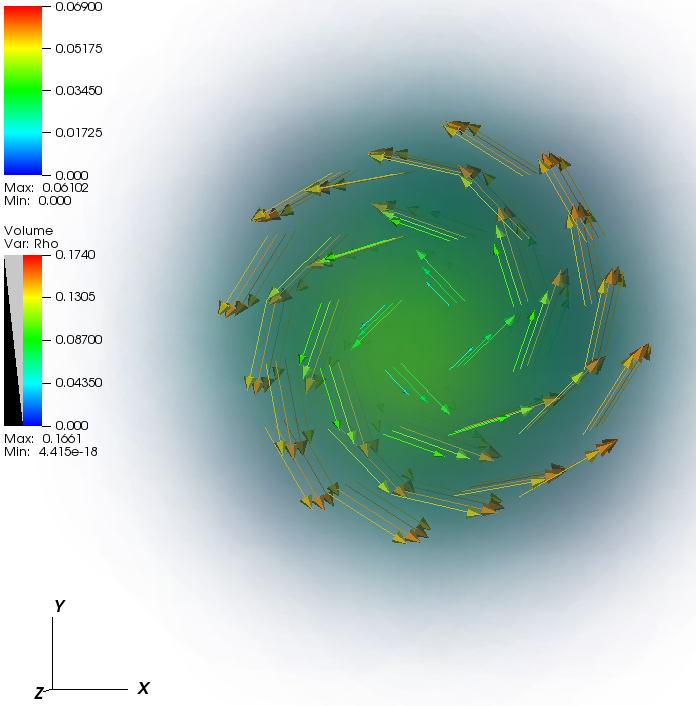}
    \includegraphics[width=0.4\textwidth]{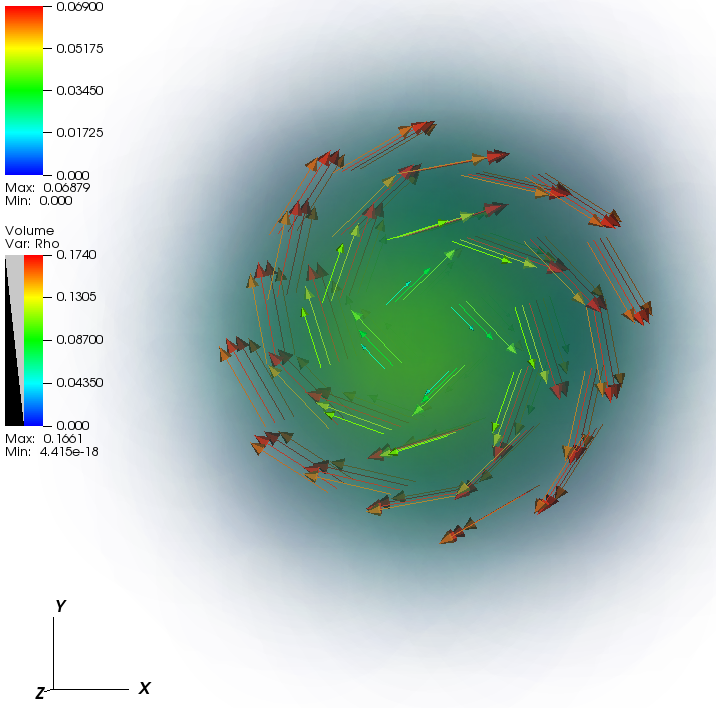}
    \caption[]{Die Stromdichte dividiert durch die Teilchendichte
      (kol\-lek\-ti\-ve Flie\ss{}\-ge\-schwin\-dig\-keit) f\"ur
      \isotope[12]{C} f\"ur $B=4{,}1\cdot10^{17}\,$G mit der
      Teilchendichte im Hintergrund f\"ur Neutronen (links) und
      Protonen (rechts). Diese Abbildung wurde mit
      VisIt~\cite{HPV:VisIt} erstellt.}
    \label{fig_0_10}
  \end{center}
\end{figure}

Als N\"achstes wollen wir auf die Spin- und die Stromdichte anhand von
\isotope[12]{C} eingehen. In Abbildung~\ref{fig_0_10} ist die
normierte Stromdichte (kol\-lek\-ti\-ve
Flie\ss{}\-ge\-schwin\-dig\-keit) dargestellt. Das Magnetfeld ist in
beiden F\"allen $B=4{,}1\cdot10^{17}\,$G, die linke Abbildung stellt
Neutronen und die rechte Protonen dar. Wir sehen, dass die
kol\-lek\-ti\-ve Flie\ss{}\-ge\-schwin\-dig\-keit senkrecht zum
Magnetfeld verl\"auft. Au\ss{}erdem sehen wir, dass die
kol\-lek\-ti\-ve Flie\ss{}\-ge\-schwin\-dig\-keit der Protonen und
Neutronen in verschiedene Richtungen verl\"auft. Dies liegt an den
unterschiedlichen Vor\-zei\-chen von $g_n$ und $g_p$.

\begin{figure}[!]
  \begin{center}
    \includegraphics[width=0.4\textwidth]{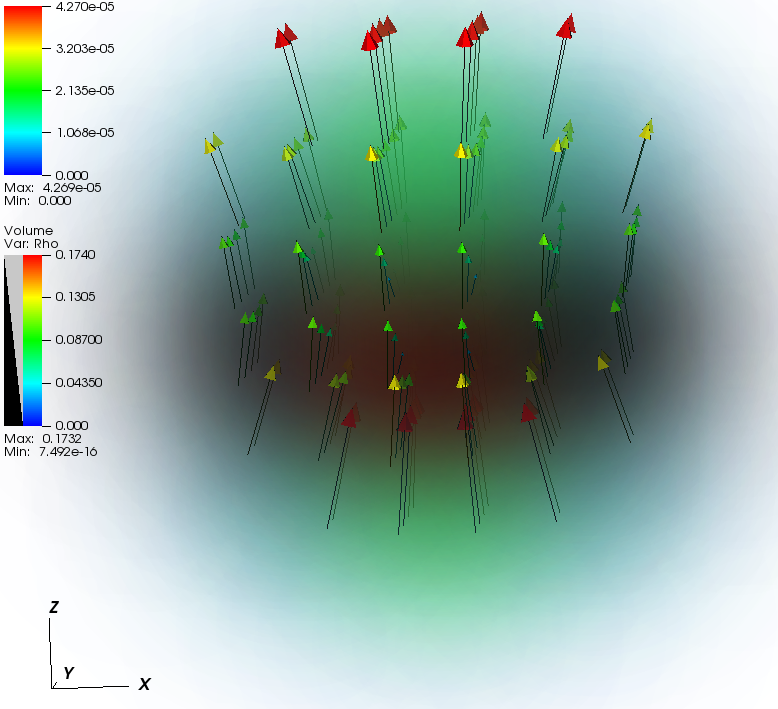}
    \includegraphics[width=0.4\textwidth]{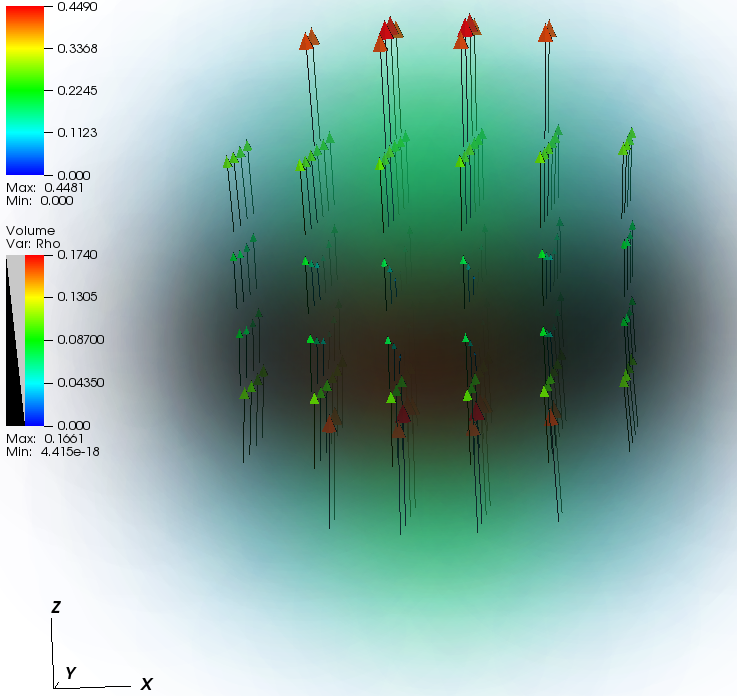}
    \caption[]{Die Spindichte dividiert durch die Teilchendichte f\"ur
      \isotope[12]{C} mit der Teilchendichte im Hintergrund f\"ur
      Protonen f\"ur verschiedene Magnetfelder:
      $B=4{,}0\cdot10^{13}\,$G (links) und $B=4{,}1\cdot10^{17}\,$G
      (rechts). Diese Abbildung wurde mit VisIt~\cite{HPV:VisIt}
      erstellt.}
    \label{fig_0_11}
  \end{center}
\end{figure}

In Abbildung~\ref{fig_0_11} sehen wir die normierte Spindichte f\"ur
Protonen bei zwei verschiedenen Magnetfeldern: Links ist das
Magnetfeld ver\-h\"alt\-nis\-m\"a\-\ss{}ig sehr klein
($B=4{,}0\cdot10^{13}\,$G) und rechts gro\ss{}
($B=4{,}1\cdot10^{17}\,$G). Hier sehen wir die Ausrichtung des Spins
bei zunehmendem Magnetfeld; bei kleinem Magnetfeld ist der Spin
aufgrund der dominanten Spin-Bahn-Kopplung wenig ausgerichtet,
wohingegen die Ausrichtung bei starkem Magnetfeld deutlich zu sehen
ist.

\subsubsection*{Verformung}
\begin{figure}[!]
  \begin{center}
    \includegraphics[width=0.32\textwidth]{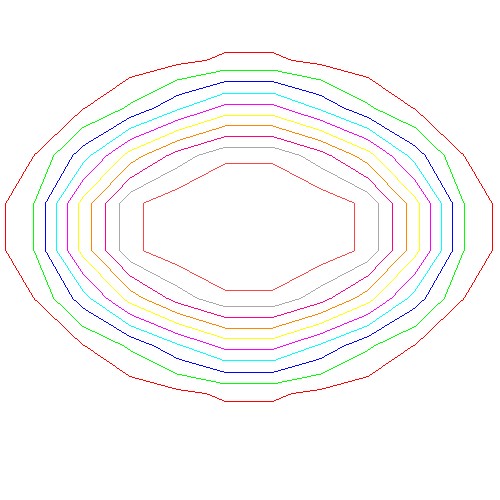}
    \includegraphics[width=0.32\textwidth]{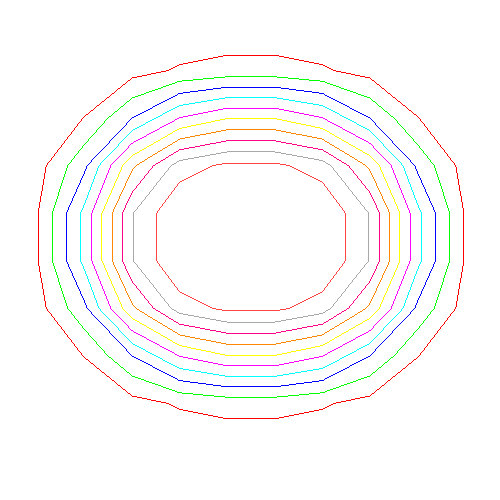}
    \includegraphics[width=0.32\textwidth]{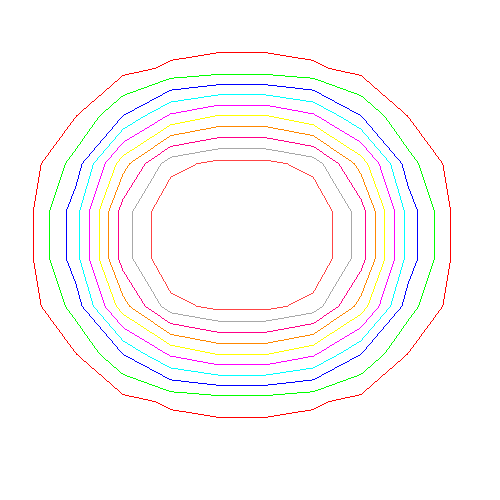}
    \caption[]{Der verformte Atomkerne \isotope[20]{Ne} f\"ur $B=0$,
      $B=2{,}4\cdot10^{17}\,$G und $B=4{,}9\cdot10^{17}\,$G. Diese
      Abbildung wurde mit VisIt~\cite{HPV:VisIt} erstellt.}
    \label{fig_0_12}
  \end{center}
\end{figure}

In Abbildung~\ref{fig_0_12} sehen wir die Verformung von
\isotope[20]{Ne}, das Magnetfeld nimmt von links nach rechts zu. Bei
$B=0$ ist der Atomkern stark verformt, mit zunehmendem Magnetfeld
nimmt die Verformung ab.

\subsection*{Neutronenmaterie in starken magnetischen Feldern}
Neutronensternmaterie kann in erster N\"aherung als reine
Neutronenmaterie behandelt werden~\cite{2009PhRvC..80d5802G}, weil der
Anteil von Protonen und Elektronen und schweren Barionen nicht mehr
als $5\%$-$10\%$ der Gesamtdichte des Systems ausmacht. Daher spielt
Neutron-Neutron-Paarung eine wichtige Rolle in der Physik der inneren
Kruste eines Neutronensterns. Au\ss{}erdem spielt sie eine wichtige
Rolle f\"ur Neutronen-reiche Atomkerne in der N\"ahe der {\it Drip
  Line} (Kerne, die keine Neutronen mehr binden
k\"onnen.)~\cite{PhysRevC.73.044309}. Es gibt ein paar
ph\"anomenologische Hinweise auf Neutronen-Suprafluidit\"at in
Neutronensternen. Bekannte Beispiele sind Periodenspr\"unge (englisch:
glitches) in dem Rotationsverhalten einiger Pulsare und das
K\"uh\-lungs\-ver\-hal\-ten des j\"ungsten bekannten Neutronensterns
in Kassiopeia A~\cite{2006pfsb.book..135S}.

\begin{figure}[!]
  \begin{center}
    \includegraphics[width=0.8\textwidth]{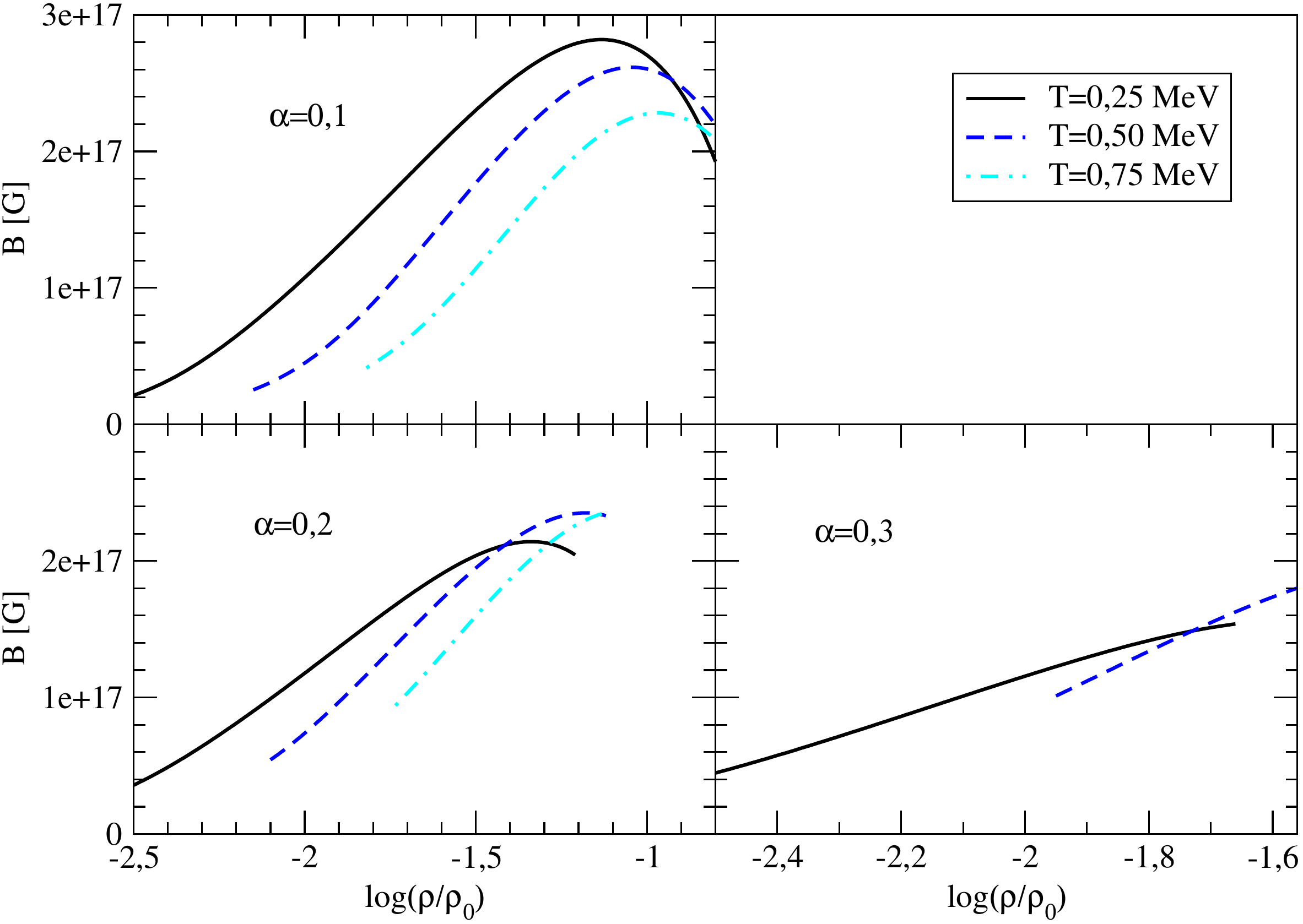}
    \caption[]{Das Magnetfeld, das ben\"otigt wird, um eine bestimmte
      Spin-Asymmetrie (Polarisation) zu erzeugen, als Funktion der
      Dichte. In jedem Feld ist eine bestimmte Polarisation
      fixiert. Verschiedene Temperaturen sind mit verschiedenen Farben
      dargestellt.}
    \label{fig_0_13}
  \end{center}
\end{figure}

Neben dem oben erw\"ahnten Pha\-sen\-dia\-gramm und den
mi\-kro\-sko\-pi\-schen Funktionen haben wir den Einfluss des
Magnetfeldes auf die Spin-Asym\-me\-trie (Polarisation)
untersucht. Au\ss{}erdem haben wir die magnetische Energie mit der
Temperatur verglichen. In Abbildung~\ref{fig_0_13} ist das
Ma\-gnet\-feld, das f\"ur eine bestimmte Polarisation ben\"otigt wird,
als Funktion der Dichte dargestellt. Verschiedene Felder zeigen
verschiedene Werte der Polarisation, verschiedene Farben stehen f\"ur
verschiedene Temperaturen. Wir sehen, dass das ben\"otigte Magnetfeld
in der Regel f\"ur steigende Po\-la\-ri\-sa\-ti\-on, steigende Dichte
oder fallende Temperatur steigt.

\begin{figure}[!]
  \begin{center}
    \includegraphics[width=0.8\textwidth]{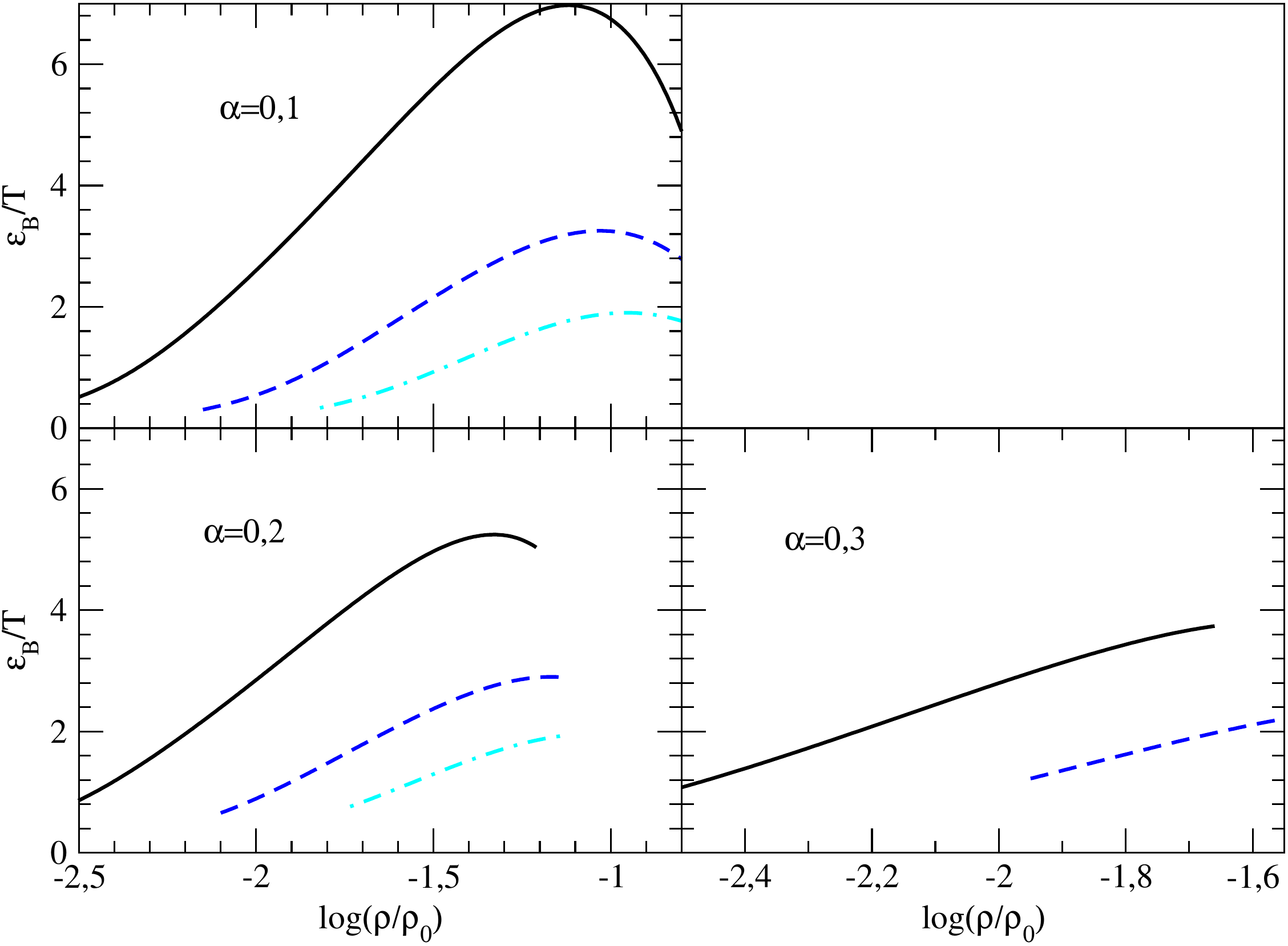}
    \caption[]{Die ma\-gne\-ti\-sche Energie dividiert durch die
      Temperatur als Funktion der Dichte f\"ur verschiedene
      Temperaturen und Polarisationen. Der Farbcode ist der gleiche
      wie in Abbildung~\ref{fig_0_13}.}
    \label{fig_0_14}
  \end{center}
\end{figure}

In Abbildung~\ref{fig_0_14} sehen wir die magnetische Energie
$\varepsilon_B$ mit
\begin{eqnarray}
  \varepsilon_B&=&\vert\tilde\mu_nB\vert\label{eq_0_06}
\end{eqnarray}
dividiert durch die Temperatur $T$. Wir sehen, dass $\varepsilon_B$ in
der Regel gr\"o\ss{}er ist als $T$. Formel~\ref{eq_0_06} wird in
Abschnitt~\ref{sec_3_2} n\"aher erkl\"art.


\section*{Schlu\ss{}folgerungen}
In Kapitel~\ref{chap_1} untersuchen wir Kernmaterie bei niedrigen
Dichten, tiefen Temperaturen und nicht verschwindender
Isospin-Asymmetrie. Hierbei erhalten wir ein reichhaltiges
Phasendiagramm bestehend aus der translations- und
rotationssymmetrischen BCS Phase, einem BEC bestehend aus
Neu\-tron-Proton-Dimeren und den exotischen Phasen LOFF und PS. Wir
erhalten zwei trikritische Punkte, die f\"ur einen bestimmten Wert von
Dichte, Temperatur und Asymmetrie in einem tetrakri\-tischen Punkt
zusammenfallen k\"onnen. Au\ss{}erdem existieren zwei Crossovers: bei
hohen Temperaturen von einer asymmetrischen BCS Phase zu einem von
BEC, das von einem Neutronengas umgeben ist. Bei tiefen Temperaturen
erhalten wir einen Crossover in der Phasenseparation. Wir haben
verschiedene mi\-kro\-sko\-pi\-sche Funktionen untersucht -- den Gap,
den Kernel der Gap-Gleichung, die Wellenfunktionen der Cooper-Paare,
die Besetzungszahlen und die Einteilchenenergien. Hierbei konnten wir
den \"Ubergang von einer schwach gebundenen BCS-Phase bei hohen
Dichten zu einem stark gebundenen BEC bei niedrigen Dichten
beobachten. Wir konnten auch sehen, dass sich die
mi\-kro\-sko\-pi\-schen Funktionen der LOFF Phase denen der BCS-Phase
bei verschwindender Asymmetrie ann\"ahern. Au\ss{}erdem konnten wir
eine L\"ucke um die Fermikante herum feststellen, die sich auf die
mi\-kro\-sko\-pi\-schen Funktionen auswirkt.

In Kapitel~\ref{chap_2} untersuchen wir den Einfluss von starken
Magnetfeldern auf verschiedene Atomkerne mit einem Skyrme-Hartree-Fock
(SHF) Ansatz unter Benutzung des Codes Sky3D. Starke Magnetfelder
k\"onnen z.B. in Neutronensternen realisiert werden. Die Elemente, die
wir untersuchen, kommen in Wei\ss{}en Zwergen vor, die auch starke
Magnetfelder haben k\"onnen. Wir haben drei verschiedene Atom\-kerne
betrachtet: \isotope[16]{O}, \isotope[12]{C} und \isotope[20]{Ne}. Wir
haben den Spin und den Bahndrehimpuls als Funktion des Magnetfeldes
untersucht; bei schwachen Magnetfeldern ist deren Ausrichtung aufgrund
der dominierenden Spin-Bahn-Wechselwirkung gering, bei starken
Ma\-gnet\-feldern dominiert die Kopplung von Spin- und Bahndrehimpuls
an das Magnetfeld. Bei \isotope[16]{O} haben wir eine Umbesetzung der
Energieniveaus bei starken Magnetfeldern sehen k\"onnen. Bei
\isotope[20]{Ne} konnten wir erkennen, dass die Verformung mit
zunehmendem Magnetfeld abnimmt.

In Kapitel~\ref{chap_3} untersuchen wir Neutronenmaterie und erhalten
ein Pha\-sen\-dia\-gramm f\"ur Spin-asym\-me\-t\-rische (polarisierte)
Materie, das dem aus Kapitel~\ref{chap_1} zwar sehr \"ahnelt, aber
einige Unterschiede aufweist. Dadurch, dass der Paarungskanal
schw\"acher ist, erhalten wir geringere kritische
Temperaturen. Au\ss{}erdem erhalten wir kein BEC und keine LOFF
Phase. Die Berechnungen der mi\-kro\-sko\-pi\-schen Funktionen in der
BCS-Phase sind mit denen aus Kapitel~\ref{chap_1} vergleichbar. Durch
das Ausbleiben der LOFF Phase erhalten wir eine untere kritische
Temperatur. Wir haben auch untersucht, welche Ma\-gnetfeld\-st\"arken
welche Polarisation verursachen. Au\ss{}erdem haben wir die
magnetische Energie mit der Temperatur verglichen; hierbei war die
magnetische Energie in der Regel gr\"o\ss{}er als die der Temperatur.

\section*{Perspektiven}
Die Rechnungen in Kapitel~\ref{chap_1} gehen von
Neutron-Proton-Paarung und zu\-s\"atz\-li\-chen Neutronen aus. Die
Rechnungen k\"onnten durch Einbeziehen von Clustern verbessert
werden. Des Weiteren k\"onnte man die Rechnungen aus den
Kapiteln~\ref{chap_1} und \ref{chap_3} kombinieren, indem sowohl
Isospin-Singulett Spin-Triplett Paarung als auch Isospin-Triplett
Spin-Singulett Paarung in die Rechnungen eingebaut werden. Hierbei ist
zu erwarten, dass bei einer bestimmten Isospin-Asymmetrie ein
Pha\-sen\-\"uber\-gang von Isospin-Singulett Spin-Triplett Paarung zu
Isospin-Triplett Spin-Singulett Paarung erfolgt.

Die Ergebnisse, die in Kapitel~\ref{chap_2} gezeigt werden, k\"onnen
in Zukunft auf verschiedene Weisen verbessert werden. Ein verbesserter
Hamiltonian k\"onnte verwendet werden; insbesondere k\"onnten
Spin-Spin-Wech\-sel\-wir\-kun\-gen in Hinblick auf starke magnetische
Felder interessant sein. Au\ss{}erdem k\"onnten Methoden entwickelt
werden, die die aktuellen Studien in den Be\-reich st\"arkerer
Magnetfelder oder schwererer Atomkerne ausdehnen.

\renewcommand{\figurename}{Figure}
\clearpage{\pagestyle{empty}\cleardoublepage}
\chapter*{Abstract}
This PhD thesis deals with nuclear matter and nuclei under extreme
conditions. These can occur e.g. in compact
stars. Chapter~\ref{chap_1} studies superfluid neutron-proton pairing
in isospin-asymmetric nuclear matter in the $\SD$ channel. In
chapter~\ref{chap_2} we study the influence of strong magnetic fields
on \isotope[12]{C}, \isotope[16]{O} and \isotope[20]{Ne}. Finally, we
study in chapter~\ref{chap_3} superfluid neutron-neutron pairing in
spin-asymmetric (polarized) neutron matter in the $^1S_0$ channel; a
polarization can be induced e.g. by a magnetic field.

In chapter~\ref{chap_1} we obtain a rich phase diagram for
isospin-asymmetric nuclear matter. A better understanding of this
matter can be important e.g. for low energy heavy ion collisions,
supernovae explosions or nuclei. In the outer area of nuclei the
density is low, thus an isospin-asymmetry hardly suppresses
neutron-proton pairing. We study the unpaired phase and several
superfluid phases. We study the crossover from the weakly coupled
Bardeen Cooper Schrieffer (BCS) phase at high densities to the
Bose-Einstein condensate (BEC) in the limit of strong coupling at low
densities. Moreover, we study two exotic phases: the
Larkin-Ovchinnikov-Fulde-Ferrell (LOFF) phase, at which Cooper-pairs
get a nonvanishing center-of-mass momentum. This phase exists only at
high densities. Moreover, we study a phase separation (PS) consisting
of an isospin symmetric BCS or BEC part and an isospin asymmetric
unpaired part with neutron excess. The phase separation can exist both
at high and low densities, thus we obtain a crossover in the phase
separation. The phase transition between LOFF and PS is of first
order, all other phase transitions are of second order. Furthermore,
we study the gap, the kernel of the gap equation, the Cooper-pair wave
functions, the occupation numbers and the quasiparticle dispersion
relations. In the BCS limit, we obtain a fermionic and in the BEC
limit a bosonic nature. For the LOFF phase the intrinsic features
approach those of the BCS phase at vanishing isospin asymmetry.

In chapter~\ref{chap_2} we study the effect of a strong magnetic field
on the elements \isotope[16]{O}, \isotope[12]{C} and
\isotope[20]{Ne}. These elements can occur e.g. in white dwarfs, which
can have strong magnetic fields. Furthermore, these elements can play
an important role for accreting neutron stars. For \isotope[16]{O} and
\isotope[12]{C} the single particle energies are splitted with
increasing magnetic field. Moreover, the $z$-component of the angular
momentum and the spin are aligned with the magnetic field at strong
magnetic fields, this alignment is suppressed by the spin-orbit
coupling for weak magnetic fields. In \isotope[16]{O} the energy
states are rearranged at strong magnetic fields. For the collective
flow velocity in the nuclei we obtain circular or almost circular
orbits around the axis of the magnetic field. The spin density aligns
with the magnetic field for strong mangetic fields. \isotope[20]{Ne}
is strongly deformed at vanishing magnetic fields, this deformation
decreases with increasing magnetic field.

The phase diagram for polarized neutron matter studied in
chapter~\ref{chap_3} can be of great importance for studies of neutron
matter; in particular for the inner crust of neutron stars. It can
also be important for studies on nuclei. There are some
phenomenological indications of neutron superfluidity in neutron
stars. The obtained phase diagram consists only of the unpaired phase
and the BCS phase. Since there exists no bound neutron-neutron pairs,
BEC cannot form. Since the coupling strength of the $^1S_0$ channel is
weaker than the one of the $\SD$ channel, the critical temperature is
lower than in the phase diagram analyzed in chapter~\ref{chap_1}. For
the intrinsic features we obtain similar results as for the BCS phase
in chapter~\ref{chap_1}. Moreover, we have studied the magnetic field
needed for a certain magnetization and compared its energy with the
temperature of the system. In the sector we studied, the magnetic
energy is normally greater than the temperature.

\clearpage{\pagestyle{empty}\cleardoublepage}
\thispagestyle{empty} {\renewcommand{\MakeUppercase}{}
\renewcommand{\contentsname}{Table of Contents}
\tableofcontents
\listoffigures}

\clearpage{\pagestyle{empty}\cleardoublepage}
\pagenumbering{arabic}
\setcounter{page}{1}

\chapter{BCS-BEC crossovers and unconventional phases in dilute
  nuclear matter}
\label{chap_1}

\section{Introduction}
\begin{figure}[!]
  \begin{center}
    \includegraphics[width=0.45\textwidth]{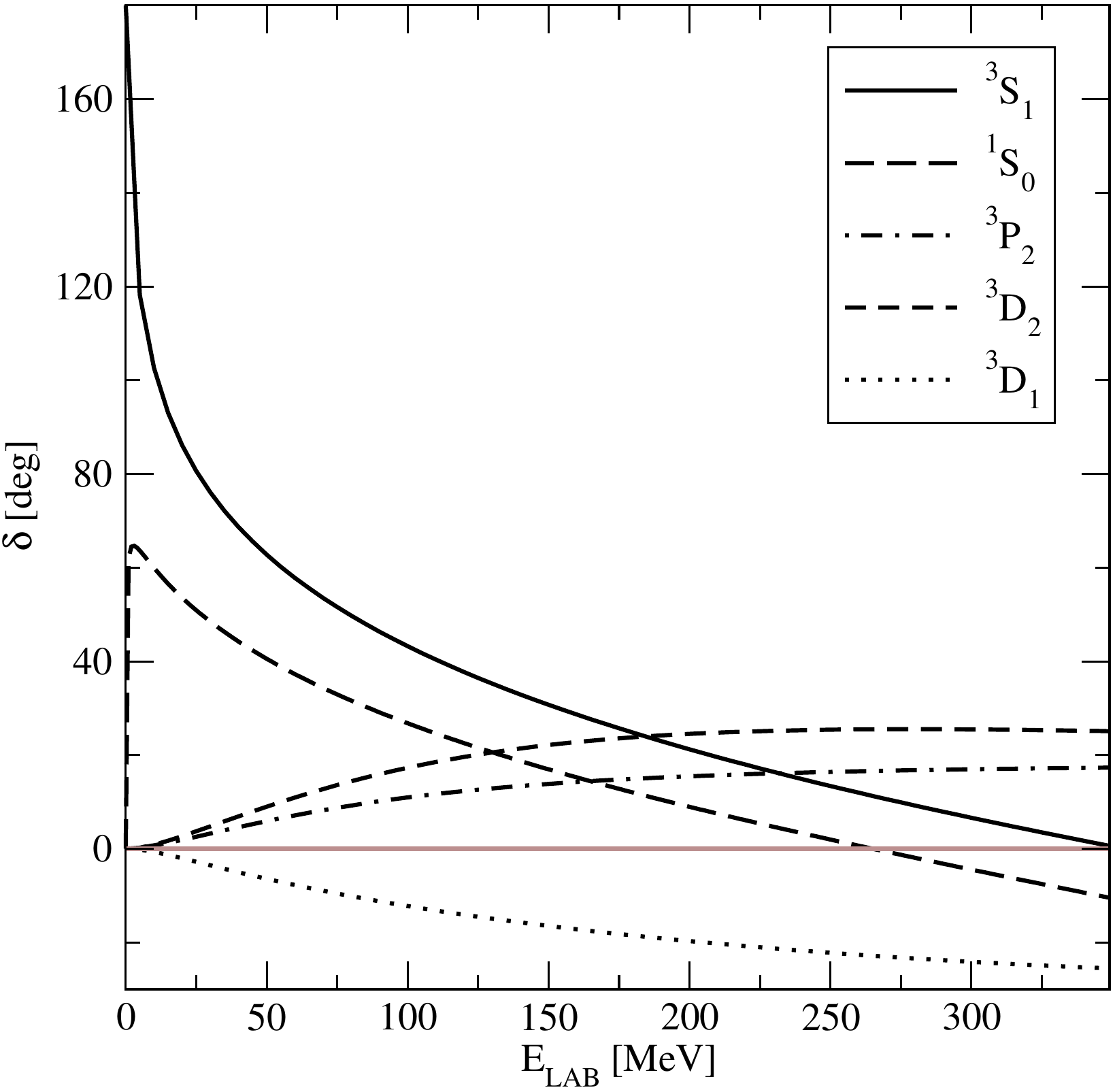}
    \hspace{0.05\textwidth}
    \includegraphics[width=0.45\textwidth]{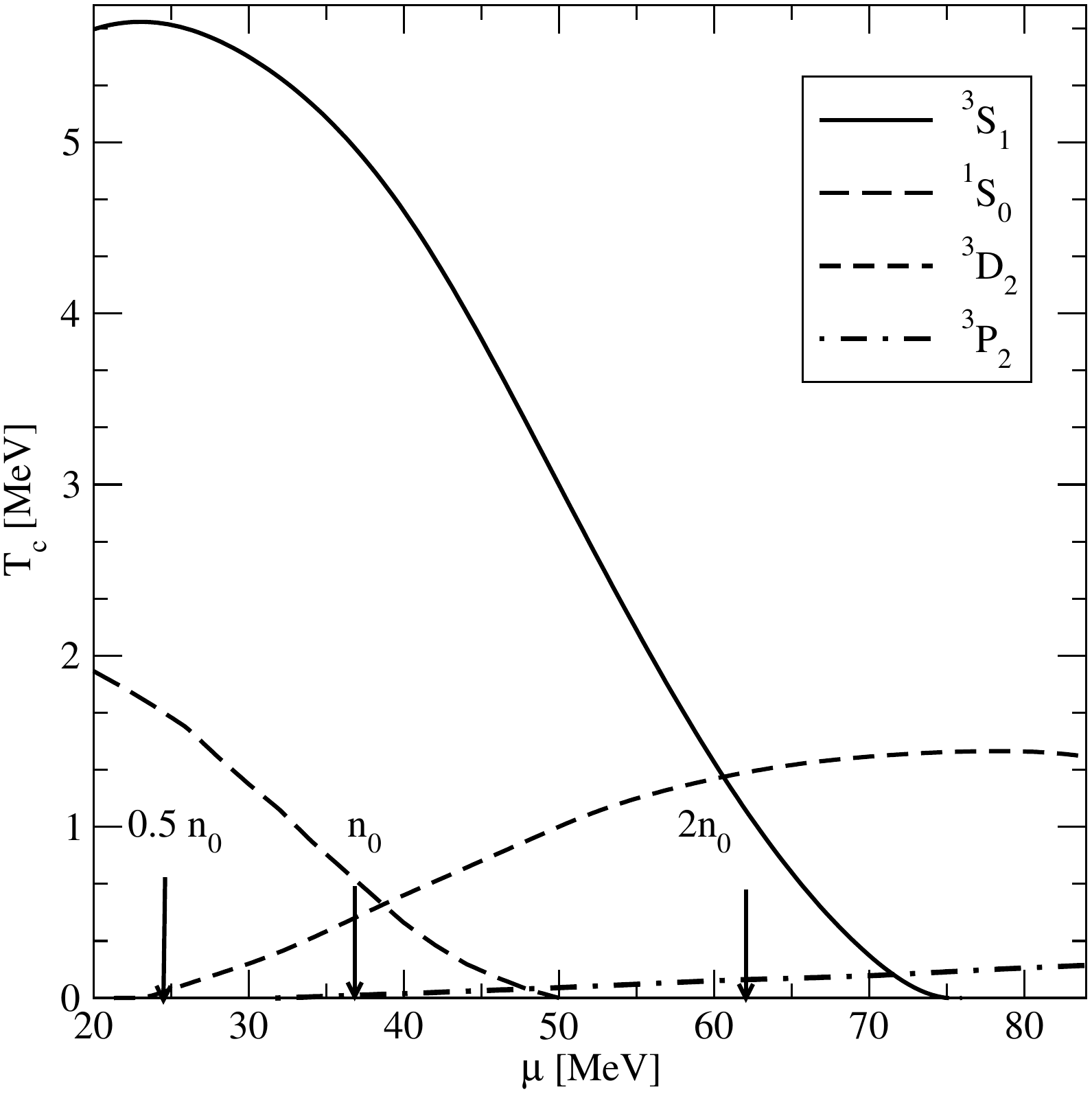}
    \caption[The elastic scattering phase shifts and the critical
    temperatures of pairing in the attractive interaction
    channels.]{a) The elastic scattering phase shifts as a function of
      the laboratory energy (left panel). The critical temperatures of
      pairing in the attractive interaction channels (right
      panel). Taken from Ref.~\cite{2007PrPNP..58..168S} (The variable
      $n_0$ of the right figure is the nuclear saturation density
      which is called $\rho_0$ in this thesis.)}
    \label{fig_1_01}
  \end{center}
\end{figure}

The vacuum two-nucleon interaction at low energies is experimentally
constrained by the phase-shift data obtained from the analysis of
elastic nucleon-nucleon collisions. The attractive part of the nuclear
interaction which is dominant at low energies leads to a formation of
nuclear clusters and the appearance of nucleonic pair condensates of
the Bardeen-Cooper-Schrieffer (BCS) type at sufficiently low
temperatures. Fig.~\ref{fig_1_01} shows the scattering phase shifts as
a function of laboratory energy for attractive channels (left panel)
and the corresponding critical temperatures for transition to
superconducting/superfluid state (right panel). The overall behavior
of nuclear matter at low density is rather complex because of the
possibility of the formation of clusters and condensates. The physics
of low density nuclear matter is relevant for astrophysics of
supernova matter and neutron stars. These settings differ in the
values of additional parameters (apart from the matter density) such
as temperature ($T$) and isospin asymmetry ($\alpha$). In supernovae
$\alpha$ is non-zero but small compared to that of cold
$\beta $-catalyzed matter in neutron stars. Under large isospin
asymmetry the neutron-proton pairing is disrupted and $^1S_0$ pairing
in the isospin-triplet, spin-singlet state of neutrons is
favored. This is the case in neutron stars. In supernova matter nearly
isospin-symmetrical matter supports $\SD$ pairing in the spin-triplet,
isospin-singlet state, because the isospin asymmetry is not large
enough to suppress the $\SD$ pairing.

Fermionic BCS superfluids, which form loosely bound Cooper-pairs in
the weak-coupling limit undergo a transition to the Bose-Einstein
condensate (BEC) of tightly bound bosonic dimers, when the pairing
strength increases
sufficiently~\cite{1985JLTP...59..195N,1969PhRv..186..456E}. In
experiments on cold atomic gases, the pairing strength can be
manipulated via the Feshbach mechanism. The transition from BCS to BEC
regime of pairing was confirmed experimentally in these systems. In
isospin-symmetric nuclear matter, this transition may occur upon
dilution of the system. If the pairing is in the $\SD$ channel the
asymptotic state of the strong-coupling limit is a BEC of
deuterons~\cite{1993NuPhA.551...45A,1995PhRvC..52..975B,1995ZPhyA.351..295S,2001PhRvC..63c8201L,2006PhRvC..73c5803S,2009PhRvC..79c4304M,2010PhRvC..81c4007H,2010PhRvC..82b4911J,2013JPhCS.413a2024S,2014JPhCS.496a2008S,2013PhRvC..88c4314P,2013PhRvC..88b5806S,2013arXiv1308.0364S,2013NuPhA.909....8S}. The
isoscalar neutron-proton ($np$) pairing is disrupted by isospin
asymmetry, which is induced by weak interactions in stellar
environments and is expected in exotic nuclei. This disruption occurs
because the mismatch in the Fermi surfaces of protons and neutrons
suppresses the pairing
correlations~\cite{2000PhRvL..84..602S}. Moreover the standard
Nozi\`eres-Schmitt-Rink theory~\cite{1985JLTP...59..195N} of the
BCS-BEC crossover must also be modified in a way that the low-density
asymptotic state becomes a gaseous mixture of neutrons and
deuterons~\cite{2001PhRvC..64f4314L}. The $\SD$ condensates can be
important in several physical backgrounds. (i)~Low-energy heavy-ion
collisions produce large amounts of deuterons in final states as
putative fingerprints of $\SD$
condensation~\cite{1995PhRvC..52..975B}. (ii)~Large nuclei may feature
spin-aligned $np$ pairs, as evidenced by recent experimental
findings~\cite{2011Natur.469...68C} on excited states in $^{92}$Pd;
moreover, exotic nuclei with extended halos provide a locus for
$n$-$p$ Cooper pairing. (iii)~Directly relevant to the parameter
ranges covered in this chapter are the observations that supernova and
hot proto-neutron-star matter at sub-saturation densities have low
temperature and low isospin asymmetry, and that the deuteron fluid is
a substantial
constituent~\cite{2010PhRvC..81a5803T,2009PhRvC..80a5805H}.

Two relevant energy scales which are important for this chapter are
the magnitude of the shift $\delta\mu = (\mu_n - \mu_p)/2$ of the
chemical potentials of neutrons $\mu_n$ and protons $\mu_p$ from their
common value $\mu_0$ at isospin symmetry and the pairing gap
$\Delta_0$ in the $\SD$ channel at $\delta\mu=0$. With increasing
isospin asymmetry, i.e., with $\delta\mu$ increasing from zero to
values of the order for $\Delta_0$, several unconventional phases may
emerge. One of these is a neutron-proton condensate with Cooper-pairs
which have a nonzero center-of-mass (CM)
momentum~\cite{2001PhRvC..63b5801S,2003PhRvC..67a5802M,2009PhRvC..79c4304M}. This
phase is the analogue of the Larkin-Ovchinnikov-Fulde-Ferrell (LOFF)
phase in electric
superconductors~\cite{1965LO,1964PhRv..135..550F}. Another possible
phase is the phase-separation (PS) consisting of an isospin symmetric
BCS part and an isospin asymmetric unpaired part. This phase was first
proposed in cold atomic gases~\cite{2003PhRvL..91x7002B}. As an
alternative to the LOFF phase we could include üthe deformed Fermi
surface (DFS) phase. In contrast to the LOFF phase it is
translationally invariant but it breaks the rotational
symmetry~\cite{2002PhRvL..88y2503M,2003PhRvC..67a5802M}. However,
these two phases have many properties in common and we concentrate
only on the LOFF phase. At large isospin asymmetry, where $\SD$
pairing is strongly suppressed, a BCS-BEC crossover may also occur in
the isotriplet $^1S_0$ pairing channel, notably in neutron-rich
systems and halo
nuclei~\cite{PhysRevC.73.044309,2007PhRvC..76f4316M,2008PhRvC..78a4306I,PhysRevC.79.054305,2009PhRvC..79e4002A,2010PhLB..683..134S,2011PhRvC..84f7301S,2012PhRvC..86a4305S}. From
the experimental phase shifts we can conclude that the pairing force
in the $\SD$ channel is stronger than in the $^1S_0$
channel. Isotriplet, spin-triplet pairing is prohibited by the Pauli
principle; accordingly, isotriplet pairing occurs only in the
spin-singlet channel. Since isosinglet, spin-triplet pairing is
favored over isotriplet spin-singlet pairing for not very high
asymmetries, we neglect isotriplet pairing. For large asymmetries,
isosinglet pairing is strongly suppressed and therefore pairing takes
place mostly in the isotriplet spin-singlet channel. Thus, to
conclude, for large asymmetries we expect pairing in the $^1S_0$ state
of neutron-neutron and proton-proton pairs, whereas at low asymmetries
the $\SD$ pairing between neutrons and protons dominates.

This chapter describes and extends the research presented in
Ref.~\cite{2012PhRvC..86f2801S} and \cite{2014PhRvC..90f5804S}. In the
first paper, the concepts of unconventional $\SD$ pairing and the
crossover were unified in a model of isospin-asymmetrical nuclear
matter by including some of the phases mentioned above. A phase
diagram for superfluid nuclear matter was constructed over wide ranges
of density, temperature, and isospin asymmetry. For this purpose the
coupled equations for the gap and the densities of the constituents
(neutrons and protons) were solved for the ordinary BCS state, its
low-density strong-coupling counterpart the BEC state, and two exotic
phases which may occur at finite isospin asymmetry: the phase with
finite Cooper-pair momentum (LOFF phase) and the PS phase, which
separates the matter into an unpaired part and an isospin symmetric
BCS part (PS-BCS) in the high-density weak-coupling limit or an
isospin symmetric BEC part (PS-BEC) in the low-densities
strong-coupling limit, respectively.

The basic parameters of the superfluid phases, such as the pairing gap
and the energy density have been studied widely for the BCS-BEC
crossover and in unconventional phases as for example the LOFF
phase. However, some {\it intrinsic features} which characterize the
condensate are less well known. These are for example the Cooper-pair
wave functions, the occupation probabilities of particles, the
coherence length, and related quantities. Certainly, for a deeper
understanding of the transitions from BCS to LOFF as well as from BCS
to BEC, an understanding of the evolution of these properties during
these transitions provide important insights into the mechanisms
underlying the emergence of new phases as well as into their
nature. In our second paper~\cite{2014PhRvC..90f5804S} we studied the
intrinsic properties of the condensate for the case of the $\SD$
condensate, thereby extending our first study in this
series~\cite{2012PhRvC..86f2801S}. As a representative of the
unconventional phases we choose the LOFF phase. In the case of the PS
phase, one of the constituents is the isospin-symmetrical BCS phase
and the other is the normal isospin-asymmetrical phase. Therefore, the
intrinsic features of the superfluid component of the PS phase are
identical to those of the BCS phase and there is no need to discuss
the intrinsic properties of the PS phase separately.

In order to induce a BCS-BEC crossover in the $\SD$-condensate we vary
the density of matter, which is a control parameter. The energies
which are relevant for scattering of two nucleons in the medium
essentially depend on their Fermi energies and therefore on the
density of the medium. Therefore, the nuclear interaction strength
also changes with density. There are two effects enforcing the BCS-BEC
crossover: a progressive dilution of the system and a concomitant
increase in the interaction strength in the $\SD$ channel at lower
energies. In~\cite{2014PhRvC..90f5804S}, we additionally varied the
isospin asymmetry for generating a mismatch in the Fermi surfaces of
paired fermions, and we changed the temperature to access the entire
density-temperature-asymmetry plane. In ultracold atomic gases the
BCS-BEC crossover is experimentally achieved by changing the effective
interaction strengths via the Feshbach mechanism, whereas the mismatch
of Fermi surfaces is accomplished by trapping different amounts of
atoms in different hyperfine states.

This chapter is structured as follows. In Sec.~\ref{sec_1_2} we
present the theory of asymmetric nuclear matter formulated in terms of
the imaginary-time finite-temperature Green's functions. In
Sec.~\ref{sec_1_3} we discuss the phase diagram of asymmetric nuclear
matter (Subsec.~\ref{subsec_1_3_1}), the temperature and asymmetry
dependence of the gap in the BCS and LOFF phases
(Subsec.~\ref{subsec_1_3_2}), the occupation numbers and chemical
potentials (Subsec.~\ref{subsec_1_3_3}), the effects of finite
momentum in the LOFF phase (Subsec.~\ref{subsec_1_3_4}), the kernel of
the gap equation across the BCS-BEC crossover and within the LOFF
phase (Subsec.~\ref{subsec_1_3_5}), the Cooper-pair wave functions
throughout the BCS-BEC crossover (Subsec.~\ref{subsec_1_3_6}), and the
occupation numbers and quasiparticle dispersion relations
(Subsec. \ref{subsec_1_3_7} and \ref{subsec_1_3_8},
respectively). This chapter is closed with a summary of the results in
Sec.~\ref{sec_1_4}.

\section{Theory}
\label{sec_1_2}
In the Nambu-Gorkov basis, the Greens function of the superfluid is
given by
\begin{eqnarray}
  \label{eq_1_01}
  i\mathscr{G}_{12} = i\left(
  \begin{array}{cc}
    G_{12}^{+} & F_{12}^{-}\\
    F_{12}^+ & G_{12}^{-}
  \end{array}
               \right) = \left(
               \begin{array}{cc}
                 \langle T_\tau\psi_1\psi_2^+\rangle & \langle T_\tau\psi_1\psi_2\rangle \\
                 \langle T_\tau\psi_1^+\psi_2^+\rangle & \langle T_\tau\psi_1^+\psi_2\rangle
               \end{array}
                                                         \right)\,,
\end{eqnarray}
with $G_{12}^{+}\equiv G^{+}_{\alpha\beta}(x_1,x_2)$ etc.,
$x=(t,\vecr)$ is the continuous space-time variable, Greek indices
label the discrete spin and isospin variables and $T_\tau$ is the
time-ordering operator for imaginary time. The operators in
Eq.~\eqref{eq_1_01} can be viewed as bi-spinors with
$\psi_{\alpha}=(\psi_{n\uparrow},\psi_{n\downarrow},\psi_{p\uparrow},\psi_{p\downarrow})^T$. The
indices $n$ and $p$ label the isospin and the indices $\uparrow$ and
$\downarrow$ label the spin.

The matrix in Eq.~\eqref{eq_1_01} obeys the familiar Dyson equation
with the formal solution
\begin{eqnarray}
  \label{eq_1_2}
  \left(\mathscr{G}_{0,13}^{-1}-\Xi_{13} \right)
  \mathscr{G}_{32} = \delta_{12}\,,
\end{eqnarray}
with $\Xi$ being the self-energy. Summation and integration over
repeated indices is implicit. In the next step we need to transform
Eq.~\eqref{eq_1_2} into momentum space, where it becomes an algebraic
equation. We cannot assume translational invariance for our purposes
and therefore we introduce center-of-mass (CM) coordinates
$\tilde r=(x_1-x_2)$ and $R=(x_1+x_2)/2$, with $\vecR$ denoting the
three vector component of $R$. $k=(ik_\nu,\veck)$ is the associated
relative momentum, whose zero component takes discrete values of
$k_\nu=(2\nu+1)\pi T$ (Matsubara frequencies), where
$\nu\in\mathbb{Z}$ and $T$ is the temperature. Here $\vecQ$ is the
three-momentum in the CM system. We first perform a variable
transformation to CM coordinates
\begin{eqnarray}
  &&iG_{12}^{+}=iG^{+}_{\alpha\beta}(x_1,x_2)
     =iG^{+}_{\tau\sigma,\tau'\sigma'}(\bm x_1,\bm x_2,\tilde t)\\
  &=&\langle T\psi_1\psi_2^+\rangle =\langle
      T\psi_{\tau\sigma}(\bm
      x_1,0)\psi_{\tau'\sigma'}^+(\bm x_2,\tilde t)\rangle\\
  &=&\left\langle T\psi_{\tau\sigma}\left(\bm
      R+\frac{\tilde{\bm
      r}}2,0\right)\psi_{\tau'\sigma'}^+\left(\bm
      R-\frac{\tilde{\bm r}}2,\tilde
      t\right)\right\rangle\,,
\end{eqnarray}
with $\tilde t=t'-t$. Afterwards we perform a Fourier transformation
with respect to the relative four-coordinate and the CM
three-coordinate. Here we first do the Fourier transformation with
respect to the three-coordinates:
\begin{eqnarray}
  G^{+}_{\tau\sigma,\tau'\sigma'}(\bm k,\bm Q,\tilde t)
  &=&\frac1{(2\pi)^3}\int d^3\bm R d^3\tilde{\bm r}\cdot
      e^{-i(\tilde{\bm r}\cdot\bm k+\bm R\cdot\bm Q)}\nonumber\\
  && \times
     G^{+}_{\tau\sigma,\tau'\sigma'}(\bm x_1,\bm x_2,\tilde t)\,,
\end{eqnarray}
and then we perform the Fourier transformation with respect to the
zero component of the relative momentum:
\begin{eqnarray}
  &&G^{+}_{\tau\sigma,\tau'\sigma'}(\bm k,\bm Q,\tilde
     t) =\frac1\beta\sum_\nu e^{-ik_\nu
     t}G^{+}_{\tau\sigma,\tau'\sigma'}(ik_\nu,\bm k,\bm Q)\,.
\end{eqnarray}
The other components of $i\mathscr{G}_{12}$ can be Fourier transformed
in an analogous manner to obtain $\mathscr{G}(k,\bm Q)$.

Thus the Fourier image of Eq.~\eqref{eq_1_2} is written as
\begin{eqnarray}
  \left[\mathscr{G}_0(k,\vecQ)^{-1}-\Xi(k,\vecQ) \right]\mathscr{G}(k,\vecQ) = {\bf{1}}_{8\times 8}\,.
\end{eqnarray}
The normal propagators of particles and holes are diagonal in both
spaces, i.e., $(G^+,G^{-}) \propto \delta_{\alpha\alpha'}$; thus the
off-diagonal elements of $\mathscr{G}_0^{-1}$ are zero. The
nonvanishing components in the Nambu-Gorkov space are:
\begin{equation} [\mathscr{G}_0^{-1}(ik_{\nu},\veck,\vecQ)]_{11} =
  -[\mathscr{G}_0^{-1}(-ik_{\nu}, \veck, -\vecQ)]_{22} =
  G_{0}^{-1}(ik_{\nu},\veck,\vecQ)
\end{equation}
with
\begin{equation}
  \label{eq_1_10} {G}_0^{-1}(k,\vecQ) ={\rm diag}( ik_{\nu} - \epsilon_{n\uparrow}^+, ik_{\nu} - \epsilon_{n\downarrow}^+, ik_{\nu} - \epsilon_{p\uparrow}^+ , ik_{\nu} - \epsilon_{p\downarrow}^+)\,.
\end{equation}
Here we have
\begin{equation}
  \epsilon_{n/p,\,\uparrow/\downarrow}^{\pm} = \frac1{2m^*}\left(\veck\pm\frac{\vecQ}2\right)^2-\mu_{n/p}\,,
  \label{eq_1_11}
\end{equation}
which we separate into symmetrical and antisymmetrical parts with
respect to the time-reversal operation into
\begin{eqnarray}
  \epsilon_{n,\uparrow/\downarrow}^\pm = E_S -\delta\mu\pm E_A\,,\\
  \epsilon_{p,\uparrow/\downarrow}^\pm = E_S+\delta\mu \pm E_A\,,
\end{eqnarray}
with
\begin{eqnarray}
  E_S &=&\left(Q^2/4+k^2\right)/2m^*-\bar\mu\,,\\
  E_A &=& \veck\cdot \vecQ /2m^*\,,
\end{eqnarray}
with $\bar\mu \equiv (\mu_n+\mu_p)/2$. Here $E_S$ is the symmetric
part of the quasiparticle spectrum which does not depend on the angle
between $\veck$ and $\vecQ$, whereas $E_A$ is the antisymmetric part
of quasiparticle spectrum which depends on the angle. The self-energy
effects can be taken into account via the effective mass $m^*$, which
we compute using the Skyrme force:
\begin{align}
  \frac{m}{m^*}=&\left[1-(m/p) \partial_p \Xi_{11} \vert_{p=p_F}\right] =\left[1+\frac{\rho\cdot m}{8\hbar^2}(3t_1+5t_2)\right]\,,\\
                &t_1=395\,\mathrm{MeV}\,\mathrm{fm}^5\,,\quad t_2=-95\,\mathrm{MeV}\,\mathrm{fm}^5\,,\quad m=939\,\mathrm{MeV}\,,\nonumber
\end{align}
where $m$ is the bare mass and $p_F$ is the Fermi momentum. We use the
SkIII parameterization of the Skyrme
interaction~\cite{1987PhRvC..35.1539S}. In our calculations we ignore
the small mismatch between neutrons and protons. Had we kept the
mismatch, we would obtain
\begin{align}
  \frac{m_{n/p}}{m^*_{n/p}}=&\left[1+\frac{\rho\cdot m_{n/p}}{2\hbar^2}(t_1+t_2) +\frac{\rho\cdot m_{n/p}}{8\hbar^2}(t_2-t_1)(1\pm\alpha)\right]\,,
\end{align}
with $\alpha=(\rho_n-\rho_p)/(\rho_n+\rho_p)$ being the density
asymmetry. This mismatch changes $E_{S/A}\to E_{S/A}(1\pm\delta_m)$
and $\delta\mu\to\delta\mu+\mu\delta_m$, with
$\delta_m = (m^*_n-m^*_p)/(m^*_n+m^*_p)\ll1$. In our analysis of this
chapter, we obtain $0\le\vert \delta_m\vert\le0.06$. Because the upper
bound that is reached for the largest asymmetries is small we can
neglect the missmatch, as stated above.

The quasiparticle spectra in Eq.~\eqref{eq_1_10} are written in a
general reference frame moving with the CM momentum of Cooper-pairs
$\vecQ$ with respect to a laboratory frame at rest. The spectrum of
quasiparticles is two-fold degenerate. The SU(4) Wigner symmetry of
the unpaired state is broken down to spin SU(2). The phase shifts in
the isoscalar and isotriplet $S$-waves differ, thus this symmetry is
always approximate. The isosinglet pairing is stronger than isotriplet
pairing in bulk nuclear matter.

The nucleon-nucleon scattering data (see Fig.~\ref{fig_1_01}) shows
that the dominant attractive interaction in low-density nuclear matter
is in the $\SD$ partial wave. Thus isosinglet spin-triplet pairing in
the $\SD$ channel dominates the pairing at low densities and not too
large asymmetries. Accordingly, we have the following relation for the
anomalous propagators:
$(F^+_{12},F^-_{12})\propto (-i\tau_y) \otimes \sigma_x$, with
$\tau_i$ and $\sigma_i$ being the Pauli matrices in isospin and spin
spaces. This implies that in the quasiparticle approximation, the
self-energy $\Xi$ has only off-diagonal elements in the Nambu-Gorkov
space. This implies that
$\Xi_{12} = \Xi_{21}^{+} = i\Delta_{\alpha\beta}$, with
$\Delta_{14}= \Delta_{23} =-\Delta_{32} = - \Delta_{41} \equiv
\Delta$,
where $\Delta$ is the (scalar) pairing gap in the $\SD$ channel.

With specifications above we obtain for $\Xi$ and $\mathscr{G}_0^{-1}$
the following matrix structure
\begin{eqnarray}
  \Xi_{12}&=&
              \begin{pmatrix}
                0&0&0&-i\Delta\\0&0&-i\Delta&0\\0&i\Delta&0&0\\i\Delta&0&0&0
              \end{pmatrix}\\
  \Rightarrow\Xi&=&
                    \begin{pmatrix}
                      \bf{0}&
                      \begin{matrix}
                        0&0&0&-i\Delta\\0&0&-i\Delta&0\\0&i\Delta&0&0\\i\Delta&0&0&0
                      \end{matrix}\\
                      \begin{matrix}
                        0&0&0&-i\Delta\\0&0&-i\Delta&0\\0&i\Delta&0&0\\i\Delta&0&0&0
                      \end{matrix}
                      &\bf{0}
                    \end{pmatrix}\,,
\end{eqnarray}
\begin{eqnarray}
  \begin{split}
    \mathscr{G}_0^{-1}&=&{\rm diag}\left(ik_{\nu} -
      \epsilon_{n\uparrow}^+,\,ik_{\nu}-
      \epsilon_{n\downarrow}^+,\,ik_{\nu} -
      \epsilon_{p\uparrow}^+,\,ik_{\nu} -
      \epsilon_{p\downarrow}^+,\right.\\ &&\left.ik_{\nu} +
      \epsilon_{n\uparrow}^-,\,ik_{\nu}+
      \epsilon_{n\downarrow}^-,\,ik_{\nu} +
      \epsilon_{p\uparrow}^-,\,ik_{\nu} +
      \epsilon_{p\downarrow}^-\right)\,.
  \end{split}
\end{eqnarray}

Since we have
$\epsilon_{n/p\uparrow}^\pm=\epsilon_{n/p\downarrow}^\pm$, we do not
lose information by reducing the $8\times 8$ equation
$(\mathscr{G}_0^{-1}-\Xi)\cdot\mathscr{G}=\bm1$ to an equation written
in terms of $4\times4$ matrices:
\begin{eqnarray}
  &&\begin{pmatrix}
    ik_{\nu} - \epsilon_{n}^+&0&0&i\Delta\\
    0& ik_{\nu} - \epsilon_{p}^+ &-i\Delta&0\\
    0&i\Delta &ik_{\nu} + \epsilon_{n}^-&0\\
    -i\Delta&0&0&ik_{\nu} + \epsilon_{p}^-
  \end{pmatrix} \nonumber\\
  &&\cdot\begin{pmatrix}
    \begin{matrix}
      G^+_n&0\\0&G^+_p
    \end{matrix} &
    \begin{matrix}
      0&F^-_{np}\\F^-_{pn}&0
    \end{matrix}\\
    \begin{matrix}
      0&F^+_{np}\\F^+_{pn}&0
    \end{matrix} &
    \begin{matrix}
      G^-_n&0\\0&G^-_p
    \end{matrix} &
  \end{pmatrix}=
                   \begin{pmatrix}
                     1&0&0&0\\0&1&0&0\\0&0&1&0\\0&0&0&1
                   \end{pmatrix}\,,
\end{eqnarray}
or explicitly
\begin{eqnarray}
  &&(ik_{\nu}-\epsilon_{n}^+)G^+_n+i\Delta
     F^+_{pn}=1\,,\\
  &&(ik_{\nu}+\epsilon_{p}^-)F^+_{pn}-i\Delta
     G^+_n=0\,,\\
  &&(ik_{\nu}-\epsilon_{p}^+)G^+_p-i\Delta
     F^+_{np}=1\,,\\
  &&(ik_{\nu}+\epsilon_{n}^-)F^+_{np}+i\Delta
     G^+_{p}=0\,,\\
  &&(ik_{\nu}+\epsilon_{n}^-)G^-_n+i\Delta
     F^-_{pn}=1\,,\\
  &&(ik_{\nu}-\epsilon_{p}^+)F^-_{pn}-i\Delta G^-_n=0\,,\\
  &&(ik_{\nu}+\epsilon_{p}^-)G^-_p-i\Delta
     F^-_{np}=1\,,\\
  &&(ik_{\nu}-\epsilon_{n}^+)F^-_{np}+i\Delta
     G^-_p=0\,.
\end{eqnarray}

These equations are solved in terms of normal and anomalous Green's
functions:
\begin{eqnarray}
  G_{n/p}^{\pm} &=&
                    \frac{ik_{\nu}\pm\epsilon_{p/n}^{\mp}}{(ik_{\nu}-E^+_{\mp/\pm})(ik_{\nu}+E^-_{\pm/\mp})}\,,\\
  F_{np}^{\pm} &=&
                   \frac{-i\Delta}{(ik_{\nu}-E^+_{\pm})(ik_{\nu}+E^-_{\mp})}\,,\\
  F_{pn}^{\pm} &=&
                   \frac{i\Delta}{(ik_{\nu}-E^+_{\mp})(ik_{\nu}+E^-_{\pm})}\,.
\end{eqnarray}
The poles of the propagators define the four possible branches of the
quasiparticle spectra, which are given by
\begin{equation}
  E_{r}^{a} = \sqrt{E_S^2+\Delta^2} + r\delta\mu +a
  E_A\,,
  \label{eq_1_33}
\end{equation}
with $a,r\in\{+,-\}$. Here $E_{r}^{a}$ accommodates the disruptive
effects such as the shift in the chemical potentials as well as the
effects of finite momentum $\vecQ$ which can compensate for the
mismatch. The latter effects can be viewed as a shift between the
centers of the Fermi spheres of protons and neutrons due to the CM
momentum $\vecQ$. At angles with $\cos\theta>0$ (where $\theta$ is the
angle between $\veck$ and $\vecQ$), the branches with $a=r$ are
located further away from each other than in ordinary BCS paring and,
conversely, the branches with $a\neq r$ are shifted closer
together. In this case, the shift due to the Cooper-pair momentum
works against the shift due to the asymmetry.

For the following calculations we need the Matsubara summations over
frequencies in the Green's function $G_{n/p}^{\pm}$ and
$F_{np/pn}^{\pm}$. They are calculated in appendix \ref{app_1}. The
result of the summations is given by
\begin{eqnarray}
  \frac1\beta\sum_\nu G_{n/p}^{\pm} &=&\frac12\left(1\pm\frac{E_S}{\sqrt{E_S^2+\Delta^2}}\right)f(E^+_{\mp/\pm})\nonumber\\
                                    &&+\frac12\left(1\mp\frac{E_S}{\sqrt{E_S^2+\Delta^2}}\right)(1-f(E^-_{\pm/\mp}))\,,\\
  \frac1\beta\sum_\nu F_{np}^{\pm} &=& \frac{i\Delta}{2\sqrt{E_S^2+\Delta^2}}\left(1-f(E^+_{\pm})-f(E^-_{\mp})\right)\,,\label{eq_1_35}\\
  \frac1\beta\sum_\nu F_{pn}^{\pm} &=&\frac{-i\Delta}{2\sqrt{E_S^2+\Delta^2}}\left(1-f(E^+_{\mp})-f(E^-_{\pm})\right)\,.\label{eq_1_36}
\end{eqnarray}

We introduce the following equation for the gap:
\begin{eqnarray}
  \Delta(\veck,\vecQ) &=& \frac1{4\beta}\int\!\!\frac{d^3k'}{(2\pi)^3}\sum_\nu V(\veck,\veck')\nonumber\\
                      &&\times {\rm Im} [F_{np}^+(k'_\nu,\veck',\vecQ)+F_{np}^-(k'_\nu,\veck',\vecQ)\nonumber\\
                      &&-F_{pn}^+(k'_\nu,\veck',\vecQ)-F_{pn}^-(k'_\nu,\veck',\vecQ)]\,,
\end{eqnarray}
with $V(\veck,\veck')$ being the neutron-proton interaction potential
and $f(E)=1/[\exp(E/T)+1]$. Using the Matsubara summations of
Eq.~\eqref{eq_1_35} and Eq.~\eqref{eq_1_36} and performing the partial
wave expansion we obtain:
\begin{eqnarray}
  \Delta_l(Q) &=& \frac{1}{4}\sum_{a,r,l'} \int\!\!\frac{d^3k'}{(2\pi)^3} V_{l,l'}(k,k')\nonumber\\
              &&\times \frac{\Delta_{l'}(k',Q)}{2\sqrt{E_{S}^2(k',Q)+\Delta^2(k',Q)}}[1-2f(E^a_r)]\,,\label{eq_1_38}
\end{eqnarray}
with $V_{l,l'}(k,k')$ being the interaction in the $\SD$ partial wave
and $\Delta^2=3/(8\pi)\sum_l\Delta_l^2$.

For the densities of neutrons and protons in any of the superfluid
states we obtain:
\begin{eqnarray}
  \rho_{n/p}(\vecQ)&=& \int \frac{d^3k}{(2\pi)^3}\cdot\frac1\beta\sum_\nu\left[(G_{n/p,\uparrow}^+(k_\nu,\veck,\vecQ) +G_{n/p,\downarrow}^+(k_\nu,\veck,\vecQ)\right]\nonumber\\
                   &=&2\int \frac{d^3k}{(2\pi)^3}\cdot\frac1\beta\sum_\nu G_{n/p}^+(k_\nu,\veck,\vecQ)\\
                   &=&\int \frac{d^3k}{(2\pi)^3}\cdot\left(1+\frac{E_S}{\sqrt{E_S^2+\Delta^2}}\right)f(E^+_\mp) \nonumber\\
                   &&+\left(1-\frac{E_S}{\sqrt{E_S^2+\Delta^2}}\right)(1-f(E^-_\pm))\,. \label{eq_1_40}
\end{eqnarray}

The grand canonical potential is given by:
\begin{eqnarray}
  \Omega(\Delta,\vecQ)&=&\frac3{4\pi}\sum_l\int\frac{d^3k}{(2\pi)^3}\Delta_l(\veck)\phi_l(\veck)\nonumber\\
                      &&-\sum_{a,r}\int\frac{d^3k}{(2\pi)^3}\left[\frac{E^a_r(\veck)-E_S(\veck)}2\right.\nonumber\\
                      &&+\left.T\ln\left(1+e^{-\beta E^a_r(\veck)}\right)\right]\,,\label{eq_1_41}
\end{eqnarray}
where
\begin{eqnarray}
  \Delta_l(\veck)&=&\sum_{l'}\int\frac{d^3k'}{(2\pi)^3}V_{l,l'}(\veck,\veck')\phi_{l'}(\veck')\,,
\end{eqnarray}
where the $\phi_{l}(\veck)$ function is given by
\begin{eqnarray}
  \phi_{l}(\veck) = \frac{1}{4}\sum_{a,r} \frac{\Delta_{l}(k,Q)}{2\sqrt{E_{S}^2(k,Q)+\Delta^2(k,Q)}}[1-2f(E^a_r)]\,.
\end{eqnarray}
The free energy can be further related to the grand canonical
potential as follows
\begin{eqnarray}
  \tilde\Omega(\Delta,\vecQ)&=&\Omega(\Delta,\vecQ)-\Omega(0,\vecQ)+\Omega(0,0)\,,\label{eq_1_44}\\
  F(\Delta,\vecQ)&=&\tilde\Omega(\Delta,\vecQ)+\mu_n\rho_n+\mu_p\rho_p\,.\label{eq_1_45}
\end{eqnarray}

The CM momentum $Q$ is obtained in the following way: First we solve
the system of equations~\eqref{eq_1_38} and \eqref{eq_1_40}
simultaneously. Afterwards we determine the free energy according to
Eq.~\eqref{eq_1_45}. This procedure is carried out for a range of
values of $Q$ and the value corresponding to the lowest free energy is
the one chosen by the system. The case with $Q = 0$ corresponds to the
BCS state, the case with $Q\neq 0$ corresponds to the LOFF phase.

For the ordinary BCS phase and the phase-separated phase it is
sufficient to find the free energy of the superfluid (S) and the
unpaired (N) phase,
\begin{equation}
  F_S = E_S-TS_S\,,\quad F_N = E_N-TS_N\,,
\end{equation}
where $E$ is the internal energy (statistical average of the system
Hamiltonian) and $S$ is the entropy. The free energy of the PS phase
can be calculated as a linear combination of the free energy of the
superfluid and the unpaired free energy:
\begin{equation}
  \label{eq_1_47}
  \mathscr{F}(x,\alpha) = (1-x) F_S(\alpha = 0) + x F_N(\alpha \neq 0)\,, \quad (Q=0)\,,
\end{equation}
with $x$ being the filling fraction of the unpaired phase. By
construction, the superfluid (S) part is isospin symmetric, whereas
the extra neutrons are shifted to the unpaired (N) part. Thus we have
$\rho_n^{(S)}=\rho_p^{(S)}=\frac12\rho^{(S)}$ and
$\rho_{n/p} = \frac12(1-x)\rho^{(S)} + x\rho_{n/p}^{(N)}$. Thus if the
ground state is achieved with $0<x<1$ we assign the ground state to
the phase-separated phase.

Putting all these together we see that we have three superfluid phases
and the normal state, which can be classified according to their
properties as follows
\begin{eqnarray}
  \label{eq_1_48}
  \left\{
  \begin{array}{llll}
    \Delta \neq 0, & Q = 0, & x = 0, &\textrm{BCS phase,}\\
    \Delta \neq 0, & Q \neq 0, & x = 0, &\textrm{LOFF phase,}\\
    \Delta \neq 0, & Q = 0, & 0<x<1, &\textrm{PS phase,}\\
    \Delta = 0, & Q = 0, & x = 1, &\textrm{unpaired phase.}
  \end{array}
                                    \right.
\end{eqnarray}
The first line of Eq.~\eqref{eq_1_48} corresponds to the homogeneous,
translational invariant, BCS phase. The second line corresponds to the
homogeneous, translational non-invariant LOFF phase. The third line
corresponds to the phase-separated phase, where the matter is divided
into an isospin symmetric BCS phase and an unpaired phase. The latter
phase is inhomogeneous but translational invariant phase-separated
(PS) phase. The last line corresponds to the normal (unpaired) state.

\section{BCS-phase, LOFF phase and crossover to BEC}
\label{sec_1_3}
\subsection{Phase diagram}
\label{subsec_1_3_1}
Eq.~\eqref{eq_1_38} and Eq.~\eqref{eq_1_40} were solved
self-consistently for pairing in the $\SD$ channel based on the
(phase-shift equivalent) Paris
potential~\cite{1984PhRvC..30.1822H}. Thus, we choose the dominant
attractive channel at relevant energies which corresponds to the
isosinglet, spin-triplet pairing. We, however, ignore the isotriplet,
spin-singlet pairing in the $^1S_0$ channel, which can become dominant
once the $\SD$ pairing is suppressed by isospin asymmetry. Thus, at
low temperatures and high asymmetries, $^1S_0$ pairing may play an
important role. The bare force in Eq.~\eqref{eq_1_38} benchmarks the
phase diagram, which should be reproducible by any
phase-shift-equivalent interaction. However, some regions of the phase
diagram may strongly be affected by polarization of the
medium. Studies of polarization in neutron matter exemplify the
complexity of this problem: while propagator-based methods predict
suppression of the gap, quantum Monte-Carlo methods predict gaps
closer to the BCS result obtained with the bare force (for a recent
assessment, see~\cite{2009PhRvC..80d5802G}). The nuclear mean field
was modelled by a Skyrme density functional. We used two
parameterizations: the first one is the SkIII taken
from~\cite{1987PhRvC..35.1539S} and the second one is the SLy4
parameterization of Ref.~\cite{1998NuPhA.635..231C}. We found that the
results are insensitive to the choice of parameterization.

\begin{figure}[!]
  \begin{center}
    \includegraphics[width=\textwidth]{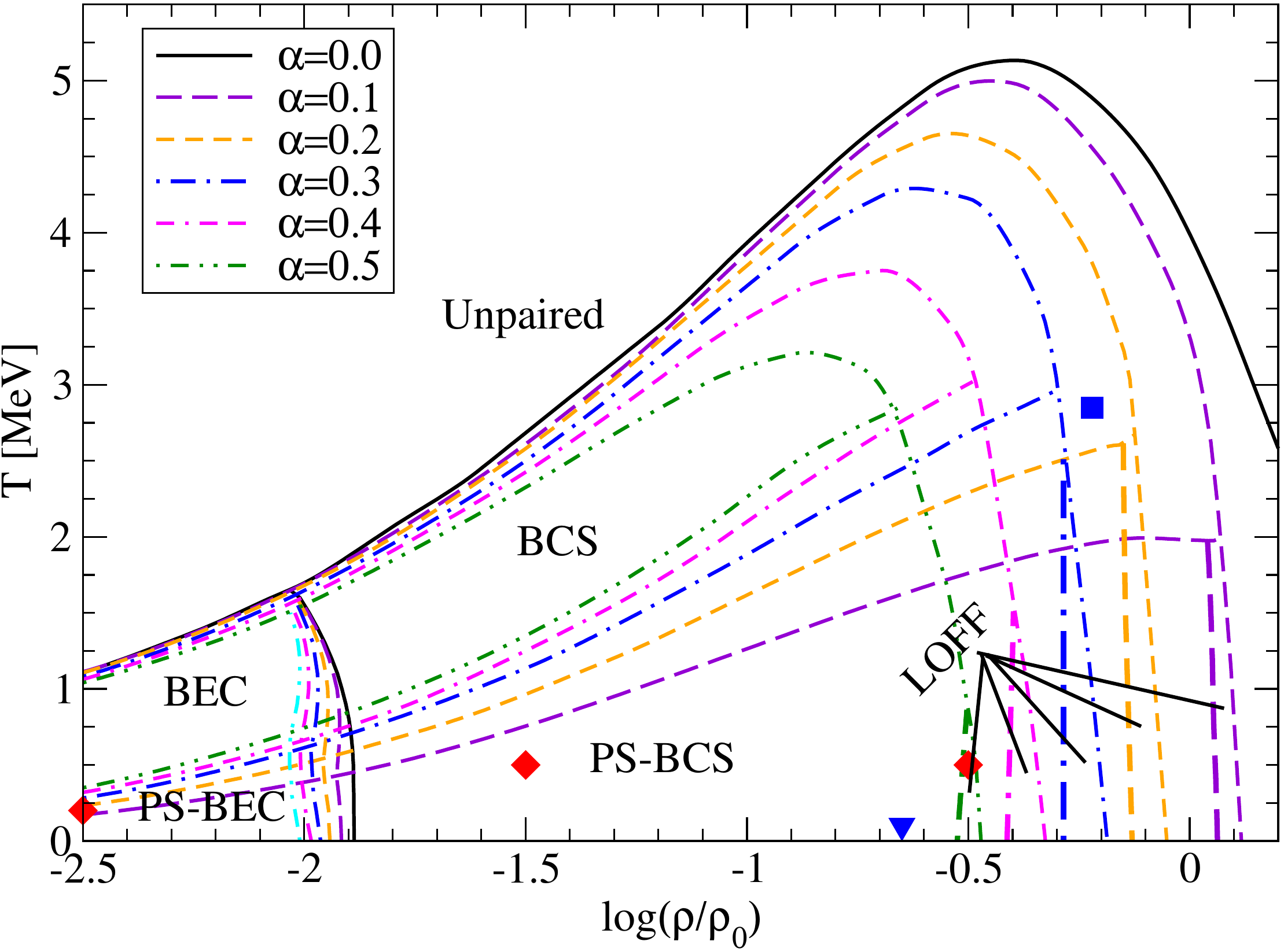}
    \caption[Phase diagram of isospin asymmetric nuclear
    matter.]{Phase diagram of dilute nuclear matter in the
      temperature-density plane for several isospin asymmetries
      $\alpha$ (from~\cite{2012PhRvC..86f2801S}). Included are four
      phases: unpaired phase, BCS (BEC) phase, LOFF phase, and PS-BCS
      (PS-BEC) phase. For each asymmetry between
      $0<\alpha<\alpha_\mathrm{LOFF}$ there are two tri-critical
      points, one of which is always a Lifshitz
      point~\cite{1980JMMM...15..387H}. For special values of
      asymmetry these two points degenerate into a single
      tetra-critical point for $\log(\rho_4/\rho_0) = -0.22$,
      $T_4 = 2.85$ MeV and $\alpha_4 = 0.255$ (shown by a square). The
      LOFF phase disappears at the point
      $\log(\rho_\mathrm{LOFF}/\rho_0) = -0.65$,
      $\alpha_\mathrm{LOFF} = 0.62$ and $T=0$ (shown by a
      triangle). The boundaries between BCS and BEC phases are
      identified by the change of sign of the average chemical
      potential $\bar\mu$. The red diamonds refer to three regions of
      the phase diagram explained later.}
    \label{fig_1_02}
  \end{center}
\end{figure}

Fig.~\ref{fig_1_02} shows the phase diagram of dilute nuclear matter
with pairing in the $\SD$ channel. We start with a discussion of the
phase transition from paired phase to unpaired phase for vanishing
asymmetry. At $T=0$, the gap has its maximal value. It decreases with
increasing temperature until it vanishes at $T_C$. The relation
between the gap and the critical temperature is given by
\begin{eqnarray}
  \Delta(T=0=\alpha)=1.76\, T_C\,.\label{eq_1_49}
\end{eqnarray}
Thus, a larger gap at vanishing temperature leads to a larger critical
temperature. Qualitative insight can be obtained from examining the
BCS weak coupling formula for the gap at zero temperature and
asymmetry
\begin{eqnarray}
  \Delta=2\varepsilon_F\cdot e^{-\frac1{NV}}\,,\label{eq_1_50}
\end{eqnarray}
with $N$ being the density of states and $V$ the strength of the
interaction and $\varepsilon_F$ the Fermi energy. The density of
states increases linearly with the Fermi momentum, whereas, according
to the phase-shift analysis, the interaction decreases as a function
of energy of colliding particles. We see that the critical temperature
increases initially due to the increase of $N$, but it becomes
suppressed in the high density limit as the attractive pairing
interactions tends to zero. This behavior is reflected in the shape we
can see in Fig.~\ref{fig_1_02}.

The phase diagram has a richer structure at non-zero isospin, as can
be seen in Fig.~\ref{fig_1_02} where the phase structure is shown for
several values of isospin asymmetry
$\alpha = (\rho_n-\rho_p)/(\rho_n+\rho_p)$, where $\rho_n$ and
$\rho_p$ are the number densities of neutrons and protons and
$\rho_0=0.16\,\mathrm{fm}^{-3}$ is the nuclear saturation
density. There are four different phases of matter in the phase
diagram (see Eq.~\eqref{eq_1_48}), which we discuss in turn:

(a) The unpaired normal phase, which is the ground state for
temperatures $T>T_{c}(\rho,\alpha)$, where $T_{c}(\rho,\alpha)$ is the
critical temperature of the superfluid phase transition for any given
asymmetry.

(b) The LOFF phase is the ground state for nonvanishing values of
$\alpha$ within the range $0<\alpha<\alpha_{\mathrm{LOFF}}$ and high
densities with $\rho>\rho_{\mathrm{LOFF}}$ and in a narrow
temperature-density strip at low temperatures with $T<T_4$. Here
$\alpha_{\mathrm{LOFF}}$ and $\rho_{\mathrm{LOFF}}$ correspond to the
point of maximal asymmetry and at the same time the minimal density
were the LOFF phase exists at $T=0$. This is shown by a blue triangle
in Fig.~\ref{fig_1_02}. $\rho_4$, $T_4$ and $\alpha_4$ belong to the
tetra-critical point, where the four phases BCS, PS-BCS, LOFF and
unpaired phase coexist. This is shown by a blue square in
Fig.~\ref{fig_1_02}. As borders for the LOFF phase we have the
triangle with $\log(\rho_{\mathrm{LOFF}}/\rho_0)=-0.65$, $T=0$ and
$\alpha_{\mathrm{LOFF}}=0.62$ and the square with
$\log(\rho_4/\rho_0)=-0.22$, $T_4=2.85$ MeV and $\alpha_4=0.255$.

(c) For nonvanishing asymmetry, the phase-separated (PS) phase is the
ground state for low temperatures and densities.

(d) The isospin-asymmetric BCS phase is the ground state at
intermediate temperatures below the transition to the unpaired phase
and above the transition to the PS phase and densities above the
crossover to a BEC.

One may, of course, pose the question of the structure of the phase
diagram in the high-density limit. At sufficiently large density, when
the chemical potentials of nucleons become of the order of the rest
mass of hyperons, the matter may become hyperon rich. This may occur
at about twice the nuclear saturation density. Furthermore, at very
high densities the interparticle distances decrease to values smaller
than the nucleon radius and the quarks bound in nucleons may deconfine
into free quarks.

The phase transitions have a very interesting shape. In addition to
the crossover lines, we see several phase transition lines, resulting
in two tri-critical points, where three phases coexist. At asymmetries
below $\alpha_4$, we have a low-density tri-critical point, where the
PS-BCS, the LOFF and the BCS phase coexist and a high-density
tri-critical point, where the LOFF, the BCS and the unpaired phase
coexist. However, at asymmetries above $\alpha_4$, we obtain a
low-density tri-critical point with PS-BCS, BCS and unpaired phase and
a high-density tri-critical point with PS-BCS, LOFF and unpaired
phase. Interestingly they degenerate into a tetra-critical point,
where PS-BCS, BCS, LOFF and unpaired phase coexist at asymmetry
$\alpha_4$.

\begin{figure}[!]
  \begin{center}
    \includegraphics[width=0.8\textwidth]{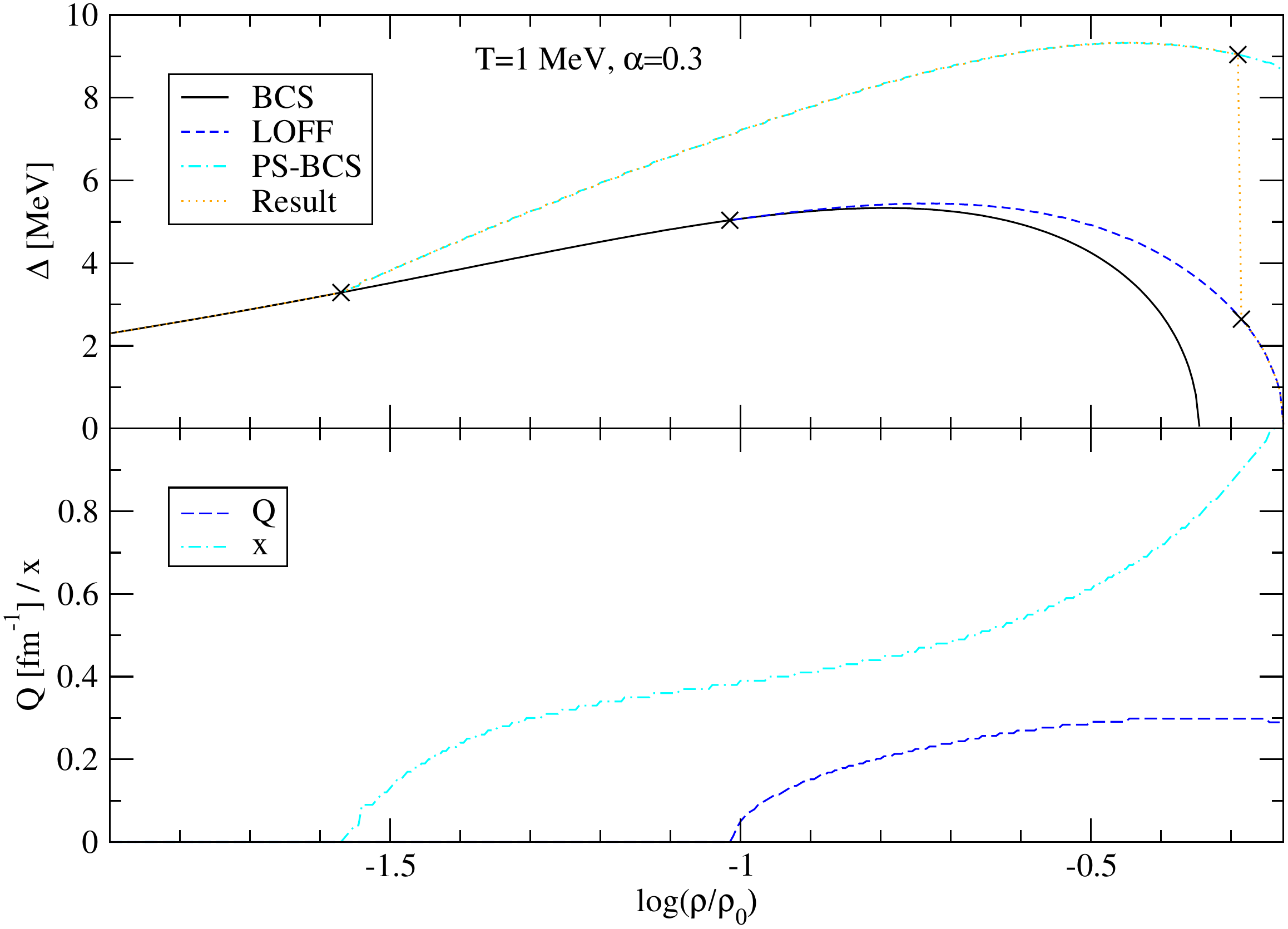}
    \caption[The gap of various phases, the LOFF momentum and the PS
    filling fraction]{Upper panel: The gap of various phases. Lower
      panel: LOFF momentum $\vecQ$ and PS filling fraction $x$.}
    \label{fig_1_03}
  \end{center}
\end{figure}

To access the order of various phase transitions (first or second
order) we examine the behavior of the gap function across the phase
diagram. This is illustrated in Fig.~\ref{fig_1_03}. In the upper
panel we present the gap at fixed temperature and asymmetry for
increasing density for three different phases. The calculated gap for
the LOFF phase does not take into account the possibility of a
PS-phase and vice versa. The BCS gap calculation ignores the
possibility of the LOFF and PS pairing. Of course, the phase realized
in nature is the one with the lowest free energy. In the lower panel
we present the LOFF momentum $\vecQ$ and the PS filling fraction $x$
with a reference to the corresponding gaps presented in the upper
panel. At low densities $\vecQ=0$ and $x=0$ and the BCS phase is the
ground state. With increasing density we find $x\neq 0$, therefore a
phase transition into the PS-BCS phase occurs which breaks the
homogeneity of the system. If we ignore the possibility of the PS
phase, a phase transition to the LOFF phase at higher density occurs;
this breaks the translational symmetry. Since both, the filling
parameter ($x$) of the PS phase and the momentum of the condensate
($\vecQ$) of the LOFF phase increase smoothly, the change in the gap
is also smooth and the phase transitions are second order. If we
increase the density, ignoring the possibility of a PS or LOFF phase,
the BCS gap vanishes smoothly. The same holds for the gap of the LOFF
phase, if we consider the possibility of the LOFF phase but ignore the
possibility of the PS phase. If we consider the PS phase but ignore
the possibility of a LOFF phase, the filling fraction $x$ increases
smoothly and we obtain a second order phase transition from BCS,
PS-BCS or LOFF to the unpaired phase. The same holds for phase
transitions from BEC, BCS, PS-BCS or LOFF to the unpaired phase with
increasing temperature. However, if we take PS-BCS and LOFF phases
into account, the free energy of the LOFF phase becomes less than the
free energy of the PS-BCS phase at a certain density. At this point
the gap does not change smoothly and therefore a first order phase
transition is expected. To summarize, we have second order phase
transitions from all superfluid phases to the unpaired phase and
between superfluid phases, with the exception of a first order phase
transition between the PS-BCS and LOFF phase (thick lines in
Fig.~\ref{fig_1_02}). The transitions from BCS to BEC and from PS-BCS
to PS-BEC are smooth crossovers.
\begin{figure}[!]
  \begin{center}
    \includegraphics[width=0.7\textwidth]{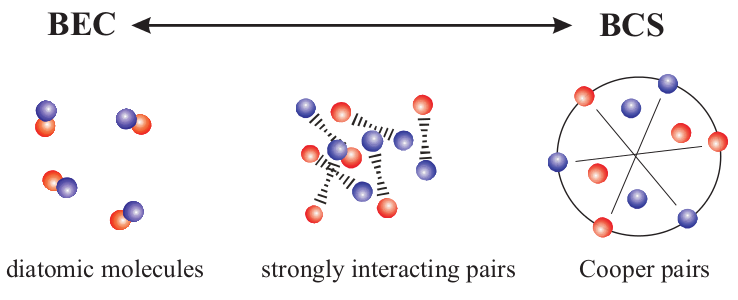}
    \caption[Illustration of the crossover.]{An illustration of the
      crossover from~\cite{2006PhDT.......107R}. On the left we see
      bound deuterons, on the right we see unbound Cooper-pairs.}
    \label{fig_1_04}
  \end{center}
\end{figure}

As mentioned above the low density limit of the phase diagram
corresponds to the strong-coupling limit where a BEC of deuterons
emerges. At intermediate temperatures we find a direct crossover from
the ordinary BCS phase to a BEC consisting of bound deuterons and free
neutrons. The situation is more complicated at low temperatures. The
crossover occurs in the presence of the PS phase. Therefore, we obtain
a crossover from the PS-BCS (which features a mixture of symmetric BCS
and an asymmetric unpaired phase) to the PS-BEC phase where the
symmetric BCS domains are replaced by a symmetric BEC of
deuterons. These transformations are not phase transitions, but smooth
crossovers, since no symmetry is broken. Therefore, the points of the
phase diagram where BCS, BEC and unpaired phases coexist cannot be
viewed as critical points. The same applies to the points where BCS,
BEC, PS-BCS, PS-BEC coexist.

\begin{figure}[!]
  \begin{center}
    \includegraphics[width=\textwidth]{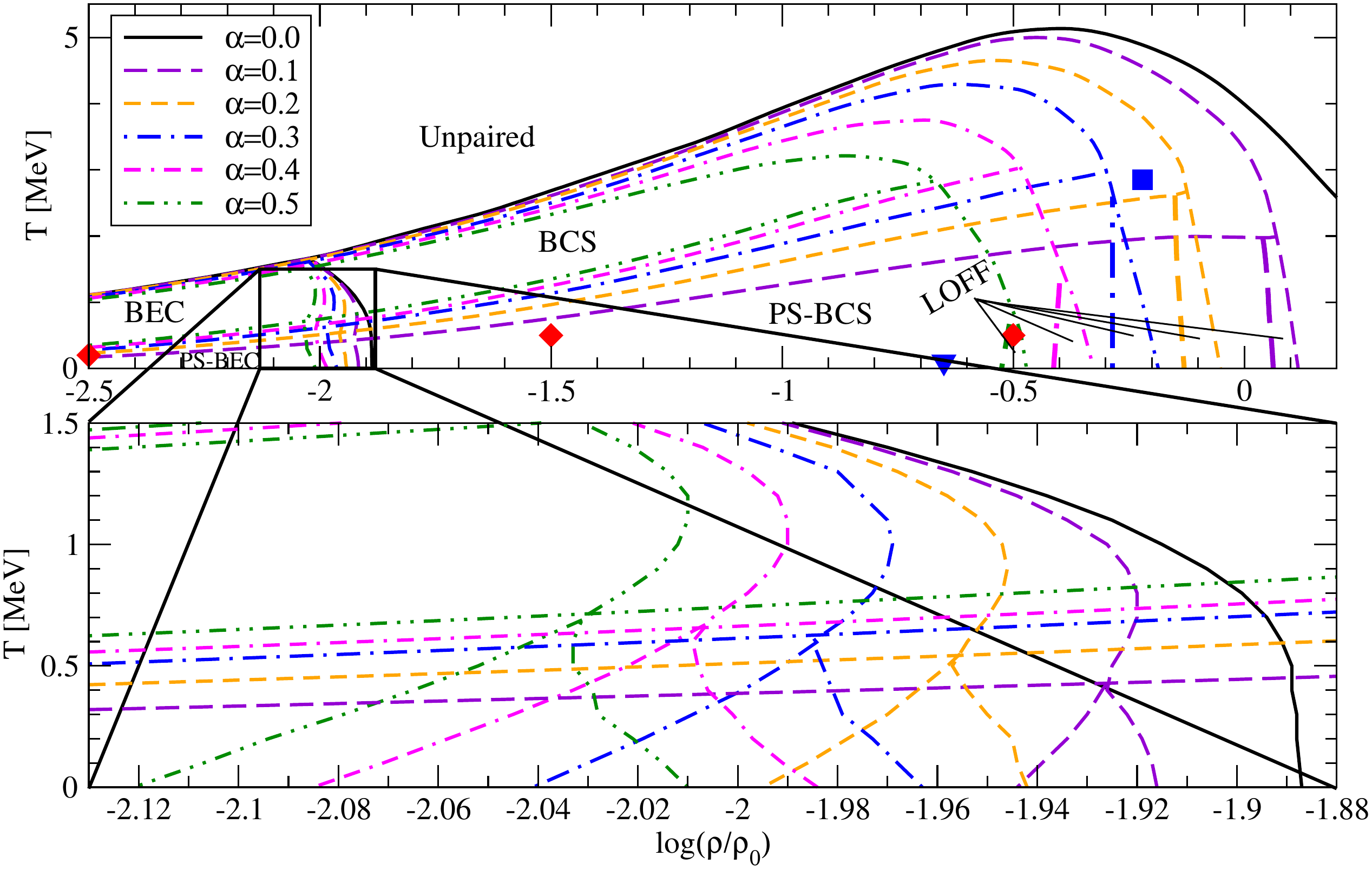}
    \caption[The zoomed crossover region of the phase diagram.]{The
      upper panel shows the complete phase diagram, as seen in
      Fig.~\ref{fig_1_02}. The lower panel zooms in the crossover
      region. The horizontal lines here show the phase transition
      lines for different asymmetries between the BCS/BEC phase (which
      exists above a given asymmetry line) to the PS phase (which
      exists below the line). The nearly vertical lines show the
      crossover from BCS to BEC regimes. The left line is the case
      without PS phase, whereas the right line is the one including
      the PS phase. Thus, the inclusion of the PS phase induces the
      BCS-BEC crossover at higher density. }
    \label{fig_1_05}
  \end{center}
\end{figure}

In the BCS limit, the size of a Cooper-pair is given by the coherence
length $\xi$ which is very large compared to the average interparticle
distance $d$. In the BEC limit the pairs are tightly bound deuterons
with $\xi\ll d$. This is illustrated schematically in
Fig.~\ref{fig_1_04}. Fig.~\ref{fig_1_05} zooms in at the crossover
region of Fig.~\ref{fig_1_02} and shows the results including and
excluding the PS phase. At higher temperatures the PS phase does not
arise and we observe an ordinary BCS-BEC crossover even in the
presence of isospin asymmetry. However, note that at sufficiently low
temperatures, the crossover density decreases with decreasing
temperature. At constant density the interparticle distance $d$ does
not change. By decreasing the temperature we have two competitive
effects affecting each other. At lower temperatures the particles have
less momentum and thus pairing can occur at lower distances, which
means that $\xi$ decreases and the crossover is shifted to higher
densities. However, by increasing asymmetry we have less protons and
thus less pairs, therefore $\xi$ increases, which means, that the
crossover is shifted to lower densities. At high temperatures, the
temperature can smear out the Fermi edges and thus the asymmetry
effect is weak. However, at low temperatures the temperature induced
smearing is weak compared to the asymmetry effect. Thus, the effect
induced by asymmetry dominates at low temperatures and high
asymmetries. Taking the PS-phase into account, we see that the
crossover density increases with decreasing temperature for
temperatures below the phase transition from BCS/BEC to
PS-BCS/PS-BEC. In the PS phase, we have an isospin symmetric BCS/BEC
domain in the matter. This means that $\xi$ is lower than in the
ordinary BCS/BEC phase and thus the crossover is shifted to higher
densities towards the $\alpha=0$ result.

In the following we will discuss the crossover in detail. For that
purpose we choose three points (marked with red diamonds in
Fig.~\ref{fig_1_02}) which correspond to the weak, strong and
intermediate couplings. Indeed, the point at $\log(\rho/\rho_0)=-0.5$
and $T=0.5$ MeV corresponds to the high-density weak-coupling region
(WCR) where we clearly have BCS pairing. For the low-density
strong-coupling region (SCR) we choose the parameter values
$\log(\rho/\rho_0)=-2.5$ and $T=0.2$ MeV as representative for the BEC
pairing. For comparison we also choose one point in between in the
intermediate-coupling region (ICR) at $\log(\rho/\rho_0)=-1.5$ and
$T=0.5$ MeV. We have chosen low values for the temperatures to make
sure that the matter is in all cases in the well developed condensate
phase.

\subsection{Temperature and asymmetry dependence of the gap:
  contrasting the BCS and LOFF phases}
\label{subsec_1_3_2}
\begin{figure}[!]
  \begin{center}
    \includegraphics[width=0.7\textwidth]{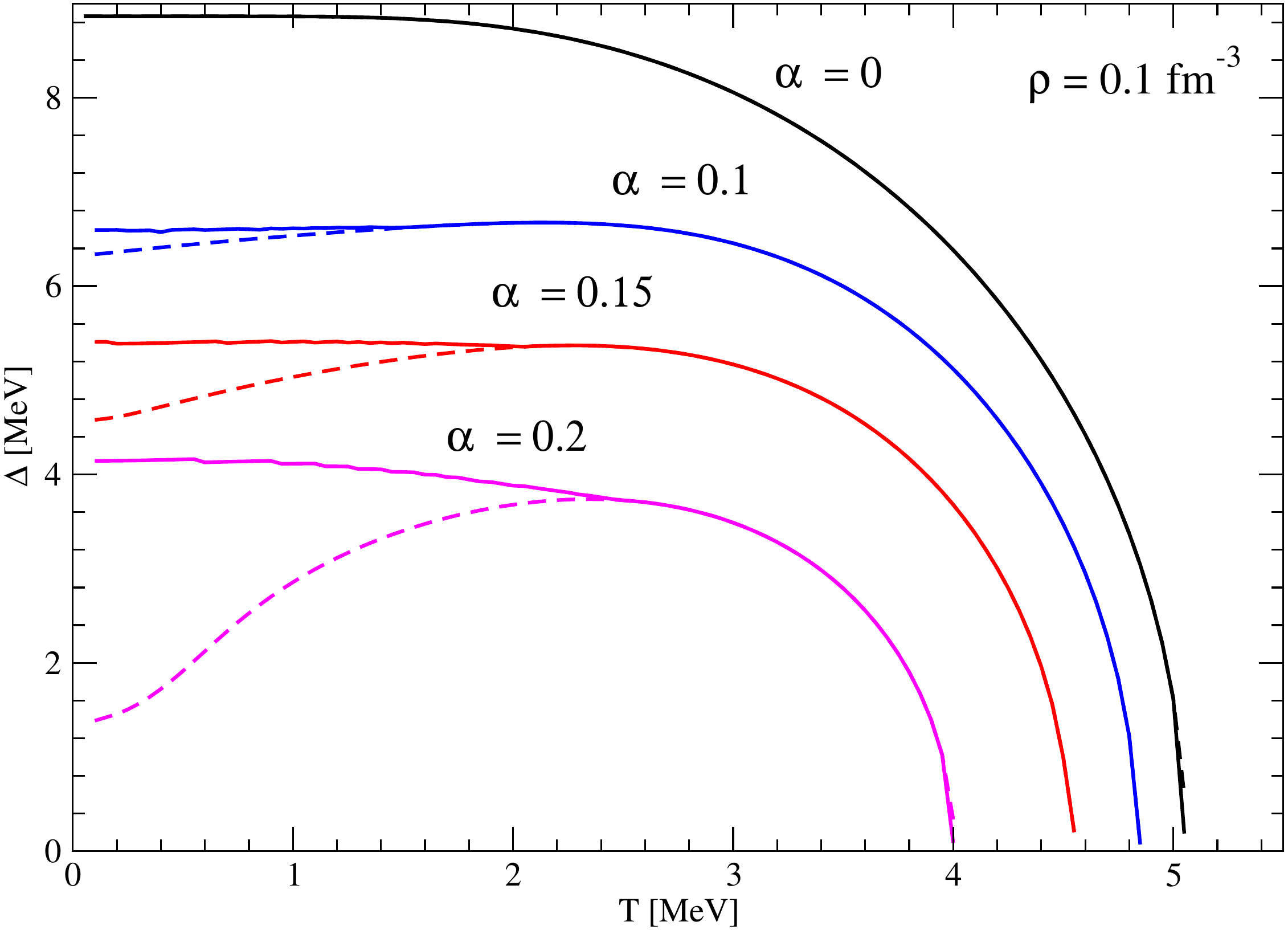}
    \caption[The gap as a function of the temperature.]{The gap as a
      function of the temperature at constant density $\rho=0.1$
      fm$^{-3}$ for several asymmetries. Solid lines allow for the
      emergence of the LOFF phase, whereas the dashed lines show only
      the BCS phase. }
    \label{fig_1_07}
  \end{center}
\end{figure}

We now turn to the discussion of the properties of individual phases
appearing in our phase diagram focusing on the key features. As a
first step in understanding the mechanism that governs the appearance
of various phases at different regimes present in the phase diagram we
now focus on the behavior of the gap function as a function of
temperature and asymmetry at constant density. We concentrate only on
the weak-coupling regime (WCR), as the behavior of the gap function in
the strong coupling regime (SCR) is self-similar to that of the
WCR. For now, we also neglect the possibility that the PS phase is the
ground state. Fig.~\ref{fig_1_07} shows the weak-coupling gap as a
function of temperature for a range of asymmetries. The plotted
results for each nonzero value of $\alpha$ reveal different regimes of
relatively low and relatively high temperature that reflect the
different behaviors of the gap when the possibility of a LOFF phase is
taken into account (solid curves) and when it is not (dashed
curves). Two branches existing at lower temperatures merge at some
point to form a single segment stretching up to the critical
temperature of phase transition. This high-temperature segment
corresponds to the BCS state, and the temperature dependence of the
gap is standard, with $d\Delta(T)/dT < 0$ and asymptotic behavior
$\Delta(\alpha, T) \sim [T_c(\alpha)(T_c(\alpha)-T)]^{1/2}$ as
$T \to T_c(\alpha)$, where $T_c(\alpha)$ is the (upper) critical
temperature. In the low-temperature region below the branching point,
there are two competing phases (BCS and LOFF), with very different
temperature dependences of the gap function. The quenching of the BCS
gap (dashed lines) as the temperature is decreased is caused by the
loss of coherence among the quasiparticles as the thermal smearing of
the Fermi surfaces disappears.
\begin{figure}[!]
  \begin{center}
    \includegraphics[width=\textwidth]{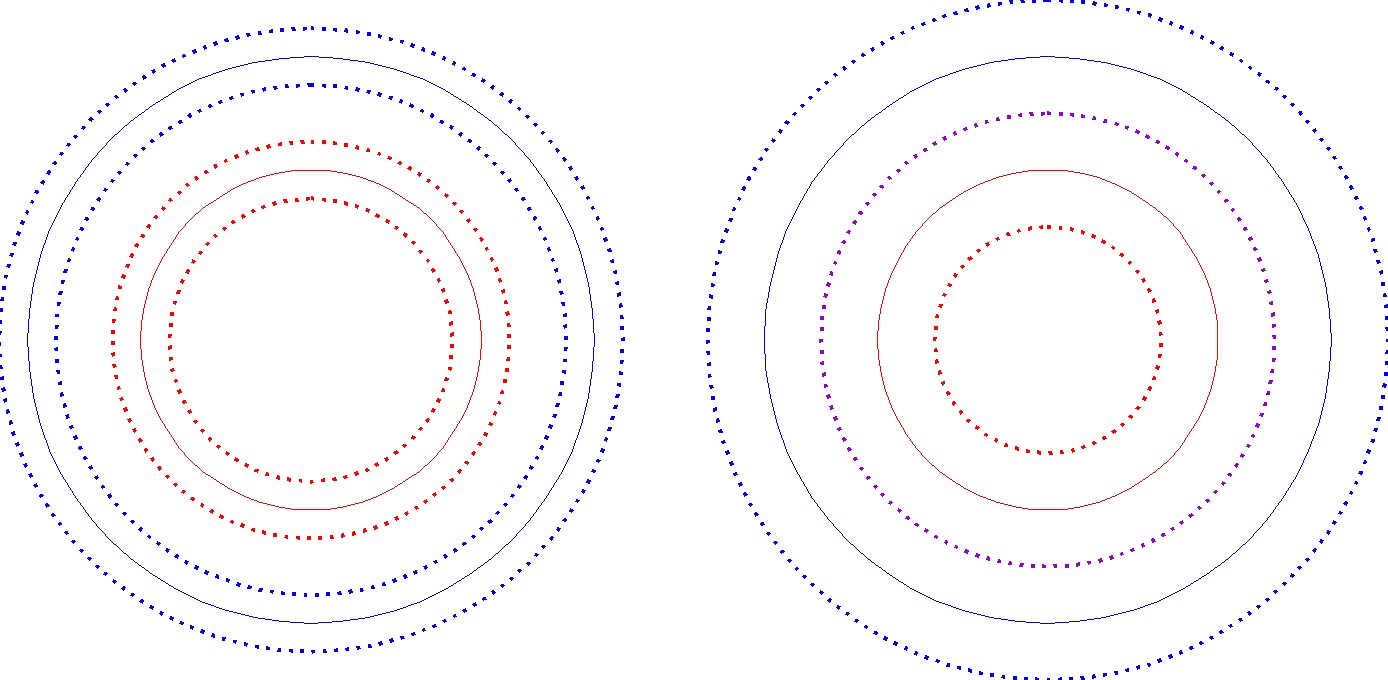}
    \caption[Illustration of temperature induced smearing of the Fermi
    surfaces.]{The proton Fermi surface is shown by a red solid line,
      the neutron Fermi surface is shown by a blue solid line. The
      dotted lines present the temperature induce smearing. The
      temperature in the right figure is bigger than in the left one.}
    \label{fig_1_06}
  \end{center}
\end{figure}

Consequently, in the low-temperature range below the branch point, the
BCS branch shows the unorthodox behavior $d\Delta(T)/dT > 0$, and for
large enough asymmetries there exists a lower critical temperature
$T_c^*$~\cite{2000PhRvL..84..602S}.

This effect is illustrate in Fig.~\ref{fig_1_06}, where the
Fermi-spheres of protons and neutrons are shown by red and blue solid
lines, respectively. The dotted concentric circles illustrate the
smearing induced by temperature. For pairing we need an overlap of the
Fermi spheres, thus the smearing of the temperature needs to overcome
the shift of the Fermi levels due to asymmetry. In the left plot the
smearing of the temperature is too low and coherence is lost. On the
right it is large enough to create an overlap. This simple picture
captures the effect of temperature on the pairing in asymmetric
systems: if temperature is high enough it restores the pairing
correlations which are otherwise suppressed by the asymmetry.

\begin{figure}[!]
  \begin{center}
    \includegraphics[width=0.7\textwidth]{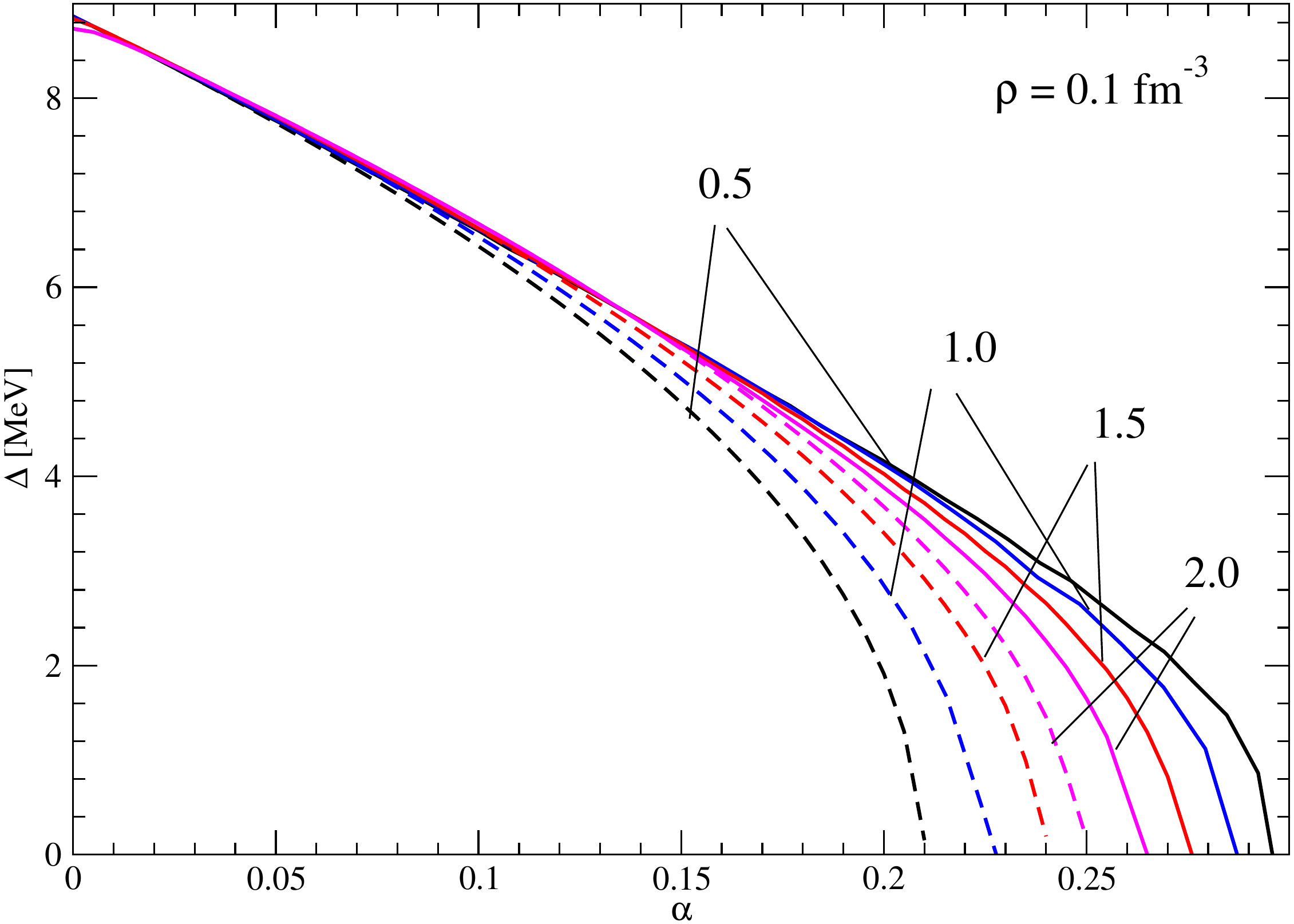}
    \caption[The gap as a function of the asymmetry.]{The gap as a
      function of the asymmetry at constant density $\rho=0.1$
      fm$^{-3}$ for several temperatures, given in MeV. Solid lines
      allow for the emergence of the LOFF phase, whereas the dashed
      lines show only the BCS phase. }
    \label{fig_1_08}
  \end{center}
\end{figure}

On the contrary, one finds $d\Delta(T)/dT < 0$ for the LOFF branch, as
is the case in ordinary (symmetrical) BCS
theory~\cite{2006PhRvB..74u4516H}. It should be mentioned that the
``anomalous" behavior of the BCS gap below the point of bifurcation
leading to the LOFF state gives rise to a number of anomalies in
thermodynamic quantities, such as negative superfluid density or
excess entropy of the superfluid~\cite{2006PhRvL..97n0404S}. These
anomalies are absent in the LOFF
state~\cite{2007IJMPE..16.2363J}. Fig.~\ref{fig_1_08} shows the
dependence of the gap function on asymmetry for several pertinent
temperatures. In accord with Fig.~\ref{fig_1_07}, there are two curves
(or segments) for each temperature: one in the low-$\alpha$ domain
where only the BCS phase exists and the other in the large-$\alpha$
domain where both BCS (dashed lines) and LOFF states (solid lines) are
possible. Clearly the LOFF solution, for which the gap extends to
larger $\alpha$ values, is favored in the latter domain.


For small $\alpha$ the gap function is linear in $\alpha$. At the
other extreme of large $\alpha$, the gap has the asymptotic behavior
$\Delta(\alpha)\sim \Delta_{00} \left(1-\alpha/\alpha_1\right)^{1/2}$,
where $\alpha_1\sim \Delta_{00}/\bar\mu$ and $\Delta_{00}$ is the
value of the gap at vanishing temperature and asymmetry. The critical
asymmetry $\alpha_2$ at which the LOFF phase transforms into the
normal phase is a decreasing function of temperature, whereas that for
termination of the BCS phase (denoted $\alpha_1$ above) increases up
to the temperature where $\alpha_1=\alpha_2$. For higher temperatures,
$\alpha_1$ decreases with temperature. Consequently, in the dominant
phase the critical asymmetry always decreases with temperature.

\subsection{Occupation numbers and chemical potentials}
\label{subsec_1_3_3}
\begin{figure}[!]
  \begin{center}
    \includegraphics[width=0.8\textwidth]{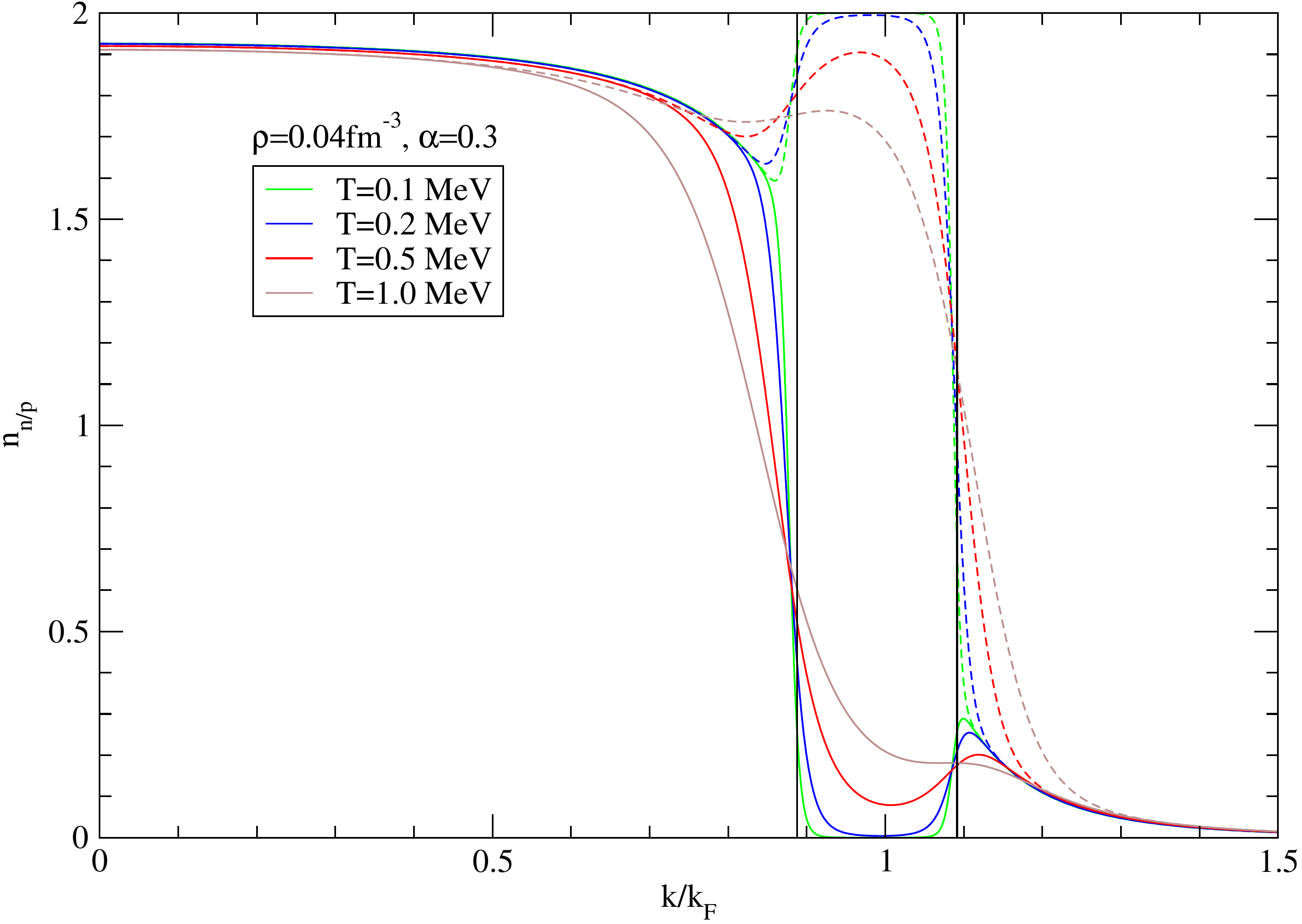}
    \caption[The neutron and proton occupation numbers for several
    temperatures in the BCS phase.]{The neutron (proton) occupation
      numbers in the BCS phase are shown with dashed (solid) lines at
      a fixed density of $\rho=0.04\,\mathrm{fm}^{-3}$ and a fixed
      asymmetry of $\alpha=0.3$. The labeling of temperatures is shown
      in the plot with various colors. The vertical black solid lines
      present the Fermi momenta of neutrons and protons,
      respectively.}
    \label{fig_1_09}
  \end{center}
\end{figure}

Next let us examine the behavior of the occupation numbers, which are
defined as integrands of the densities appearing in Eq.~\ref{eq_1_40},
i.e.,
\begin{eqnarray}
  n_{n/p}(\vecQ)
  &=&\left(1+\frac{E_S}{\sqrt{E_S^2+\Delta^2}}\right)f(E^+_\mp)
      \nonumber\\
  &&+\left(1-\frac{E_S}{\sqrt{E_S^2+\Delta^2}}\right)(1-f(E^-_\pm))\,.
\end{eqnarray}
Fig.~\ref{fig_1_09} shows the occupation numbers of neutrons and
protons respectively for a fixed density of
$\rho=0.04\,\mathrm{fm}^{-3}$ and a fixed asymmetry of $\alpha=0.3$
for several values of the temperature (see also the discussion in
Subsec.~\ref{subsec_1_3_7}). Due to the asymmetry, the Fermi surfaces
are shifted by $\delta\mu$. Because
$\rho_{n/p}=(1\pm\alpha)/2\cdot\rho$ and $k_F\propto\sqrt[3]{\rho}$,
one can define new Fermi surfaces for neutrons and protons as
$k_{F_{n/p}}/k_F=\sqrt[3]{1\pm\alpha}$, where $k_{F_{n/p}}$ are the
Fermi momenta for neutrons and protons and $k_F$ is the Fermi momentum
in isospin symmetric nuclear matter. The Fermi surfaces of neutrons
and protons are presented by vertical black solid lines in
Fig.~\ref{fig_1_09}. The prominent feature is the depletion of the
proton occupation numbers around the common Fermi surface, which is
most pronounced at low temperatures. At finite temperature this
depletion is gradually washed out. Note that at the neutron Fermi
surface, the proton occupancy increases again and these protons
contribute most to the Cooper pairing with the neutrons at their Fermi
surface. We thus have a Fermi distribution type occupation for protons
and neutrons for $k\lesssim k_{{F_p}}$ and $k\gtrsim k_{{F_n}}$
respectively with a ``breach'' in the momentum range
$k_{{F_p}}\lesssim k\lesssim k_{{F_n}}$. The effect of the temperature
smearing is demonstrated illustratively in Fig.~\ref{fig_1_06}.
\begin{figure}[!]
  \begin{center}
    \includegraphics[width=0.8\textwidth]{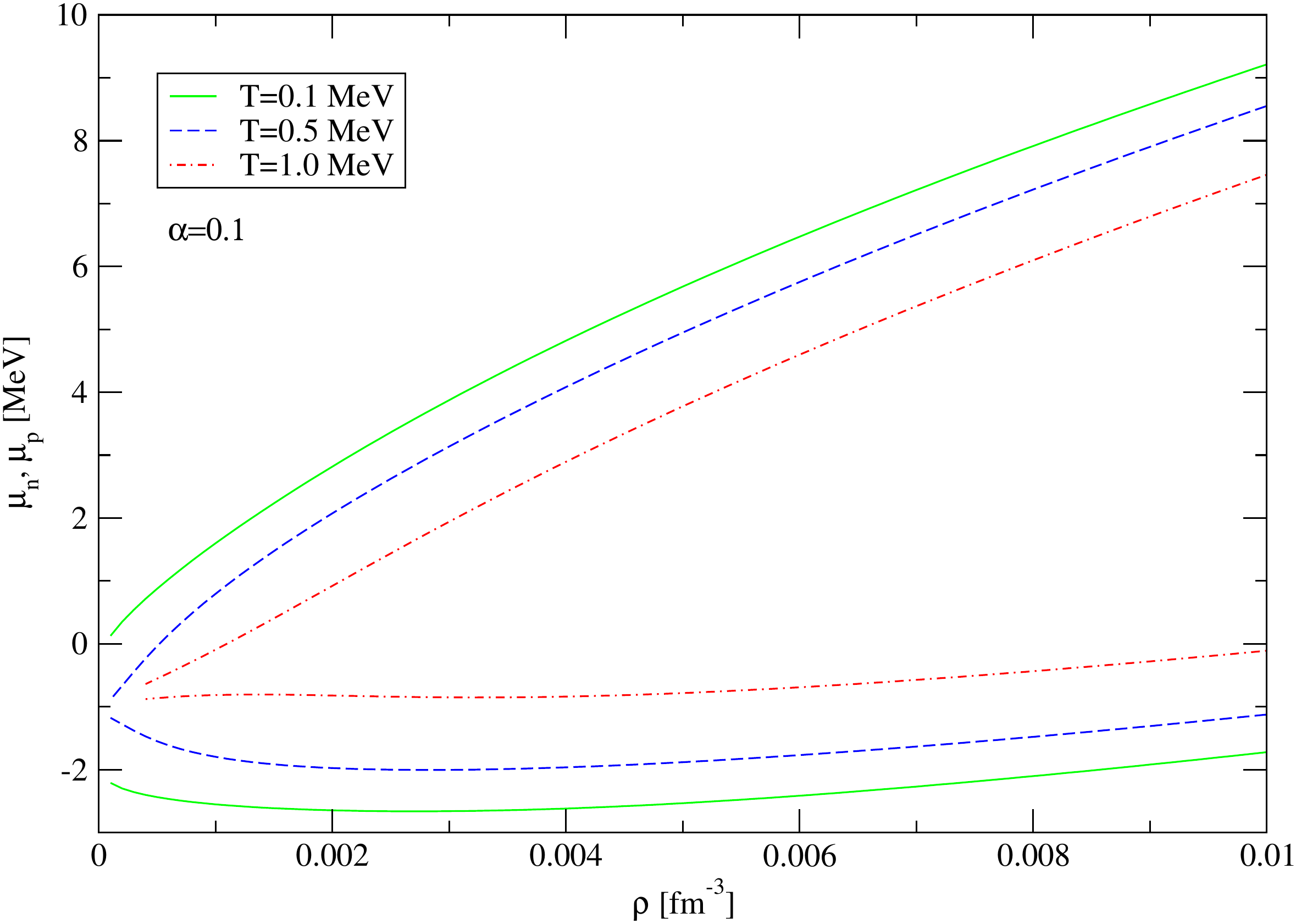}
    \caption[Chemical potentials of neutrons and protons for several
    temperatures in the BCS/BEC phase.]{Chemical potentials of
      neutrons and protons in the BCS/BEC phase as a function of the
      density at $\alpha=0.1$ for several temperatures.}
    \label{fig_1_10}
  \end{center}
\end{figure}

Fig.~\ref{fig_1_10} shows the chemical potentials of protons and
neutrons as a function of density for a fixed asymmetry of
$\alpha=0.1$ for several values of the temperature. We see that the
separation of proton and neutron chemical potentials decreases with
decreasing density and with increasing temperature. In both cases the
distributions of neutrons and protons are smeared out, Pauli blocking
is less effective and the difference of the chemical potentials is
also smeared out.

\subsection{Effects of finite momentum in the LOFF phase}
\label{subsec_1_3_4}
\begin{figure}[!]
  \begin{center}
    \includegraphics[width=0.6\textwidth]{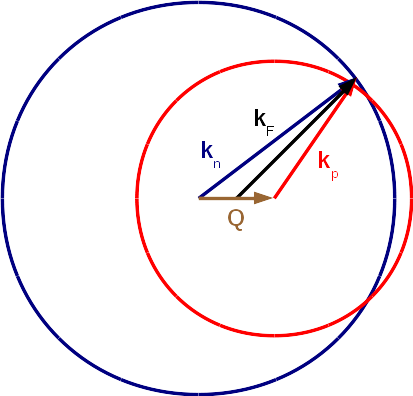}
    \caption[Illustration of the mechanism of phase-space restoration
    by the LOFF phase.]{The purpose of this figure is to illustrate
      the mechanism of phase-space restoration by the LOFF phase. The
      centers of the neutron and proton Fermi surfaces are shifted by
      $\vecQ$. The neutron and proton Fermi surfaces are shown by a
      blue or red cycle, respectively. We show the momenta of neutrons
      and protons for $\theta=45\degree$ which are constructed
      according to $\veck_{n} = \veck_F+\vecQ/2$ and
      $\veck_{p} = \veck_F-\vecQ/2$. These are drawn towards a point
      where the Fermi surfaces intersect and there is a maximal phase
      space overlap. }
    \label{fig_1_11}
  \end{center}
\end{figure}

A phase-space overlap between the members of a Cooper-pair is required
for pairing. Increasing the asymmetry shifts the Fermi momenta of
neutrons and protons apart. BCS pairing at finite asymmetry thus
requires smearing out of Fermi surfaces, which then provides the
needed phase-space overlap. The overlap is large at high temperatures
and low densities. Similar effect of restoration of phase-space
overlap can be achieved if a total Cooper-pair momentum $\vecQ$ is
allowed, as is the case in the LOFF phase. The shift of the
Fermi-surfaces due to finite $\vecQ$ which restores pairing
correlations in the limit of high densities, low temperatures and
large asymmetries, is illustrated in Fig.~\ref{fig_1_11}.

Fig.~\ref{fig_1_11} illustrates the mechanism of phase-space
restoration by the LOFF phase. In the case of high densities, low
temperatures and finite asymmetry, pairing with finite $\vecQ$ is
energetically favorable, because the negative pairing energy
compensates the positive kinetic energy of motion of Cooper-pairs. The
momenta of protons are shown in red and the ones of neutrons in
blue. The Cooper-pair momentum $\vecQ$ describes the shift of the
centers of the Fermi spheres. The relative momentum of the pairs at
the Fermi surface $\veck_F$ is shown for the angle $45\degree$. The
corresponding neutron momentum is then given as $\veck_F+\vecQ/2$ (in
blue) and that of the proton is given by $\veck_F-\vecQ/2$ (in
red). By construction the sum of the momenta is such that
$(\veck_F+\vecQ/2)+(-\veck_F+\vecQ/2)=\vecQ$. Note that we show the
case where the Fermi-surfaces intersect and the overlap is optimal for
pairing.

\begin{figure}[!]
  \begin{center} \includegraphics[width=\textwidth]{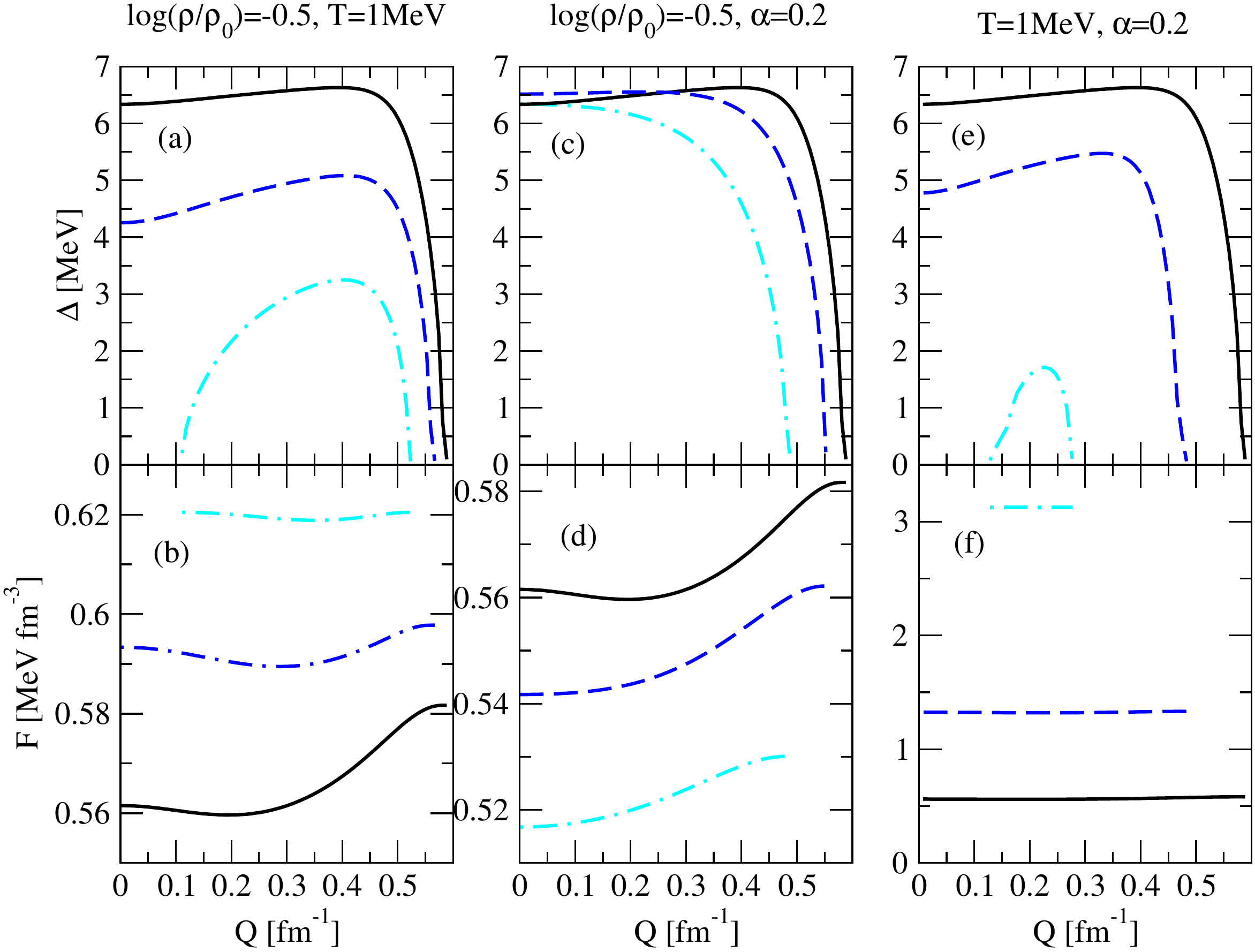} \caption[Properties
    of the nuclear LOFF phase. Pairing gaps and the corresponding free
    energies are shown as functions of the total momentum of a
    Cooper-pair.]{Properties of the nuclear LOFF phase. The upper
      panel shows the pairing gaps and the lower panel shows the free
      energies as a function of the total momentum $\vecQ$ of a
      Cooper-pair. \\
      In (a) and (b) the density is fixed at
      $\log(\rho/\rho_0) = -0.5$ and the temperature is fixed at $T=1$
      MeV, the asymmetries are:\\
      $\left.\right.\qquad\alpha=0.2$
      (black, solid),\\
      $\left.\right.\qquad\alpha=0.3$ (blue, dashed)
      and \\
      $\left.\right.\qquad\alpha=0.4$
      (cyan, dash-dotted). \\
      In (c) and (d) the density is fixed at
      $\log(\rho/\rho_0) = -0.5$, the asymmetry is fixed at $\alpha =
      0.2$, the temperatures are:\\
      $\left.\right.\qquad$1 MeV (black,
      solid),\\
      $\left.\right.\qquad$2
      MeV (blue, dashed) and\\
      $\left.\right.\qquad$3 MeV (cyan,
      dash-dotted). \\
      In (e) and (f) the temperature is fixed at $T=1$ MeV, the
      asymmetry is fixed at $\alpha = 0.2$, the densities
      are:\\
      $\left.\right.\qquad\log(\rho/\rho_0) = -0.5$ (black,
      solid),\\
      $\left.\right.\qquad\log(\rho/\rho_0) = -0.3$ (blue,
      dashed) and \\
      $\left.\right.\qquad\log(\rho/\rho_0) = -0.1$ (cyan,
      dash-dotted). }
    \label{fig_1_12}
  \end{center}
\end{figure}

Fig.~\ref{fig_1_12} shows the gap and the free energy for several
densities, temperatures and asymmetries. We see that the maximum of
the gap and the minimum of the free energy are at finite values of
$\vecQ$ at high density, high asymmetry or low temperature. In
particular, the gap at vanishing $\vecQ$ vanishes for high asymmetry
or high density. At high temperatures, low asymmetries or low
densities, we expect the translational symmetric BCS phase to be
favored over the LOFF phase. By introducing the effective chemical
potential $\tilde{\bar\mu}=(\bar\mu-Q^2/8m^*)$ we obtain
\begin{eqnarray}
  E_S &=&k^2/2m^*-\tilde{\bar\mu}.
\end{eqnarray}
Thus, the non-zero total momentum implies that the average chemical
potential of the BCS phase $\bar\mu$ is reduced.

\subsection{The kernel of the gap equation}
\label{subsec_1_3_5}
\begin{figure}[!]
  \begin{center}
    \includegraphics[width=0.7\textwidth]{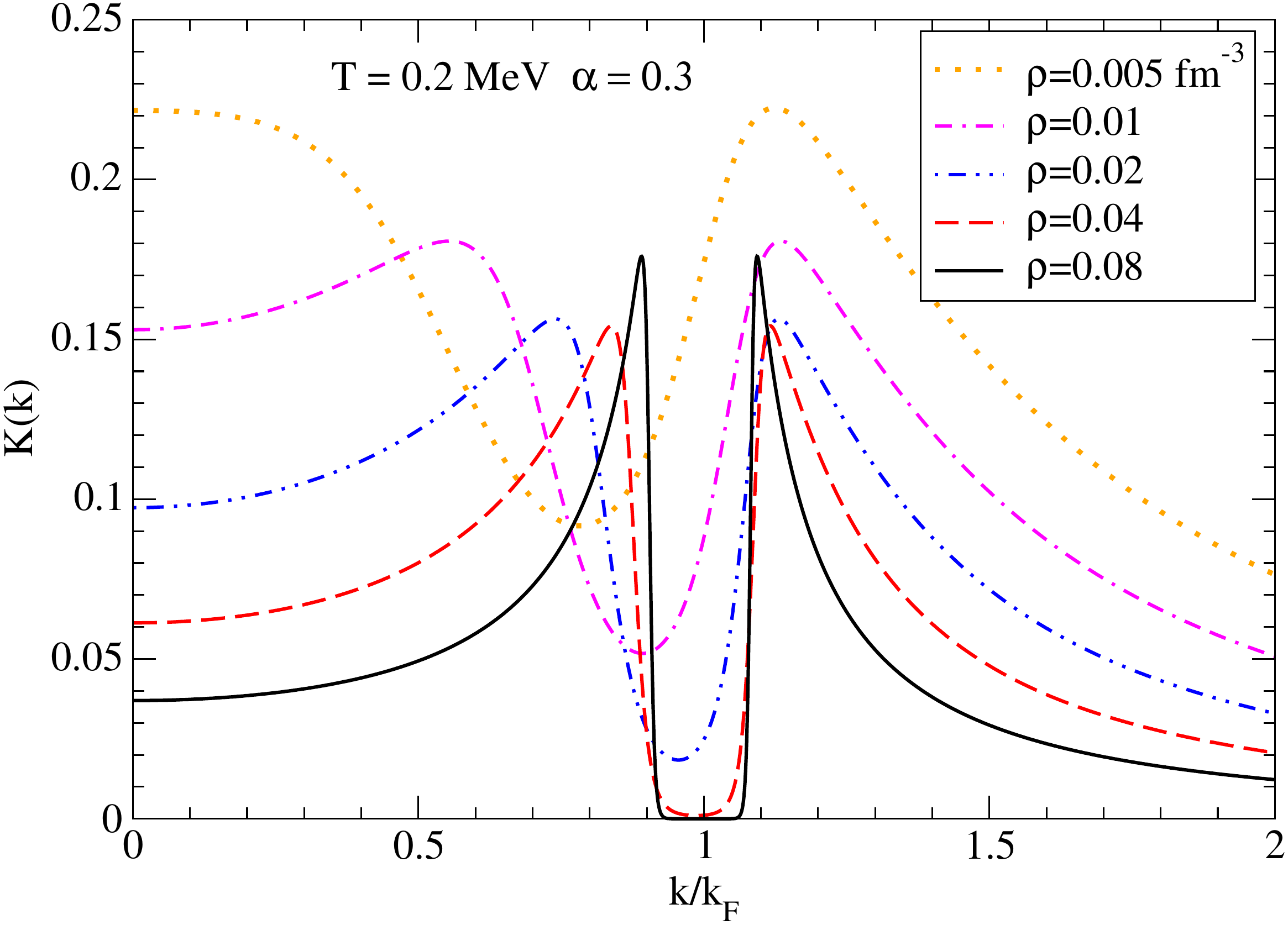}
    \caption[Dependence of the kernel on momentum for various
    densities.]{Dependence of the kernel $K(k)$ on momentum in units
      of Fermi momentum for fixed $T=0.2$ MeV, $\alpha = 0.3$, and
      various densities indicated in the plot. }
    \label{fig_1_13}
  \end{center}
\end{figure}

\begin{figure}[!]
  \begin{center}
    \includegraphics[width=0.7\textwidth]{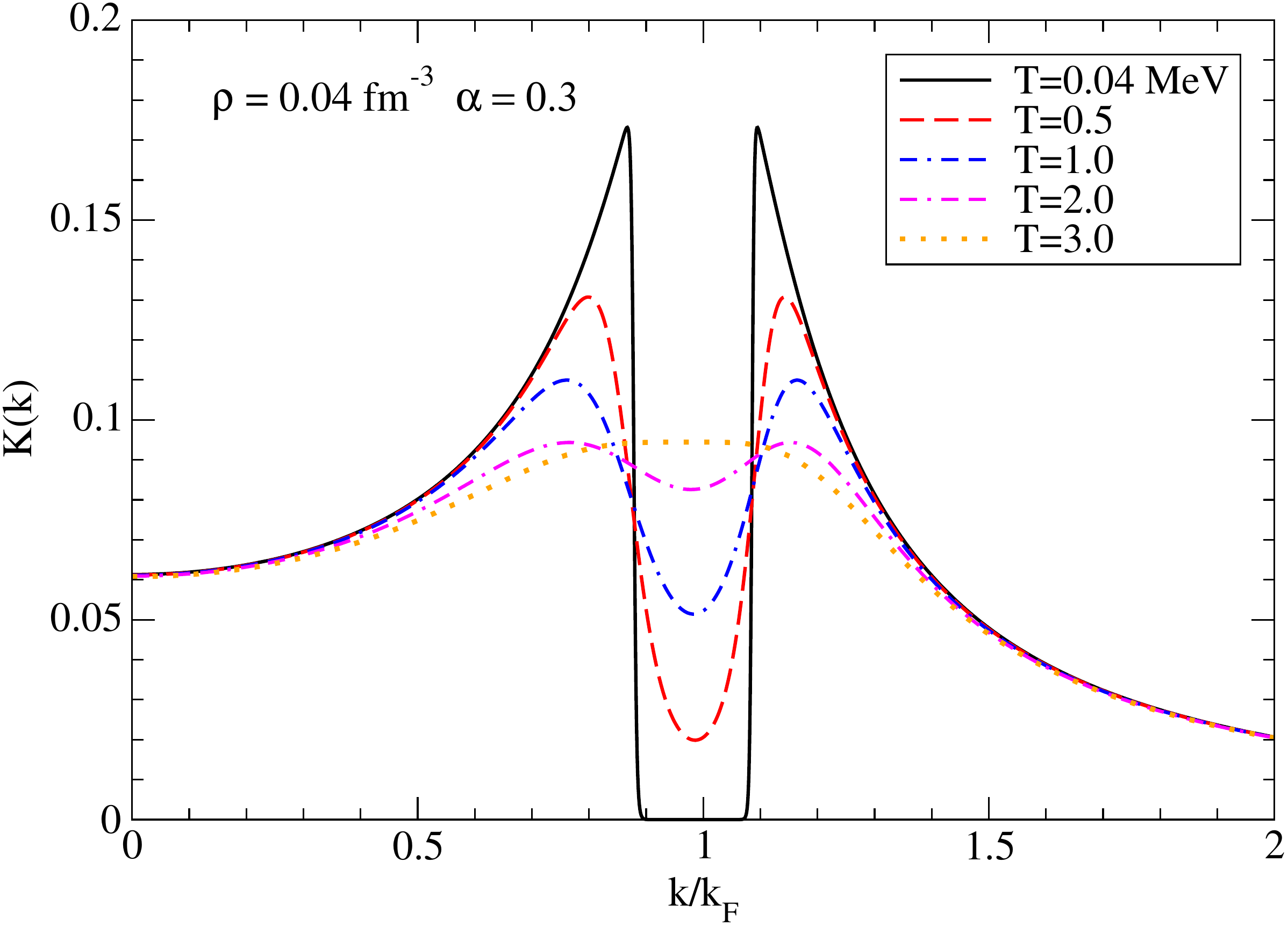}
    \caption[Dependence of the kernel on momentum for various
    temperatures.]{Dependence of the kernel $K(k)$ on momentum in
      units of Fermi momentum for fixed $\rho=0.04$ fm$^{-3}$,
      $\alpha=0.3$, and various temperatures indicated in the plot. }
    \label{fig_1_14}
  \end{center}
\end{figure}

\begin{figure}[!]
  \begin{center}
    \includegraphics[width=0.7\textwidth]{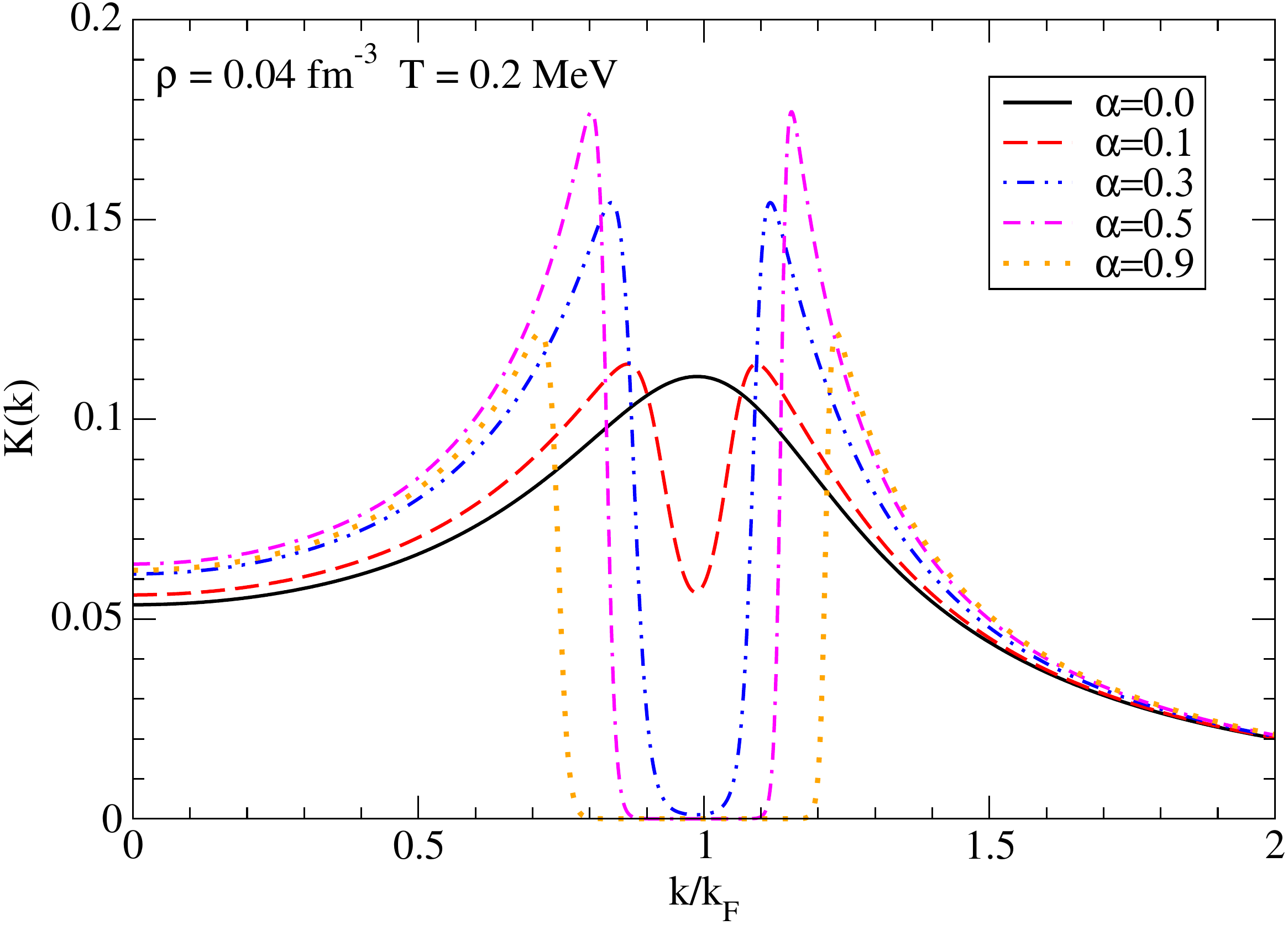}
    \caption[Dependence of the kernel on momentum for various
    asymmetries.]{Dependence of the kernel $K(k)$ on momentum in units
      of Fermi momentum for fixed $\rho=0.04$ fm$^{-3}$, $T=0.2$ MeV,
      and various values of asymmetry indicated in the plot. }
    \label{fig_1_15}
  \end{center}
\end{figure}

\begin{figure}[!]
  \begin{center}
    \includegraphics[width=0.7\textwidth]{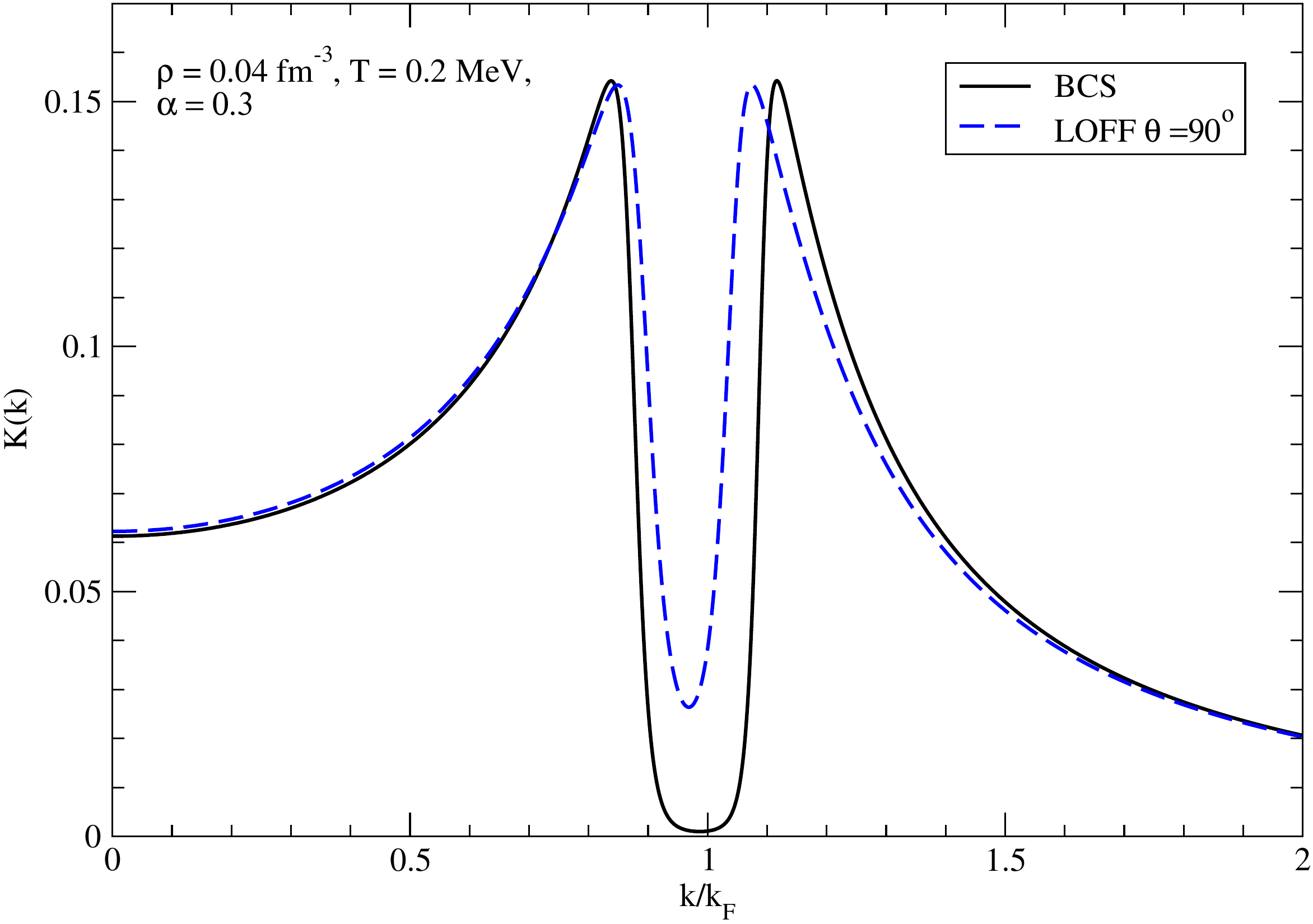}
    \caption[Dependence of the kernel on momentum in the BCS and the
    LOFF phase.]{Dependence of the kernel $K(k)$ on momentum in units
      of Fermi momentum at fixed $\rho=0.04$~fm$^{-3}$, $T=0.2$~MeV,
      and $\alpha=0.3$ for the BCS phase and the LOFF phase at
      $\theta=90\degree$, where $\theta$ is the angle formed by the CM
      and relative momenta in Eq.~\eqref{eq_1_53}.}
    \label{fig_1_16}
  \end{center}
\end{figure}

\begin{figure}[!]
  \begin{center}
    \includegraphics[width=0.7\textwidth]{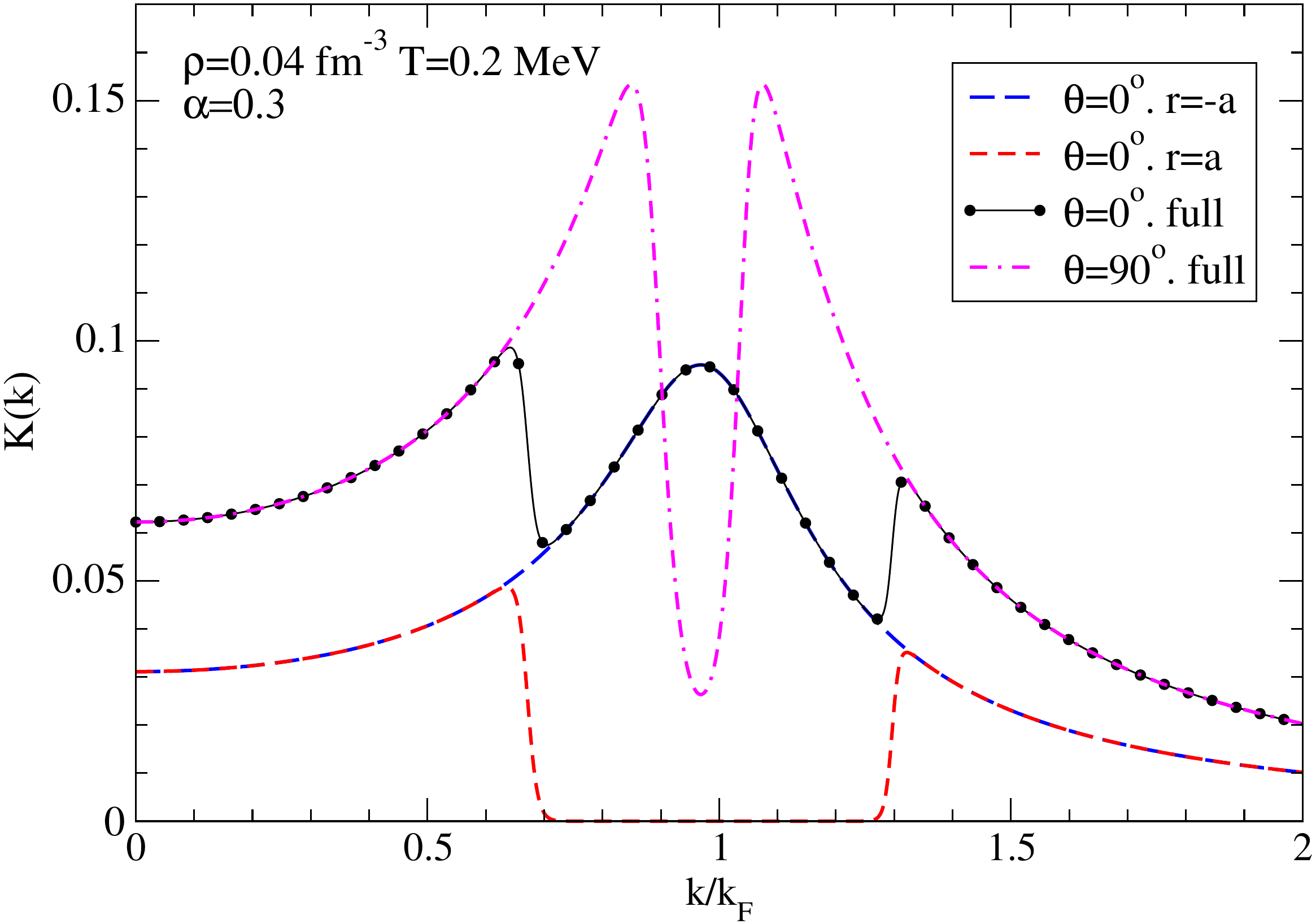}
    \caption[Dependence of the kernel on momentum in the LOFF phase
    for different angles.]{Dependence of the kernel $K(k)$ on momentum
      in units of Fermi momentum at fixed $\rho=0.04$~fm$^{-3}$,
      $T=0.2$~MeV, and $\alpha=0.3$ for the LOFF phase, where $\theta$
      is the angle formed by the CM and relative momenta in
      Eq.~\eqref{eq_1_53}. In the case $\theta= 0\degree$ the full
      result (black, solid-filled-circle line) is decomposed into
      components with $r=-a$ (blue, long dashed line) and $r=a$ (red,
      short dashed line). }
    \label{fig_1_17}
  \end{center}
\end{figure}

We start our study of the intrinsic quantities with the kernel of the
gap equation,
\begin{eqnarray}
  \label{eq_1_53} K(k,\theta) &\equiv&\sum_{a,r}\frac{P^{a}_r}{4\sqrt{E_{S}^2(k)+\Delta^2(k,Q)}}.
\end{eqnarray}
This kernel is proportional to the imaginary part of the retarded
anomalous propagator and the Pauli operator represented by
$P_r^a = 1-2f(E^a_r)$. Physically, $K(k)$ can be interpreted as the
wave function of the Cooper-pairs, since it obeys a Schr\"odinger-type
eigenvalue equation in the limit of extremely strong coupling. The
Pauli operator is a smooth function of the momentum having a minimum
at the Fermi surface, where $E_S$ vanishes in the limit of
weak-coupling. In Figs.~\ref{fig_1_13}-\ref{fig_1_17} we present the
kernel for several values of density, temperature and asymmetry as a
function of the momentum. When studying the variation with density,
temperature or asymmetry we fix the remaining quantities at the
following values $\rho=0.04$ fm$^{-3}$, $T=0.2$ MeV, and $\alpha=0.3$.
These values correspond to the BCS region in all figures where the
density is fixed. The ranges of momenta which contribute substantially
to the gap equation in different regimes of the phase diagram can be
identified from these figures. We now discuss the insights that can be
gained from these figures in some detail.

Fig.~\ref{fig_1_13} shows the function $K(k)$ at constant temperature
and asymmetry for various densities. The high densities correspond to
the BCS regime, and the low densities to the BEC regime, allowing us
to follow the evolution of this function through the BCS-BEC
crossover. In the BCS regime, $K(k)$ has two sharp maxima which are
separated by a depression of width ∼ $\delta \mu$ around the Fermi
momentum. The lower maximum is located at the Fermi momentum of
protons, whereas the upper maximum at the Fermi momentum of
neutrons. As discussed in Subsec.~\ref{subsec_1_3_7} below, this
feature originates from the Pauli operator. Because of their strong
localization in momentum space, the Cooper-pairs have an intrinsic
structure that is broad in real space, implying a large coherence
length. This is characteristic of the BCS regime. The picture is
reversed in the strong-coupling (low-density) limit, where $K(k)$ is a
broad function of momentum, corresponding to the presence of bound
states (deuterons), which are well-localized in real space. This is
characteristic of the BEC regime. In addition, as the density
decreases, the lower (proton) peak moves toward $k = 0$ and also the
minimum moves away from $k=k_F$ towards lower momenta, due to the fact
that $\bar\mu$ changes its sign from positive to negative at the
transition from the BCS to the BEC regime. As a consequence, the
prefactor of the Pauli operator $P^r_a $ peaks at $k = 0$ in the BEC
regime, rather than at the Fermi surface as in the BCS regime. In
addition, one can clearly see a smearing of the two peaks of the
kernel with decreasing density.

Fig.~\ref{fig_1_14} shows the function $K(k)$ for various
temperatures, now at constant asymmetry and constant density, such
that the system is situated in the BCS regime. At low temperatures,
$K(k)$ is seen to have two maxima separated by a depression around the
Fermi momentum, as already discussed above. Increasing the temperature
smears out the structures characteristic of the low-temperature case,
due to temperature-induced blurring of the Fermi surface. Close to
$T_c$, the temperature effects dominate over the effects of
asymmetry. Consequently, the double-peak structure disappears and the
isospin asymmetry does not affect the properties of the condensate.

Fig.~\ref{fig_1_15} shows the function $K(k)$ for various asymmetries
at constant temperature and the same density as above (thus again
implying the BCS regime). We can now follow how the double
peak-structure builds up as the asymmetry is increased. Because the
width of the depression is proportional to $\delta\mu$, it increases
with increasing isospin asymmetry, a behavior consistent with the
facts that the Fermi surfaces of neutrons and protons are pulled apart
by the isospin asymmetry, and that in the BCS regime the available
phase space is constrained to the vicinity of the corresponding Fermi
surface. (See also the Subsec.~\ref{subsec_1_3_3} for a discussion of
the Fermi momenta.)

Finally, in Figs.~\ref{fig_1_16} and \ref{fig_1_17} we show $K(k)$ for
fixed values of temperature, asymmetry, and density, in
Fig.~\ref{fig_1_16} for the BCS phase and the LOFF phase at
$\theta=90\degree$ and in Fig.~\ref{fig_1_17} for the LOFF phase at
two values of the angle formed by the relative and CM momenta, as
defined in Eq.~\eqref{eq_1_53}. It is seen from Fig.~\ref{fig_1_16}
that in the orthogonal case ($\theta = 90\degree$) the double-peak
structure present in the BCS phase remains, although the effects of
asymmetry are weaker compared to the BCS case. This is easily
understood by noting that $E_A = 0$ for $\theta = 90\degree$,
therefore finite momentum induces only a shift in the energy origin
according to $\bar\mu \to \bar \mu - Q^2/8m^*$.  The case
$\theta = 0\degree$ in Fig.~\ref{fig_1_17} exposes an interesting
feature of the LOFF phase: for a range of orientations of the CM
momentum of Cooper-pairs ($\theta \approx 0\degree$), the effects of
asymmetry are mitigated and the kernel obtains a maximum at
$k/k_F = 1$, which is a combination of the contribution from $r=-a$
which acts to enhance the pairing correlations in the vicinity of the
Fermi surface and the $r = a$ contribution which vanishes in this
region.

\subsection{The Cooper-pair wave function across the BCS-BEC
  crossover}
\label{subsec_1_3_6}
The transition to the BEC regime of strongly-coupled neutron-proton
pairs, which are asymptotically identical with deuterons, occurs at
low densities. The criterion for the transition from BCS to BEC is
that either the average chemical potential $\bar \mu$ changes its sign
from positive to negative values, or the coherence length $\xi$ of a
Cooper-pair becomes comparable to the interparticle distance, i.e.,
$\xi$ becomes of order $d\sim \rho^{-1/3}$. (In the BCS regime
$\xi \gg d$, whereas in the BEC regime $\xi \ll d$).

The coherence length can be related to the root-mean-square of the
Cooper-pair wave function, as we show below. The wave function of a
Cooper-pair is defined in terms of the kernel of the gap equation
according to
\begin{eqnarray}
  \label{eq_1_54}
  \Psi(\vecr) = \sqrt{N} \int
  \frac{d^3p}{(2\pi)^3}
  [K(\vecp,\Delta)-K(\vecp,0)]e^{i\vecp\cdot\vecr}\,,
\end{eqnarray}
which, after integration over angles, becomes
\begin{eqnarray}
  \Psi(r)&=& \frac{\sqrt{N}}{2\pi^2r} \int_0^\infty dp\,p\,[K(p,\Delta)-K(p,0)]\sin(pr)\,,
\end{eqnarray}
where $N$ is a constant determined by the normalization condition
\begin{equation}
  N\int d^3r \vert
  \Psi(\vecr)\vert^2 =
  1\,.
\end{equation}
In Eq.~\eqref{eq_1_54} we subtract from the kernel its value
$K(\vecp,0)$ in the normal state to regularize the integral, which is
otherwise divergent. Cut-off regularization of this strongly
oscillating integral is not appropriate. The mean-square radius of a
Cooper-pair is defined via the second moment of the probability
density,
\begin{equation}
  \langle
  r^2\rangle = \int d^3r\, r^2
  \vert \Psi(\vecr)\vert^2\,.
\end{equation}
The coherence length, i.e., the spatial extension of a Cooper-pair, is
then defined as
\begin{equation}
  \xi_{\rm rms} =
  \sqrt{\langle
    r^2\rangle}\,.
\end{equation}
Thus the change in the coherence length is related to the change of
the condensate wave function across the BCS-BEC crossover. The regimes
of strong and weak coupling can be identified by comparing the
coherence length to the mean interparticle distance
$d = (3/(4\pi \rho))^{1/3}$. In the BCS regime the coherence length is
given by the well-known analytical formula
\begin{eqnarray}
  \xi_a =
  \frac{\hbar^2 k_F}{\pi m^*
  \Delta}\,. \label{eq_1_59}
\end{eqnarray}
Table \ref{table_1_1} lists the analytical and root-mean-square values
of the coherence length for several densities and temperatures, chosen
to represent the different regimes WCR, ICR, and SCR, together with
the corresponding values of the mean interparticle distance. It is
seen that in the case of neutron-proton pairing, one of the criteria
for the BCS-BEC transition is fulfilled, namely, the mean distance
between the pairs becomes larger than the coherence length of the
superfluid as one goes from WCR to SCR for the numerical computed
coherence length $\xi_\mathrm{rms}$. This is not the case for the
analytical expression $\xi_a$. Thus one should rely only on the
numerical value $\xi_\mathrm{rms}$. We have verified that the average
chemical potential changes its sign accordingly, so that the second
criterion is fulfilled as well.
\begin{table}
  \begin{tabular}{ccccccc}
    \hline & log$\left(\frac{\rho}{\rho_0}\right)$ & $k_F$[fm$^{-1}$] &$T$ [MeV] & $d$ [fm] & $\xi_{\rm rms}$ [fm] & $\xi_{a}$ [fm] \\
    \hline\hline WCR & $-0.5$ &0.91& 0.5 & 1.68 & 3.17 & 1.41 \\
    ICR & $-1.5$ & 0.42& 0.5 & 3.61 & 0.94 & 1.25 \\
    SCR & $-2.5$ & 0.20 & 0.2 & 7.79 & 0.57 & 1.79 \\
    \hline\\
  \end{tabular}
  \caption{For each of the three regimes of coupling strength, corresponding values are presented for the density $\rho$ (in units of nuclear saturation density $\rho_0 = 0.16$ fm$^{-3}$), Fermi momentum $k_F$, temperature $T$, interparticle distance $d$, and coherence parameters $\xi_{\rm rms}$ and $\xi_{a}$. The values of the gap and effective mass (in units of bare mass) at $\alpha = 0$ in these three regimes are 9.39, 4.50, 1.44 MeV and 0.903, 0.989, 0.999, respectively. In the WCR, the LOFF phase is found in the vicinity of asymmetry $\alpha = 0.49$, for which $\Delta = 1.27$ MeV and $Q = 0.40$ fm$^{-1}$.}
  \label{table_1_1}
\end{table} 

\begin{figure}[!]
  \begin{center}
    \includegraphics[width=0.7\textwidth]{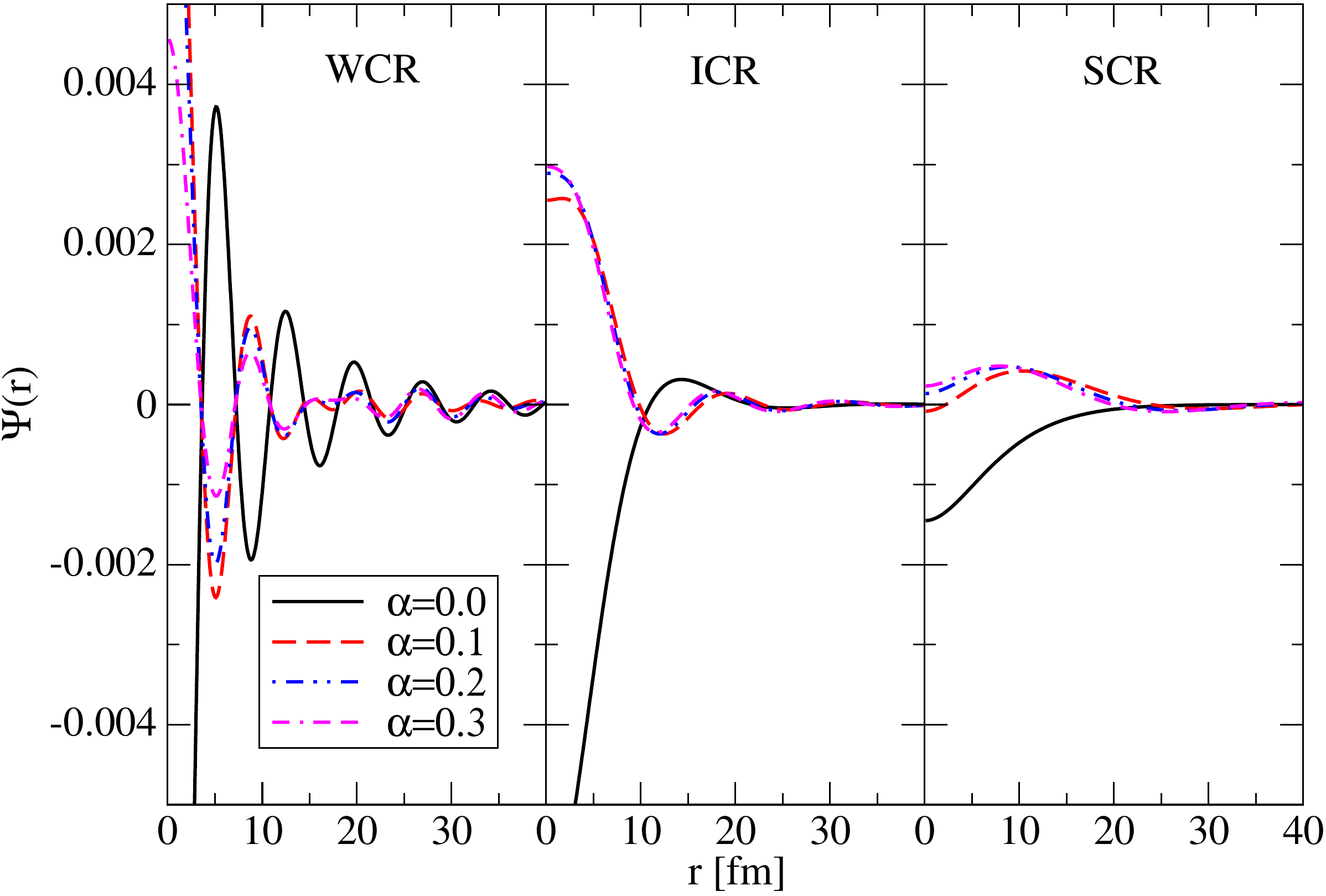}
    \caption[Dependence of $\Psi(r)$ on $r$ for the three coupling
    regimes.]{Dependence of $\Psi(r)$ on $r$ for the three coupling
      regimes and various values of asymmetry (see Table
      \ref{table_1_1} for values of density and temperature).}
    \label{fig_1_18}
  \end{center}
\end{figure}
\begin{figure}[!]
  \begin{center}
    \includegraphics[width=0.7\textwidth]{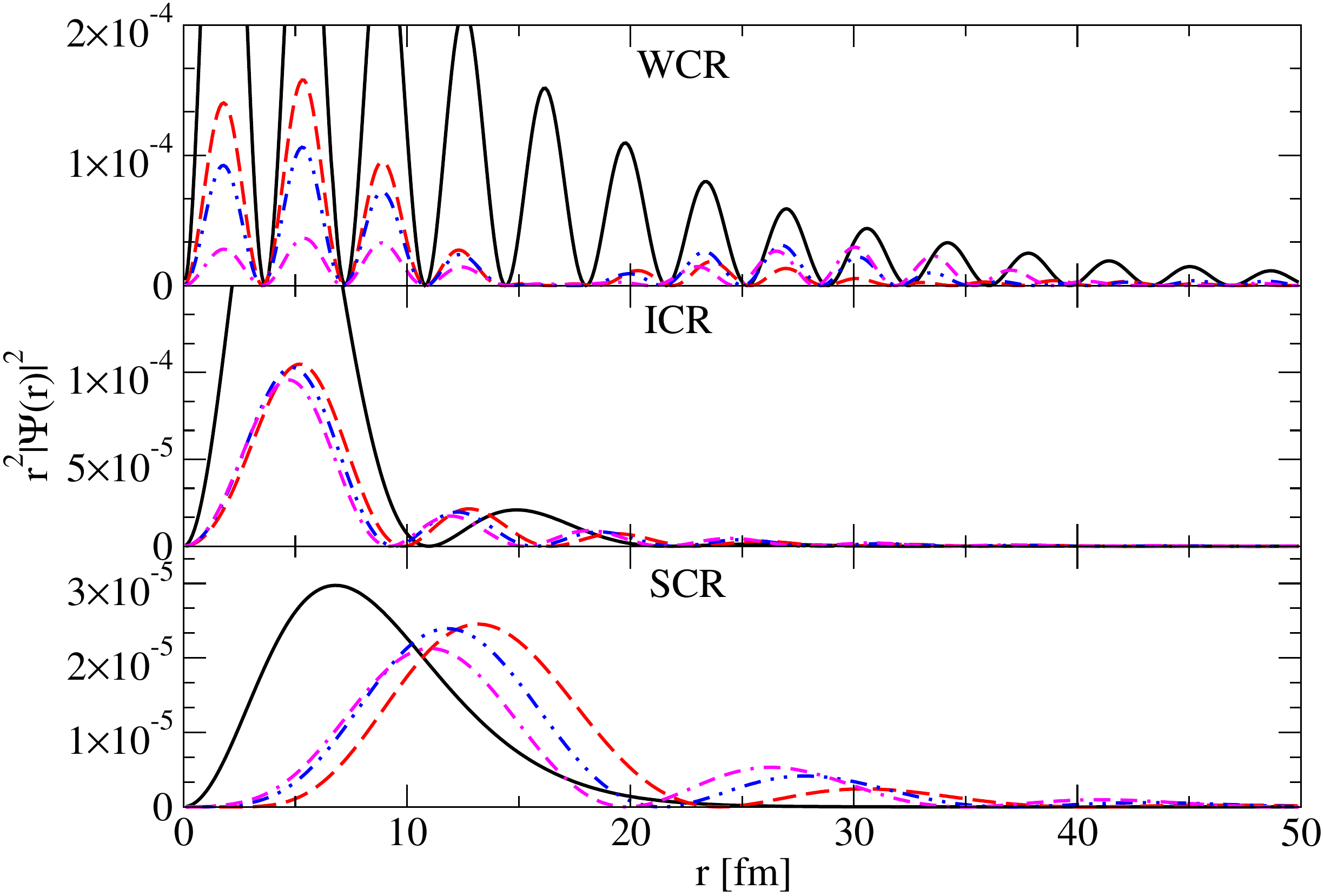}
    \caption[Dependence of $r^2|\Psi(r)|^2$ on $r$ for the three
    coupling regimes.]{Dependence of $r^2|\Psi(r)|^2$ on $r$ for the
      three coupling regimes. Conventions are the same as in
      Fig.~\ref{fig_1_18}.}
    \label{fig_1_19}
  \end{center}
\end{figure}
\begin{figure}[!]
  \begin{center}
    \begin{minipage}[t]{\textwidth}
      \begin{center}
        \includegraphics[width=0.8\textwidth]{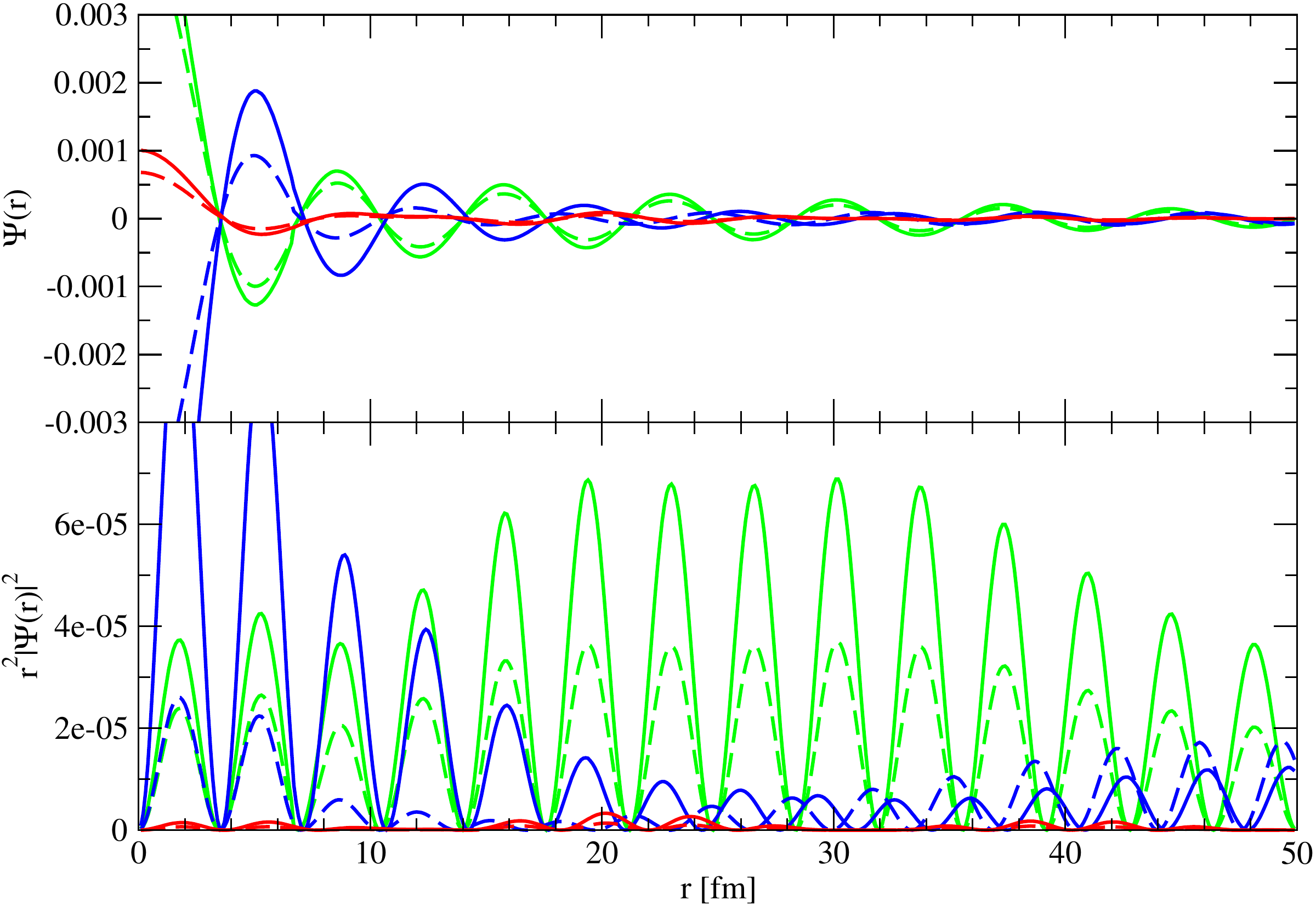}
      \end{center}
    \end{minipage}
    \newline\newline
    \begin{minipage}[t]{\textwidth}
      \begin{center}
        \includegraphics[width=0.8\textwidth]{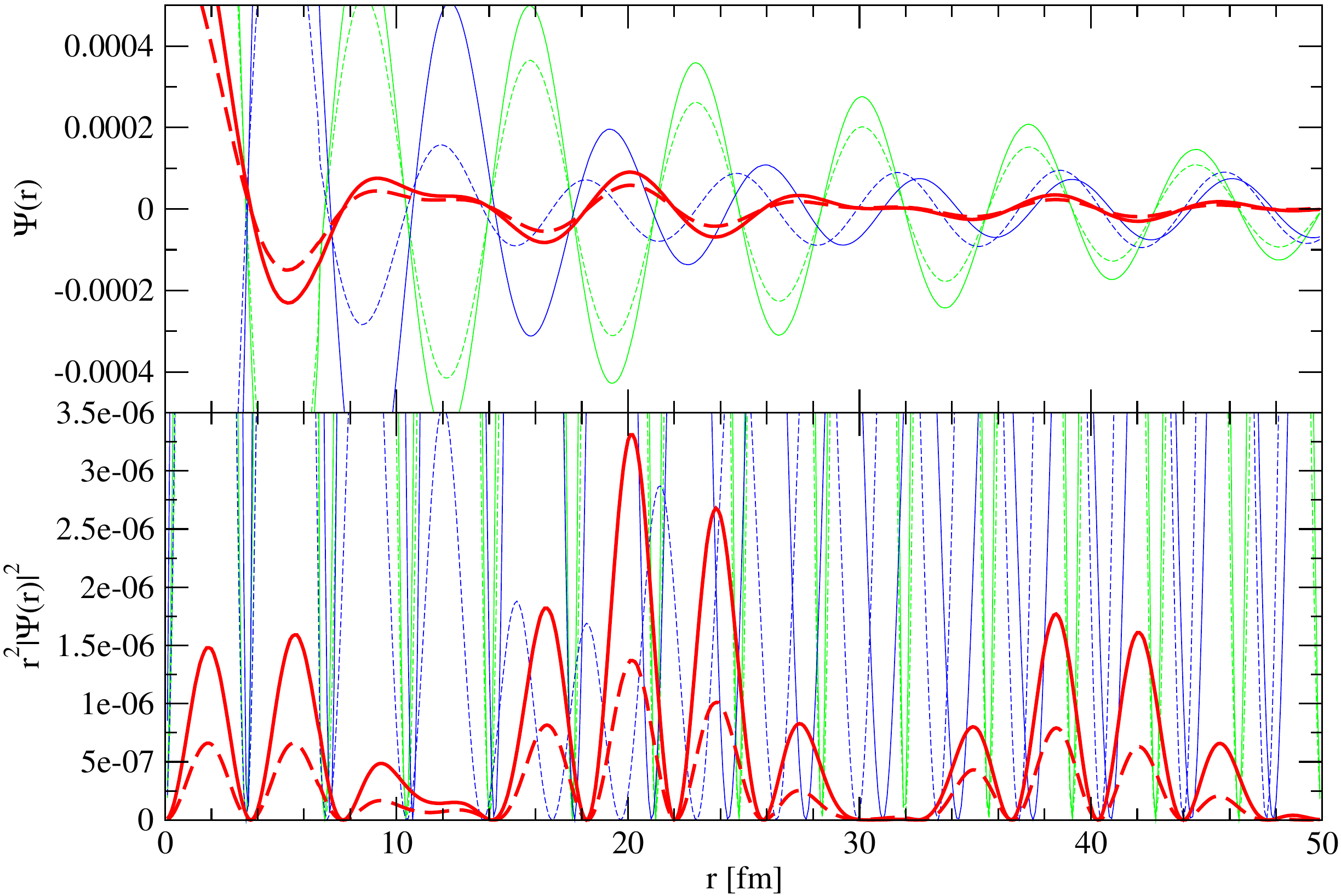}
      \end{center}
    \end{minipage}
    \caption[Dependence of $\Psi(r)$ and $r^2|\Psi(r)|^2$ on $r$ in
    the LOFF phase in the WCR for three different angles and two
    different asymmetries.]{Dependence of $\Psi(r)$ and
      $r^2|\Psi(r)|^2$ on $r$ in the WCR for three different angles
      $\theta$: green $\theta=0\degree$, blue $\theta=45\degree$ and
      red $\theta=90\degree$ for two different asymmetries: solid
      lines for $\alpha = 0.49$ and dashed lines for $\alpha=0.50$ at
      which the LOFF phase is the ground state. The figures only
      differ in the scale and show the results for $\theta=0\degree$
      and $\theta=45\degree$ (top figure) and $\theta=90\degree$
      (bottom figure).}
    \label{fig_1_20}
  \end{center}
\end{figure}

Fig.~\ref{fig_1_18} shows the wave function of Cooper-pairs as a
function of radial distance across the BCS-BEC crossover for various
densities. In weak coupling, the wave function has a well-defined
oscillatory form that extends over many periods of the interparticle
distance. Such a state conforms to the familiar BCS picture, in which
the spatial correlations are characterized by scales that are much
larger than the interparticle distance. We clearly see a decrease of
the amplitude with increasing asymmetry, which is correlated with the
observation that the asymmetry decreases the gap. For intermediate and
strong coupling the wave function is increasingly concentrated at the
origin with at most a few periods of oscillation. The strong-coupling
limit corresponds to pairs that are well localized in space within a
small radius. This regime clearly has BEC character, with the pair
correlations extending only over distances comparable to the
interparticle distance. At large distances the asymmetry does not
change the shape of the wave function significantly. However, at small
distances the changes are significant. In the SCR $\Psi(\vecr)$ has,
for vanishing asymmetry, a minimum at $r=0$ and reaches asymptotically
$\Psi(\vecr)=0$ for $r\rightarrow\infty$. Increasing the asymmetry
increases the small-distance values leading to a maximum at
nonvanishing $r$. The function $\Psi(\vecr)$ in the ICR and at
vanishing asymmetry starts at a large negative value and oscillates
only once to a maximum. At nonvanishing asymmetry $\Psi(\vecr)$ starts
at a large positive value which is followed by oscillatory
behavior. The first minimum at nonvanishing asymmetry is at lower $r$
than the first maximum at vanishing asymmetry. In the WCR we find a
regularly oscillating shape at vanishing asymmetry with the amplitude
vanishing for large $r$. For nonvanishing asymmetry we basically have
two segments. In the large $r$ segment the symmetric and asymmetric
condensate wave-functions oscillate in-phase. For small $r$ the
oscillations are counter-phase. The transition between these segments
is characterized by $\Psi(\vecr)\approx 0$. As a general trend we find
that the wave function is almost independent of the asymmetry in the
WCR. In the SCR there are substantial changes as soon as asymmetry is
switched on.
\begin{figure}[!]
  \begin{center}
    \includegraphics[width=0.8\textwidth]{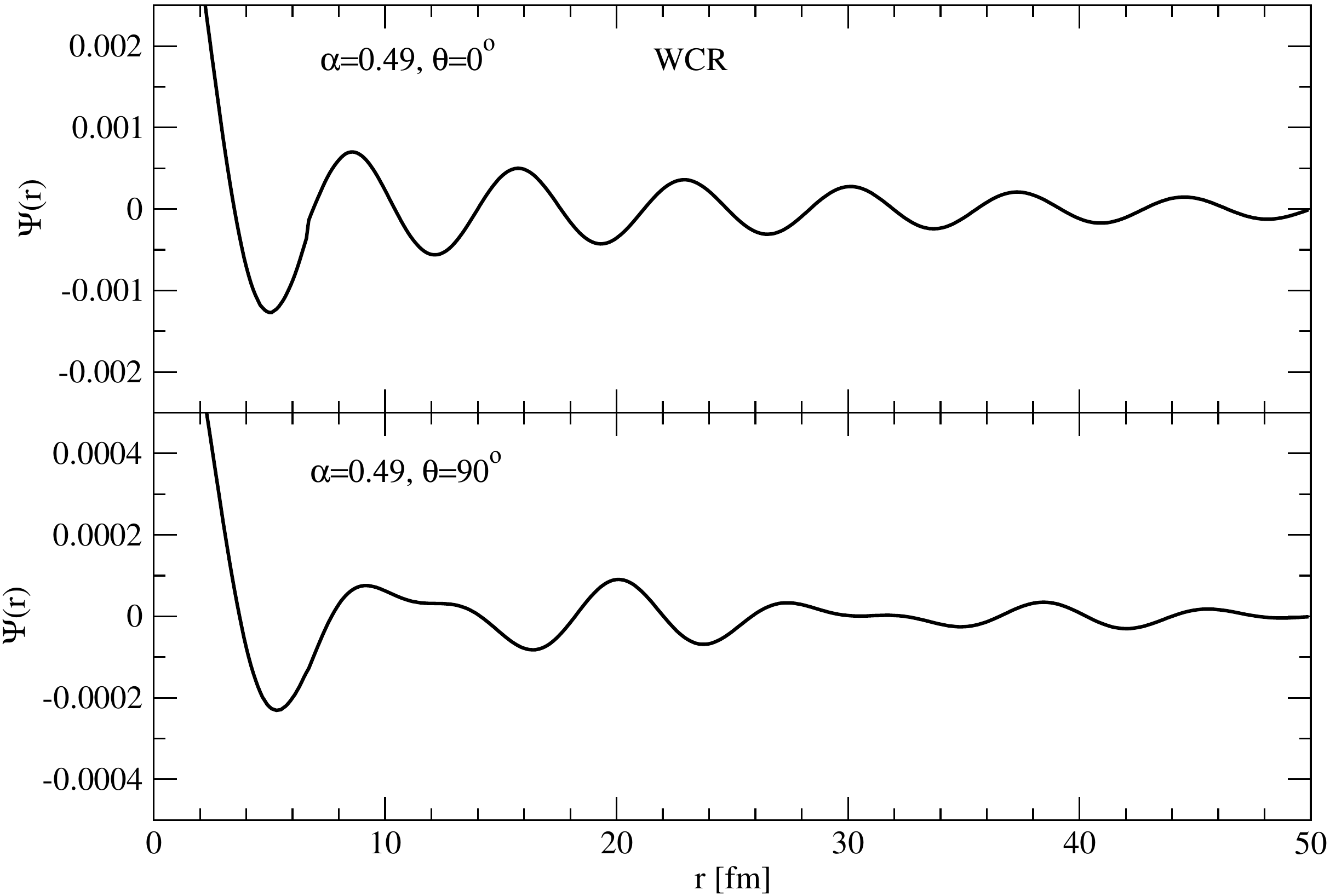}
    \caption[Dependence of $\Psi(r)$ on $r$ in the LOFF phase in the
    WCR for two different angles and one asymmetry.]{Dependence of
      $\Psi(r)$ on $r$ in the WCR for two different angles $\theta$
      and for asymmetry $\alpha = 0.49$ at which the LOFF phase is the
      ground state. (Cutout of Fig.~\ref{fig_1_20})}
    \label{fig_1_21}
  \end{center}
\end{figure}
\begin{figure}[!]
  \begin{center}
    \includegraphics[width=0.8\textwidth]{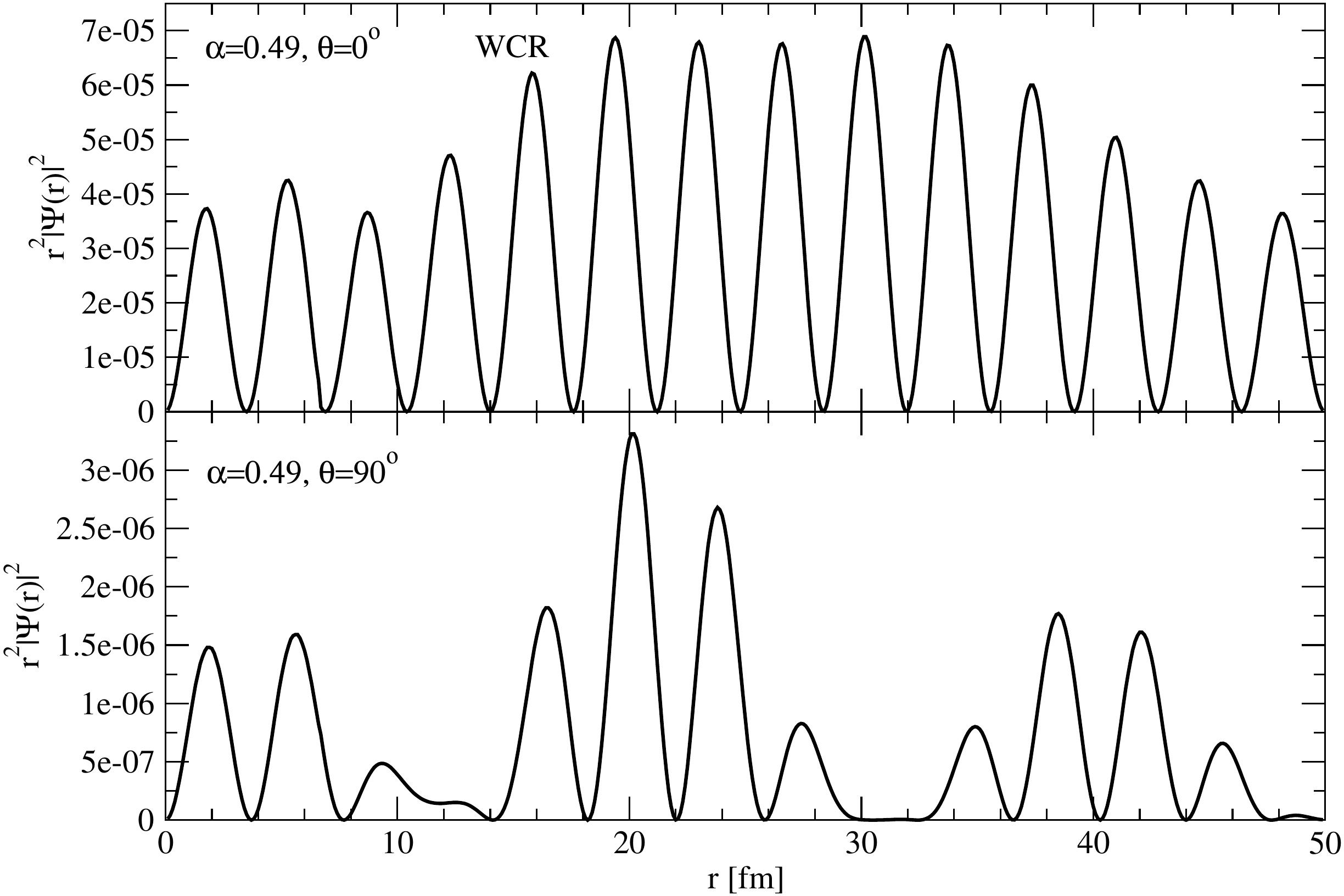}
    \caption[Dependence $r^2|\Psi(r)|^2$ on $r$ in the LOFF phase in
    the WCR for two different angles and one asymmetry.]{Dependence of
      $r^2|\Psi(r)|^2$ on $r$ in the WCR for two different angles
      $\theta$ and for asymmetry $\alpha = 0.49$ at which the LOFF
      phase is the ground state. (Cutout of Fig.~\ref{fig_1_20})}
    \label{fig_1_22}
  \end{center}
\end{figure}
\begin{figure}[!]
  \begin{center}
    \includegraphics[width=0.8\textwidth]{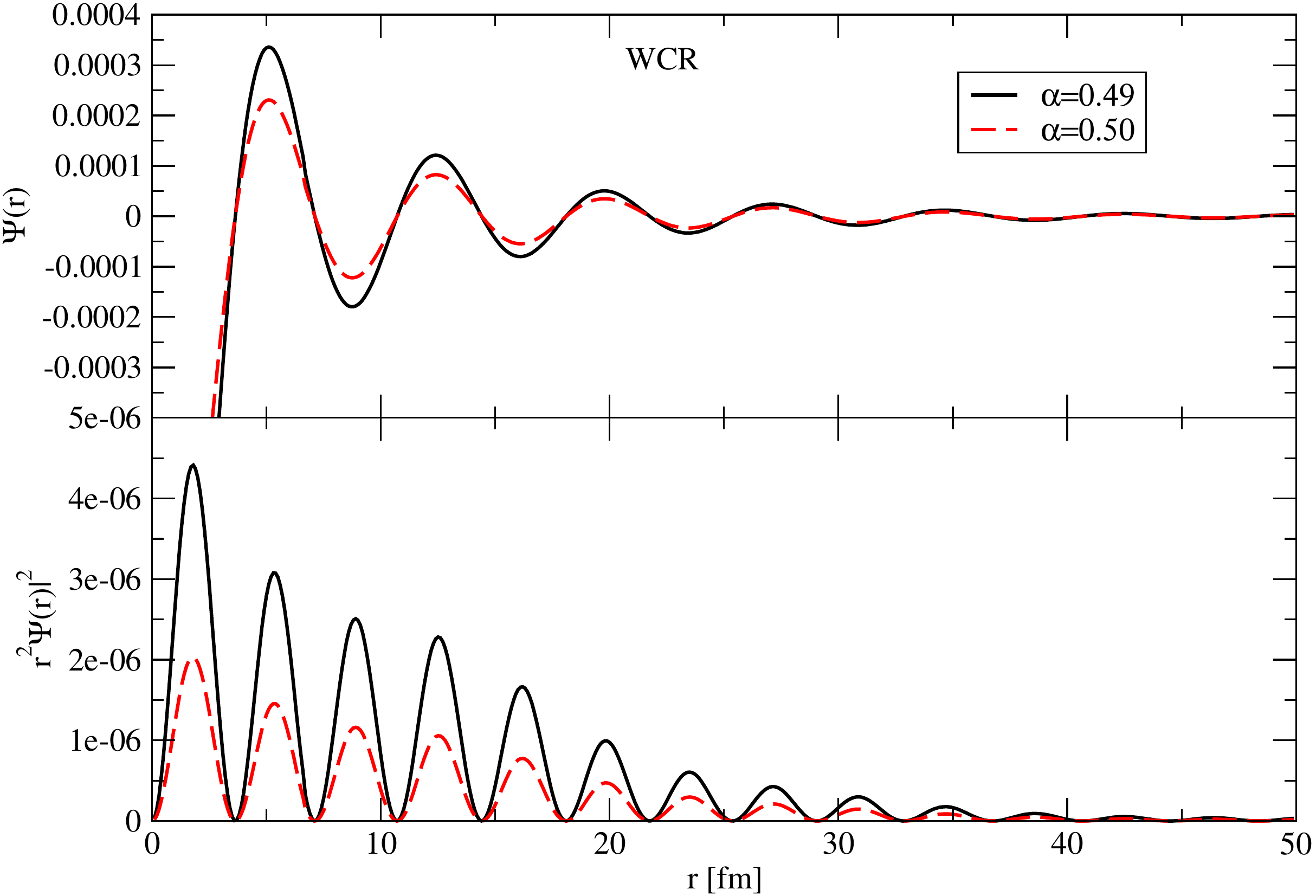}
    \caption[Dependence of $\Psi(r)$ and $r^2|\Psi(r)|^2$ on $r$ in
    the LOFF phase in the WCR integrated over the angle for two
    asymmetries.]{Dependence of $\Psi(r)$ and $r^2|\Psi(r)|^2$ on $r$
      in the WCR integrated over the angle $\theta$ for asymmetries
      $\alpha = 0.49$ and $\alpha = 0.50$ at which the LOFF phase is
      the ground state.}
    \label{fig_1_23}
  \end{center}
\end{figure}

Fig.~\ref{fig_1_19}, complementary to Fig.~\ref{fig_1_18}, displays
the quantity $r^2\vert \Psi(\vecr)\vert^2$.  The spatial correlation
in the SCR is dominated by a single peak corresponding to a tightly
bound state close to the origin. The existence of residual
oscillations indicates that there is no unique bound state formed at
such coupling, but the tendency towards its formation is clearly
seen. We find that in the weak coupling there is perfect match in the
maxima of the function $r^2\vert \Psi(\vecr)\vert^2$ for all
asymmetries (which was less visible in Fig.~\ref{fig_1_18} where the
oscillations are counter-phase). An oscillatory structure appears in
the ICR as a fingerprint of the transition from the BEC to the BCS
regime. In this case there are similar changes of maxima and zeros,
which is also the case in SCR. At low and high asymmetries the
strong-coupling peaks are well defined, whereas at intermediate
asymmetries the weight of the function is distributed among several
peaks. By increasing the coupling from WCR via ICR to SCR we see a
change of the shape from a heavily oscillating wave function of
unbound Cooper-pairs fixed in momentum space and spread in real space
in the WCR to a single peak dominated wave function of bound deuterons
fixed in real space and spread in momentum space, pronounced best for
$\alpha=0$.

Figs.~\ref{fig_1_20} to \ref{fig_1_23} demonstrate the same
quantities, i.e., $\Psi(\vecr)$ and $r^2\vert \Psi(\vecr)\vert^2$ for
the case of the LOFF phase computed at the WCR point of the phase
diagram (as specified in Table \ref{table_1_1}). In
Fig.~\ref{fig_1_20} a broad range of angles and asymmetries is shown,
whereas Figs.~\ref{fig_1_21} and \ref{fig_1_22} show only the
essential features. Fig.~\ref{fig_1_23} presents the angle integrated
quantities. At this point the LOFF phase is the ground state of the
matter at asymmetry $\alpha = 0.49$ ($\delta\mu=6.45$ MeV), where
$\Delta = 1.27$ MeV and $Q = 0.40$ fm$^{-1}$ and at $\alpha = 0.50$
($\delta\mu=6.51$ MeV), where $\Delta = 0.84$ MeV and $Q = 0.40$
fm$^{-1}$. For slightly lower asymmetries ($\alpha \le 0.48$) the
system is in the PS phase, whereas for $\alpha > 0.50$ the gap is
vanishingly small, the system being in the normal state. Unfortunately
it is not possible to carry out a direct comparison between BCS and
LOFF, since the gap of the BCS vanishes at $\alpha=0.37$ in the
WCR. In the case $\theta=0\degree$ and $\theta=45\degree$ the perfect
oscillatory behavior seen in $\Psi(\vecr)$ in the BCS case is
replicated, as in this case the finite momentum of the condensate does
not contribute to the spectrum of the Cooper-pairs. This shape is also
seen for the angle integrated result, because contributions of angles
around $\theta\approx0\degree$ are dominant. Regarding
$r^2\vert \Psi(\vecr)\vert^2$ at $\theta=0\degree$ we see that the
maxima first increase in value with a maximum at about $r=25$ fm and
afterwards decrease. This is due to the phase, which can be chosen
freely. Thus the physics at $\theta=45\degree$ and $\theta=0\degree$
does not differ significantly. In the case $\theta=90\degree$
$\Psi(\vecr)$ is distorted in the LOFF phase by the presence of a
second oscillatory mode with the period $2\pi/Q$ in addition to the
first mode, with the period $2\pi/k_F$. The additional periodic
structure is more pronounced in the quantity
$r^2\vert \Psi(\vecr)\vert^2$, where the rapid oscillations are
modulated with a period $\approx 16$~fm. Moreover we see a large
decrease of the amplitude at $\theta=90\degree$ compared to
$\theta\approx0\degree$ because of the low phase-space overlap.

\subsection{Occupation numbers across the BCS-BEC crossover}
\label{subsec_1_3_7}
\begin{figure}[!]
  \begin{center}
    \includegraphics[width=0.7\textwidth]{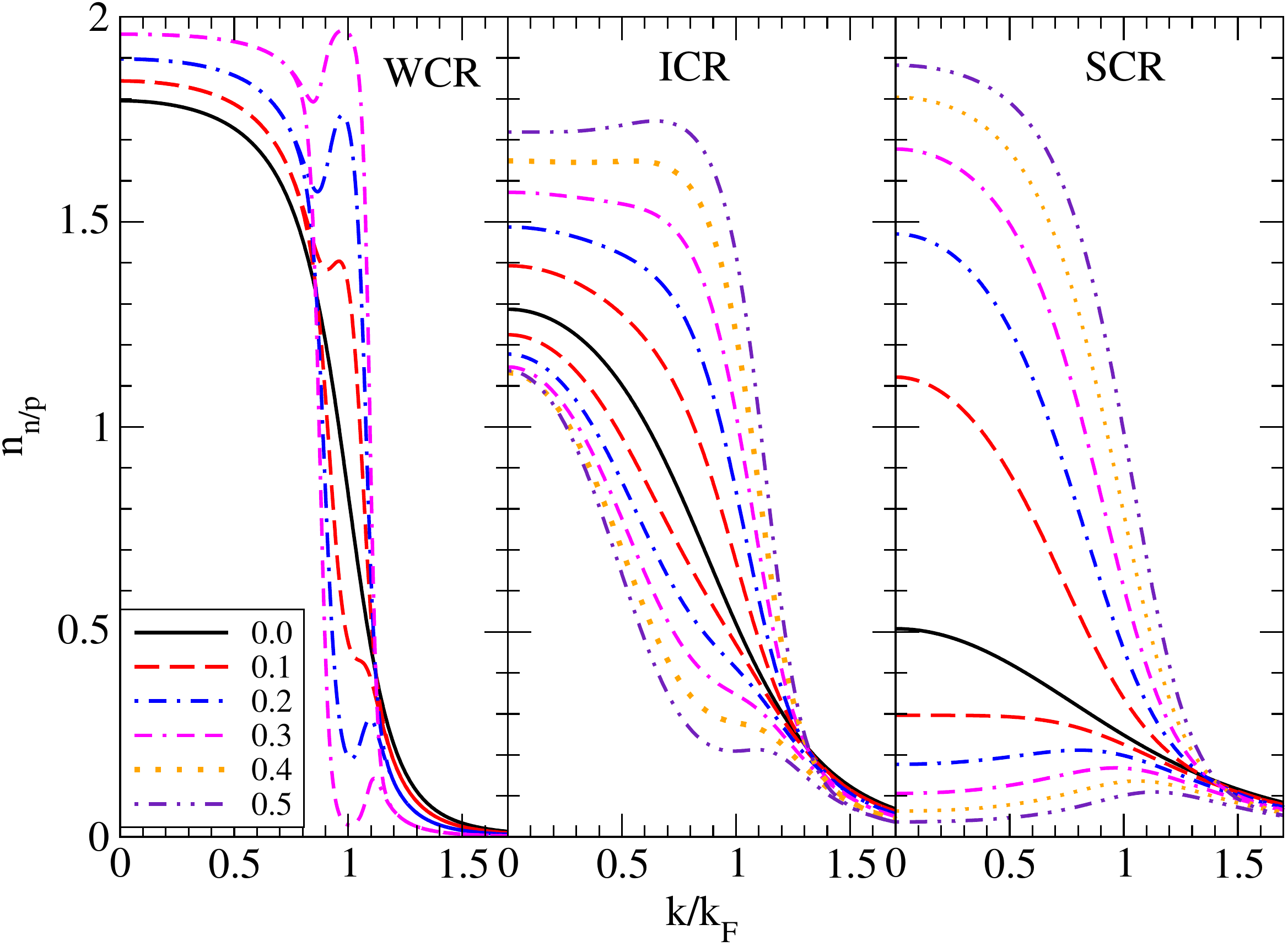}
    \caption[Dependence of the neutron and proton occupation numbers
    on momentum for the three coupling regimes.]{Dependence of the
      neutron and proton occupation numbers on momentum $k$ (in units
      of Fermi momentum) for the three coupling regimes and various
      asymmetries indicated in the legend.}
    \label{fig_1_24}
  \end{center}
\end{figure}
\begin{figure}[!]
  \begin{center}
    \includegraphics[width=0.7\textwidth]{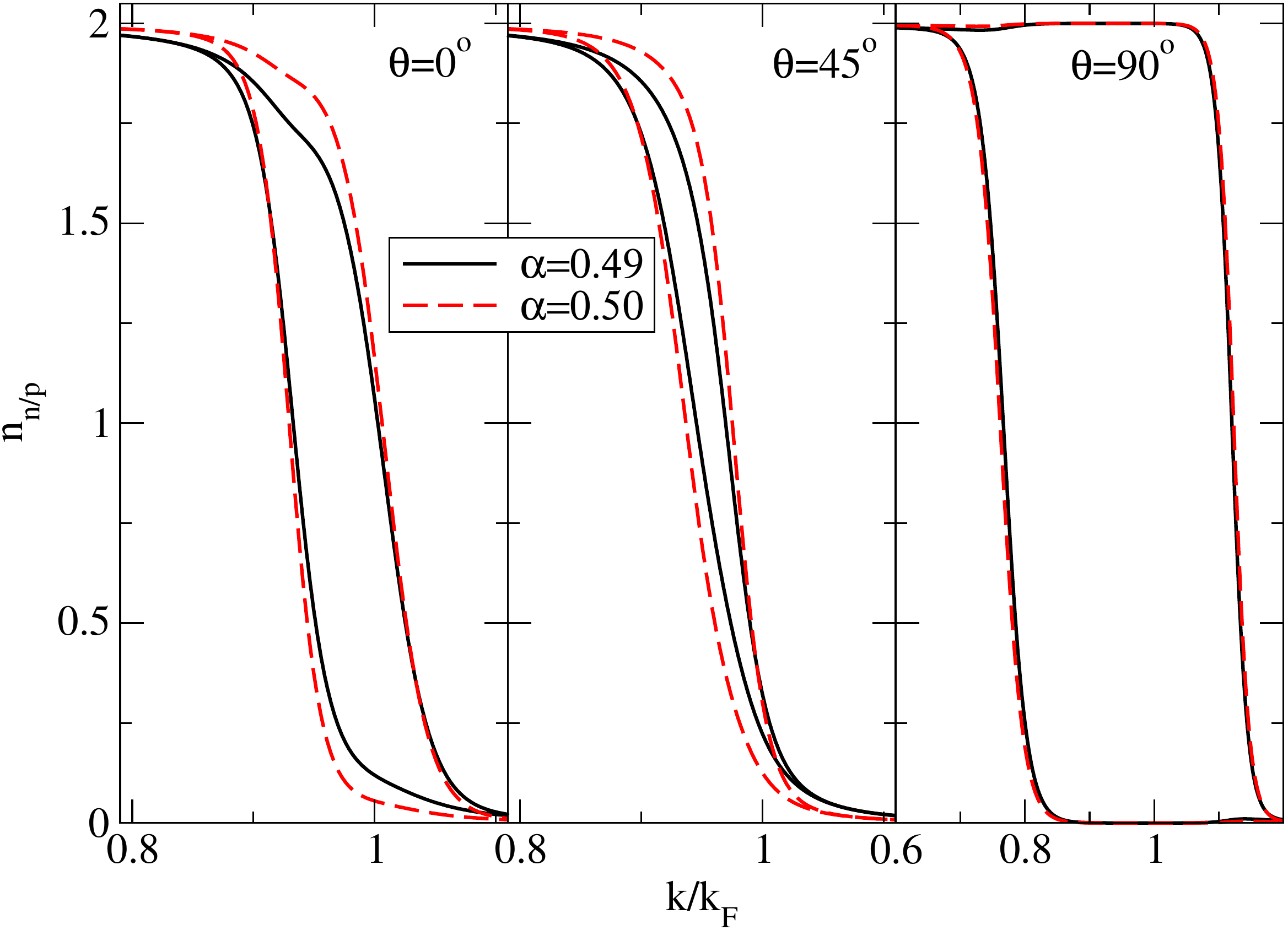}
    \caption[Dependence of the neutron and proton occupation numbers
    on momentum in the LOFF phase in the WCR for two asymmetries and
    three angles.]{Dependence of the neutron and proton occupation
      numbers on momentum $k$ (in units of Fermi momentum) in the WCR
      for two asymmetries where the LOFF phase is the ground
      state. The three angles indicated refer to the neutron
      occupation numbers. The proton occupation numbers are plotted
      for angles $180\degree-\theta$. Note: For $\theta=0\degree$ the
      neutron curves are the left ones.}
    \label{fig_1_25}
  \end{center}
\end{figure}

The integrand of Eq.~\eqref{eq_1_40} defines the occupation numbers
$n_{n/p}(k)$ of the neutrons and protons, given by
$n_{n/p}(k)=n_{n/p,\uparrow}(k)+n_{n/p,\downarrow}(k)$. The maximal
value of $n_{n/p}(k)$ is therefore two. The Cooper-pairs with total
momentum $\vecQ$ represent pairs with individual momenta
$\veck+\vecQ/2$ (neutrons) and $-\veck+\vecQ/2$ (protons).

Regarding Eq.~\eqref{eq_1_40} and Eq.~\eqref{eq_1_33} we see that the
energy shift due to the Cooper-pair momentum has the opposite
direction of the energy shift due to the mismatch of the Fermi spheres
and therefore promotes pairing for $\theta<90\degree$ in case of
neutrons and $\theta>90\degree$ in case of protons. To achieve a
better comparison, we therefore depict the proton occupation numbers
at $180\degree-\theta$. These quantities are shown in different
coupling regimes of the BCS-BEC crossover in Fig.~\ref{fig_1_24}. In
the WCR (leftmost panel) the occupation numbers of protons exhibit a
``breach'' \cite{2003PhRvL..91c2001G} or ``blocking region'' for large
asymmetries, i.e., the minority component is entirely expelled from
the blocking region ($n_p=0$), while the majority component is
maximally occupied ($n_n/2=1$). In the small-$\alpha$ limit the
occupation numbers are clearly fermionic (with some diffuseness due to
the temperature), in that all single-particle states below a certain
mode (the Fermi momentum at $T=0$) are almost filled, while all states
above are nearly empty. We have verified that in the high-temperature
limit the breach is filled in, the occupation numbers becoming smooth
functions of momentum; consequently the low-momentum modes are less,
this can be seen in Fig.~\ref{fig_1_09}. Since the densities of
neutrons and protons are different, Fermi momenta are shifted to
$k_{F_{n/p}}/k_F=\sqrt[3]{1\pm\alpha}$ as explained in
Subsec.~\ref{subsec_1_3_3}. For uncoupled particles, we would expect
two independent Fermi distributions with different $k_{F_{n/p}}$ at
finite asymmetry. However, we see a special behavior around the Fermi
surface also explained in Subsec.~\ref{subsec_1_3_3}.

In the ICR (middle panel) the fermionic nature of the occupation
numbers is lost. The low-momentum modes are not fully populated and,
accordingly, high-momentum modes are more heavily occupied. A Fermi
surface cannot be identified because of the smooth population of the
modes. Moreover, a breach no longer appears for the parameters
chosen. It is also to be noted that for large asymmetries
$\alpha \ge 0.4$, the momentum dependence of the occupation numbers
becomes non-monotonic; for the minority component this is a precursor
of the change in the topology of the Fermi surface under increase of
coupling strength. Furthermore, this non-monotonic behavior could be
interpreted as a relict of the effect at the neutron Fermi sphere
explained for the WCR.

The SCR (rightmost panel) can be identified with the BEC phase of
strongly coupled pairs. At large asymmetries the distribution of the
minority component undergoes a topological change. First there
develops an empty strip within the distribution function, which is
reorganized at larger momenta into a distribution in which the modes
are populated starting from a certain nonzero value. Thus, the Fermi
sphere occupied by the minority component in the weakly coupled BCS
limit evolves into a shallow shell structure in the strongly coupled
Bose-Einstein-condensed limit. This behavior was already revealed in
the case of the $\SD$ condensate in
Ref.~\cite{2001PhRvC..64f4314L}. In this shallow shell structure, the
occupation number of the minority component is approximately equal to
the occupation number of the majority component, which promotes
pairing.

Fig.~\ref{fig_1_25} depicts the occupation numbers in the WCR at
asymmetries corresponding to a LOFF-phase ground state for three fixed
angles $\theta = 0\degree$, $45\degree$, and $90\degree$. In the case
$\theta = 90\degree$ we have $E_A = 0$, and the LOFF spectrum differs
from the asymmetrical BCS spectrum only by a shift in the energy
origin, $\bar\mu \to \bar \mu - Q^2/8m^*$.  Therefore the occupation
numbers do not depart qualitatively from their BCS behavior; moreover,
the ``breach'' is clearly seen. The occupation numbers of protons and
neutrons are very rarely correlated as one would expect for unpaired
particles. This fits to the fact that there is no pairing in the BCS
phase at this asymmetry. For $\theta = 45\degree$ the difference
between the occupation numbers disappears, i.e., the superfluid
behaves as if it were isospin symmetric. This result follows from the
fact that the nonzero CM momentum of the LOFF phase compensates for
the mismatch of the Fermi spheres and restores the coherence needed
for pairing. In the case $\theta = 0\degree$ the effect of $E_A$
attains its maximal value, but the occupation numbers are intermediate
between those of the two cases previously addressed. This is due to
the fact that the overlap between the spectra of neutron and proton
quasiparticle branches is better for $\theta = 45\degree$ than for
$\theta = 0\degree$, in which case the quasiparticle spectra
``overshoot'' the optimal overlap (see the discussion in the following
section and Fig.~\ref{fig_1_11}). Also in quark matter, matter in the
LOFF phase at high asymmetry behaves as isospin symmetric at
$\theta=45\degree$~\cite{2009PhRvD..80g4022S}.

\subsection{Quasiparticle spectra}
\label{subsec_1_3_8}
\begin{figure}[!]
  \begin{center}
    \includegraphics[width=0.7\textwidth]{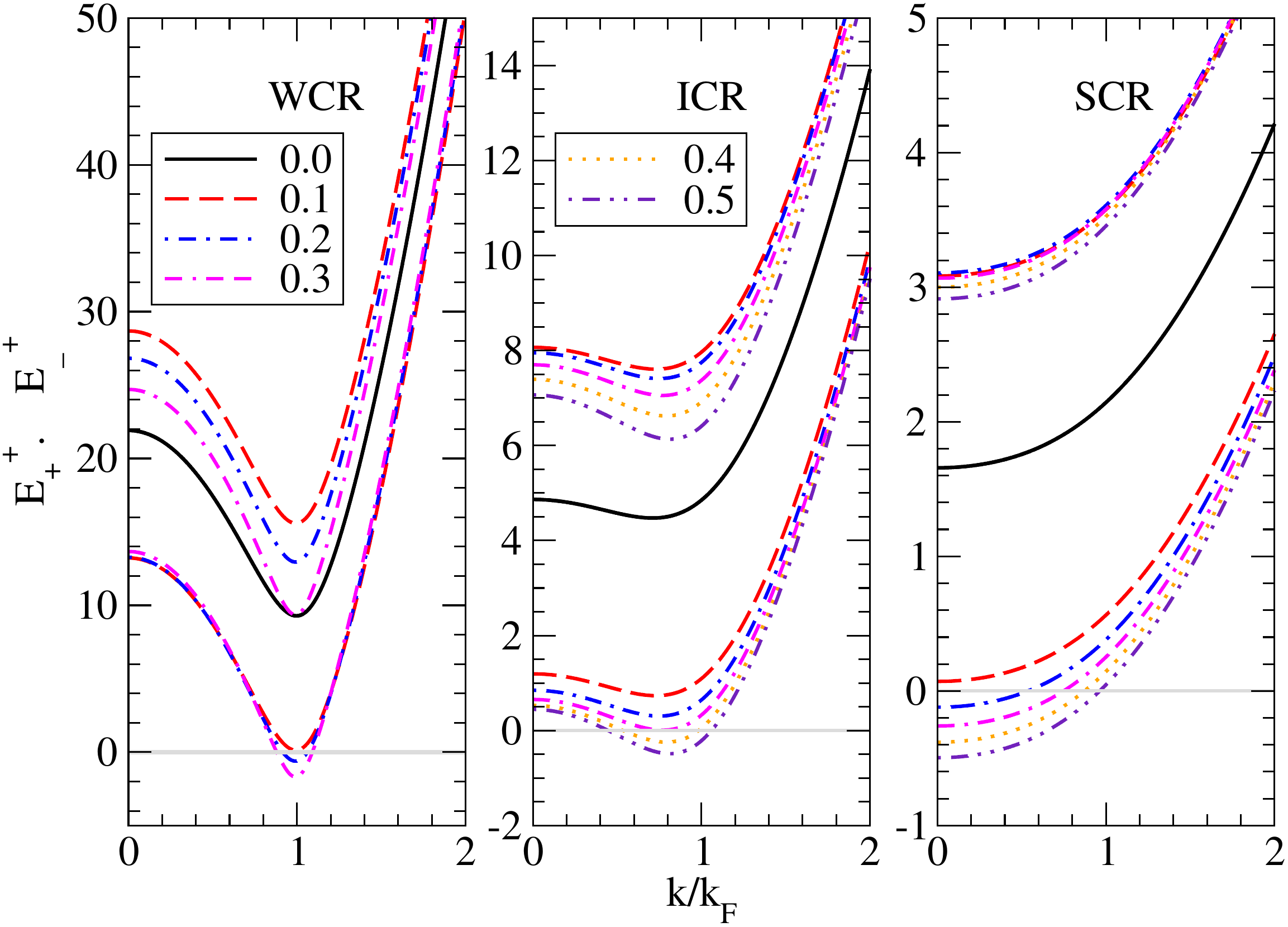}
    \caption[Dispersion relations for quasiparticle spectra in the
    case of the BCS condensate, as functions of momentum.]{Dispersion
      relations for quasiparticle spectra in the case of the BCS
      condensate, as functions of momentum in units of Fermi
      momentum. For each asymmetry, the upper branch corresponds to
      $E^+_+$, and the lower to the $E^+_-$ solution. }
    \label{fig_1_26}
  \end{center}
\end{figure}

Finally, let us consider the dispersion relations for quasiparticle
excitations about the $\SD$ condensate. We first examine in some
detail the spectra $E^a_{\pm}$ in the BCS case defined in
Eq.~\eqref{eq_1_33}, which are then independent of the sign of $a$ and
we take $a = +$.
\begin{eqnarray}
  E_{\pm}^{-} =
  E_{\pm}^{+} =
  \sqrt{E_S^2+\Delta^2} \pm
  \delta\mu\,.
\end{eqnarray}
These are shown in Fig.~\ref{fig_1_26} for the three coupling regimes
of interest. In the isospin-symmetric BCS case, the dispersion
relation has a minimum at $E^+_{+} = E^+_{-} = \Delta$ for $k= k_F$
due to:
\begin{eqnarray}
  E_{\pm}^{+}(k)
  &=&
      \sqrt{\left(\frac{k^2}{2m^*}-\bar\mu\right)^2+\Delta^2}\,,\qquad\bar\mu=\mu_n=\mu_p=\frac{k_F^2}{2m^*}\,.
      \label{eq_1_61}
\end{eqnarray}
For finite asymmetries one has
$E^+_{\pm} = \sqrt{E_S^2+\Delta^2} \pm \delta\mu$; hence the minima of
the dispersion relations of neutron and proton quasiparticles are
given by an asymmetry-dependent gap value modified by the shift in
chemical potential, i.e., $\Delta(\alpha)\pm\delta\mu$. For protons
this leads to a gapless spectrum, which does not require a finite
minimum energy for excitation of two modes (say $k_1$ and $k_2$) for
which the dispersion relation intersects the zero-energy axis. This
phenomenon is well known as {\it gapless superconductivity}. In a
gaped BCS phase, the energy levels of the pairing particles are
beneath the Fermi surface. Therefore, they can not scatter with other
particles. However, in the gapless BCS, the pairing protons are at the
Fermi surface and can therefore scatter with other particles. In
nuclear matter e.g. of supernovae, this can change the properties of
matter, i.e. the neutrino transport. Also in quark matter of cooling
neutron stars this can affect the neutrino transport and have a
remarkable effect on the cooling curve~\cite{2013A&A...555L..10S}. The
momentum interval $k_1\le k\le k_2$ corresponds to the interval in
Fig.~\ref{fig_1_24} where the occupation numbers of majority and
minority components separate and the ``breach'' in the occupation of
the minority component becomes prominent. Moreover it is the interval
in which the kernel has a very low value, see
Subsec.~\ref{subsec_1_3_5}.

Consider now the SCR, in which case we are dealing with a gas of
deuterons and free neutrons. Due to the negative average chemical
potential, the minimum of $E^+_\pm$ is shifted to $k=0$, as can be
seen from Eq.~\eqref{eq_1_61}. In the symmetrical limit (i.e.\ when
only deuterons are present), the dispersion relation has a minimum at
the origin that corresponds to the (average) chemical potential, which
asymptotically approaches half the binding energy of a deuteron in
vacuum~\cite{2001PhRvC..64f4314L}. The effect of asymmetry is to shift
the average chemical potential downwards and to introduce the
separation $\delta\mu$ in the quasiparticle spectra.

Since the minimum is now at the origin, there is only one mode for
which the dispersion relation crosses zero at a finite $k$. The
dispersion relations in the ICR experience a transition from the WCR
to the SCR, such that their key features resemble those of the WCR,
but with a shallower minimum and a larger momentum interval
$[k_1,k_2]$ over which the excitation spectrum becomes gapless.
\begin{figure}[!]
  \begin{center}
    \includegraphics[width=0.7\textwidth]{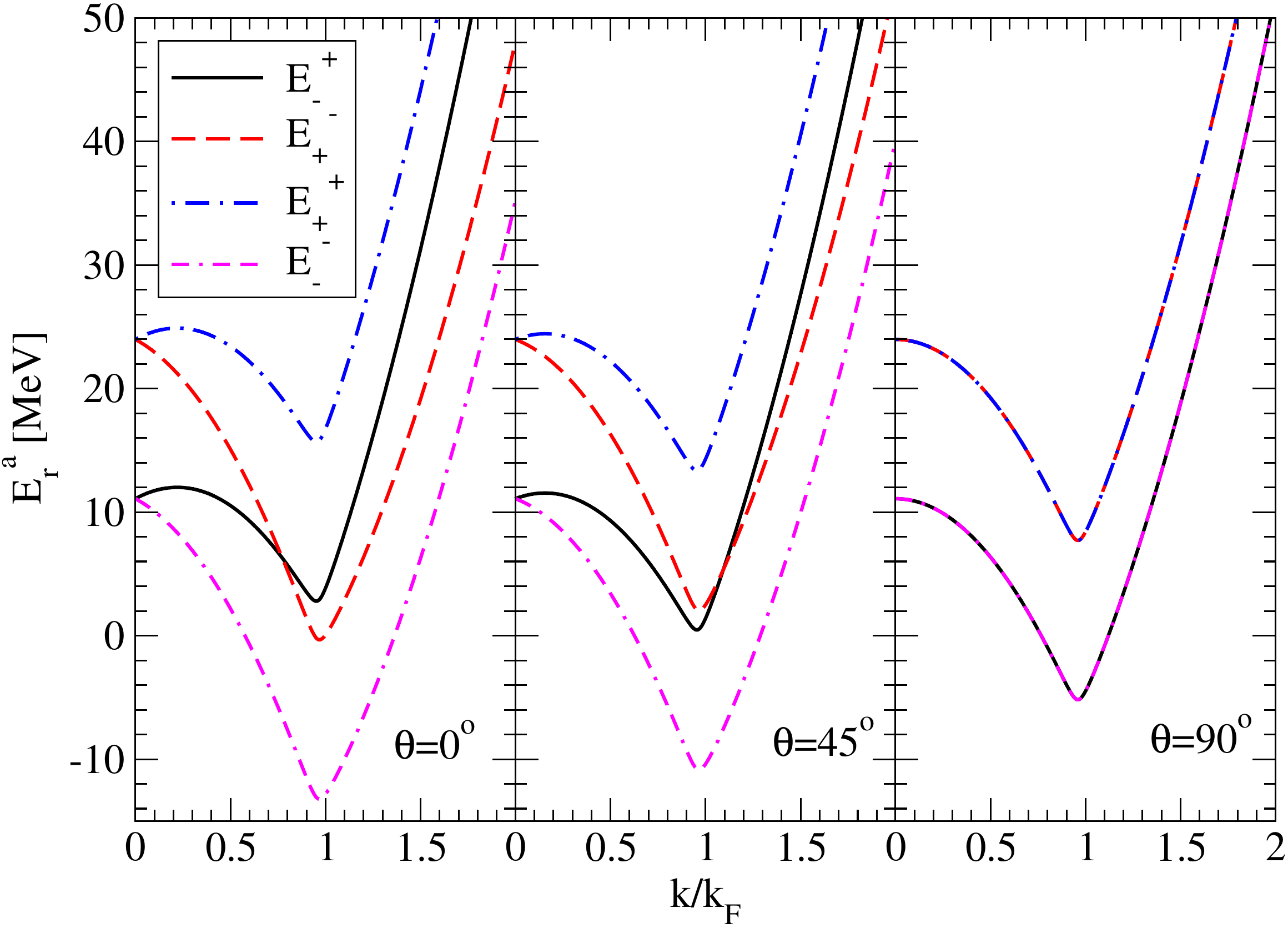}
    \caption[Dispersion relations for quasiparticle spectra in the
    LOFF phase in the WCR, as functions of momentum for three
    angles.]{Dispersion relations for quasiparticle spectra in the
      LOFF phase in the WCR, as functions of momentum for three angles
      and $\alpha = 0.49$. }
    \label{fig_1_27}
  \end{center}
\end{figure}

The dispersion relations for quasiparticles in the LOFF phase for
special angles $\theta$ are shown in Fig.~\ref{fig_1_27} in the WCR
and for $\alpha=0.49$ corresponding to the LOFF phase as ground
state. In this case, we show all four branches of quasiparticle
spectrum. Consistent with the earlier discussion of
Figs.~\ref{fig_1_11} and \ref{fig_1_25} for $\theta =90\degree$, the
LOFF phase resembles the BCS phase and there is a large mismatch
between the spectra of protons and neutrons. In this case the branches
$a = +$ and $a=-$ are degenerated. For other angles we see again that
the nonzero CM momentum mitigates the asymmetry and brings the
quasiparticle spectra closer together, i.e., the LOFF phase resembles
the symmetrical BCS phase for the two branches with $a \neq r$ for
$\theta<90\degree$. This is particularly clear for
$\theta = 45\degree$, in which case two of the four dispersion
relations coincide in the vicinity of the Fermi momentum. It is clear
that the optimal mitigation of the isospin mismatch by the finite
moment does not need to be for $\theta = 0\degree$, but can occur at
some angle $0\degree\le \theta\le 90\degree$; it is seen that for
$\theta = 0\degree$ the branches cross and, hence, ``overshoot'' the
optimal compensation.

\begin{figure}[!]
  \begin{center}
    \includegraphics[width=\textwidth]{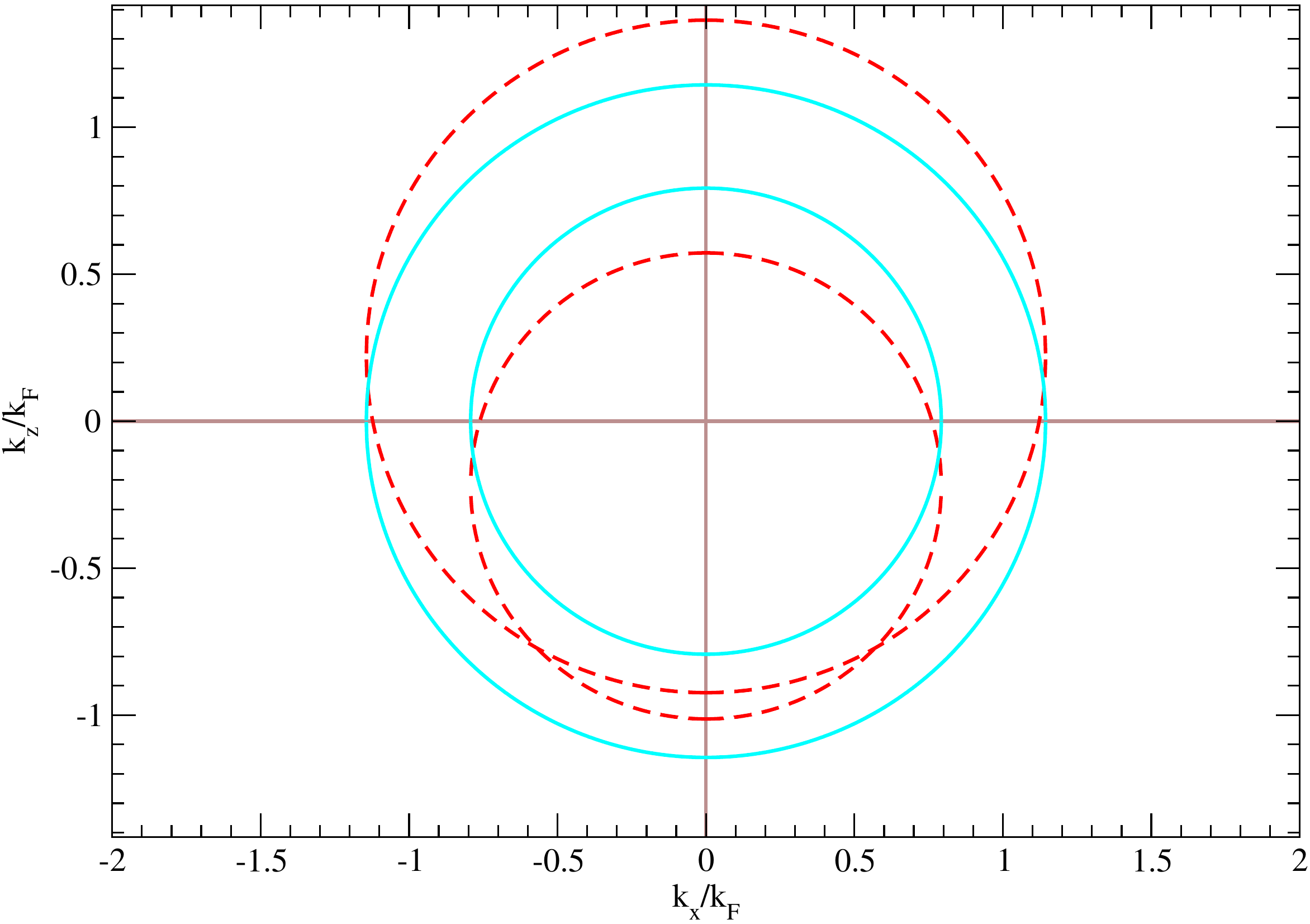}
    \caption[Illustration of Fermi surfaces in the LOFF phase.]
    {Illustration of Fermi surfaces in the LOFF phase. It is
      characterized by the following values of parameters:
      $\alpha = 0.49$, $\delta\mu=6.45$~MeV, $\Delta = 1.27$~MeV and
      $Q = 0.40$~fm$^{-1}$. The dashed lines are the actual results
      for the LOFF phase; by the calculations of the solid lines we
      use the effective mass and chemical potentials of the LOFF phase
      in the WCR with $\alpha=0.49$, but we set $\vecQ=0$.}
    \label{fig_1_28}
  \end{center}
\end{figure}

The restoration of the coherence (Fermi-surface overlap) in the LOFF
phase can be illustrated by looking at the solutions of
$\epsilon_{n/p,\uparrow/\downarrow}^\pm = 0$ [see Eq.~\eqref{eq_1_11}]
which define the Fermi-surface in the limit $\Delta\to 0$ but
$Q\neq 0$. Solutions for $\epsilon_{n}^-=0$ and $\epsilon_{p}^+=0$
with $\vecQ=Q\bm e_z$ are illustrated in Fig.~\ref{fig_1_28} in two
cases $Q = 0$ and $Q\neq 0$. In both cases we calculated with the
effective mass and the chemical potentials we obtained for the LOFF
phase in the WCR at $\alpha=0.49$. In the first case the Fermi
surfaces are concentric spheres which have no intersection. In the
second case the non-zero CM leads to an intersection of the
Fermi-spheres; in these regions of intersection the pair-correlations
are restored to the magnitude characteristic to the BCS phase. Of
course, the CM momentum costs positive kinetic energy, which must be
smaller than the negative condensation energy for LOFF phase to be
stable.

\section{Conclusion}
\label{sec_1_4}
Low-density nuclear matter is predicted to feature a rich phase
diagram at low temperatures and nonzero isospin asymmetry. The phase
diagram contains at least the following phases: the translationally
and rotationally symmetric, but isospin-asymmetrical BCS phase, the
BEC phase containing neutron-proton dimers, the current carrying
Larkin-Ovchinnikov-Fulde-Ferrell phase, and associated phase-separated
phases.

Our analysis of these phases can be summarized as follows:
\begin{itemize}
\item The phase diagram of nuclear matter composed of these phases has
  two tri-critical points in general, one of which is a Lifshitz
  point. These can combine in a tetra-critical point for a special
  combination of density, temperature, and isospin
  asymmetry. Tri-critical points exist only for
  $0<\alpha<\alpha_\mathrm{LOFF}$. The phase diagram contains two
  types of crossovers from the asymmetrical BCS phase to the BEC of
  deuterons and an embedded neutron gas: a transition between the
  homogeneous BCS-BEC phases at relatively high temperatures and
  between the heterogeneous BCS-BEC phases at low temperatures. We
  have shown that the LOFF phase exists only in a narrow strip in the
  high-density, low-temperature domain and at nonzero asymmetries.

\item The crossovers of BCS-BEC type are smooth and are characterized
  by lines in the temperature-density plane that do not change much
  with isospin asymmetry. These lines were obtained by examining the
  sign of the average chemical potential.

\item Detailed analysis of key intrinsic quantities, including the
  kernel of the gap equation along with the Cooper-pair wave function
  and its probability density, clearly establishes that in the BCS
  limit one deals with a coherent state, whose wave function
  oscillates over many periods with a wavelength characterized by the
  inverse Fermi momentum $k_F^{-1}$. In the opposite limit the wave
  function is well-localized around the origin, indicating that one is
  then dealing with a Bose condensate of strongly bound states, namely
  deuterons. For high densities and high asymmetries, when matter is
  stable in the LOFF phase, an oscillation emerges belonging to the
  inverse Cooper-pair momentum $Q^{-1}$, additional to the one
  belonging to $k_F^{-1}$.

\item The analysis of the kernel of the wave function, the occupation
  probabilities of neutrons and protons, and the quasiparticle
  dispersion relations demonstrates the prominent role played by the
  Pauli-blocking region (called ``the breach'')
  \cite{2003PhRvL..91c2001G} that appears in these quantities. In the
  BCS phase and the low-temperature limit of the weak-coupling regime
  (WCR), the blocking region embraces modes in the range
  $k_1\le k\le k_2$ around the Fermi surface. In this modal region, it
  has been found that (a) the minor constituents (protons) are
  extinct; (b) there are no contributions to the kernel of the gap
  equation from these modes; and (c) the end of points of this region
  correspond to the onset of gapless modes that can be excited without
  any energy cost. The LOFF phase appearing in this regime
  substantially mitigates the blocking mechanism by allowing for
  nonzero CM momentum of the condensate. As a consequence, all the
  intrinsic quantities studied are much closer to those of the
  isospin-symmetric BCS state.

\item We have traced the evolution of the targeted intrinsic
  properties into the strong-coupling regime (SCR) as the system
  crosses over from the BCS condensate to a BEC of deuterons plus a
  neutron gas. In the SCR the long-range coherence of the condensate
  is lost. The dispersion relations change their form from a spectrum
  having a minimum at the Fermi surface to a spectrum that is minimal
  at $k=0$, as would be expected for a BEC, independent of isospin
  asymmetry. With increasing isospin asymmetry, the proton dispersion
  relation acquires points with zero excitation energy in this
  regime. The occupation numbers reach a maximum for finite $k$ and
  reflect a change of topology at large asymmetries: the filled
  ``Fermi sphere'' becomes an empty ``core.''
\end{itemize}

The present investigation of BCS-BEC crossovers with inclusion of
unconventional phases, such as the LOFF phase and the heterogeneous
phase-separated phase, could be useful in the studies of
spin/flavor-imbalanced fermionic systems in ultracold atomic gases,
for recent studies see,
e.g.,~\cite{2007AnPhy.322.1790S,2008RvMP...80.1215G,2011JPhCS.321a2028S},
dense quark matter,
e.g.,~\cite{2009PhRvD..80g4022S,2010PhRvD..82e6006M,2014PhRvD..89c6009M,2010PhRvD..82g6002F,2012NuPhA.875...94K},
and other related quantum systems.

\clearpage{\pagestyle{empty}\cleardoublepage}

\chapter{Hartree-Fock}
\label{chap_2}

\section{Introduction}
\label{sec_2_1} This chapter describes matter in strong magnetic
fields which may occur in compact
stars~\cite{1986bhwd.book.....S,1991ApJ...383..745L,2000ApJ...537..351B,2013PhRvD..88b5008S,2010arXiv1005.4995S,2009PhRvD..79l3001S,2012PhRvC..86e5804C}. After
the hydrogen burning process, a star develops into a red giant or a
red supergiant, depending on its mass. After the state of a planetary
nebular or a supernova, respectively, the star finely results in a
compact star, namely white dwarf, neutron star or black hole. Matter
in stellar objects can occur in the form of ordinary baryonic matter
either in confined (hadronic) or deconfined (quark-gluon) state. It
can also occur in the form of strange matter if the hypothesis that
the strange matter is the absolute ground state of matter is
true~\cite{2013PhRvD..88b5008S}. In this chapter we do not discuss
strange matter; its physics is described for example in
Refs.~\cite{2013PhRvD..88b5008S,2010arXiv1005.4995S,2009PhRvD..79l3001S}. The
ordinary baryonic matter in strong magnetic fields has been studied
extensively in the literature, see
e.g. Ref.~\cite{2000ApJ...537..351B}. The surface magnetic field of a
white dwarf is $B\approx10^{6}-10^{8}\,$G and the surface magnetic
field of a neutron star is
$B\approx10^{12}\,$G~\cite{1991ApJ...383..745L}. Neutron stars with
strong magnetic fields (known as magnetars) with a surface magnetic
field of $B\approx 10^{14}-10^{15}\,$G have been observed. It is
further conjectured that magnetic fields of the order of
$B \approx 10^{18}\,$G can exist in the interiors of
magnetars~\cite{2000ApJ...537..351B,2013PhRvD..88b5008S,2010arXiv1005.4995S,2009PhRvD..79l3001S,1991ApJ...383..745L}. Due
to the virial theorem, the magnetic field can not exceed
$B\approx10^{18}\,$G in the interior of a neutron star and
$B\approx10^{12}\,$G in the interior of a white
dwarf~\cite{1991ApJ...383..745L}. It is estimated that the magnetic
field in self-bound compact stars made of strange quark matter cannot
exceed $B\approx10^{20}\,$G~\cite{PhysRevC.82.065802}. Moreover,
strong magnetic fields ($B\approx10^{16}-10^{17}\,$G) have been
considered in newly-born neutron
stars~\cite{2012PhRvC..86e5804C}. Magnetic fields of order
$B\gtrsim10^{17}\,$G can substantially affect the composition of the
outer crust of a neutron star~\cite{2012PhRvC..86e5804C}. A discussion
of the composition of the outer crust can be found in
Ref.~\cite{1986bhwd.book.....S,2007ASSL..326.....H,2008LRR....11...10C}. A
recent study of the composition of crustal matter in relation to the
magnetic field can be found in Ref.~\cite{2012PhRvC..86e5804C}. One
source of the changes in the composition of the crust in strong
magnetic field is the Landau quantization of electron orbits. It has
been found that strong magnetic fields favor more isospin-symmetric
nuclei in the outer crust of a neutron star. Macroscopically strong
magnetic fields act to make the crust of a magnetar more massive than
its non-magnetized counterpart~\cite{2012PhRvC..86e5804C}. Due to the
lower mass of the progenitor stars of white dwarfs, white dwarfs
consist of lighter elements than the neutron star crust; massive white
dwarfs are conjectured to consist largely of carbon and
oxygen~\cite{1986bhwd.book.....S}. The third element we study in this
chapter, neon, can also be found in white dwarfs. These relatively
light elements can also occur in accreting neutron stars.

In this introduction~\ref{sec_2_1} we give an overview of the
Hartree-Fock theory, which is used to compute the properties of nuclei
in strong magnetic fields. The discussion follows mostly
Refs.~\cite{2011JPhG...38c3101E} and~\cite{2014CoPhC.185.2195M}. In
Section~\ref{sec_2_2} we give an overview of the code
Sky3D~\cite{2014CoPhC.185.2195M} which was used in numerical
computations. In Section~\ref{sec_2_3} we describe the extension of
the Hartree-Fock theory to include magnetic fields. The corresponding
modifications to the code are described. Further we present our
results and discussion.

Nuclear systems can be described at different levels of sophistication
and precision. At the most fundamental level to describe strongly
interacting matter one needs to start from the quantum chromodynamics
(QCD). However, it is more rational to start from the nucleon-nucleon
interaction to describe bound states of nucleons which form the
nuclei. The methods which rely on the basic nucleon-nucleon
interaction fitted to the empirical data as a starting point are
called ab initio methods. Ab initio calculations are numerically very
expensive. The ultimate goal of an ab initio theory is to start from
basic microscopic interactions and to predict the properties of a
complex compound. This program, however, is not yet realized, although
there are no fundamental obstacles. Indeed, for molecules and solids
where the basic Coulomb interaction is well known such a program works
well. To circumvent the difficulties in the case of finite nuclei or
nuclear matter, approximations like restricting to the low-momentum
regime or chiral perturbation theory have been applied with increasing
success. Within these approximations, many-body methods are exploited
to obtain the fully correlated nuclear state, although often one
restricts to the mean-field state (Slater determinant or BCS state).

On the other extreme is the macroscopic nuclear liquid-drop model
(LDM). It parameterizes the energy of a nucleus in terms of its bulk
properties. The parameters of the model are obtained from extensive
fits to a large pool of ground-state data of nuclei. The necessary
quantum effects, which are not explicit in the LDM, can be added for
an improvement of the model. In-between these methods there is the
so-called microscopic-macroscopic (mic-mac) method, which combines the
single particle model and LDM approaches. The mic-mac method relies on
a large amount of preconceived knowledge, in particular on the
expected nuclear mean field. It has disadvantages caused by
uncertainties when it is extrapolated to the unknown regime of exotic
nuclei and it is restricted to ground-state properties.

Another class of ``intermediate'' approaches is based on the
self-consistent mean-field (SCMF). They work at the microscopic level,
but they employ effective interactions. Like the vast majority of
microscopic models that describe many-body systems, they use a
description in terms of single-particle (s.p.) wave functions. SCMF
models generate the optimal one-body potential corresponding to the
s.p. wave functions starting from various types of zero-range or
finite-range {\it effective interactions}. Widely used interactions
are the Skyrme (zero-range) interaction and the Gogny (finite range)
interaction. The effective interaction, as opposed to the bare
nucleon-nucleon interaction is ``soft'' and therefore the Hartree-Fock
(HF) theory can be straighforwardly applied. One thus obtains the
s.p. wave functions variationally for a given effective interaction,
which is calibrated to reproduce the empirical data. The great
strength of SCMF is that they reach a high-quality description of
ground-state properties, excitations and large-amplitude
dynamics. (Covariant versions of SCMF methods based on meson picture
of interaction between nuclei and relativistic mean-field (RMF)
approach have also been used with success to describe nuclei.)

Correlations in nuclei can be divided roughly into short-range,
long-range and collective correlations. The short-range correlations
describe the hard repulsive core within the range $r\leq 0.5\,$fm. The
long-range correlations act over long distances, which are
characterized by coherence lengths that are larger than the
interparticle distance. Collective correlations refer to e.g. the
center-of-mass motion or rotation of a nucleus. The short- and
long-range correlations are fully active in the nuclear volume, their
effects can be expressed via smoothly varying functions of densities
and currents. This can be summarized in an effective energy-density
functional or in an effective interaction. Short- and long-range
correlations have been a priori built into the mean-field model. They
should not be computed again with effective interactions. Collective
correlations on the other hand cannot be transferred to a simple
effective functional and need a posteriori treatment.

\section{Overview of the TDHF Code}
\label{sec_2_2}

\subsection{Introduction}
The calculations reported in this chapter were done with the TDHF code
Sky3D~\cite{2014CoPhC.185.2195M}. Below we mostly follow
Ref.~\cite{2014CoPhC.185.2195M} but we discuss in some detail the
effects of the magnetic field. These were included in the code in
manner similar to the inclusion of the angular momentum in cranking
term described in Ref.~\cite{2008PhRvC..77d1301G}. In addition we have
included the effects of the interaction between the magnetic field and
the spin of the nucleons.

The code Sky3D contains a useful selection of Skyrme forces. However
it does not contain all terms that have been included in some recent
works which aim a high precision description of nuclear
systems. Therefore the Skyrme interaction used in Sky3D is useful for
a semiquantitative description, where high precision of the Skyrme
force is not decisive. (These additional terms can be added to the
code without much difficulty).

The code Sky3D solves the static or dynamic equations of motion using
Skyrme-like forces on a three-dimensional Cartesian grid. Certain
symmetries are assumed when imposing isolated or periodic boundary
conditions. Consequently the nucleonic wave function spinors are
always periodic functions, while it is possible to choose an isolated
charge distribution for the Coulomb potential. Due to the possibility
of choosing periodic boundary conditions and due to the highly
flexible initialization, the code is also suitable for astrophysical
nuclear matter applications. All spatial derivatives in the code are
calculated with the finite Fourier transform method. For the static
Hartree-Fock equations a damped gradient iteration method is used and
for the time-dependent Hartree-Fock (TDHF) equations an expansion of
the time-development operator is employed. It is possible to place any
number of initial nuclei into the mesh at arbitrary positions with any
velocities. It is also possible to include pairing in the BCS
approximation for the static case. However, due to the absence of some
time-odd terms in the implementation of the Skyrme interaction,
calculations may be restricted only to even-even nuclei. Altogether
the code Sky3D can be used (within the limitations of mean-field
theory) for a wide variety of applications in nuclear structure,
collective excitations, and nuclear reactions.

An overview of the code Sky3D is included in
Ref.~\cite{2014CoPhC.185.2195M} to offer the possibility to include
additional physics or special analysis of the results. In this chapter
and in appendix~\ref{app_2} we give an overview of the code and add
the extensions which are needed for the inclusion of magnetic fields.

\subsection{Physics implemented in the code}
\label{subsec_2_2_2}
Here we give a brief discussion of the physics implemented in the code
which is important for the studies of this chapter. See
appendix~\ref{app_2} and Ref.~\cite{2014CoPhC.185.2195M} for further
details.

In the mean-field theory, the many-body system is described in terms
of a set of single-particle (s.p.) wave functions. With these
s.p. wave functions local densities and currents can be defined, see
appendix~\ref{sec_b_1}.

The code Sky3D solves the mean-field equations based on the widely
used Skyrme energy functional. The energy-density functional contains
an expansion in a number of derivatives, i.e., it corresponds to a
low-momentum expansion of many-body theory (see
Ref.~\cite{2011JPhG...38c3101E}). The energy functional as implemented
in the code can be written as
\begin{eqnarray}
  E_\mathrm{tot} &=& T + (E_0 + E_1 + E_2 + E_3 +
                     E_\mathrm{ls})\nonumber\\
                 &&+ E_\mathrm{Coulomb} + E_\mathrm{pair} +
                    E_\mathrm{corr}\,,\label{eq_2_01}
\end{eqnarray}
where the terms which arise from the Skyrme force are collected within
the parenthesis. The various terms of Eq.~\eqref{eq_2_01} are defined
in appendix~\ref{sec_b_2}.

The variation of the energy-density functional discussed above with
$\partial_{\psi_\alpha^*}E=\hat h\psi_\alpha$ leads to the mean-field
Hamiltonian $\hat h$. It is given by
\begin{eqnarray}
  \hat h_q&=&U_q(\vecr)-\nabla\cdot\left[B_q(\vecr)\nabla\right]+\mathrm i\vecW_q\cdot\left(\vecsigma\times\nabla\right)+\vecS_q\cdot\vecsigma\nonumber\\
          && -\tfrac{\mathrm i}2\left[\left(\nabla\cdot\vecA_q\right)+2\vecA_q\cdot\nabla\right]\,,\label{eq_2_02}
\end{eqnarray}
with $q\in\{p,n\}$ specifying the isospin. The terms are discussed in
appendix~\ref{sec_b_4}, using the force coefficients of
appendix~\ref{sec_b_3}. Note that because protons are charged, the
Coulomb potential acting between protons should be added to the
potential.

With the Hamiltonian operator the eigenvalues of the system can be
computed through the Schr\"odinger equation
\begin{eqnarray}
  \hat h\psi_\alpha=\varepsilon_\alpha\psi_\alpha\,,\label{eq_2_03}
\end{eqnarray}
with $\hat h$ being the mean-field Hamiltonian of Eq.~\eqref{eq_2_02}
and $\varepsilon_\alpha$ being the single-particle energy of state
$\alpha$. This equation follows upon variation with respect to
single-particle wave-function $\psi_\alpha$. In the code Sky3D,
pairing can be included in the BCS approximation. See
appendix~\ref{sec_b_5} for more details on the static calculations
without pairing. For more details also on static Hartree-Fock
including pairing and on time-dependent Hartree-Fock
see~\cite{2014CoPhC.185.2195M}.

As an output, the code provides several observables, e.g. the total
deformation $\beta$, the triaxiality $\gamma$ and the r.m.s radii
$r_\mathrm{rms}^{q}$. See appendix~\ref{sec_b_6} for further details
on the observables.

The calculations of Sky3D run on a three dimensional regular Cartesian
grid. The number of grid points and the distance between the grid
points can be chosen by the user. The code uses a fast Fourier
transform and therefore periodic boundary conditions, except for the
Coulomb force.

A particular strength of the code Sky3D is the possibility of a
flexible initialization. For the calculations of this chapter, we used
the harmonic oscillator initialization. Here the user can implement
the radii of the harmonic oscillator states in each spatial
direction. Moreover, it is possible to choose the numbers of neutrons
and protons and it is also possible to include unoccupied neutron and
proton states. When using unoccupied states, there are more nucleon
states calculated than the actual existing ones. This can lead to a
faster convergence. Initially the harmonic oscillator states are
filled up. For certain set-ups, e.g., if we consider magnetic fields,
some normally higher harmonic oscillator states may be energetically
favored, but not taken into account by using only occupied states. In
this case it can happen that the code converges to two different
configurations, whether we use unoccupied states or not; or it can
happen that the code is stuck in one configuration for a long time,
before it converges to the lower energy state without the additional
unoccupied states.

\section{Hartree-Fock with magnetic field}
\label{sec_2_3}
\subsection{Introduction}
\label{subsec_2_3_1}
Now we turn to the central problem of this chapter -- the
determination of the properties of nuclei in strong magnetic
fields. To introduce the magnetic field the Hamiltonian
\eqref{eq_2_02} is modified as
\begin{subequations}
  \label{eq_2_04}
  \begin{eqnarray}
    \hat h_{\mathrm{mod},\,q} &=&\hat h_q+ \hat h_{\mathrm{mag},\,q}\,,\\
    \hat h_{\mathrm{mag},\,q} &=&- \left(\vecl\cdot\delta_{q,p} + g_q\frac{\vecsigma}2\right) \cdot \tilde\vecB_q\,,
  \end{eqnarray}
\end{subequations}
with $q\in\{p,n\}$ specifying the isospin, $\vecsigma$ is the spin
Pauli matrix and $\vecl$ is the (dimensionless) orbital angular
momentum related to the spin $\vecS$ and the orbital angular momentum
$\vecL$ via $\vecS=\hbar\vecsigma/2$ and $\vecL=\hbar\vecl$, where
$g_n=-3.82608544$ and $g_p=5.584694712$ are the Land\'e $g$-factors of
neutrons and protons and
\begin{eqnarray}
  \tilde{\vecB_q}=\frac{e\hbar}{2\,m_qc}\vecB\,,
\end{eqnarray}
where $\vecB$ being the magnetic field. The Kronecker Delta is due to
the fact that neutrons are charge neutral and thus do not couple to
the orbital angular momentum. Despite their charge neutrality, they
couple to the spin because of the charge distribution of the
constituent quarks. If dipole modes are considered, effective currents
for neutrons and protons should be introduced; neutrons get a negative
effective charge and the proton charge is reduced~\cite{1995GM}. Due
to the interactions of the nucleons within the nucleus, the $g$
factors are modified, and one should include quenching factors as it
is done in~\cite{2011JPhG...38c3101E}. These quenching factors are
known for random-phase approximation (RPA), but not for our studies,
therefore we can not adopt them. However, without these quenching
factors, the results are only qualitative.

Using a constant magnetic field in $\bm e_z$ direction
($\vecB=B_z\bm e_z$) allows us to simplify the equations to the form
\begin{subequations}
  \begin{eqnarray}
    \hat h_{\mathrm{mod},\,q}&=&\hat h_q- \left(l_z\cdot\delta_{q,p} + g_q\frac{\sigma_z}2\right) \cdot \tilde B_{q,z}\,,\\
    \hat h_{\mathrm{mod},\,q}&=&\hat h_q+ \left(\mathrm i(x\partial_y-y\partial_x)\delta_{q,p} + \frac{g_q}2
                                 \begin{pmatrix}
                                   -1&0\\0&1
                                 \end{pmatrix}
                                            \right) \cdot \tilde{B}_{q,z}\,.
  \end{eqnarray}
\end{subequations}
The additional terms due to the magnetic field appearing in
Eq.~\eqref{eq_2_04} are implemented in the module Meanfield of the
code.

\subsection{Clebsch-Gordan coefficients}
The definition of the Clebsch-Gordan coefficients follows the book of
Greiner and Maruhn~\cite{1995GM}. We need the general formula
\begin{eqnarray}
  \left|JMls\right>&=&\sum_{m_lm_s}\left|lm_lsm_s\right>(lsJ|m_lm_sM)\,,
\end{eqnarray}
hereby $l$ is the orbital angular momentum, $s$ the spin and $J$ the
total angular momentum and $m_l$, $m_s$ and $M$ are the $z$-components
of $l$, $s$ or $J$, respectively. The following conditions need to be
fulfilled:
\begin{subequations}
  \begin{eqnarray}
    m_l+m_s&=&M\,,\\ |l-s|&\leq&J\leq l+s\,.
  \end{eqnarray}
\end{subequations}
In our case we deal with nucleons, which have $s=1/2$ and thus have
$m_s=\pm1/2$. For $l$ we have non-negative integer numbers, for $m_l$
we obtain integer numbers, which can be positive or negative or
zero. Thus we have non-negative half-integer numbers for $J$ and we
obtain half-integer numbers, which can be positive or negative for
$M$. For clarity we introduce the following conventions:
\begin{itemize}
\item $s$ is always $1/2$, therefore it is skipped.
\item $m_s$ is always $\pm1/2$. To prevent confusion with other terms,
  it is written in the following way:
  \begin{itemize}
  \item $m_s=\uparrow$ for $m_s=+1/2$
  \item $m_s=\downarrow$ for $m_s=-1/2$
  \end{itemize}
\item $l$ is always positive, it is always written without sign.
\item $m_l$ is always written with sign to prevent confusion with $l$.
\item For $J$ and $M$ the same conventions as for $l$ and $m_l$ are
  introduced. There is no confusion between $J$, $M$ and $l$, $m_l$,
  because the former are half-integer and the latter are integer
  numbers.
\end{itemize} Thus, to summarize our convention,

\begin{tabular}{cll} $s$ & & skipped \\
  $m_s$ & & $\uparrow$ or $\downarrow$ \\
  $l$ & without sign & integer number \\
  $m_l$ & with sign & integer number \\
  $J$ & without sign & half-integer number \\
  $M$ & with sign & half-integer number \\
\end{tabular}

\newpage Now we can introduce the Clebsch-Gordan coefficients for
$s_{1/2}$, $p_{3/2}$ and $p_{1/2}$ states
\begin{subequations}
  \begin{eqnarray}
    \left|J,M,l\right>
    &=&\sum_{m_l,m_s}\left|l,m_l,m_s\right>\nonumber\\
    &&\times\left(l,J\middle|m_l,m_s,M\right)\,,\\
    s_{1/2},M=-\tfrac12:\quad\left|\tfrac12,-\tfrac12,0\right>
    &=&\left|0,+0,\downarrow\right>\left(0,\tfrac12\middle|+0,\downarrow,-\tfrac12\right)\,,\\
    s_{1/2},M=+\tfrac12:\quad\left|\tfrac12,+\tfrac12,0\right>
    &=&\left|0,+0,\uparrow\right>\left(0,\tfrac12\middle|+0,\uparrow,+\tfrac12\right)\,,\\
    p_{3/2},M=-\tfrac32:\quad\left|\tfrac32,-\tfrac32,1\right>
    &=&\left|1,-1,\downarrow\right>\left(1,\tfrac32\middle|-1,\downarrow,-\tfrac32\right)\,,\\
    p_{3/2},M=-\tfrac12:\quad\left|\tfrac32,-\tfrac12,1\right>
    &=&\left|1,-1,\uparrow\right>\left(1,\tfrac32\middle|-1,\uparrow,-\tfrac12\right)\nonumber\\
    &&+\left|1,+0,\downarrow\right>\left(1,\tfrac32\middle|+0,\downarrow,-\tfrac12\right)\,,\\
    p_{3/2},M=+\tfrac12:\quad\left|\tfrac32,+\tfrac12,1\right>
    &=&\left|1,+0,\uparrow\right>\left(1,\tfrac32\middle|+0,\uparrow,+\tfrac12\right)\nonumber\\
    &&+\left|1,+1,\downarrow\right>\left(1,\tfrac32\middle|+1,\downarrow,+\tfrac12\right)\,,\\
    p_{3/2},M=+\tfrac32:\quad\left|\tfrac32,+\tfrac32,1\right>
    &=&\left|1,+1,\uparrow\right>\left(1,\tfrac32\middle|+1,\uparrow,+\tfrac32\right)\,,\\
    p_{1/2},M=-\tfrac12:\quad\left|\tfrac12,-\tfrac12,1\right>
    &=&\left|1,-1,\uparrow\right>\left(1,\tfrac12\middle|-1,\uparrow,-\tfrac12\right)\nonumber\\
    &&+\left|1,+0,\downarrow\right>\left(1,\tfrac12\middle|+0,\downarrow,-\tfrac12\right)\,,\\
    p_{1/2},M=+\tfrac12:\quad\left|\tfrac12,+\tfrac12,1\right>
    &=&\left|1,+0,\uparrow\right>\left(1,\tfrac12\middle|+0,\uparrow,+\tfrac12\right)\nonumber\\
    &&+\left|1,+1,\downarrow\right>\left(1,\tfrac12\middle|+1,\downarrow,+\tfrac12\right)\,,
  \end{eqnarray}
\end{subequations}

which by inserting the Clebsch-Gordan coefficients reduce to
\begin{subequations}
  \label{eq_2_10}
  \begin{eqnarray}
    s_{1/2},M=-\tfrac12:\quad\left|\tfrac12,-\tfrac12,0\right>&=&\left|0,+0,\downarrow\right>\,,\label{eq_2_10a}\\
    s_{1/2},M=+\tfrac12:\quad\left|\tfrac12,+\tfrac12,0\right>&=&\left|0,+0,\uparrow\right>\,,\label{eq_2_10b}\\
    p_{3/2},M=-\tfrac32:\quad\left|\tfrac32,-\tfrac32,1\right>&=&\left|1,-1,\downarrow\right>\,,\label{eq_2_10c}\\
    p_{3/2},M=-\tfrac12:\quad\left|\tfrac32,-\tfrac12,1\right>&=&\tfrac1{\sqrt3}\left|1,-1,\uparrow\right>\nonumber\\
                                                              &&+\sqrt{\tfrac23}\left|1,+0,\downarrow\right>\,,\label{eq_2_10d}\\ p_{3/2},M=+\tfrac12:\quad\left|\tfrac32,+\tfrac12,1\right>&=&\sqrt{\tfrac23}\left|1,+0,\uparrow\right>\nonumber\\
                                                              &&+\tfrac1{\sqrt3}\left|1,+1,\downarrow\right>\,,\label{eq_2_10e}\\ p_{3/2},M=+\tfrac32:\quad\left|\tfrac32,+\tfrac32,1\right>&=&\left|1,+1,\uparrow\right>\,,\label{eq_2_10f}\\
    p_{1/2},M=-\tfrac12:\quad\left|\tfrac12,-\tfrac12,1\right>&=&-\sqrt{\tfrac23}\left|1,-1,\uparrow\right>\nonumber\\ &&+\tfrac1{\sqrt3}\left|1,+0,\downarrow\right>\,,\label{eq_2_10g}\\
    p_{1/2},M=+\tfrac12:\quad\left|\tfrac12,+\tfrac12,1\right>&=&-\tfrac1{\sqrt3}\left|1,+0,\uparrow\right>\nonumber\\ &&+\sqrt{\tfrac23}\left|1,+1,\downarrow\right>\,.\label{eq_2_10h}
  \end{eqnarray}
\end{subequations}

\newpage We can now calculate the $z$-components of the orbital
angular momentum and the spin with the following definition
\begin{eqnarray}
  \left<\mathcal{O}(J,M,l)\right>=\left<J,M,l\middle|\mathcal{O}(J,M,l)\middle|J,M,l\right>\,,
\end{eqnarray}
and using the following relations
\begin{subequations}
  \label{eq_2_12}
  \begin{alignat}{2}
    \left<l,m_l\middle|\frac{L_z}\hbar\middle|l,m_l'\right>=&m_l\delta_{m_l,m_l'}\,,&\quad
    \left<l,m_l\middle|l,m_l'\right>=&\delta_{m_l,m_l'}\,,\\
    \left<m_s\middle|\frac{S_z}\hbar\middle|m_s'\right>=&m_s\delta_{m_s,m_s'}\,,&\quad
    \left<m_s\middle|m_s'\right>=&\delta_{m_s,m_s'}\,.
  \end{alignat}
\end{subequations}

For the orbital angular momentum we obtain
\begin{subequations}
  \begin{eqnarray}
    \left<\frac{L_z(J,M,l)}{\hbar}\right>
    &=&\left<J,M,l\middle|\frac{L_z}{\hbar}\middle|J,M,l\right>\\
    &=&\sum_{m_l,m_l',m_s,m_s'}\left<l,m_l,m_s\middle|\frac{L_z}\hbar\middle|l,m_l',m_s'\right>\nonumber\\
    &&\times\left(l,J|m_l,m_s,M\right)\left(l,J|m_l',m_s',M\right)\nonumber\\
    &=&\sum_{m_l,m_s}\left<l,m_l,m_s\middle|\frac{L_z}\hbar\middle|l,m_l,m_s\right>\left(l,J|m_l,m_s,M\right)^2\nonumber\\
    &=&\sum_{m_l,m_s}\left(l,J|m_l,m_s,M\right)^2\left<l,m_l\middle|\frac{L_z}\hbar\middle|l,m_l\right>\left<m_s\middle| m_s\right>\nonumber\\
    \Rightarrow\left<\frac{L_z(J,M,l)}{\hbar}\right>
    &=&\sum_{m_l,m_s}m_l\cdot\left(l,J|m_l,m_s,M\right)^2\,.
  \end{eqnarray}
\end{subequations}

In an analogous manner we evaluate the spin component
\begin{subequations}
  \begin{eqnarray}
    \left<\frac{S_z(J,M,l)}{\hbar}\right>
    &=&\left<J,M,l\middle|\frac{S_z}{\hbar}\middle|J,M,l\right>\\
    \Rightarrow\left<\frac{S_z(J,M,l)}{\hbar}\right>
    &=&\sum_{m_l,m_s}m_s\cdot\left(l,J|m_l,m_s,M\right)^2\,.
  \end{eqnarray}
\end{subequations}

\begin{figure}[!]
  \begin{center}
    \includegraphics[width=0.8\textwidth]{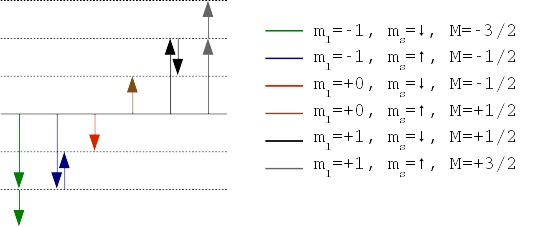}
    \caption[Spin and orbit angular momentum quantum numbers of $s$-
    and $p$-states.] {The quantum numbers for different states. The
      orbital angular momentum and spin of states are given by these
      quantum numbers or by superpositions of them. The limit of weak
      magnetic fields can be seen in Eq.~\eqref{eq_2_10}, for
      increasing magnetic field the prefactors of the mixed states
      differ. The limit of strong magnetic fields can be seen in
      Eq.~\eqref{eq_2_19}.}
    \label{fig_2_01}
  \end{center}
\end{figure}

\newpage Performing the summation and calculating the Clebsch-Gordan
coefficients gives
\begin{subequations}
  \label{eq_2_15}
  \begin{alignat}{3}
    s_{1/2},M=&-\tfrac12: &\quad \left<L_z\right>=&+0\cdot\hbar\,,
    &\quad \left<S_z\right>=&-\tfrac12\cdot\hbar\,,\\
    s_{1/2},M=&+\tfrac12: &\quad \left<L_z\right>=&+0\cdot\hbar\,,
    &\quad \left<S_z\right>=&+\tfrac12\cdot\hbar\,,\\
    p_{3/2},M=&-\tfrac32: &\quad \left<L_z\right>=&-1\cdot\hbar\,,
    &\quad \left<S_z\right>=&-\tfrac12\cdot\hbar\,,\\
    p_{3/2},M=&-\tfrac12: &\quad
    \left<L_z\right>=&-\tfrac13\cdot\hbar\,, &\quad
    \left<S_z\right>=&-\tfrac16\cdot\hbar\,,\\ p_{3/2},M=&+\tfrac12:
    &\quad \left<L_z\right>=&+\tfrac13\cdot\hbar\,, &\quad
    \left<S_z\right>=&+\tfrac16\cdot\hbar\,,\\ p_{3/2},M=&+\tfrac32:
    &\quad \left<L_z\right>=&+1\cdot\hbar\,, &\quad
    \left<S_z\right>=&+\tfrac12\cdot\hbar\,,\\ p_{1/2},M=&-\tfrac12:
    &\quad \left<L_z\right>=&-\tfrac23\cdot\hbar\,, &\quad
    \left<S_z\right>=&+\tfrac16\cdot\hbar\,,\\ p_{1/2},M=&+\tfrac12:
    &\quad \left<L_z\right>=&+\tfrac23\cdot\hbar\,, &\quad
    \left<S_z\right>=&-\tfrac16\cdot\hbar\,.
  \end{alignat}
\end{subequations}
Let us now have a closer look at these coefficients. In all states we
obtain $M=m_l+m_s=(\left<L_z\right>+\left<S_z\right>)/\hbar$ as a good
quantum number. However we only obtain $\left<L_z\right>=\hbar\,m_l$
and $\left<S_z\right>=\hbar\,m_s$ as a good quantum number for states
with pure Clebsch-Gordan coefficients. Hereby ``pure'' refers to
Clebsch-Gordan coefficients where one $\left|J,M,L\right>$ state
refers to one $\left|l,m_l,m_s\right>$ state and ``mixed'' refers to
those, where $\left|J,M,L\right>$ are formed by superpositions of
$\left|l,m_l,m_s\right>$ states. For the states with mixed
Clebsch-Gordan coefficients, this differs due to the {\it l-s}
coupling. In these states, $\left<L_z\right>$ and $\left<S_z\right>$
are superpositions of one state with $\vert{m_l}\vert=l$ and one with
$\vert{m_l}\vert<l$. The quantum numbers $m_l$ and $m_s$ and the
resulting $M=m_l+m_s$ are shown in Fig.~\ref{fig_2_01} for different
states. We have one solid line at $M=0$ and dashed lines with
intervals of $1/2$. The arrows corresponding to $m_l$ start at the
origin and the ones corresponding to $m_s$ start at the end points of
the arrows corresponding to $m_l$ to present $M$.

We can now also calculate the energy difference of these states
according to Eq.~\eqref{eq_2_04} with the relations of
\eqref{eq_2_12}:
\begin{subequations}
  \begin{eqnarray}
    \Delta E_q(J,M,l)&=&\left<J,M,l\middle|\hat h_{\mathrm{mag},\,q}\middle|J,M,l\right>\,,\\
    \Delta E_q(J,M,l)&=&\left<J,M,l\middle|-\tilde B_q\middle(\frac{L_z}\hbar\cdot\delta_{q,p}\right.\nonumber\\
                     &&\left.+ g_q\cdot\frac{S_z}{\hbar}\middle)\middle|J,M,l\right>\,.
  \end{eqnarray}
\end{subequations}

This can be written as
\begin{subequations}
  \begin{eqnarray}
    \Delta E_q(J,M,l)
    &=&\left<J,M,l\middle|-\tilde B_q\middle(\frac{L_z}\hbar\cdot\delta_{q,p} + g_q\cdot\frac{S_z}{\hbar}\middle)\middle|J,M,l\right>\\
    &=&-\sum_{m_l,m_l',m_s,m_s'}\left<l,m_l,m_s\middle|\tilde B_q\middle(\frac{L_z}\hbar\cdot\delta_{q,p} \right.\nonumber\\
    &&+\left. g_q\cdot\frac{S_z}{\hbar}\middle)\middle|l,m_l',m_s'\right>\nonumber\\
    &&\times\left(l,J|m_l,m_s,M\right)\left(l,J|m_l',m_s',M\right)\nonumber\\
    &=&-\sum_{m_l,m_s}\left<l,m_l,m_s\middle|\tilde B_q\middle(\frac{L_z}\hbar\cdot\delta_{q,p} \right.\nonumber\\
    &&+\left. g_q\cdot\frac{S_z}{\hbar}\middle)\middle|l,m_l,m_s\right>\left(l,J|m_l,m_s,M\right)^2\nonumber\\
    &=&-\sum_{m_l,m_s}\left(l,J|m_l,m_s,M\right)^2\nonumber\\
    &&\times\left(\left<l,m_l\middle|\middle<m_s\middle|\tilde B_q\cdot\frac{L_z}\hbar\cdot\delta_{q,p}\middle|l,m_l\middle>\middle|m_s\right>\right.\nonumber\\
    &&+\left.\left<l,m_l\middle|\middle<m_s\middle|\tilde B_q\cdot g_q\cdot\frac{S_z}\hbar\middle|l,m_l\middle>\middle|m_s\right>\right)\nonumber\\
    &=&-\sum_{m_l,m_s}\left(l,J|m_l,m_s,M\right)^2\left(\left<l,m_l\middle|\tilde B_q\cdot\frac{L_z}\hbar\cdot\delta_{q,p}\middle|l,m_l\right>\right.\nonumber\\
    &&\times\left<m_s\middle|m_s\right> +\left.\left<m_s\middle|\tilde B_q\cdot g_q\cdot\frac{S_z}\hbar\middle|m_s\right>\left<l,m_l\middle|l,m_l\right>\right)\nonumber\\
    &=&-\sum_{m_l,m_s}\left(l,J|m_l,m_s,M\right)^2\tilde B_q\left(\delta_{q,p}\left<l,m_l\middle|\frac{L_z}\hbar\middle|l,m_l\right>\right.\nonumber\\
    &&\left.+g_q\left<m_s\middle|\frac{S_z}\hbar\middle|m_s\right>\right)\nonumber\\
    \Rightarrow\Delta E_q(J,M,l)
    &=&-\sum_{m_l,m_s}\left(l,J|m_l,m_s,M\right)^2\tilde B_q\nonumber\\
    &&\times\left(m_l\cdot\delta_{q,p}+g_q\cdot m_s\right)\,. \label{eq_2_17b}
  \end{eqnarray}
\end{subequations}

Using Eq.~\eqref{eq_2_17b} and Eq.~\eqref{eq_2_10} leads to:
\begin{subequations}
  \label{eq_2_18}
  \begin{alignat}{3}
    s_{1/2},M=&-\tfrac12:&\Delta E_n=&\tfrac12\tilde
    B_n\cdot{g_n}\,, & \Delta E_p=&\tilde B_p\cdot\tfrac{g_p}2\,,\\
    s_{1/2},M=&+\tfrac12:&\Delta E_n=&-\tfrac12\tilde B_n\cdot{g_n}\,,
    & \Delta E_p=&-\tilde B_p\cdot\tfrac{g_p}2\,,\\
    p_{3/2},M=&-\tfrac32:&\Delta E_n=&\tfrac12\tilde B_n\cdot{g_n}\,,
    & \Delta E_p=&\tilde B_p\left(1+\tfrac{g_p}2\right)\,,\\
    p_{3/2},M=&-\tfrac12:&\Delta E_n=&\tfrac16{\tilde
      B_n}\cdot{g_n}\,, & \Delta E_p=&\tfrac{\tilde
      B_p}3\left(1+\tfrac{g_p}2\right)\,,\\
    p_{3/2},M=&+\tfrac12:&\Delta E_n=&-\tfrac16{\tilde
      B_n}\cdot{g_n}\,, & \Delta E_p=&-\tfrac{\tilde
      B_p}3\left(1+\tfrac{g_p}2\right)\\ p_{3/2},M=&+\tfrac32:&\Delta
    E_n=&-\tfrac12\tilde B_n\cdot{g_n}\,, & \Delta E_p=&-\tilde
    B_p\left(1+\tfrac{g_p}2\right)\,,\\ p_{1/2},M=&-\tfrac12:&\Delta
    E_n=&-\tfrac16{\tilde B_n}\cdot{g_n}\,, & \Delta
    E_p=&\tfrac23\tilde B_p\left(1-\tfrac{g_p}4\right)\,,\\
    p_{1/2},M=&+\tfrac12:&\Delta E_n=&\tfrac16{\tilde
      B_n}\cdot{g_n}\,, & \Delta E_p=&-\tfrac23\tilde
    B_p\left(1-\tfrac{g_p}4\right)\,.
  \end{alignat}
\end{subequations}

\subsection{Results}
\label{subsec_2_3_3} All calculations used for the following analysis
were done on a grid with 32 grid points in each direction and a
distance between the grid points of 1.0~fm. For the convergence
parameters $\delta$ and $E_0$ of Eq.~\eqref{eq_b_5} we chose
$\delta=0.4$ and $E_0=100$ as recommended
in~\cite{2014CoPhC.185.2195M}. For testing, we varied the number of
grid points, the distance between the grid points and the convergence
parameters, leading to different computational times and different
levels of convergence, but not to different physical
results. Moreover, we used a fragment initialization with one fragment
for testing. For the radii of the harmonic oscillator states we chose
3.0~fm, 3.2~fm and 3.1~fm for the $x$-, $y$- and $z$-direction,
respectively. We chose the Skyrme force
SV-bas~\cite{2009PhRvC..79c4310K} without pairing. The boundary
condition for the Coulomb force was implemented with isolated boundary
conditions. To achieve a better convergence, we used unoccupied
states. We studied nuclei with equal amount of neutrons and protons:
\isotope[12]{C}, \isotope[16]{O} and \isotope[20]{Ne} and chose the
same number of neutron and proton unoccupied states each. We chose 8
occupied and 8 unoccupied states for \isotope[16]{O}, 10 occupied and
10 unoccupied for \isotope[20]{Ne} and 6 occupied and 4 unoccupied for
\isotope[12]{C}. See subsection~\ref{subsec_2_2_2},
appendix~\ref{app_2} and Ref.~\cite{2014CoPhC.185.2195M} for further
discussion on the terms in this paragraph.

The magnetic field was chosen in $\bm e_z$ direction. For testing,
also other directions were calculated, leading to the same physical
results. For stronger magnetic fields, convergence was not always
achieved. Within the code, we used natural units for the magnetic
field of $\sqrt{\mathrm{MeV}\, \mathrm{fm}^{-3}}=4.00\cdot10^{16}\,$G.
We started at $B=0$, increased it slightly to
$B=0.001\sqrt{\mathrm{MeV}\, \mathrm{fm}^{-3}}=4.00\cdot10^{13}\,$G.
Afterwards we used multiples of
$0.25\sqrt{\mathrm{MeV}\, \mathrm{fm}^{-3}}=1.00\cdot10^{16}\,$G to
increment the magnetic field. First, we computed 2000 iterations at
$B=0$, then 2000 at $B=0.001\sqrt{\mathrm{MeV}\, \mathrm{fm}^{-3}}$
and then 4000 at each multiple of
$0.25\sqrt{\mathrm{MeV}\, \mathrm{fm}^{-3}}$, as long as we reached
convergence. For \isotope[12]{C}, a calculation starting at $B=0$ did
not give good converging results for stronger magnetic
fields. However, a calculation starting at
$B=6\sqrt{\mathrm{\mathrm{MeV}\, \mathrm{fm}^{-3}}}$ did. With these
steps, we got better convergence for stronger magnetic
fields. However, for certain strengths of the magnetic field, no
convergence was achieved. Since we got a change of the occupation for
protons in \isotope[16]{O}, we changed the step size in this region,
see below for discussion.

To access the shape and size of the nuclei we examined the parameters
$r_\mathrm{rms} \equiv r_\mathrm{rms}^{(\mathrm{total})}$ of
Eq.~\eqref{eq_b_7h} and $\beta,\,\gamma$ of Eq.~\eqref{eq_b_7g}. Here
$\gamma=0\degree$ refers to a prolate deformed nucleus,
$\gamma=60\degree$ refers to an oblate deformed nucleus and angles
between $\gamma=0\degree$ and $\gamma=60\degree$ refer to a
deformation in a state between prolate and oblate. If $\beta=0$ the
nucleus is spherical (undeformed) independent of $\gamma$. For
non-zero $\beta$ nuclei are deformed \cite{1995GM}.

Moreover, we calculated the current and spin densities. We evaluated
them separately for neutrons and protons normalised with the particle
density. The current density over the particle density results in the
collective flow velocity. Since the particle density approaches $0$
outside of the nucleus, we had to do a cut off and displayed these
quantities only in the region with $\rho_{n/p}>0.01\,$fm$^{-3}$. All
figures for the velocity (Figs.~\ref{fig_2_08}, \ref{fig_2_10} and
\ref{fig_2_12}) and for the spin (Figs.~\ref{fig_2_09}, \ref{fig_2_11}
and \ref{fig_2_13}) are done from the same perspective,
respectively. For a specific nucleus, magnetic field and quantity
(current or spin density) we used the same scaling for the neutron and
proton quantity. To show the position of these vectors relative to the
nucleus, we added the particle density ($\rho=\rho_n+\rho_p$) as
background. The scaling of this particle density is always the same
for both quantities (current and spin density) for both isospins for
all magnetic fields, but differs for different nuclei. These figures
were created with VisIt~\cite{HPV:VisIt}. A note on the corresponding
coding of the vectors: In each figure, the magnitudes of the vectors
are specified by the length of the vector and by its color. The color
changes from red for high values via yellow, green and cyan to blue
for low values.

\subsubsection{Effects of the magnetic field on \isotope[16]{O}}
\begin{figure}[!]
  \begin{center}
    \includegraphics[width=\textwidth]{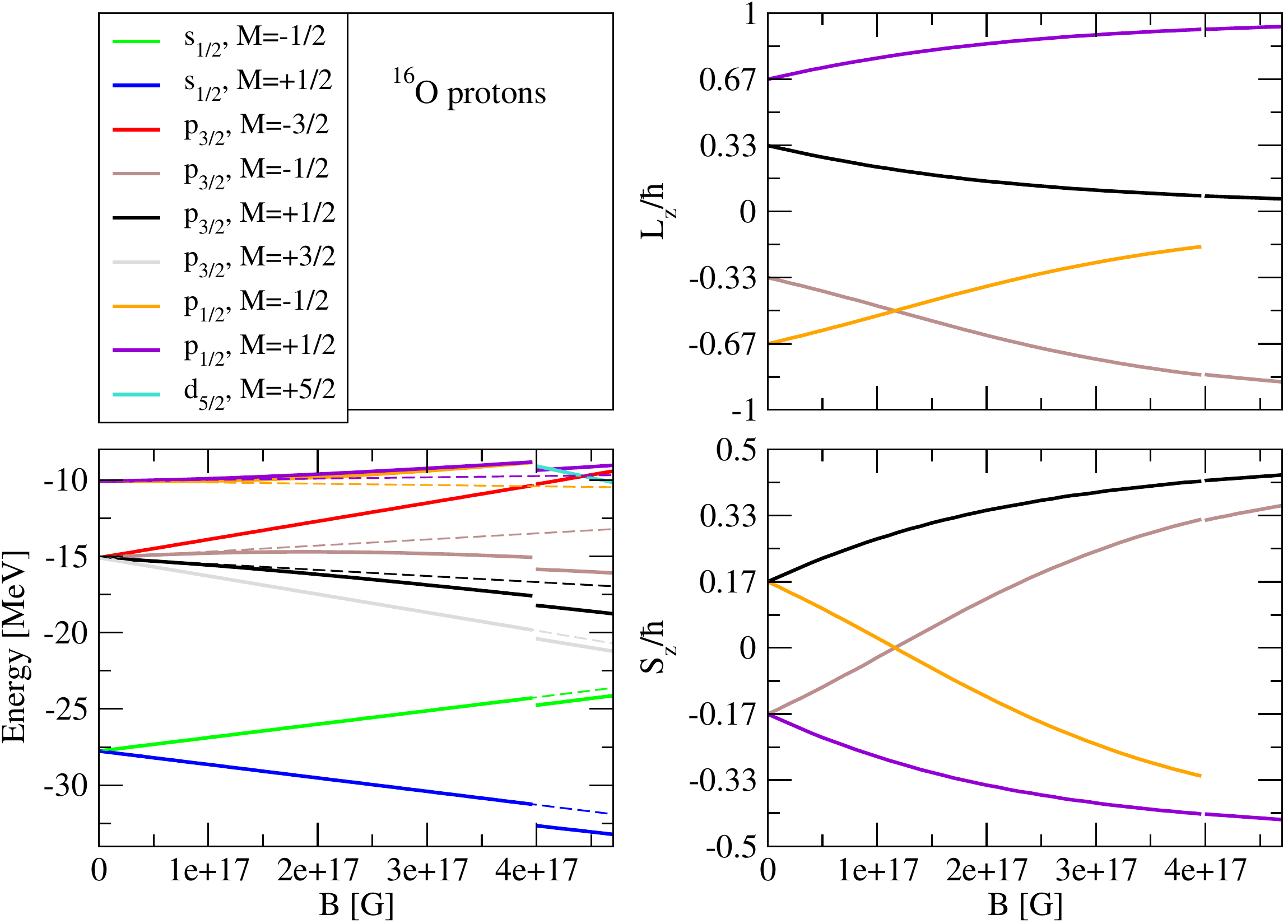}
    \caption[The energy levels, $\left<L_z\right>$ and
    $\left<S_z\right>$ as functions of the magnetic field for protons
    in $^{16}${O}.] {The energy levels (analytical (dashed) and
      numerical (solid)), $\left<L_z\right>$ and $\left<S_z\right>$ as
      functions of the magnetic field for protons in \isotope[16]{O}.}
    \label{fig_2_02}
  \end{center}
\end{figure}

\begin{figure}[!]
  \begin{center}
    \includegraphics[width=\textwidth]{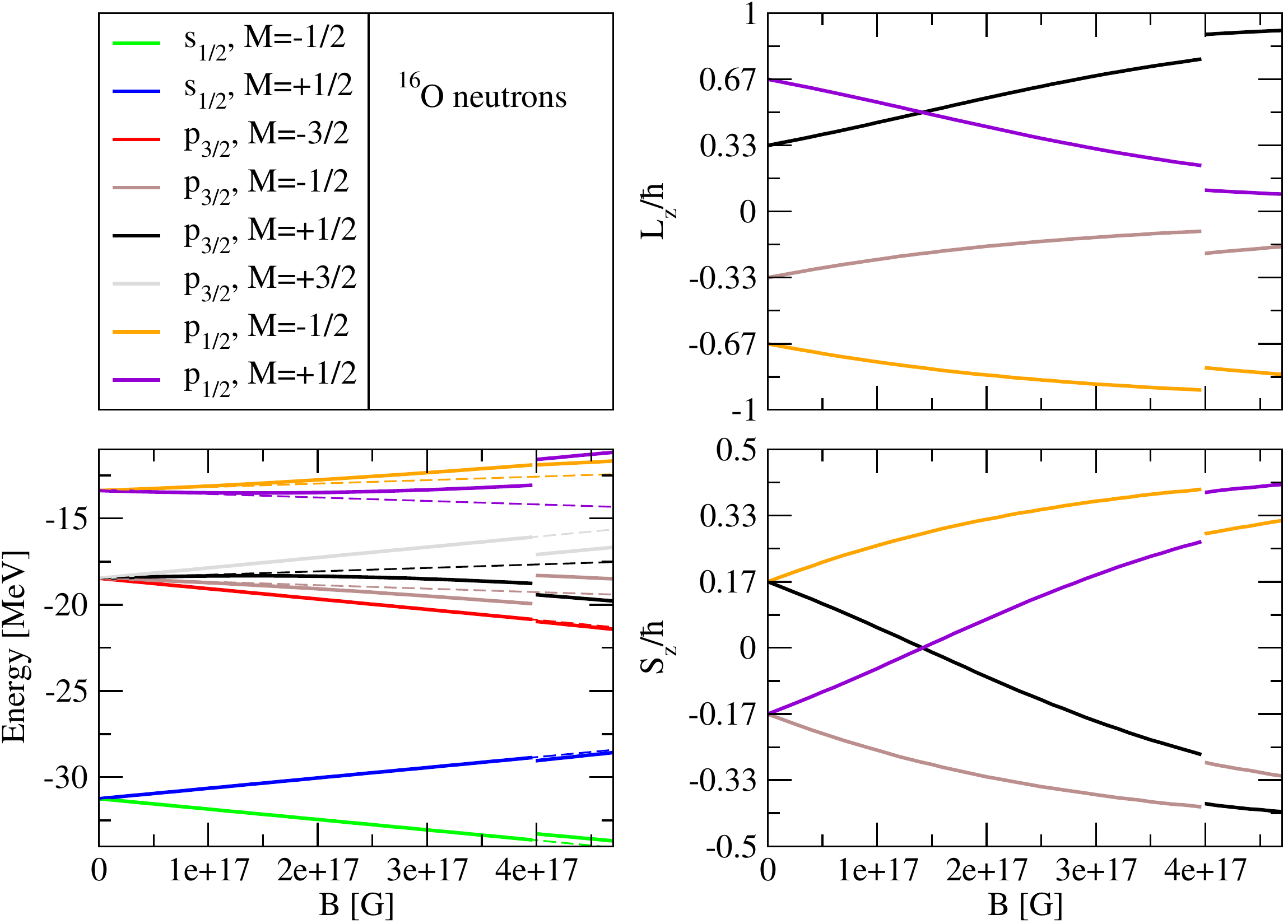}
    \caption[The energy levels, $\left<L_z\right>$ and
    $\left<S_z\right>$ as functions of the magnetic field for neutrons
    in $^{16}${O}.] {The energy levels (analytical (dashed) and
      numerical (solid)), $\left<L_z\right>$ and $\left<S_z\right>$ as
      functions of the magnetic field for neutrons in
      \isotope[16]{O}.}
    \label{fig_2_03}
  \end{center}
\end{figure}

We evaluated the effect of the magnetic field on different
nuclei. First we want to report our results for \isotope[16]{O}. Here
we calculated the energy levels and the $z$-components of orbital
angular momentum $\left<L_z\right>$ and spin $\left<S_z\right>$ of
neutrons and protons as functions of the magnetic field. The results
are shown in Figs.~\ref{fig_2_02} and~\ref{fig_2_03}. For the states
defined in Eqs.~\eqref{eq_2_10a}, \eqref{eq_2_10b}, \eqref{eq_2_10c}
and \eqref{eq_2_10f} we obtain integer or half-integer numbers for
$\left<L_z\right>/\hbar$ or $\left<S_z\right>/\hbar$ which are
identical to the quantum numbers $m_l$ or $m_s$, respectively,
independent of the magnetic field. However, for the states with mixed
Clebsch-Gordan coefficients defined in Eqs.~\eqref{eq_2_10d},
\eqref{eq_2_10e}, \eqref{eq_2_10g} and \eqref{eq_2_10h},
$\left<L_z\right>$ and $\left<S_z\right>$ change as functions of the
magnetic field. In Figs.~\ref{fig_2_02} and~\ref{fig_2_03}
$\left<L_z\right>$ and $\left<S_z\right>$ are only shown for those
state, where $\left<L_z\right>$ and $\left<S_z\right>$ change as
functions of the magnetic field.

We now take a closer look at these states. For all magnetic fields, we
obtain $M=m_l+m_s=(\left<L_z\right>+\left<S_z\right>)/\hbar$ as a good
quantum number. However, for the orbital angular momentum and the spin
there are two ranges of values for the magnetic field. In the limit of
weak magnetic fields the angular momentum and spin are coupled via the
{\it l-s} coupling. Then $\left<L_z\right>$ and $\left<S_z\right>$ are
given according to Eq.~\eqref{eq_2_15}. Because of the {\it l-s}
coupling the vectors of $\vecL$ and $\vecS$ are not aligned with the
magnetic field separately. The influence of the weak magnetic field on
the system is described by the Zeeman effect. In the regime of strong
magnetic fields the {\it l-s} coupling is ineffective, i.e., the
orbital angular momentum $l$ and the spin $s$ couple separately to the
magnetic field. In this case, the mixed states of Eq.~\eqref{eq_2_10}
reach asymptotically the following non mixed states:
\begin{subequations}
  \label{eq_2_19}
  \begin{eqnarray}
    p_{3/2},M=-\tfrac12:&\quad&\left|1,+0,\downarrow\right>\,,\\
    p_{3/2},M=+\tfrac12:&\quad&\left|1,+1,\downarrow\right>\,,\\
    p_{1/2},M=-\tfrac12:&\quad&\left|1,-1,\uparrow\right>\,,\\
    p_{1/2},M=+\tfrac12:&\quad&\left|1,+0,\uparrow\right>\,.
  \end{eqnarray}
\end{subequations}
The influence of the strong magnetic fields on the system is described
by the Paschen-Back effect.

For magnetic fields lower than
$B_{\isotope[16]{O},\,c}=4.0\cdot10^{17}\,$G, the lowest modes of the
harmonic oscillator, the $1s$ and $1p$ states, are filled up. The
shape of the nucleus is spherically symmetrical, as one would expect
for a doubly magic nucleus. A comparison between our numerical and
analytical results is given in Figs.~\ref{fig_2_02}
and~\ref{fig_2_03}. The analytical results for non-zero $B$ are
obtained in the following way: We first take the numerical solution
for $B=0$ and then use analytical formulae given by
Eq.~\eqref{eq_2_18} to obtain the splitting of the energies for
non-zero $B$.

For the states with pure Clebsch-Gordan coefficients (all $s$ states
and $p$ states with $M=\pm3/2)$, we obtain a good agreement between
the numerical and analytical results for the energy levels if the
magnetic field $B<B_{\isotope[16]{O},\,c}$. For the states with mixed
Clebsch-Gordan coefficients the analytical results differ
significantly from the numerical ones. For
$B>B_{\isotope[16]{O},\,c}$, we find a change in the occupation for
protons: the $d_{5/2},M=+5/2$ state becomes occupied instead of the
$p_{1/2},M=-1/2$ state. Therefore, in Fig.~\ref{fig_2_02}, the lines
corresponding to $p_{1/2},M=-1/2$ stop at
$B_{\isotope[16]{O},\,c}$. Since $\left<L_z\right>/\hbar=m_l=2$ and
$\left<S_z\right>/\hbar=m_s=1/2$ assume constant values for
$d_{5/2},M=+5/2$ for all magnetic fields we do not show these in
Fig.~\ref{fig_2_02} to keep it clear. However the corresponding energy
values, which are not constant, are shown. This redistribution in the
energy states affects the other proton and neutron states as well as
it slightly deforms the nucleus, i.e., the nucleus looses its
spherical symmetry.

For magnetic fields larger than
$B_{\isotope[16]{O},\,e}=4.7\cdot10^{17}\,$G we do not find
convergence. We expect that there are further redistributions of
energy states above this value of magnetic field which we leave for
future studies.

\begin{figure}[!]
  \begin{center}
    \includegraphics[width=0.49\textwidth]{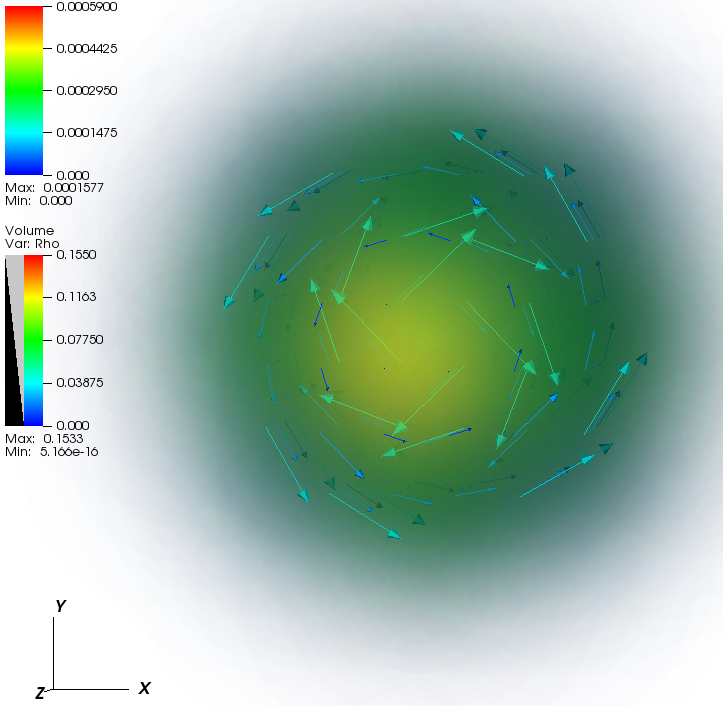}
    \includegraphics[width=0.49\textwidth]{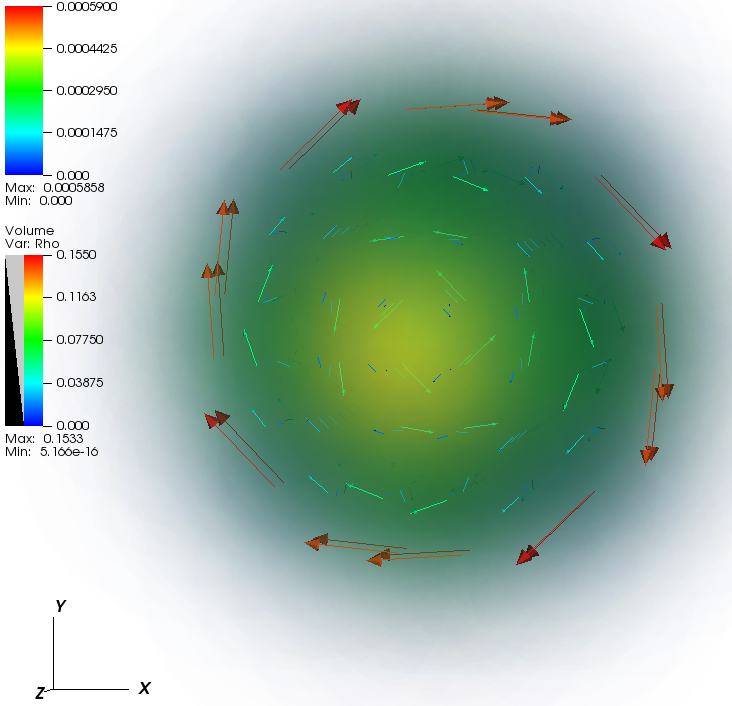}
    \includegraphics[width=0.49\textwidth]{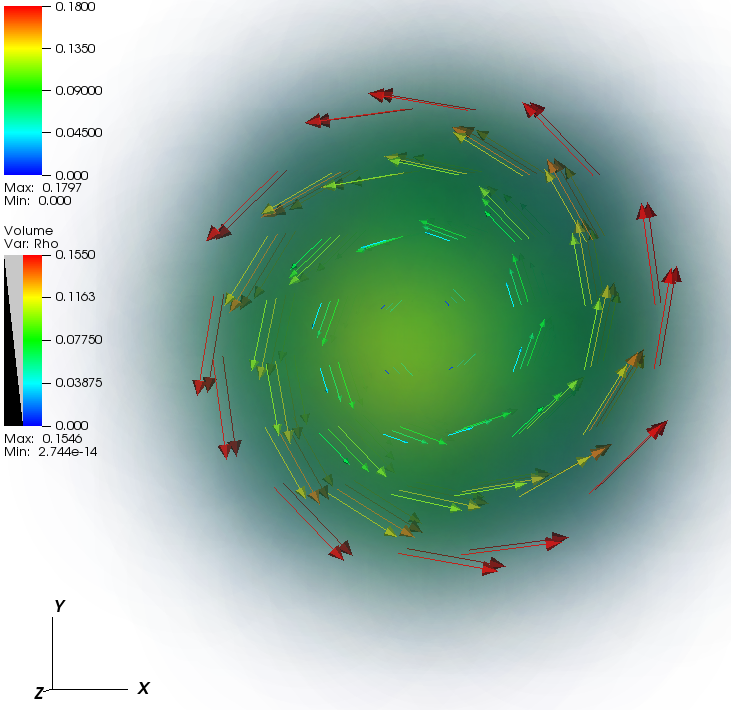}
    \includegraphics[width=0.49\textwidth]{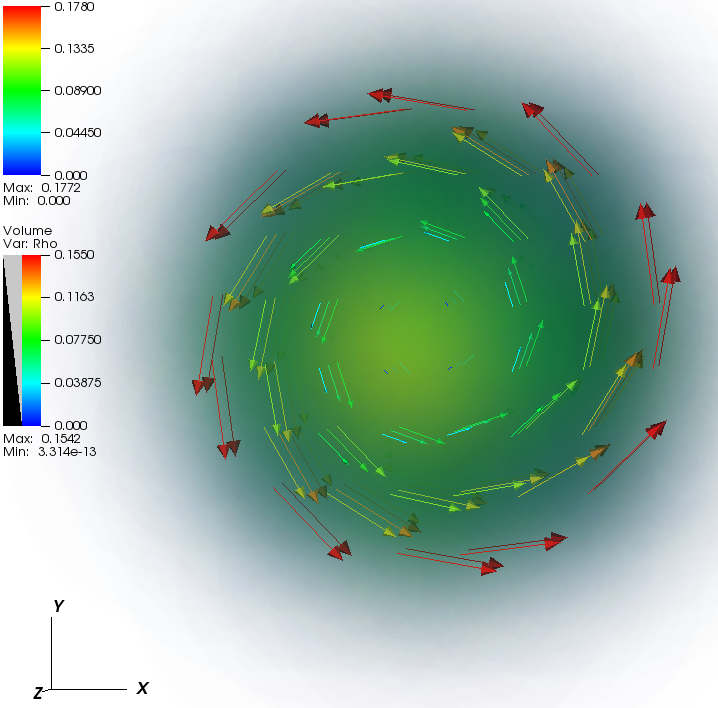}
    \caption[The collective flow velocity for neutrons and protons for
    $^{16}${O} for different values of the magnetic field.] {The
      current density divided by the particle density (collective flow
      velocity) for neutrons (top left) and protons (right and bottom)
      for \isotope[16]{O} for different values of the magnetic
      field. Top: $B=3.9\cdot10^{17}\,$G, bottom left:
      $B=4.1\cdot10^{17}\,$G, bottom right: $B=4.7\cdot10^{17}\,$G. As
      background the particle density is added. This figure was
      created with VisIt~\cite{HPV:VisIt}.}
    \label{fig_2_08}
  \end{center}
\end{figure}

\begin{figure}[!]
  \begin{center}
    \includegraphics[width=0.49\textwidth]{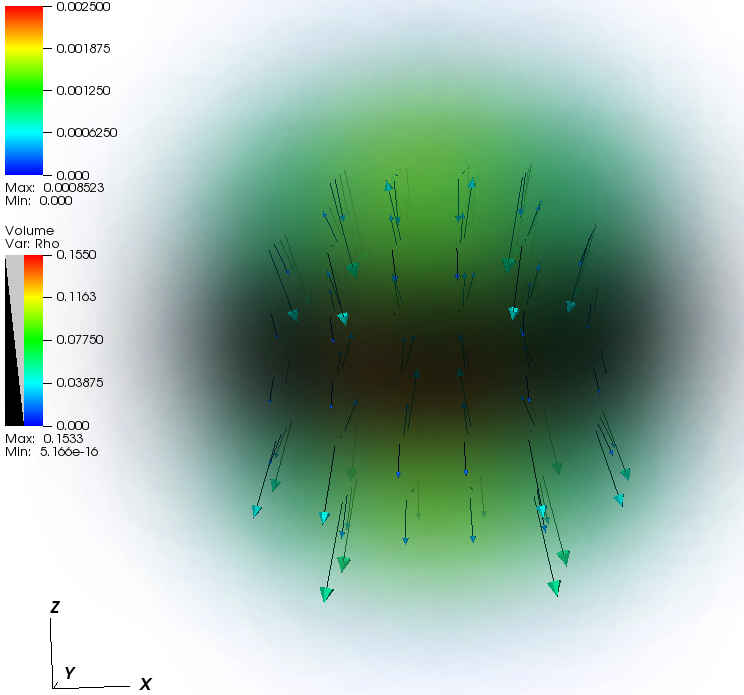}
    \includegraphics[width=0.49\textwidth]{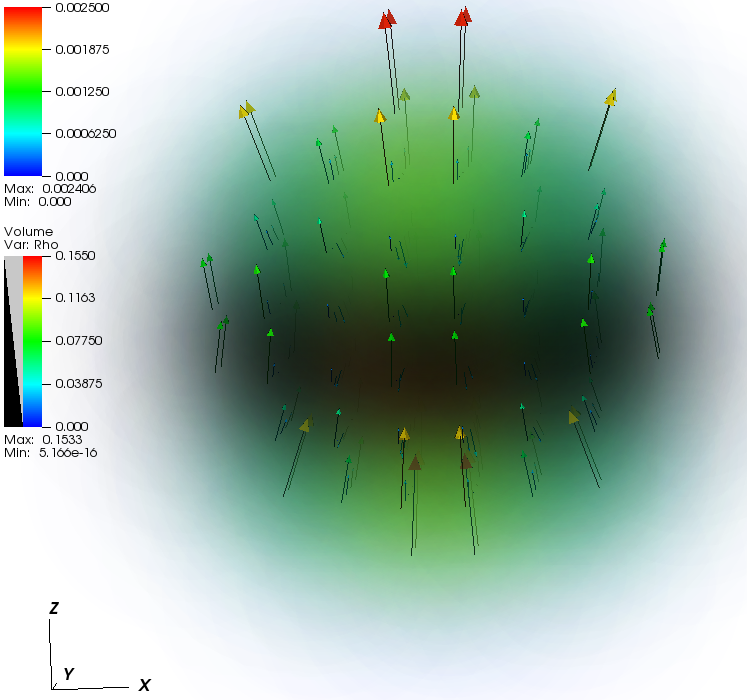}
    \includegraphics[width=0.49\textwidth]{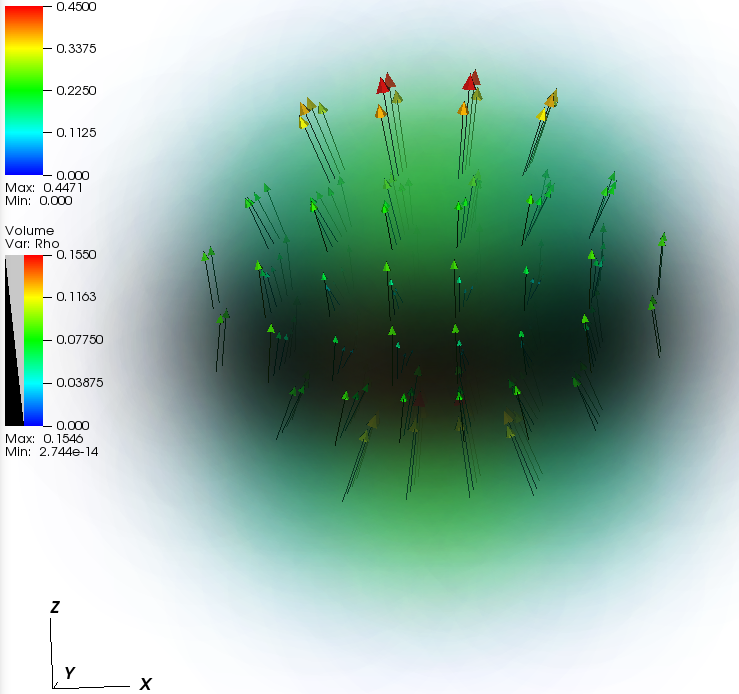}
    \includegraphics[width=0.49\textwidth]{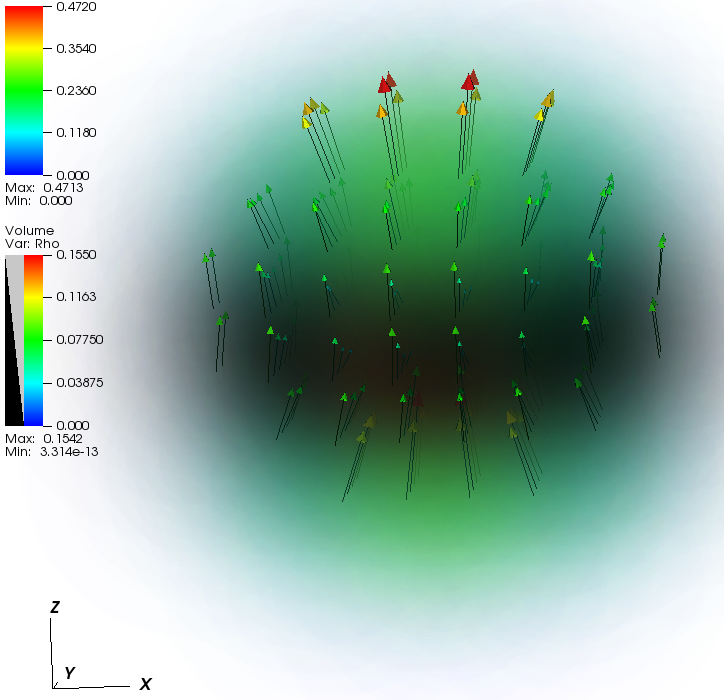}
    \caption[The spin density divided by the particle density for
    neutrons and protons for $^{16}${O} for different values of the
    magnetic field.] {The spin density divided by the particle density
      for \isotope[16]{O} with the same arrangement and magnetic
      fields as in Fig.~\ref{fig_2_08}. This figure was created with
      VisIt~\cite{HPV:VisIt}.}
    \label{fig_2_09}
  \end{center}
\end{figure}

Next we want to look at the current and the spin densities shown in
Fig.~\ref{fig_2_08} and Fig.~\ref{fig_2_09}, respectively. For
stronger magnetic fields, only the proton values are shown, because
the neutron values are much smaller. In the top panels the magnetic
field is $B=3.9\cdot10^{17}\,$G; neutron quantities are shown on the
left and proton quantities on the right. In the bottom panels proton
quantities are shown for stronger magnetic fields: on the left for
$B=4.1\cdot10^{17}\,$G and on the right for $B=4.7\cdot10^{17}\,$G. We
chose one value slightly below $B_{\isotope[16]{O},\,c}$, one slightly
above and finally $B_{\isotope[16]{O},\,e}$. Below the redistribution,
the current and spin densities of neutrons and protons are roughly of
the same order of magnitude, but in opposite direction. Above the
redistribution, we see an alignment of the proton quantities. A
comparison of the values of neutrons and protons yields that the
proton values are much higher, more than one order of magnitude. Since
the neutron quantities are much lower, no meaningful statement can be
made. The current density is perpendicular to the magnetic field. For
the protons we see a current at the surface of the nucleus, whereas
the current inside is relatively low. In the inner parts we
approximately obtain a rigid body rotation. In Fig.~\ref{fig_2_02} we
see that the energy levels of the $p$ and $d$ states are very close,
whereas the $s$ states are below. Therefore we can get a mixing of
states resulting in a current at the surface.

Finally, we want to consider the shape and the size of the
\isotope[16]{O} nucleus for non-zero magnetic fields. For
$B<B_{\isotope[16]{O},\,c}$, the lowest states of the harmonic
oscillator are filled and the shape of the nucleus is spherical. Its
radius is $r_\mathrm{rms}=2.69\,$fm. For
$B_{\isotope[16]{O},\,c}<B<B_{\isotope[16]{O},\,e}$ the nucleus is
deformed with deformation parameters $\beta=0.1$ and
$\gamma=60\degree$, which imply that the deformation is oblate. The
mean radius is $r_\mathrm{rms}=2.72\,$fm in this case.

\subsubsection{Effects of the magnetic field on \isotope[12]{C}}
\begin{figure}[!]
  \begin{center}
    \includegraphics[width=\textwidth]{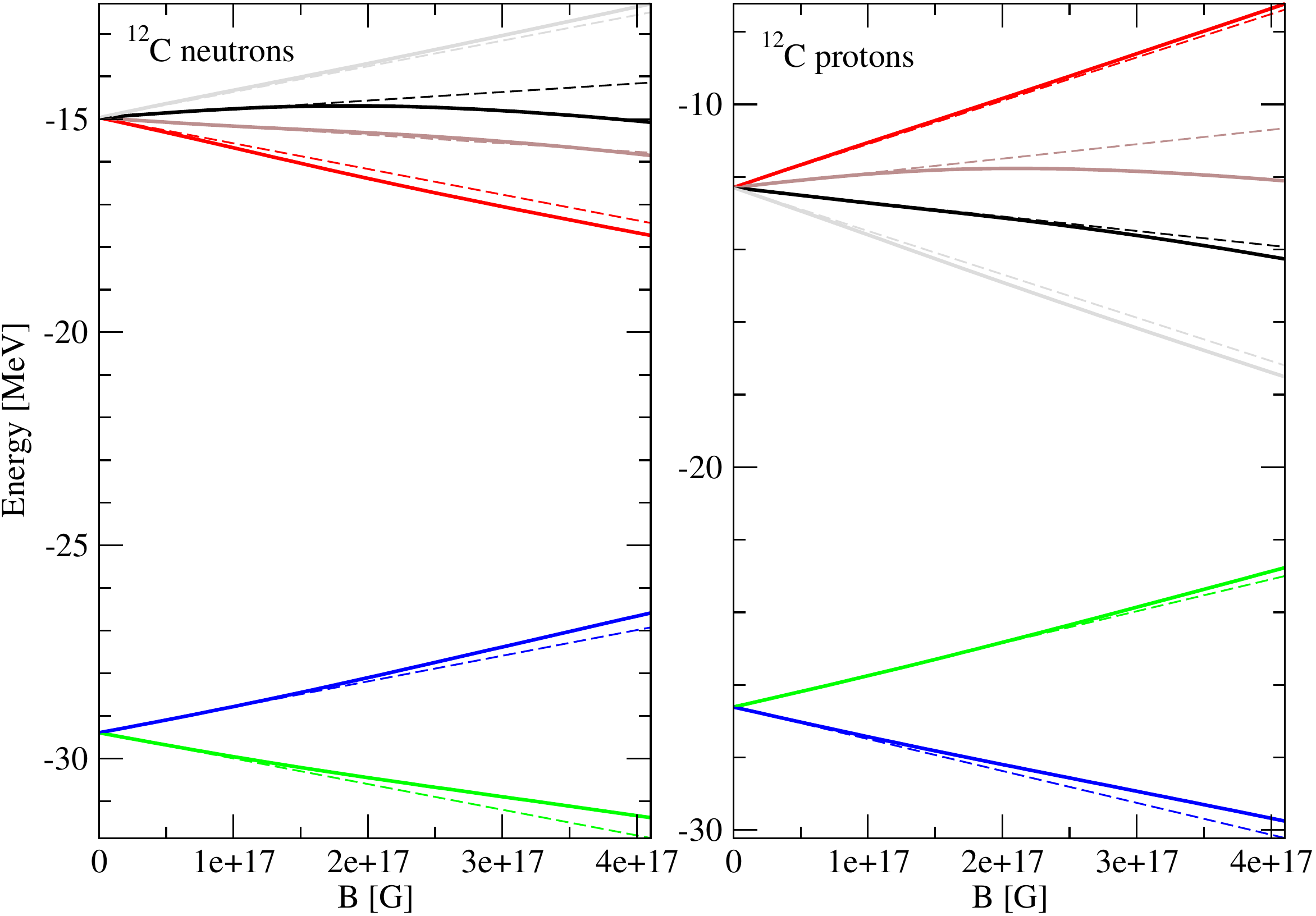}
    \caption[The energy levels as functions of the magnetic field for
    neutrons and protons in $^{12}$C.] {The energy levels (analytical
      (dashed) and numerical (solid)) as functions of the magnetic
      field for neutrons and protons in \isotope[12]{C}. The color
      code is analog to Figs.~\ref{fig_2_02} and~\ref{fig_2_03}.}
    \label{fig_2_04}
  \end{center}
\end{figure}

We have repeated the computations done for \isotope[16]{O} also for
the nucleus \isotope[12]{C}. In Fig.~\ref{fig_2_04} we present the
energy levels. The difference between the analytical and the numerical
results is much greater in this case compared to the \isotope[16]{O}
nucleus.

\begin{figure}[!]
  \begin{center}
    \includegraphics[width=\textwidth]{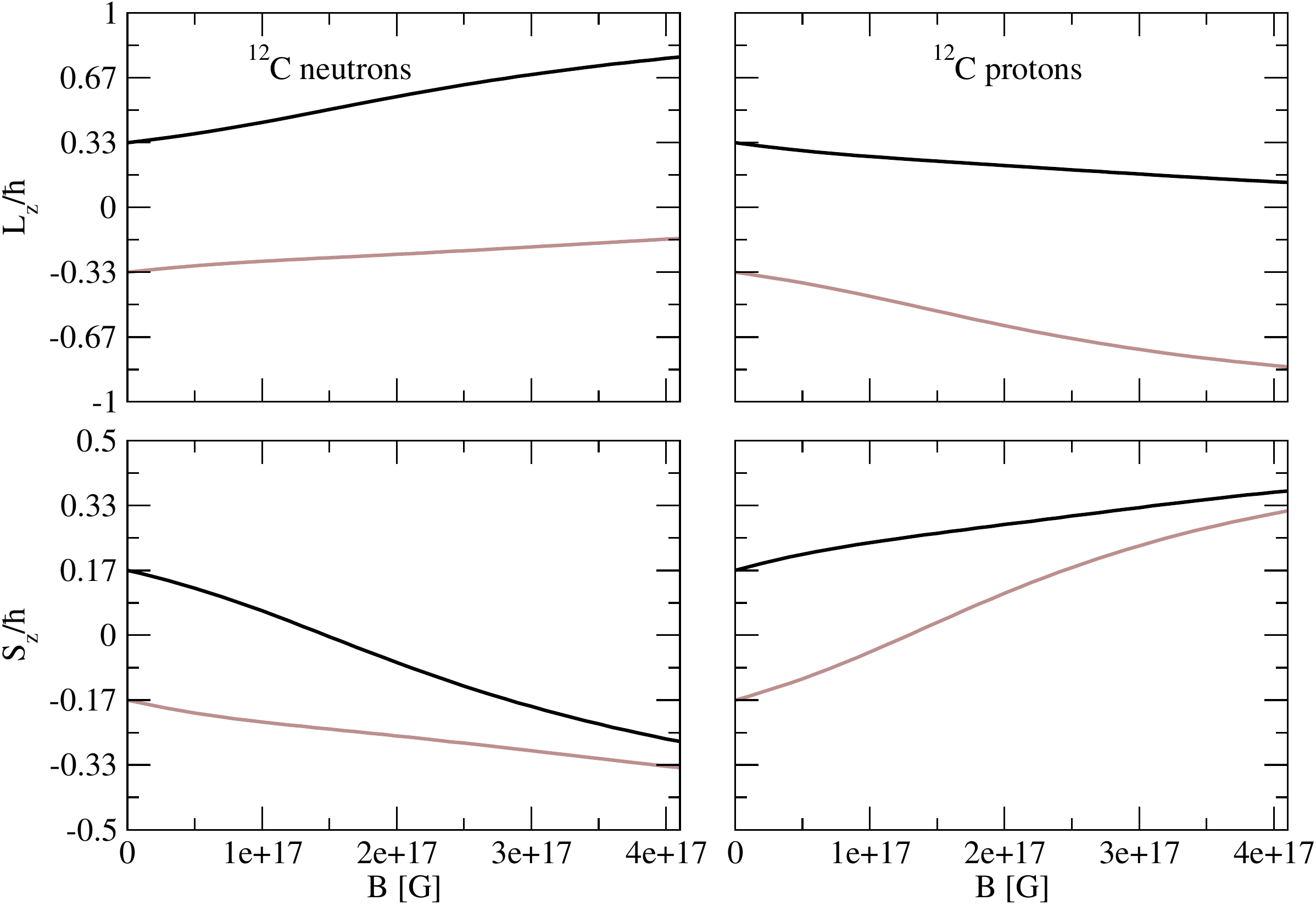}
    \caption[The angular momentum and the spin as functions of the
    magnetic field for neutrons and protons in $^{12}$C.] {The angular
      momentum $\left<L_z\right>$ and the spin $\left<S_z\right>$ as
      functions of the magnetic field for neutrons and protons in
      \isotope[12]{C}. The color code is analog to
      Figs.~\ref{fig_2_02} and~\ref{fig_2_03}.}
    \label{fig_2_05}
  \end{center}
\end{figure}

In Fig.~\ref{fig_2_05} we show only those $\left<L_z\right>/\hbar$ and
$\left<S_z\right>/\hbar$ components of \isotope[12]{C} which differ
from (half-)integer values. These are again those with mixed
Clebsch-Gordan coefficients. They behave in the same way as those of
\isotope[16]{O}.

The shape of the nucleus does not change much. At $B=0$, it is
spherical symmetric with $r_\mathrm{rms}=2.47\,$fm, increasing
slightly to 2.51\,fm for $B_{\isotope[12]{C},\,e}=4.1\cdot10^{17}\,$G
which is defined in analogy to $B_{\isotope[16]{O},\,e}$. Increasing
the magnetic field also results in a smooth deformation from $\beta=0$
at $B=0$ to $\beta=0.071$ at $B=B_{\isotope[12]{C},\,e}$. The
deformation is always oblate with $\gamma=60\degree$.

\begin{figure}[!]
  \begin{center}
    \includegraphics[width=0.49\textwidth]{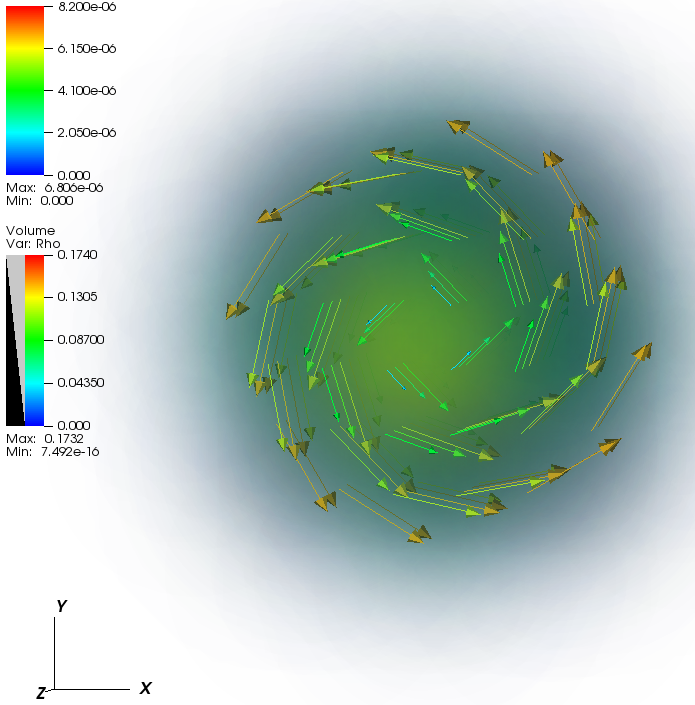}
    \includegraphics[width=0.49\textwidth]{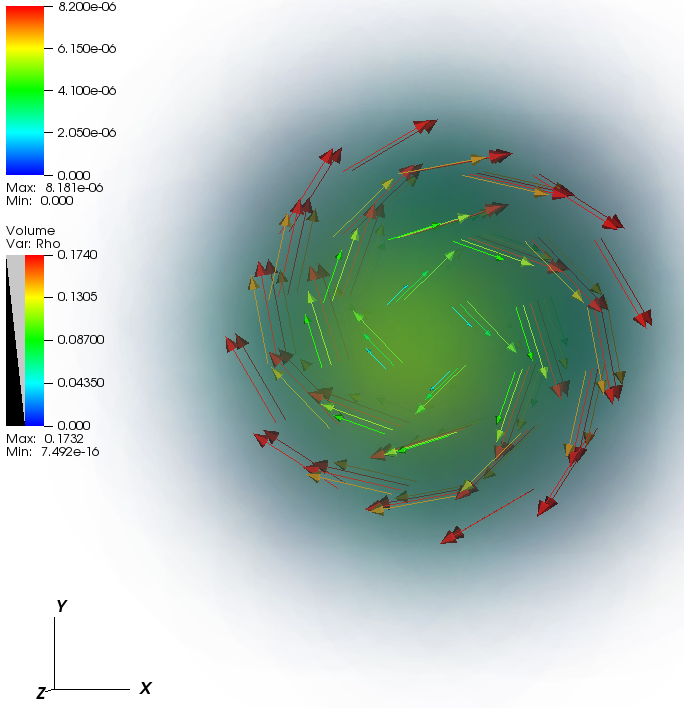}
    \includegraphics[width=0.49\textwidth]{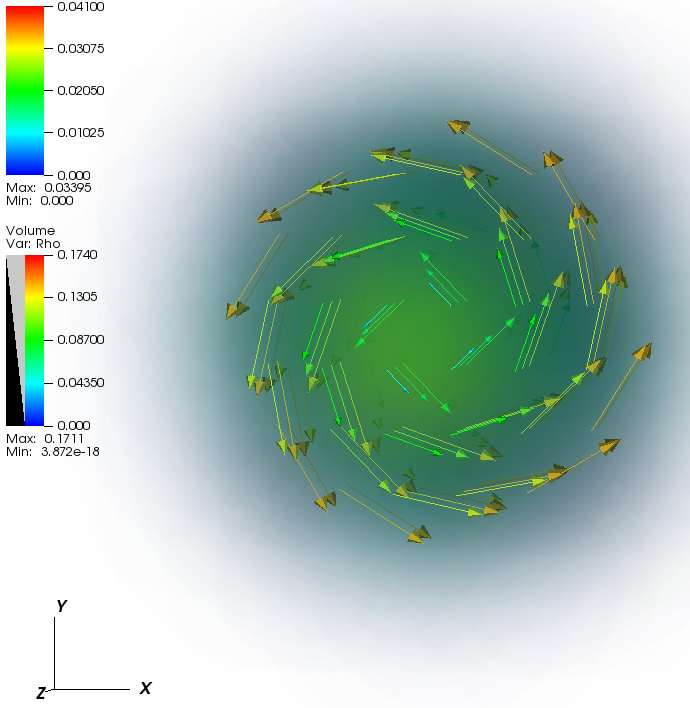}
    \includegraphics[width=0.49\textwidth]{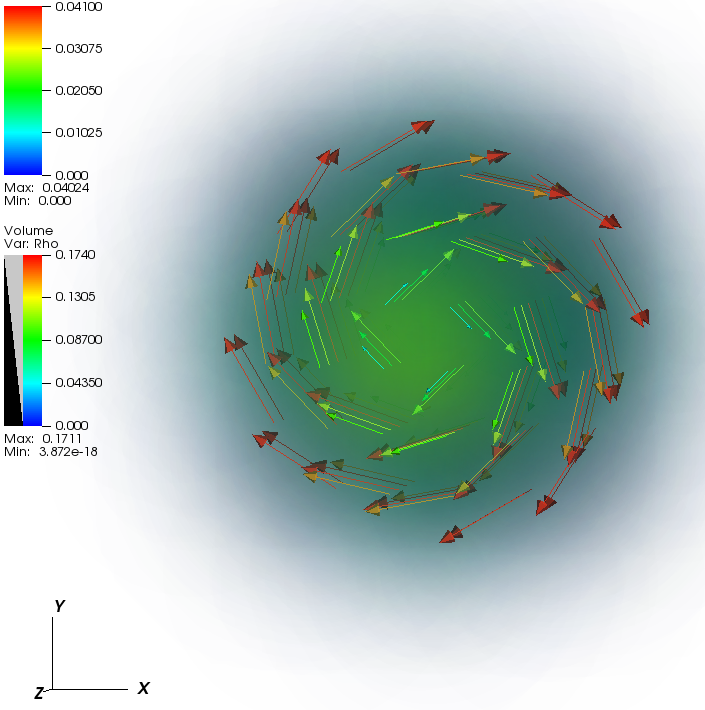}
    \includegraphics[width=0.49\textwidth]{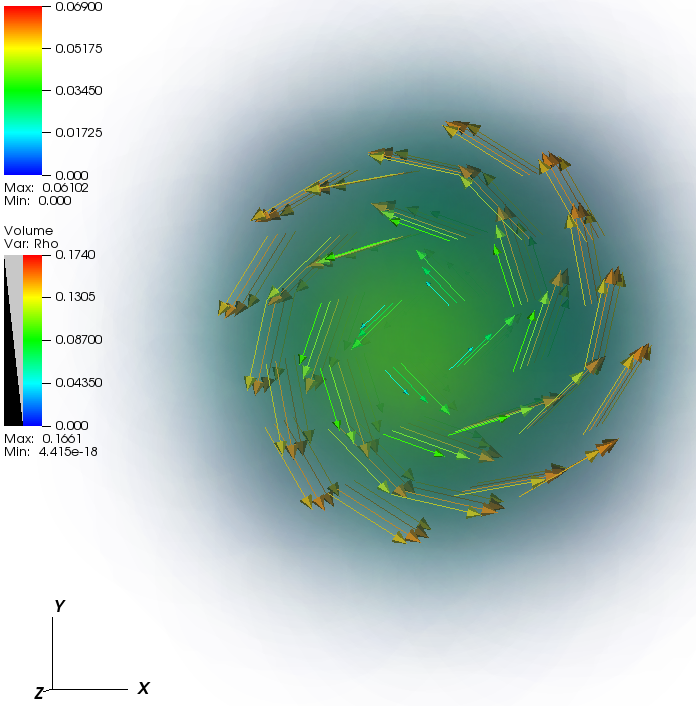}
    \includegraphics[width=0.49\textwidth]{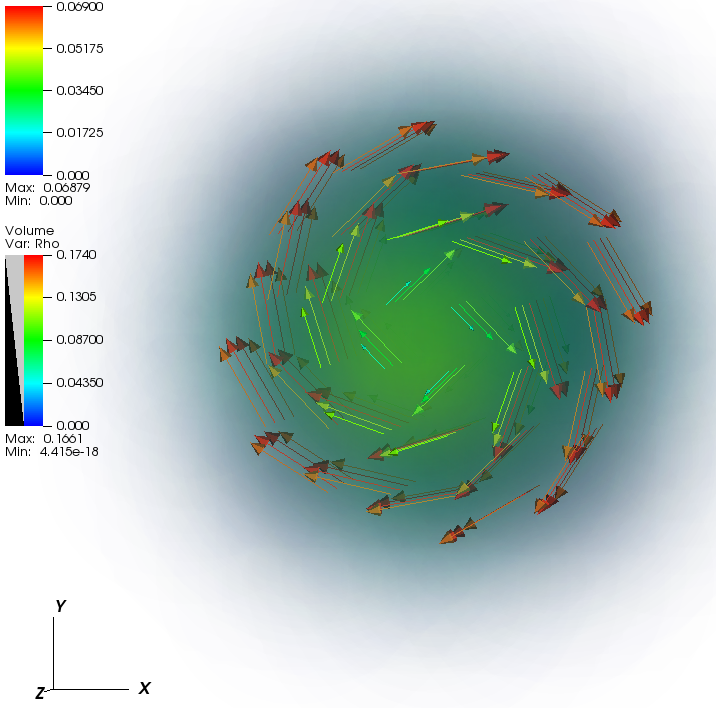}
    \caption[The collective flow velocity for neutrons and protons for
    $^{12}$C for different values of the magnetic field.] {The current
      density divided by the particle density (collective flow
      velocity) for neutrons (left) and protons (right) for
      \isotope[12]{C} for different values of the magnetic field. Top:
      $B=4.0\cdot10^{13}\,$G, middle: $B=2.0\cdot10^{17}\,$G, bottom:
      $B=4.1\cdot10^{17}\,$G. As background the particle density is
      added. This figure was created with VisIt~\cite{HPV:VisIt}.}
    \label{fig_2_10}
  \end{center}
\end{figure}

\begin{figure}[!]
  \begin{center}
    \includegraphics[width=0.49\textwidth]{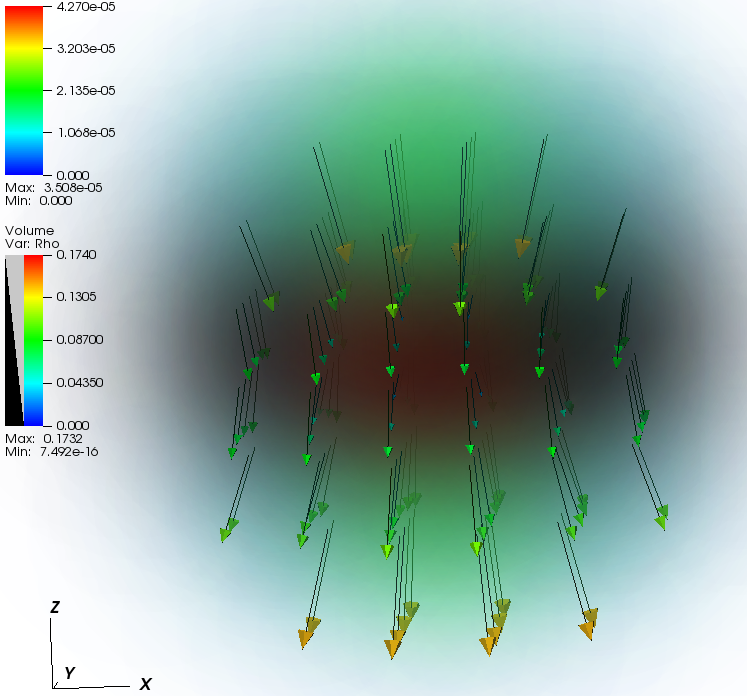}
    \includegraphics[width=0.49\textwidth]{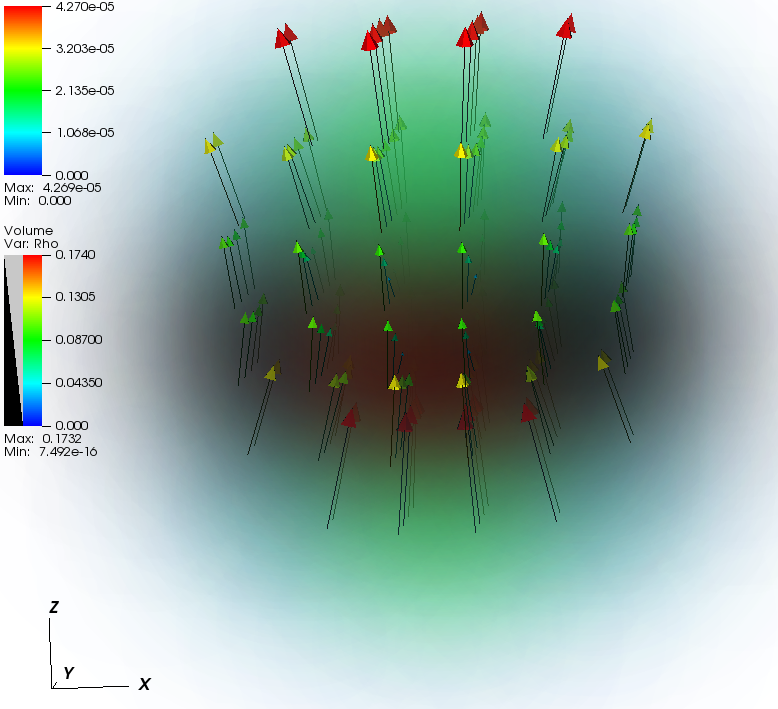}
    \includegraphics[width=0.49\textwidth]{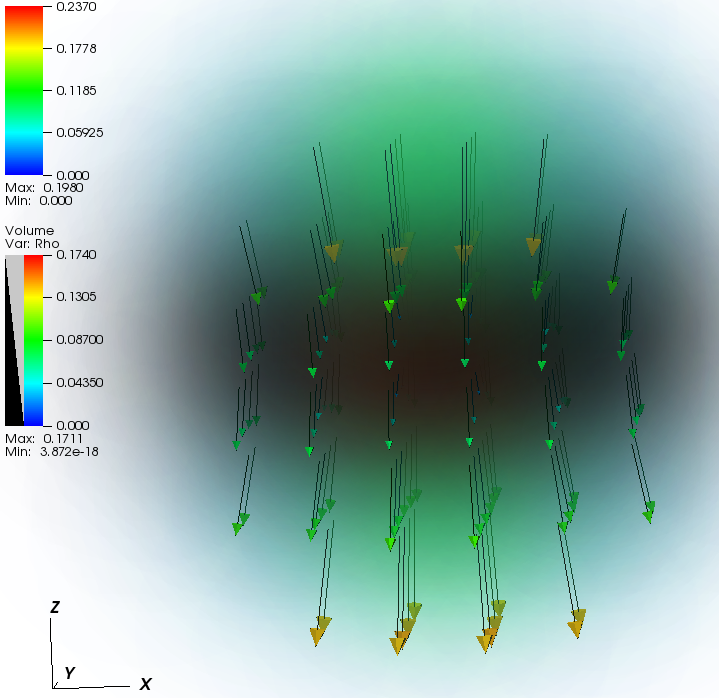}
    \includegraphics[width=0.49\textwidth]{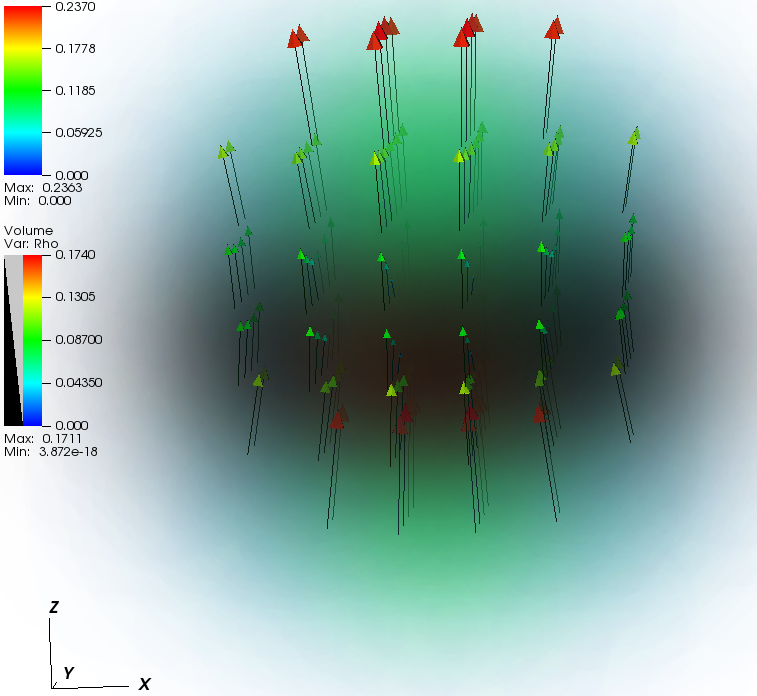}
    \includegraphics[width=0.49\textwidth]{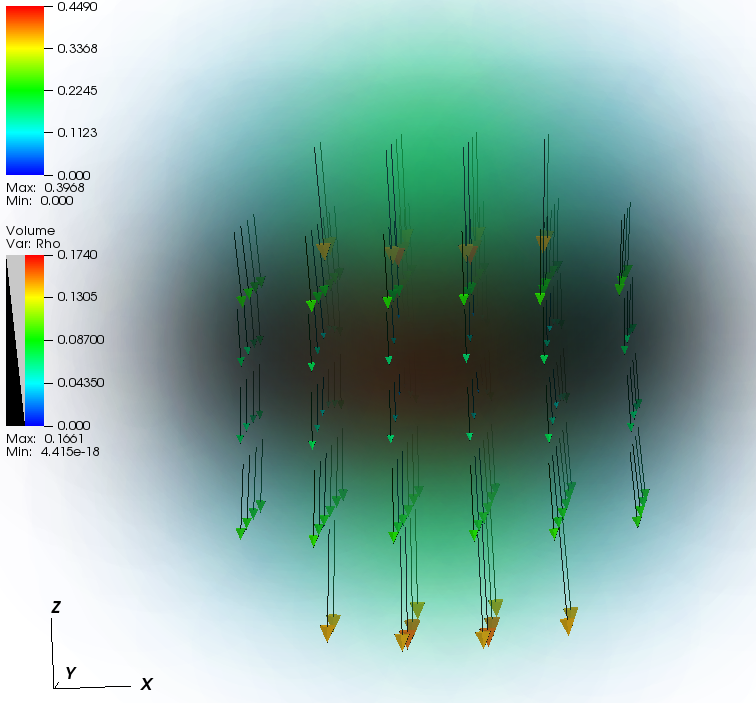}
    \includegraphics[width=0.49\textwidth]{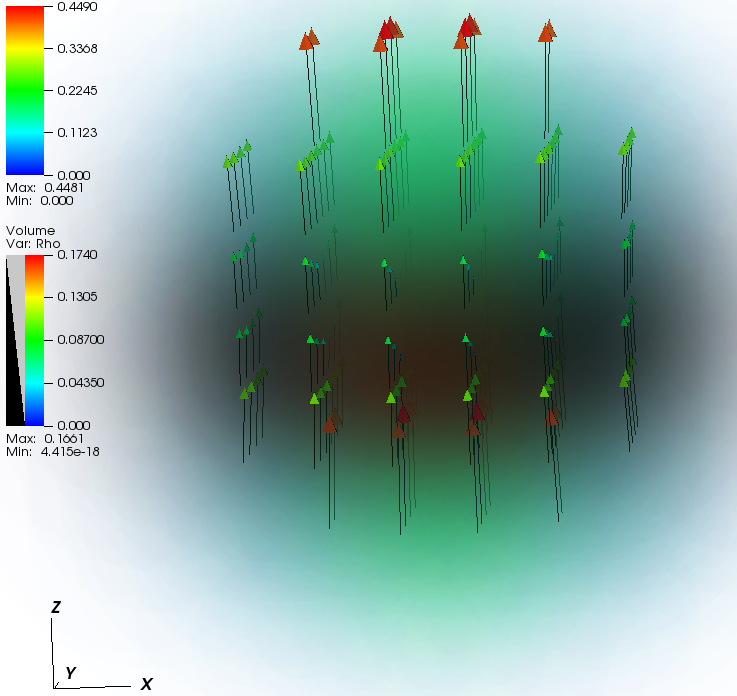}
    \caption[The spin density divided by the particle density for
    neutrons and protons for $^{12}$C for different values of the
    magnetic field.] {The spin density divided by the particle density
      for \isotope[12]{C} with the same arrangement and magnetic
      fields as in Fig.~\ref{fig_2_10}. This figure was created with
      VisIt~\cite{HPV:VisIt}.}
    \label{fig_2_11}
  \end{center}
\end{figure}

We also evaluated the current and spin densities in
Figs.~\ref{fig_2_10} and \ref{fig_2_11} analogously to
\isotope[16]{O}. We chose three magnetic fields. An infinitesimal
magnetic field, $B_{i}=4.0\cdot10^{13}\,$G,
$B_{\isotope[12]{C},\,h}=2.0\cdot10^{17}\,$G and
$B_{\isotope[12]{C},\,e}$. Hereby
$B_{\isotope[12]{C},\,h}\approx1/2\cdot B_{\isotope[12]{C},\,e}$.
$B_{i}$ is large enough for an orientation on the magnetic field, but
too small to have other significant effects. For both quantities, the
absolute values are approximately equal at each magnetic field for
protons and neutrons, but the direction is opposite. The values
increase with increasing magnetic field. We see that the current is
concentrated at the surface for neutrons and protons, as we obtained
for the protons of \isotope[16]{O}. For the spin we clearly see the
change from the Zeeman effect to the Paschen-Back effect: For weak
magnetic fields the {\it l-s} coupling is dominant, whereas for strong
magnetic fields the spin is aligned with the $z$-axis.

To summarize, the \isotope[12]{C} nucleus in a magnetic field behaves
similarly to \isotope[16]{O} for $B<B_{\isotope[16]{O},\,c}$. We do
not find any redistribution of energy states in this case up to the
strongest converging magnetic field. However, we expect
redistributions to occur also for \isotope[12]{C}. For \isotope[12]{C}
we find a similar behavior as for \isotope[16]{O} regarding the energy
levels and the $z$-components of the s.p. angular momentum and
spin. The alignment of spin density was more pronounced for
\isotope[12]{C} than for \isotope[16]{O}.

\subsubsection{Effects of the magnetic field on \isotope[20]{Ne}}
Among the nuclei considered in this study (\isotope[12]{C},
\isotope[16]{O} and \isotope[20]{Ne}) the nucleus \isotope[20]{Ne} is
the one which is farthest away from the closed shell
structure. Indeed, \isotope[16]{O} is a double magic nucleus which has
all $1s$ and $1p$ states filled. The \isotope[12]{C} is also a nucleus
which has all $1s$ and all $1p_{3/2}$ states filled. If the states are
filled according to the harmonic oscillator, in addition to the $1s$
and $1p$ states two $1d_{5/2}$ states should be filled in the case of
\isotope[20]{Ne} nucleus.

We do not find states with half-integer numbers for
$(\left<L_z\right>+\left<S_z\right>)/\hbar$. This nucleus is deformed
and the symmetry axis of the nucleus is not equal to the axis of the
magnetic field. Therefore the $z$-components are not good quantum
numbers. Since we have no good quantum numbers for the
$(\left<L_z\right>+\left<S_z\right>)/\hbar$ states, we do not present
energy states and $\left<L_z\right>$ and $\left<S_z\right>$ for
\isotope[20]{Ne} in contrast to the evaluation of the other nuclei.

\begin{figure}[!]
  \begin{center}
    \includegraphics[width=\textwidth]{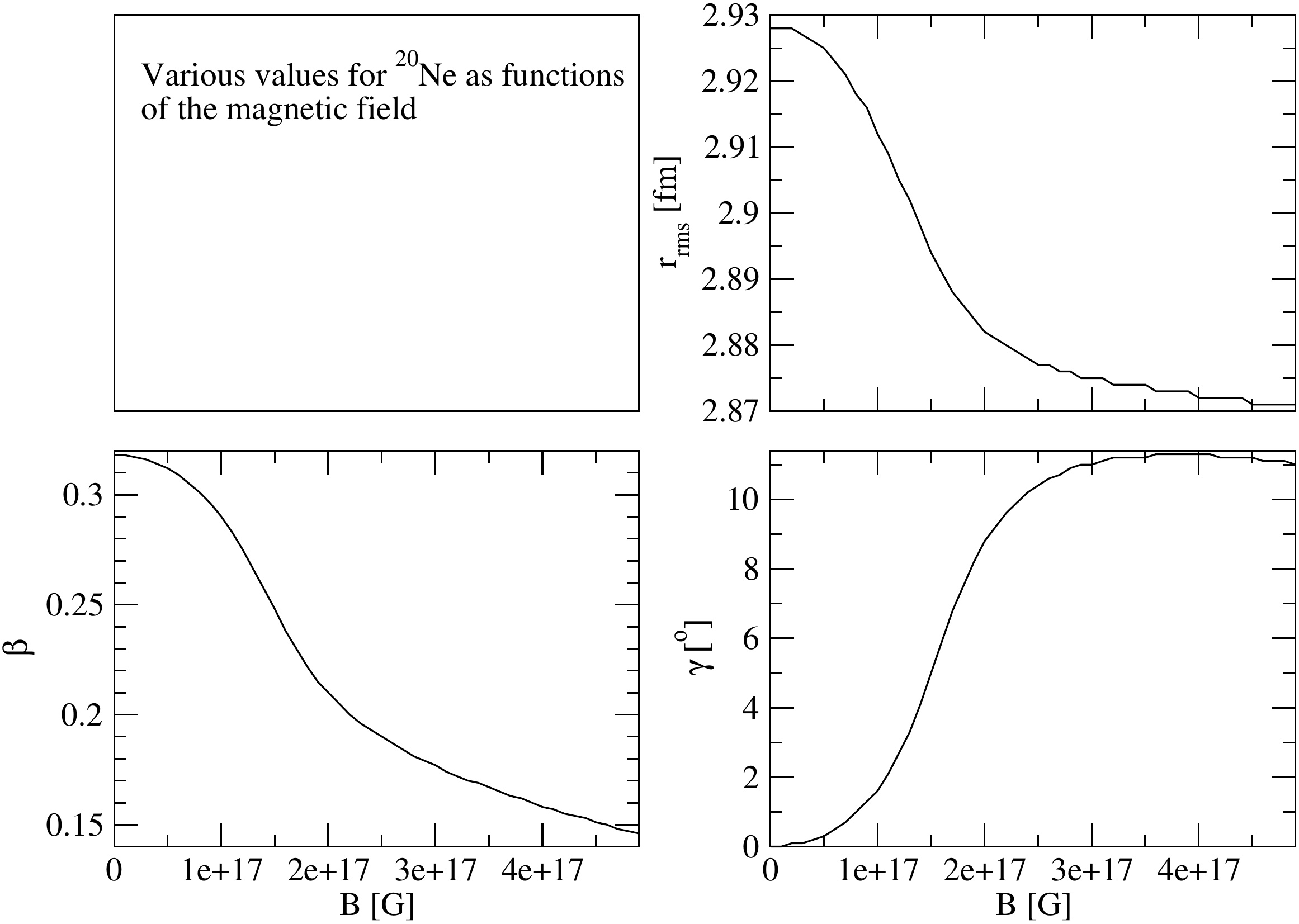}
    \caption{The $r_\mathrm{rms}$, $\beta$ and $\gamma$ as functions
      of the magnetic field for \isotope[20]{Ne}.}
    \label{fig_2_06}
  \end{center}
\end{figure}

\begin{figure}[!]
  \begin{center}
    \includegraphics[width=0.32\textwidth]{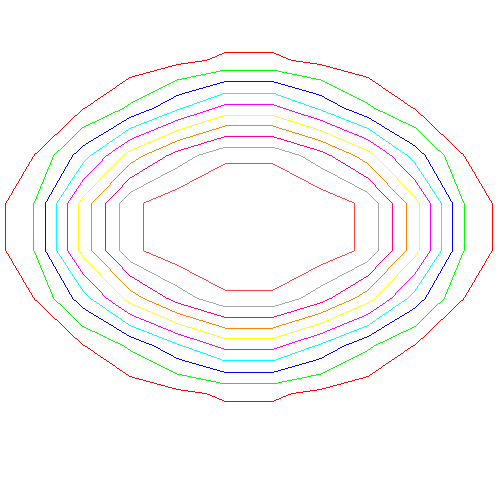}
    \includegraphics[width=0.32\textwidth]{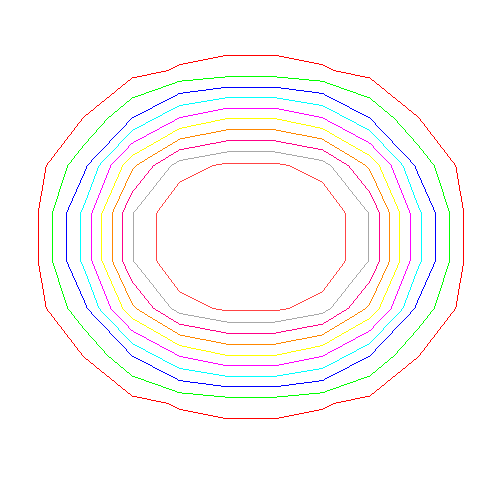}
    \includegraphics[width=0.32\textwidth]{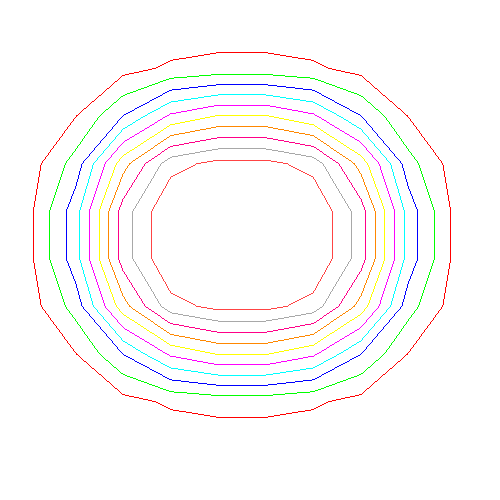}
    \caption[Illustration of the deformed $^{20}$Ne nucleus for
    different values of the magnetic field.] {The deformed
      \isotope[20]{Ne} nucleus for $B=0$, $B=2.4\cdot10^{17}\,$G and
      $B=4.9\cdot10^{17}\,$G. This figure was created with
      VisIt~\cite{HPV:VisIt}.}
    \label{fig_2_07}
  \end{center}
\end{figure}

In contrast to \isotope[16]{O}, we do not have two sectors, but we
have continuous deformations as smooth functions of the magnetic
field. As for \isotope[16]{O}, we regard $r_\mathrm{rms}$, $\beta$ and
$\gamma$ for evaluating the size and shape of the nucleus, see
Figs.~\ref{fig_2_06} and~\ref{fig_2_07}. Fig.~\ref{fig_2_07} was
created with VisIt~\cite{HPV:VisIt}. Fig.~\ref{fig_2_06} shows some
parameters as functions of the magnetic fields. Fig.~\ref{fig_2_07}
shows the deformed nucleus for $B=0$,
$B_{\isotope[20]{Ne},\,h}=2.4\cdot10^{17}\,$G and
$B_{\isotope[20]{Ne},\,e}=4.9\cdot10^{17}\,$G, hereby
$B_{\isotope[20]{Ne},\,h}$ and $B_{\isotope[20]{Ne},\,e}$ are defined
in analogy to $B_{\isotope[12]{C},\,h}$ and
$B_{\isotope[12]{C},\,e}$. The radius $r_\mathrm{rms}$ slightly
decreases from 2.93\,fm to 2.87\,fm. However, $\beta$ decreases from
0.32 to 0.15 to a value being less than half of the original one. This
denotes a continuous significant change in the deformation, which can
be seen well in Figs.~\ref{fig_2_06} and~\ref{fig_2_07}. $\gamma$
starts at $0\degree$ for $B=0$, but increases asymptotically to
$11\degree$, denoting a change from a purely prolate deformed nucleus
to a mainly prolate deformed one.

\begin{figure}[!]
  \begin{center}
    \includegraphics[width=0.49\textwidth]{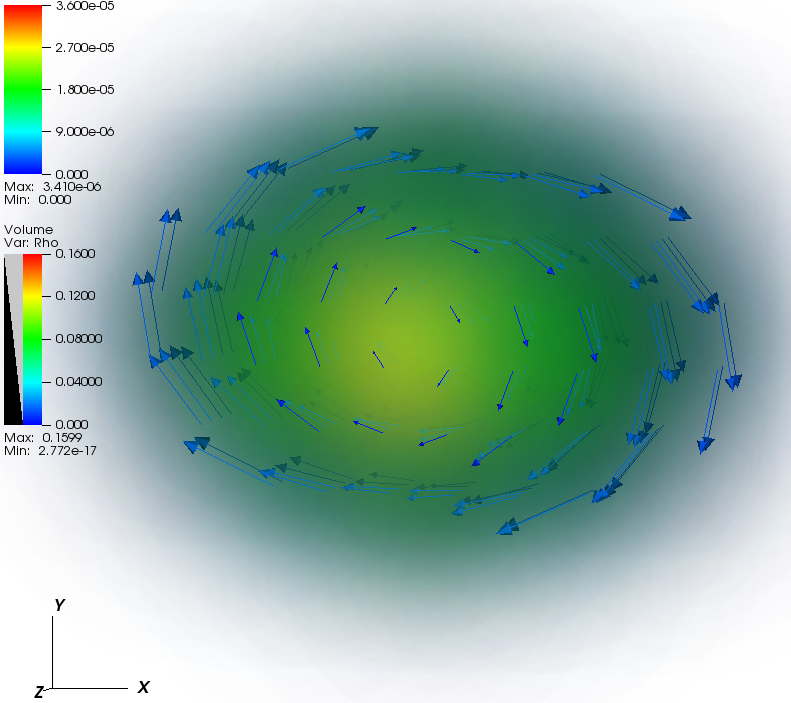}
    \includegraphics[width=0.49\textwidth]{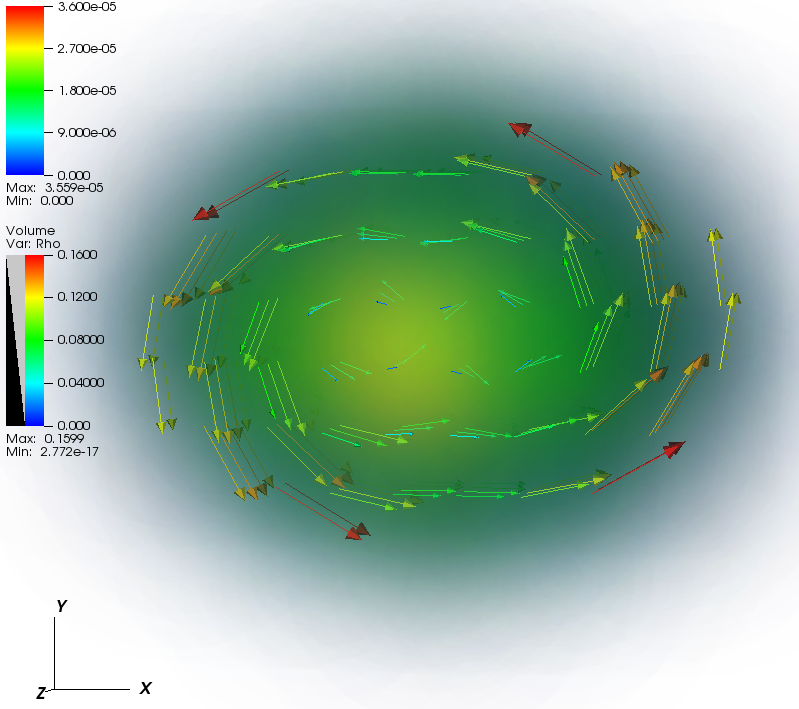}
    \includegraphics[width=0.49\textwidth]{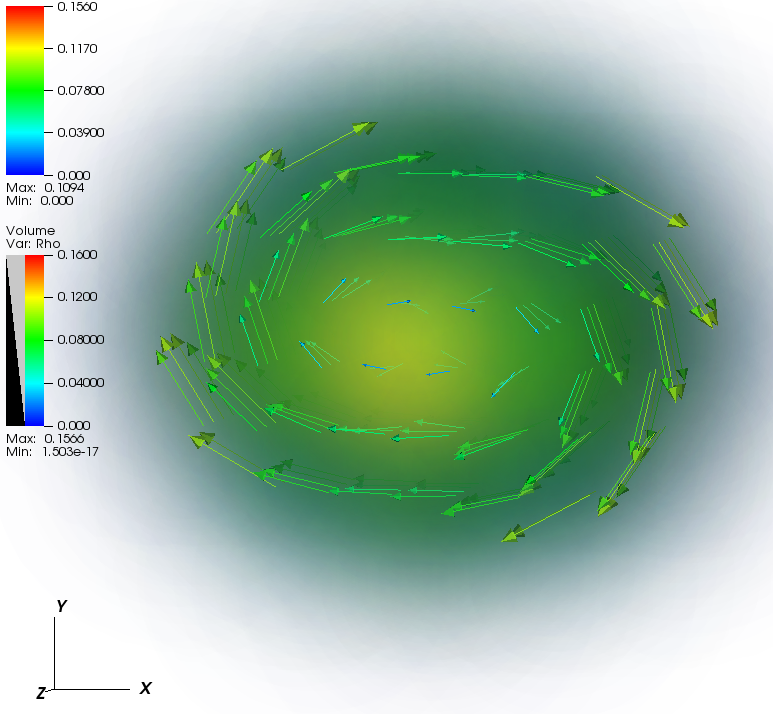}
    \includegraphics[width=0.49\textwidth]{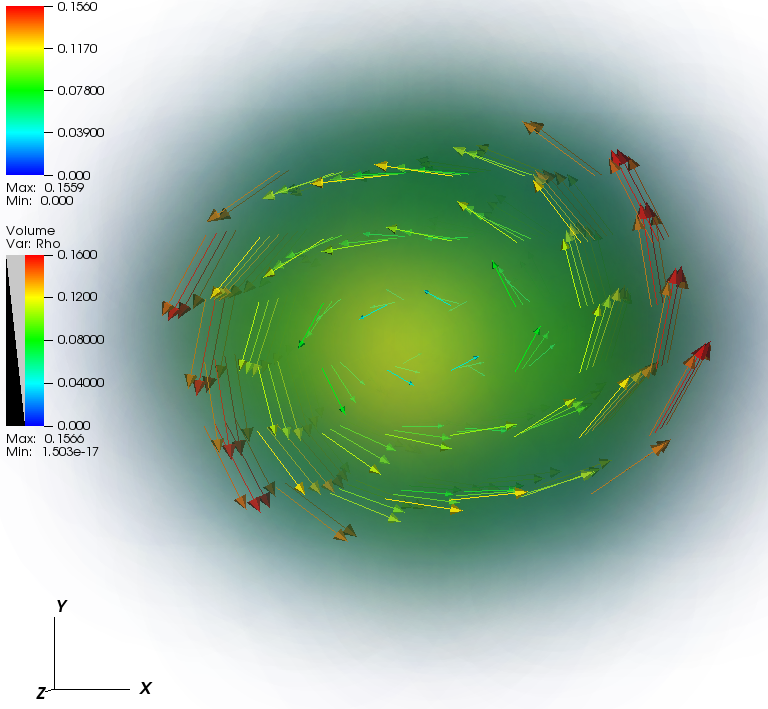}
    \includegraphics[width=0.49\textwidth]{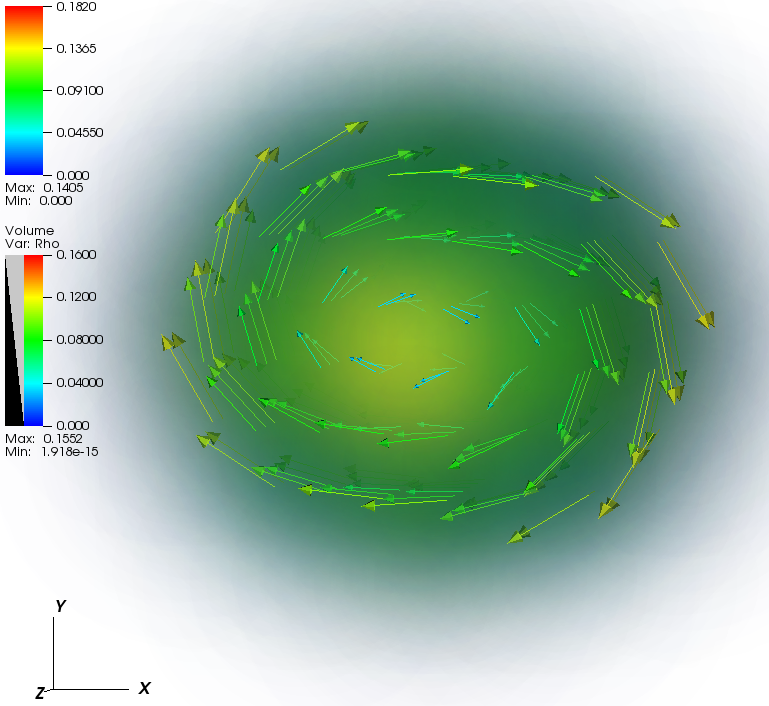}
    \includegraphics[width=0.49\textwidth]{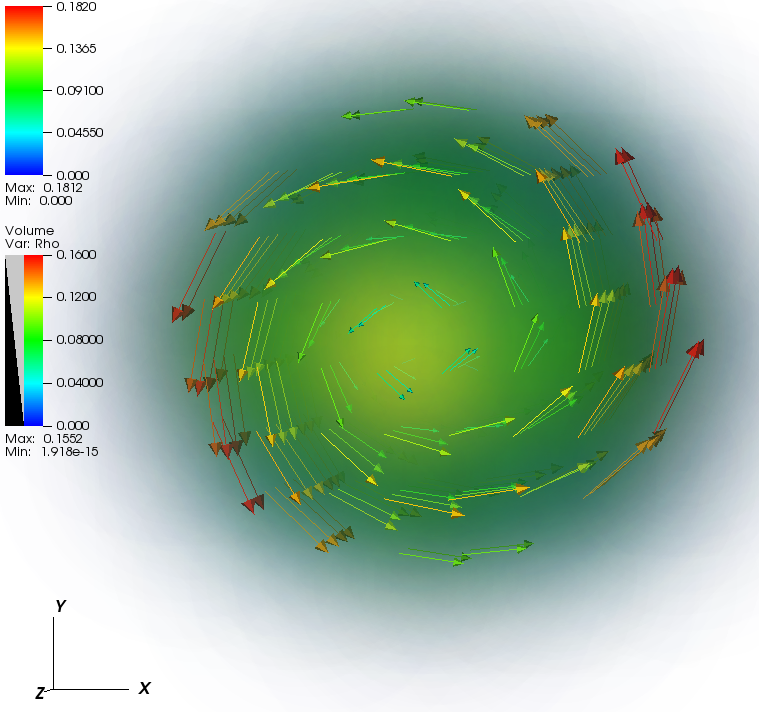}
    \caption[The collective flow velocity for neutrons and protons for
    $^{20}$Ne for different values of the magnetic field.] {The
      current density divided by the particle density (collective flow
      velocity) for neutrons (left) and protons (right) for
      \isotope[20]{Ne} for different values of the magnetic
      field. Top: $B=4.0\cdot10^{13}\,$G, middle:
      $B=2.4\cdot10^{17}\,$G, bottom: $B=4.9\cdot10^{17}\,$G. As
      background the particle density is added. This figure was
      created with VisIt~\cite{HPV:VisIt}.}
    \label{fig_2_12}
  \end{center}
\end{figure}

\begin{figure}[!]
  \begin{center}
    \includegraphics[width=0.49\textwidth]{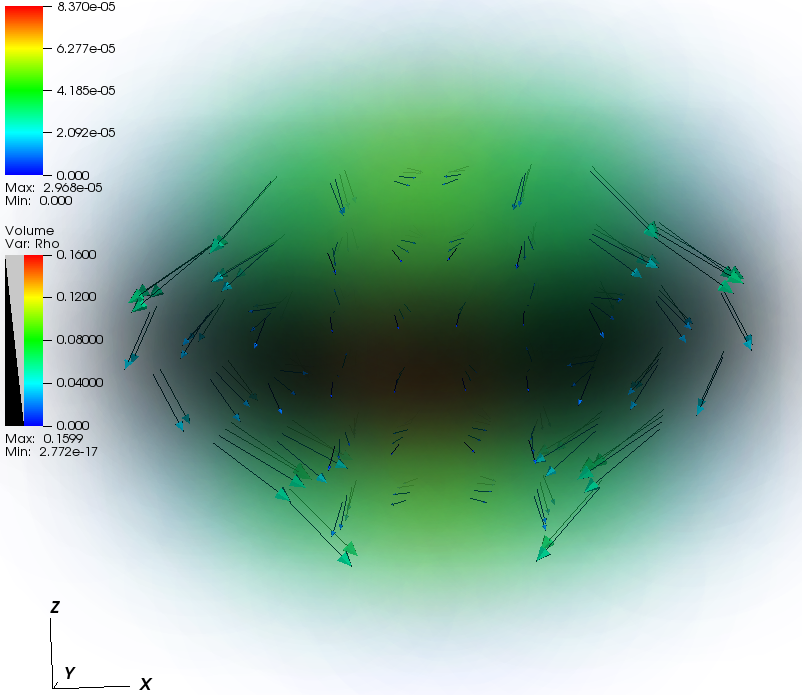}
    \includegraphics[width=0.49\textwidth]{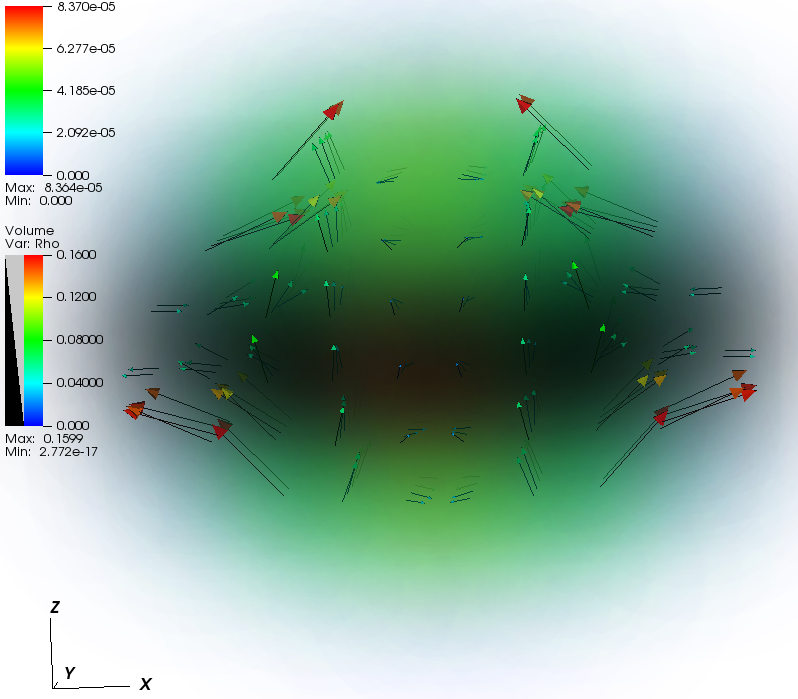}
    \includegraphics[width=0.49\textwidth]{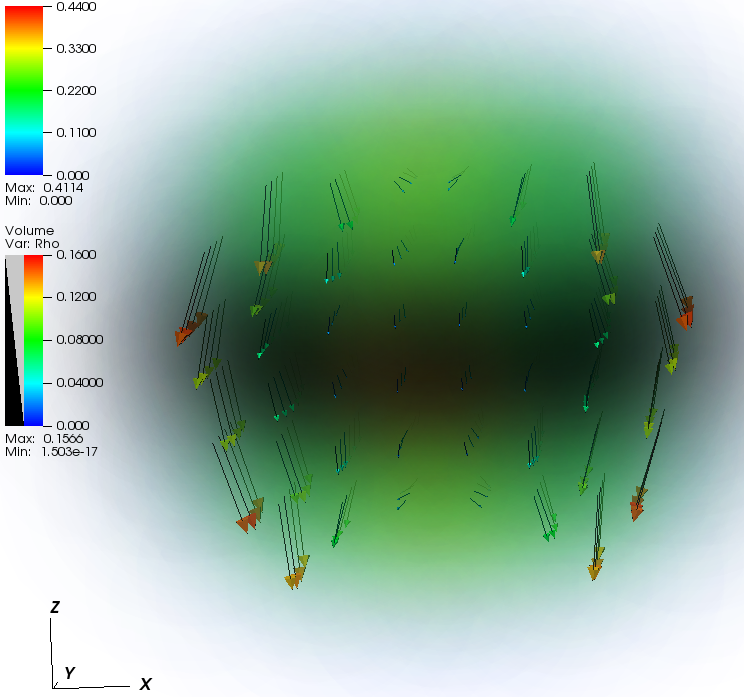}
    \includegraphics[width=0.49\textwidth]{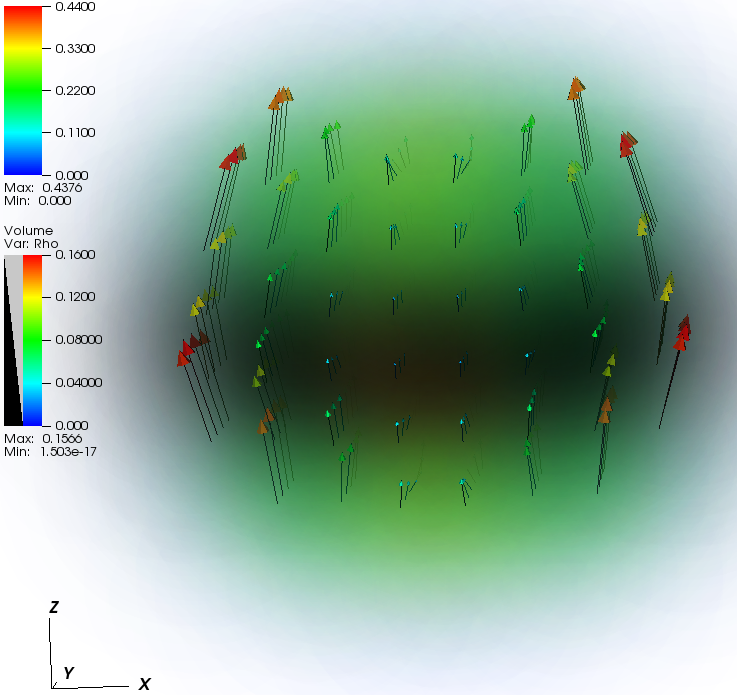}
    \includegraphics[width=0.49\textwidth]{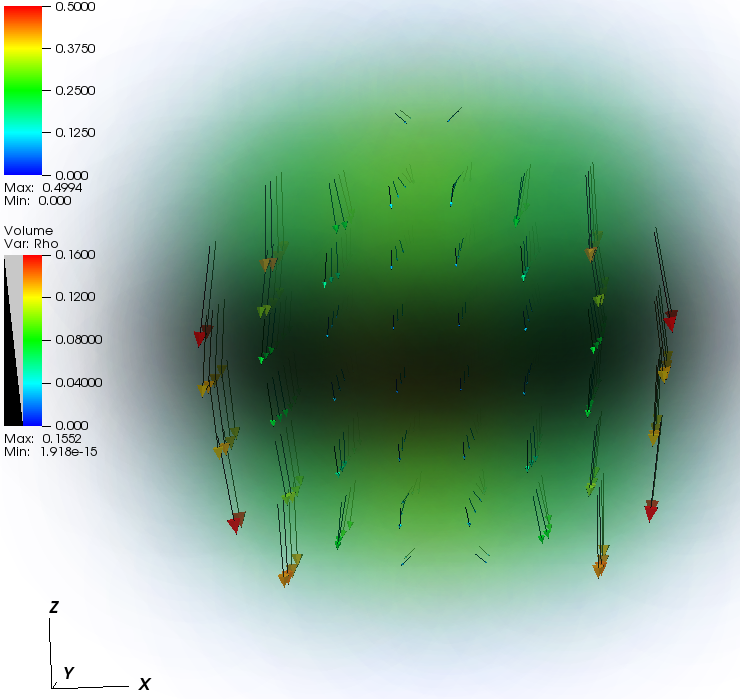}
    \includegraphics[width=0.49\textwidth]{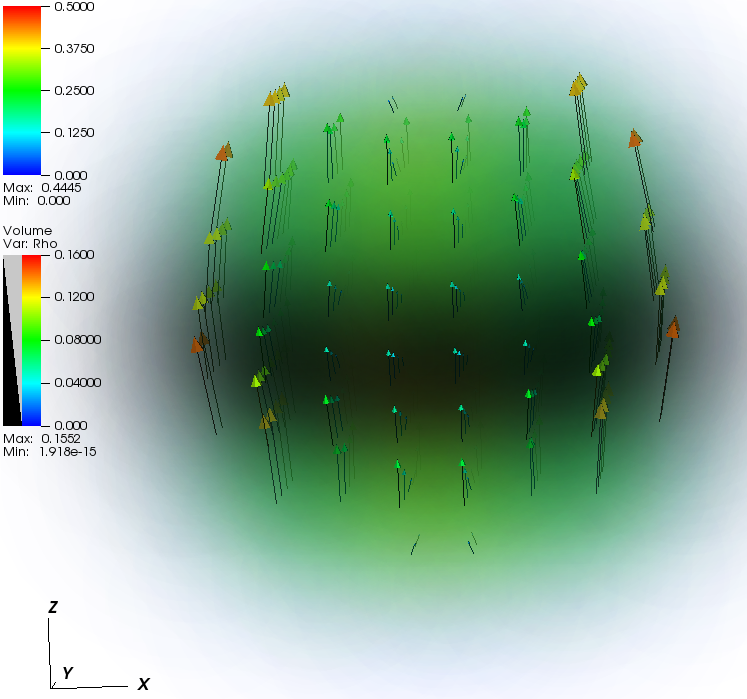}
    \caption[The spin density divided by the particle density for
    neutrons and protons for $^{20}$Ne for different values of the
    magnetic field.] {The spin density divided by the particle density
      for \isotope[20]{Ne} with the same arrangement and magnetic
      fields as in Fig.~\ref{fig_2_12}. This figure was created with
      VisIt~\cite{HPV:VisIt}.}
    \label{fig_2_13}
  \end{center}
\end{figure}

Analogously to \isotope[12]{C} we evaluated the current and spin
densities in Figs.~\ref{fig_2_12} and \ref{fig_2_13} for $B_{i}$,
$B_{\isotope[20]{Ne},\,h}$ and $B_{\isotope[20]{Ne},\,e}$. Apart from
the current at infinitesimal magnetic field, the absolute values of
both quantities are again approximately equal at each magnetic field
for protons and neutrons and the direction is again opposite. The
values increase with increasing magnetic field. Again the current is
concentrated at the surface. The change from the Zeeman effect to the
Paschen-Back effect is again clearly seen in the spin. However, in
this highly deformed nucleus, we see two main axis for spin
alignment. In all figures concerning \isotope[20]{Ne}, we see a
stronger change of values at weaker magnetic fields, explaining why
figures for $B_{\isotope[20]{Ne},\,h}$ look more similar to those of
$B_{\isotope[20]{Ne},\,e}$ than to those of $B=0$ or $B_i$.

The current distribution could give a clearer understanding of the
deformation. At vanishing magnetic field, the nucleus is deformed with
nucleons at low velocity. Increasing the magnetic field increases the
collective flow velocity. Hence the orbits of the nuclei approach to
circular orbits resulting in a less deformed nucleus at stronger
magnetic fields.

\subsubsection{Accuracy considerations}
The code Sky3D offers several variables for judging the
convergence. E.g. the average uncertainties in the single particle
energies $\Delta\varepsilon_1$ and $\Delta\varepsilon_2$ defined as
\begin{subequations}
  \label{eq_2_20}
  \begin{eqnarray}
    \Delta\varepsilon_1&=&\sqrt{\left<\psi\middle|\hat{h}^2\middle|\psi\right>-\left<\psi\middle|\hat{h}\middle|\psi\right>^2}\,,\\
    \Delta\varepsilon_2&=&\sqrt{\left<\hat{h}\psi\middle|\hat{h}\psi\right>-\left<\psi\middle|\hat{h}\middle|\psi\right>^2}\,.
  \end{eqnarray}
\end{subequations}
These uncertainties have to be low. Moreover, the change in the total
energy has to be low, but the uncertainties of Eq.~\eqref{eq_2_20} are
more important~\cite{2014CoPhC.185.2195M}.

For all results of this chapter, we obtain
$\Delta\varepsilon_1<10^{-3}$ and $\Delta\varepsilon_2<10^{-3}$. For
\isotope[16]{O} we obtain $\Delta\varepsilon_1<10^{-4}$ and
$\Delta\varepsilon_2<10^{-4}$ for $B<B_{\isotope[16]{O},\,c}$. For
\isotope[12]{C} we obtain $\Delta\varepsilon_1<10^{-4}$ and
$\Delta\varepsilon_2<10^{-4}$ for $B\leq2.9\cdot10^{17}\,$G. For all
results of this chapter we obtain $\Delta\varepsilon_2<10^{-4}$ for
\isotope[12]{C}, whereas $\Delta\varepsilon_1$ increases above
$10^{-4}$.

\subsection{Outlook}
To achieve convergence on strong magnetic fields, we used several
methods, explained in~\ref{subsec_2_3_3}. Hereby we improved the
convergence and got the results presented in this chapter. Besides
these techniques, we changed the Skyrme force and we introduced the
spin-spin term in the Hamiltonian analogue
to~\cite{2011JPhG...38c3101E}. However, these attempts were not yet
successful. For future works, it will be interesting to develop all
these techniques and to study effects for strengths of the magnetic
field where our present methods fail to converge. In addition, it will
be interesting to study nuclei which are likely to occur in magnetars,
the nuclei we studied for this work are not likely to occur in
magnetars~\cite{2012PhRvC..86e5804C}. Moreover, it will be interesting
to study the effects of the magnetic field on different terms of the
Hamiltonian, which we neglected, e.g. the spin-spin interaction.

\section{Conclusion}
We have studied the effects of strong magnetic fields on different
nuclei using a Skyrme-Hartree-Fock (SHF)
approach~\cite{2011JPhG...38c3101E,2014CoPhC.185.2195M} calculated
with the code Sky3D~\cite{2014CoPhC.185.2195M}. The strong magnetic
fields can be realized e.g. in neutron
stars~\cite{2000ApJ...537..351B,2013PhRvD..88b5008S,2010arXiv1005.4995S,2009PhRvD..79l3001S}. The
elements we study occur in white dwarfs~\cite{1986bhwd.book.....S},
which also have strong magnetic fields~\cite{1991ApJ...383..745L}.

We studied the effects of strong magnetic fields on different nuclei
and our results can be summarized as follows:
\begin{itemize}
\item \isotope[16]{O} is a double magic nucleus. For
  $B<B_{\isotope[16]{O},\,c}$ it has the lowest states of the harmonic
  oscillator filled and it is spherical symmetric. We find a change
  from a Zeeman effect to a Paschen-Back effect dominated region by
  increasing the magnetic field. For magnetic fields above
  $B_{\isotope[16]{O},\,c}$, we find a rearrangement of energy states
  and a slight deformation of the nucleus. The current and spin
  densities of neutrons and protons differ a lot above
  $B_{\isotope[16]{O},\,c}$.
\item \isotope[12]{C} is an isospin symmetric nucleus which is in many
  ways analogous to \isotope[16]{O}. Also the lowest states of the
  harmonic oscillator are filled and the change from a Zeeman effect
  to a Paschen-Back effect dominated region is clearly seen. The
  current and spin densities of neutrons and protons differ in
  direction but not much in magnitude.
\item \isotope[20]{Ne} is an isospin symmetric nucleus. At vanishing
  magnetic field it is strongly deformed in a purely prolate
  deformation. Increasing the magnetic field decreases the strength of
  the deformation and changes the shape from purely prolate to mainly
  prolate. In general we obtain equal features for current and spin
  densities as for \isotope[12]{C}, but due to the deformation it
  shows more interesting features.
\end{itemize}

Future work may extend the present studies in different directions, in
particular as follows:
\begin{itemize}
\item An improved Hamiltonian can be considered. In particular
  including the spin-spin interaction may be interesting in the
  context of strong magnetic fields.
\item Extension of the present studies to stronger magnetic
  fields. For that purpose the techniques which have been used in this
  study must be improved or modified.
\end{itemize}
\clearpage{\pagestyle{empty}\cleardoublepage}

\chapter{BCS pairing in neutron matter}
\label{chap_3}

\section{Introduction}
In this chapter we continue the discussion of pairing in nuclear
systems, started in chapter~\ref{chap_1}. In particular, we study here
the pairing in pure neutron matter in strong magnetic fields (strong
magnetic fields we have already discussed in
chapter~\ref{chap_2}). The order of magnitude of these fields is
within the range $10^{16}-10^{17}$~G, which was also relevant for our
discussion of nuclei in the previous chapter. We will show the
analogies between isospin asymmetric pairing in neutron-proton systems
and spin-polarized (spin asymmetric) neutron matter and contrast them
by pointing out the key differences.

Neutron-neutron pairing was studied extensively in the past, see for
example~\cite{2009PhRvC..79c4304M,2013PhRvC..88c4314P,PhysRevC.73.044309,2007PhRvC..76f4316M,2008PhRvC..78a4306I,2009PhRvC..80d5802G,2006pfsb.book..135S}.
Neutron-neutron pairing comes into play in systems where the isospin
asymmetry is large enough to suppress the dominant $\SD$ pairing. The
isospin triplet pairing in the $\SD$ channel is prohibited by Pauli
blocking in pure neutron matter. Therefore, the dominant pairing
channel is an isospin singlet state, which at low-energies
(low-densities) is in the $^1S_0$ channel. We may then expect that a
spin-polarization, e.g., induced by a magnetic field will suppress the
pairing significantly. At large energies (high densities) the dominant
pairing channel in neutron matter is the $^3P_2$-$^3F_2$ pairing
channel, which corresponds to a spin-1 condensate of
neutrons~\cite{2006pfsb.book..135S}. In this case the spin-polarizing
effect of the magnetic field on the internal structure of the spin-1
pairs is non-destructive.

Because in the vacuum the two-neutron system is unbound, diluting
neutron matter does not lead automatically to a state with strongly
bound two-neutron gas, which may then form a Bose
condensate. Therefore, in general one cannot expect a Bose condensed
regime of neutron-pairs in the low-density
limit~\cite{2009PhRvC..79c4304M}. Nevertheless, as shown in
Refs~\cite{PhysRevC.73.044309,2007PhRvC..76f4316M,2009PhRvC..79c4304M}
a BCS-BEC crossover region may exist also for neutron matter, when the
neutron matter is diluted. This, in principle, occurs in full analogy
with the $\SD$ pairing, with the exception that the asymptotical state
of the system at low-densities is non-interacting neutron gas, instead
of a Bose condensate of neutron-pairs.

In a first approximation neutron star matter can be treated as pure
neutron matter~\cite{2009PhRvC..80d5802G}, because the fraction of
protons and electrons (and other heavier baryons) does not exceed
$5\%$-$10\%$ of the total density of the system. Thus, it is well
known that neutron-neutron pairing plays an important role in the
physics of the inner crust of a neutron star. It plays also a
significant role for neutron-rich nuclei near the drip
line~\cite{PhysRevC.73.044309}. This type of pairing may also occur
for halo neutrons in halo nuclei, such as,
e.g. $^{11}$Li~\cite{PhysRevC.73.044309}. There are some
phenomenological indications of neutron superfluidity in neutron
stars. Prominent examples are glitches in the rotational behavior of
some pulsars, as well as the cooling behavior of the youngest known
neutron star in Cassiopeia A~\cite{2006pfsb.book..135S}.

\section{Theory}
\label{sec_3_2} In this section we adapt the formalism developed in
chapter~\ref{chap_1} to the case of neutron matter. We will continue
to use the separable version of the Paris potential. However, now we
would like to extend the discussion to arbitrary rank $n$ separable
potentials. This extension applies for both uncoupled channels, such
as the $^1S_0$ channel as well as for coupled channels as the $\SD$
channel.

We start with some basic definition following
Ref.~\cite{1988NuPhA.481..294J}. The gap equation for an uncoupled
neutron-neutron channel can be written as
\begin{eqnarray}
  \Delta_l(\veck)=\sum_{l'} \int\frac{d^3k'}{(2\pi)^3} V_{l,l'}(\veck,\veck')\phi_{l'}(\veck')\label{eq_3_01}\,,
\end{eqnarray}
where we use the same notations as in chapter~\ref{chap_1} and we
define
\begin{eqnarray}
  \phi_{l}(\veck)=\frac{1}{4}\sum_{a,r}\frac{\Delta_{l}(\veck)}{2\sqrt{E_{S}^2(\veck)+\Delta^2(\veck)}}[1-2f(E^a_r(\veck)]\,.
\end{eqnarray}
Here the neutron spin-up and spin-down spectra are given by
\begin{eqnarray}
  E_{r}^{a} &=& \sqrt{E_S^2+\Delta^2} + r\delta\mu +a E_A\,,
\end{eqnarray}
where
\begin{eqnarray}
  E_S =\left(Q^2/4+k^2\right)/2m^*-\bar\mu\,,\qquad E_A
  = \veck\cdot \vecQ /2m^*\,,
\end{eqnarray}
and
\begin{eqnarray}
  \delta\mu =
  \tfrac12(\mu_\uparrow - \mu_\downarrow)\,,\qquad
  \bar\mu = \tfrac12(\mu_\uparrow+\mu_\downarrow)\,,
\end{eqnarray}
where $\delta\mu$ describes the spin-polarization of neutron matter in
terms of chemical potentials of spin-up $\mu_\uparrow$ and spin-down
$\mu_\downarrow$ chemical potentials. The symmetric part of the
spectrum $E_S$ contains a kinetic energy shift due to the finite
momentum and an ``averaged'' chemical potential $\bar\mu$. The
angle-dependent part of the spectrum $E_A$ is the kinematical factor
that allows for LOFF phase. It describes the mutual shift in the
Fermi-surfaces of spin-up and spin-down neutrons which allows for
partial overlap of the Fermi-surfaces at the cost of additional
kinetic energy appearing in $E_S$. Thus we see that for $\delta\mu=0$
one recovers the spin-unpolarized limit, whereas for $E_A=0$ the LOFF
phase is excluded from consideration.

The condensation energy of neutron matter is given by
\begin{align}
  E_\mathrm{cond}=&\frac{g_\mathrm{deg}}2 N_\Omega\sum_{l,l'}\int\frac{d^3k}{(2\pi)^3}
                    \int\frac{d^3k'}{(2\pi)^3} V_{l,l'}(\veck,\veck') \phi_l(\veck)\phi_{l'}(\veck')\label{eq_3_06}\,,
\end{align}
where $g_\mathrm{deg}$ is the degeneration factor; $g_\mathrm{deg} =2$
for neutron matter and $g_\mathrm{deg} =4$ for symmetric nuclear
matter. $N_\Omega$ is the gap averaging factor. For the $^1S_0$
pairing channel we obtain $N_\Omega = 1$. For the $\SD$ pairing
channel, which can occur e.g. for neutron-proton pairing, we obtain
$N_\Omega = 3/(8\pi)$.

It will be convenient in the following to describe the polarization of
matter also in terms of the partial densities of spin-up and spin-down
particles
\begin{eqnarray}
  \alpha=\frac{\rho_\uparrow-\rho_\downarrow} {\rho_\uparrow+\rho_\downarrow}\,.
\end{eqnarray}
Finally, we note that we use the same Skyrme functional based
expression for the effective mass of neutrons as was used in
chapter~\ref{chap_1}
\begin{eqnarray}
  \frac{m}{m^*}&=&\left[1+\frac{\rho\cdot m}{8\hbar^2}(3t_1+5t_2)\right]\,.
\end{eqnarray}
We will neglect the effects of spin-polarization on the effective mass
of neutrons. This is a small effect compared to the ones that are
taken into account (i.e. the splitting in the chemical potentials of
spin-up and down particles.)

Now we proceed to implement a separable potential in the preceding
equations. Following Ref.~\cite{1984PhRvC..30.1822H} we define a
separable potential as
\begin{eqnarray}
  V_{l,l'}(k,k')=-2\pi^2\sum_{i=1}^n\sum_{j=1}^n\lambda_{ij}\cdot
  g_{li}(k)g_{l'j}(k')\,.
\end{eqnarray}
To solve the gap equation we carry out angle averaging in the
denominator of the gap equation by substituting
\begin{eqnarray}
  \Delta^2(k)&=&N_\Omega\sum_l\Delta_l^2(k)\, .
\end{eqnarray}
The separable potential allows one to make the following ansatz for
the gap function
\begin{eqnarray}
  \Delta_l(k)&=&\sum_{i=1}^nc_i\cdot g_{li}(k)\,,
\end{eqnarray}
where $c_i$ are constants. Then the gap equation takes the form
\begin{eqnarray}
  \sum_{i=1}^nc_i\cdot g_{li}(k) &=&-2\pi^2\cdot\frac14\sum_{a,r}\sum_{l'}\sum_{i=1}^ng_{li}(k)\sum_{j=1}^n\lambda_{ij}\int\frac{d^3k'}{(2\pi)^3} g_{l'j}(k')\nonumber\\
                                 &&\times\frac{\Delta_{l'}(k')}{2\sqrt{E_S^2(k')+\Delta^2(k')}}\cdot[1-2f(E^a_r(k'))]\,.\nonumber
\end{eqnarray}
This equation will be fulfilled for arbitrary sets of functions
$g_{li}(k)$ only if for each term in the $i$ sum we have
\begin{eqnarray}
  c_i&=&-2\pi^2\cdot\frac14\sum_{a,r}\sum_{l}\sum_{j=1}^n\lambda_{ij}\int\frac{d^3k}{(2\pi)^3}
         g_{lj}(k)\newline\\
     &&\times\frac{\Delta_{l}(k)}{2\sqrt{E_S^2(k)+\Delta^2(k)}}\cdot[1-2f(E^a_r(k))]\nonumber\\
     &=&-2\pi^2\sum_l\sum_{j=1}^n\lambda_{ij}\int\frac{d^3k}{(2\pi)^3}
         g_{lj}(k)\phi_l(k)\,.
\end{eqnarray}
Thus the problem reduces to solving a set of non-linear algebraic
equations.

In the following step we compute the condensation energy for separable
interactions. For that purpose with insert Eq.~\eqref{eq_3_01} in
Eq.~\eqref{eq_3_06} and find
\begin{eqnarray}
  E_\mathrm{cond}&=&\frac{g_\mathrm{deg}}2N_\Omega\sum_{l}\int\frac{d^3k}{(2\pi)^3}\Delta_l(k)\phi_l(k)\,.
\end{eqnarray}
For a rank~1 potential we obtain the solution of the gap equation as
\begin{eqnarray}
  c&=&-2\pi^2\sum_l\lambda\int\frac{d^3k}{(2\pi)^3}
       g_{l}(k)\phi_l(k)\,,\qquad \Delta_l(k)=c\cdot g_{l}(k)\,.
\end{eqnarray}
The condensation energy is then given in terms of the $c$-coefficient
\begin{eqnarray}
  E_\mathrm{cond}&=&\frac{g_\mathrm{deg}}2 \cdot\frac{N_\Omega\cdot c}{\lambda} \sum_{l}\lambda \int\frac{d^3k}{(2\pi)^3}g_{l}(k)\phi_l(k)\nonumber\\
                 &=&-\frac{g_\mathrm{deg}}2\cdot\frac{N_\Omega\cdot c^2}{2\pi^2\cdot\lambda}\,.
\end{eqnarray}

In analogy to Eq.~\eqref{eq_1_41} we can define the grand canonical
potential as
\begin{eqnarray}
  \Omega(\Delta,\vecQ)
  &=&\frac{g_\mathrm{deg}}2N_\Omega
      \sum_{l}\int\frac{d^3k}{(2\pi)^3}\Delta_l(\veck)\phi_l(\veck)\nonumber\\
  &&-\frac{g_\mathrm{deg}}4\sum_{a,r}\int\frac{d^3k}{(2\pi)^3}\left[\frac{E^a_r(\veck)-E_S(\veck)}2\right.\nonumber\\
  &&+\left.T\ln\left(1+e^{-\beta
     E^a_r(\veck)}\right)\right]
\end{eqnarray}
and in analogy to Eq.~\eqref{eq_1_44} we can define
\begin{eqnarray}
  \tilde\Omega(\Delta,\vecQ)
  &=&\Omega(\Delta,\vecQ)-\Omega(0,\vecQ)+\Omega(0,0)\,.
\end{eqnarray}
In analogy to Eq.~\eqref{eq_1_45} the associated free energy of the
system is given by
\begin{eqnarray}
  F(\Delta,\vecQ) &=&\tilde\Omega(\Delta,\vecQ)
                      +\mu_\uparrow\rho_\uparrow+\mu_\downarrow\rho_\downarrow\,.\label{eq_3_19}
\end{eqnarray}

To complete our set of equations, we define the densities of the
spin-up and down particles in analogy to the densities of neutrons and
protons in chapter~\ref{chap_1}:
\begin{eqnarray}
  \rho_{\uparrow/\downarrow}(\vecQ) &=&\frac{g_\mathrm{deg}}4\int \frac{d^3k}{(2\pi)^3}\cdot\left(1+\frac{E_S}{\sqrt{E_S^2+\Delta^2}}\right)f(E^+_\mp)\nonumber\\
                                    &&+\left(1-\frac{E_S}{\sqrt{E_S^2+\Delta^2}}\right)(1-f(E^-_\pm))\,.\label{eq_3_20}
\end{eqnarray}

For the following discussion we also define the Fermi momenta of
spin-up and down neutrons at zero temperature
\begin{eqnarray}
  k_{F\uparrow/\downarrow}
  = (6\pi^2\rho_{\uparrow/\downarrow})^{1/3}\,,
\end{eqnarray}
as well as the Fermi-momentum of unpolarized (spin-symmetrical)
neutron matter
\begin{eqnarray}
  k_F = (3\pi^2\rho)^{1/3}\,,
\end{eqnarray}
where $\rho = \rho_{\uparrow}+\rho_{\downarrow}$ is the net density.

In closing of this section we related the spin-polarization to the
magnetic field. Obviously,
\begin{eqnarray}
  \delta\mu = -\tilde \mu_n B\,,
\end{eqnarray}
where the minus sign is chosen to have positive $\delta\mu$ as in
chapter~\ref{chap_1} and
\begin{eqnarray}
  \tilde\mu_n=\frac{g_n}2\cdot\mu_N=\frac{g_n}2\cdot\frac{e\hbar}{2m^*c}\,,
\end{eqnarray}
where $g_n=-3.82608544$ is the neutron $g$-factor. We will sometimes
use the energy scale that describes the spin-polarization
\begin{eqnarray}
  \varepsilon_B = \vert \delta\mu\vert = \left|\tilde\mu_nB\right| \,.\label{eq_3_25}
\end{eqnarray}

\section{Results}
Using strategies as in chapter~\ref{chap_1} the phase diagram of
spin-polarized neutron matter was calculated. The intrinsic features
of spin-polarized matter were extracted at the points in the
density-temperature diagram listed in Table \ref{table_3_1}. The
computations were carried out with the rank-3 separable Paris
potential (PEST~3) in the $^1S_0$ partial wave channel with parameters
listed in Ref.~\cite{1984PhRvC..30.1822H}. We found that there exists
neither LOFF phase nor BEC regime in spin-polarized neutron
matter. The main reason for the absence of the LOFF phase is that the
neutron condensate is more ``fragile'' than the neutron-proton
condensate. Indeed the ratio of the gap to the chemical potential is
by about factor 10 smaller in neutron matter, because for the same
density the chemical potential is larger in pure neutron matter than
in the symmetrical nuclear matter and at the same time the $^1S_0$
interaction is at all energies weaker than the $\SD$
interaction. Furthermore, we ignored the possibility of a
phase-separation in spin-polarized neutron matter. Finally, concerning
the BCS-BEC crossover our findings are as follows. No change in the
sign of the chemical potential was observed, i.e., the chemical
potential had only positive values at the lowest density studied. The
lowest value we checked was $\mu_n=0.24$ MeV for
$\log(\rho/\rho_0)=-3.57$, $T=0.05$ MeV in spin-symmetric neutron
matter. This point is located at the transition line to the unpaired
phase. This indicates a vanishing chemical potential at the low
density vanishing temperature transition. The reason that we do not
find clear BEC condensate of neutron pairs is that the no bound
neutron-neutron pairs are supported in the vacuum by the nuclear
interaction. However, it should be noted that although the chemical
potential does not change its sign, we find that $d/\xi_{a}\ll 1$ at
high density and $d/\xi_{a}\ge 1$ at low-densities, which is an
indication of a BEC precursor.

In Table~\ref{table_3_1} (which is similar to Table~\ref{table_1_1})
we present several quantities of interest for three different values
for the density at $T=0.25$ MeV at vanishing spin-polarization. The
above indicated behavior of the ratio $d/\xi_{a}$ is demonstrated.
\begin{table}
  \begin{tabular}{cccccccc}
    \hline log$\left(\frac{\rho}{\rho_0}\right)$ & $k_F$ [fm$^{-1}$] & $\Delta$ [MeV] & $m^\ast/m$ & $\mu_n$ [MeV] & $d$ [fm] & $\xi_{\rm rms}$ [fm] & $\xi_{a}$ [fm]\\
    \hline\hline $-1.0$ & 0.78 & 2.46 & 0.967 &12.94 & 2.46 & 4.87 & 4.33 \\
    $-1.5$ & 0.53 & 1.91 & 0.989 & \,\,\,5.65 & 3.61 & 3.55 & 3.71 \\
    $-2.0$ & 0.36 & 1.07 & 0.997 & \,\,\,2.49 & 5.30 & 2.36 & 4.48 \\
    \hline\\
  \end{tabular}
  \caption{The table shows the net density $\rho$ (in units of nuclear saturation density $\rho_0 = 0.16$ fm$^{-3}$), Fermi momentum $k_F$, gap $\Delta$, effective mass (in units of bare mass), chemical potential $\mu_n$, interparticle distance $d$, and coherence lengths $\xi_{\rm rms}$ and $\xi_{a}$ in unpolarized neutron matter at fixed $T=0.25$ MeV. The pairing is in the $^1S_0$ channel.}\label{table_3_1}
\end{table}

\subsection{Phase diagram}
The phase diagram was computed by solving Eqs.~\eqref{eq_3_01} and
\eqref{eq_3_20} self-consistently for pairing in the $^1S_0$
channel. These equations are analogous to Eqs.~\eqref{eq_1_38} and
\eqref{eq_1_40}. The LOFF phase was searched for by varying the
Cooper-pair momentum $\vecQ$ in Eq.~\eqref{eq_3_19} and by looking at
the minimum of the free energy. It was found that in the parameter
range considered the minimum was always at $\vecQ=0$. The reason for
this, as explained above, is the ``weakness'' of the neutron pairing
compared to the neutron-proton pairing.

\begin{figure}[!]
  \begin{center}
    \includegraphics[width=\textwidth]{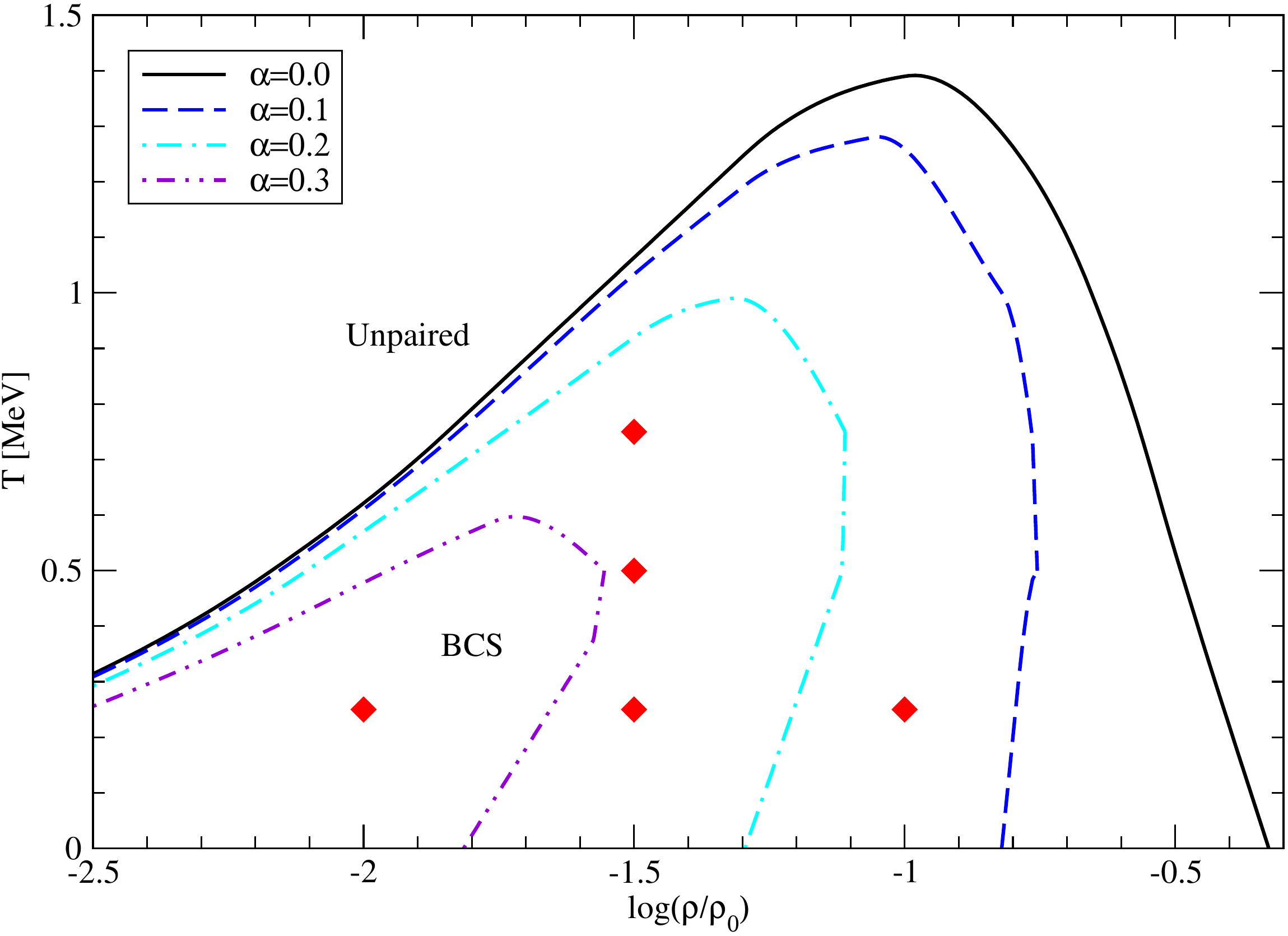}
    \caption[Phase diagram of spin polarized neutron matter.] {Phase
      diagram of neutron matter in the temperature-density plane for
      several spin-polarizations $\alpha$, induced by magnetic
      fields. Included are the BCS and the unpaired phase. The red
      diamonds refer to different points in the phase diagram at which
      we evaluated some intrinsic features. This figure is analog to
      Fig.~\ref{fig_1_02}.}
    \label{fig_3_01}
  \end{center}
\end{figure}

The resulting phase diagram of neutron matter is shown in
Fig.~\ref{fig_3_01}. It only consists of the BCS and the unpaired
phases. In general we obtain the same structure as in the case of
nuclear matter shown in Fig.~\ref{fig_1_02}. At low densities the
critical temperature increases with increasing density, for high
densities it decreases. This shape results from Eqs.~\eqref{eq_1_49}
and \eqref{eq_1_50}, see discussion in
subsection~\ref{subsec_1_3_1}. The polarization suppresses the pairing
efficiently only for high densities.

The phase diagram of spin-polarized neutron matter is less complex
because no exotic phase are present. However, an interesting feature
seen in the diagram is that the transition line to the unpaired phase
is not a single-valued function of the density in a range of
densities. The reason is well understood: at low temperature the
matter is in the unpaired phase by small separation of the Fermi
spheres as the thermal smearing of the Fermi-surfaces is ineffective
to produce phase-space overlap. Increasing the temperature at fixed
density and polarization makes the smearing of the Fermi surfaces more
effective resulting in an overlap and thus a restoration of the BCS
phase (see subsections~\ref{subsec_1_3_3} and \ref{subsec_1_3_2} for
further discussion).

\subsection{Intrinsic features}
We now proceed to study some intrinsic features of the isospin-triplet
$^1S_0$ neutron condensate, as was done for the $\SD$ condensate in
chapter~\ref{chap_1}. Because there are many similarities in the
relevant quantities (such as the pairing gap, the kernel, the wave
functions, the occupation numbers and the quasiparticle spectrum) we
will not carry out a complete discussion. We will try to point out the
main differences and the most prominent features.

\subsubsection{The Gap}
\begin{figure}[!]
  \begin{center}
    \includegraphics[width=0.8\textwidth]{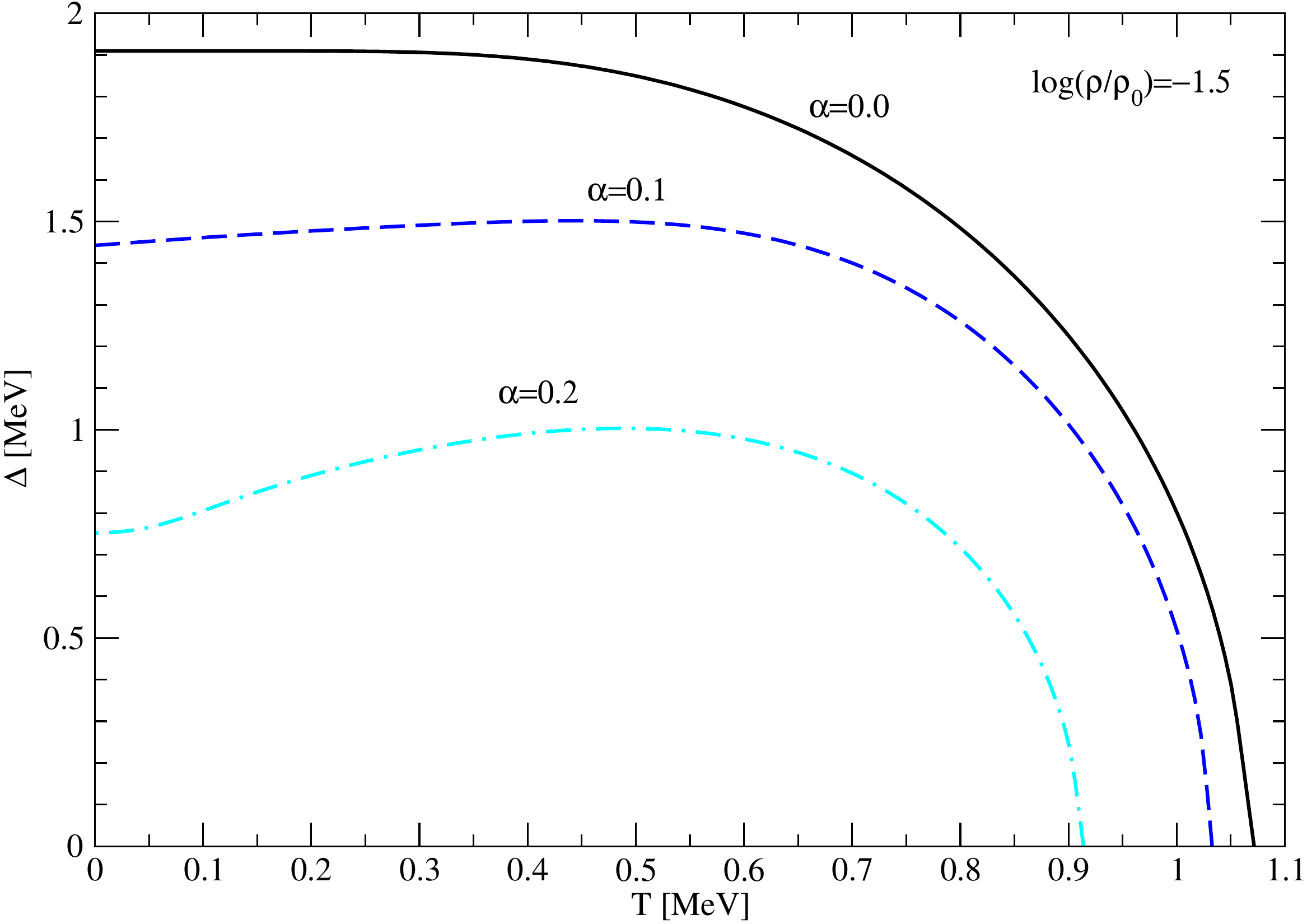}
    \caption[The gap as a function of the temperature.]{The gap as a
      function of the temperature at constant density
      $\log(\rho/\rho_0)=-1.5$ for several polarizations. This figure
      is analog to Fig.~\ref{fig_1_07}.}
    \label{fig_3_02}
  \end{center}
\end{figure}

\begin{figure}[!]
  \begin{center}
    \includegraphics[width=0.8\textwidth]{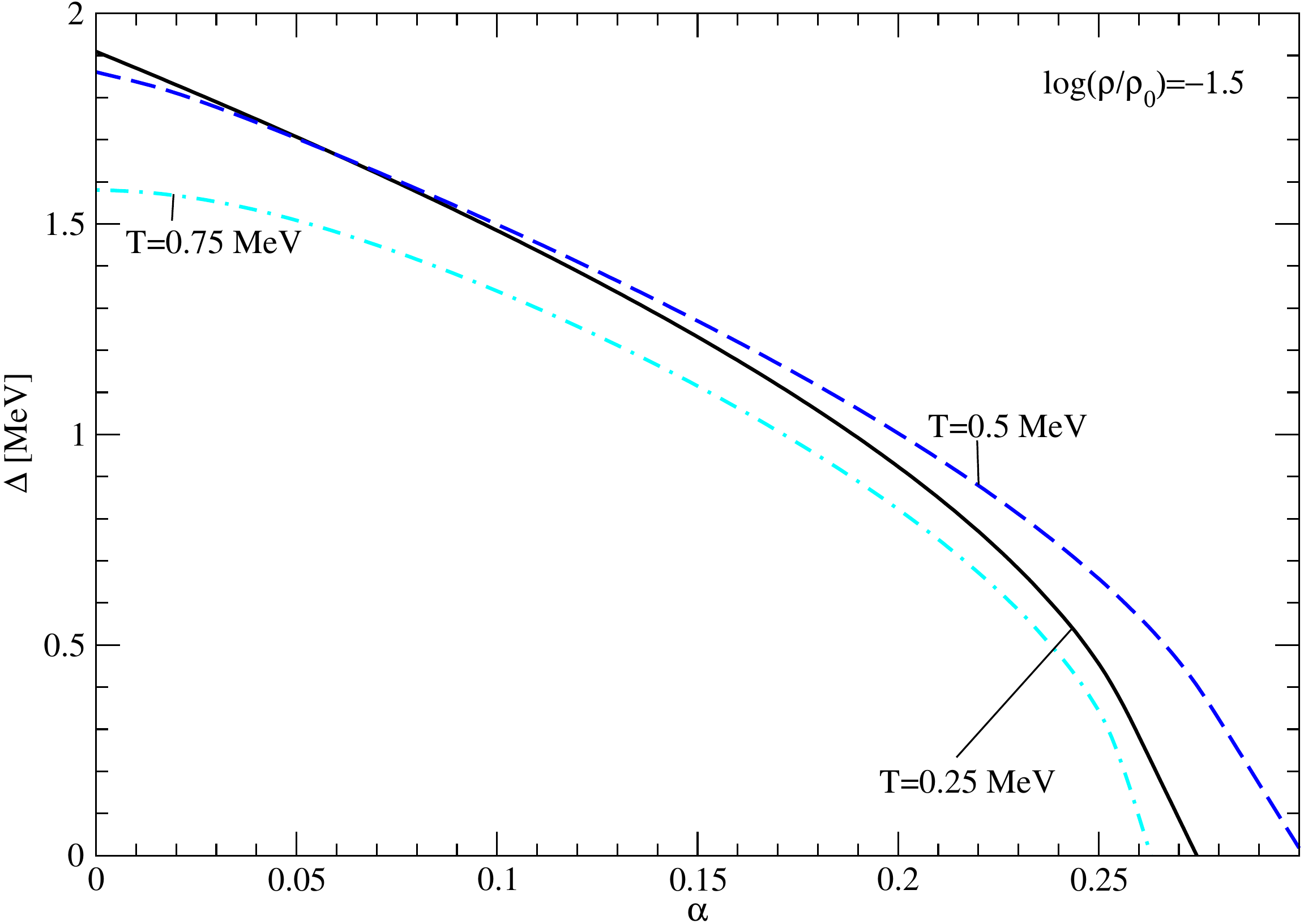}
    \caption[The gap as a function of the polarization.]{The gap as a
      function of the polarization at constant density
      $\log(\rho/\rho_0)=-1.5$ for several temperatures. This figure
      is analog to Fig.~\ref{fig_1_08}.}
    \label{fig_3_03}
  \end{center}
\end{figure}

In Figs.~\ref{fig_3_02} and~\ref{fig_3_03} we present the gap at fixed
density $\log(\rho/\rho_0)=-1.5$ (these figures are analog to
Figs.~\ref{fig_1_07} and \ref{fig_1_08}). In Fig.~\ref{fig_3_02} the
gap is shown as a function of the temperature for various values of
the polarization. For zero polarization, i.e., in the case of
symmetrical BCS state, the value of the gap is largest because of the
perfect overlap of Fermi surfaces of spin-up and down particles. The
temperature dependence of the gap corresponds to the standard BCS
behavior. Increasing the polarization has two effects: first, due to
the separation of the Fermi surfaces the gap decreases and so does
$T_C$. Second, it shifts the maximum of the gap from $T=0$ to
nonvanishing temperatures. This shift, for large enough polarizations
can lead to the appearance of a lower critical temperature (see
subsection~\ref{subsec_1_3_2} for a discussion of this effect).

In Fig.~\ref{fig_3_03} the gap is shown as a function of the
polarization for various temperatures. For $\alpha = 0$ the
temperature increase decreases the gap, as it should according to the
BCS theory. The crossing of constant temperature curves at finite
$\alpha$ reflects the fact that larger temperatures favor pairing in
polarized systems, because they increase the overlap between the Fermi
surfaces. Of course, this effect is counterbalanced by the destruction
of the superfluid state by high enough temperatures. These arguments
are reflected in the figure where at high enough polarizations the
increase of temperature from $T=0.25$ MeV to $T=0.5$ MeV increases the
gap, but the increase of the temperature from $T=0.5$ MeV to $T=0.75$
MeV again decreases the gap. See subsection~\ref{subsec_1_3_2} for a
discussion of this behavior in the $\SD$ condensate.

\subsubsection{The kernel of the gap equation}
\begin{figure}[!]
  \begin{center}
    \includegraphics[width=0.8\textwidth]{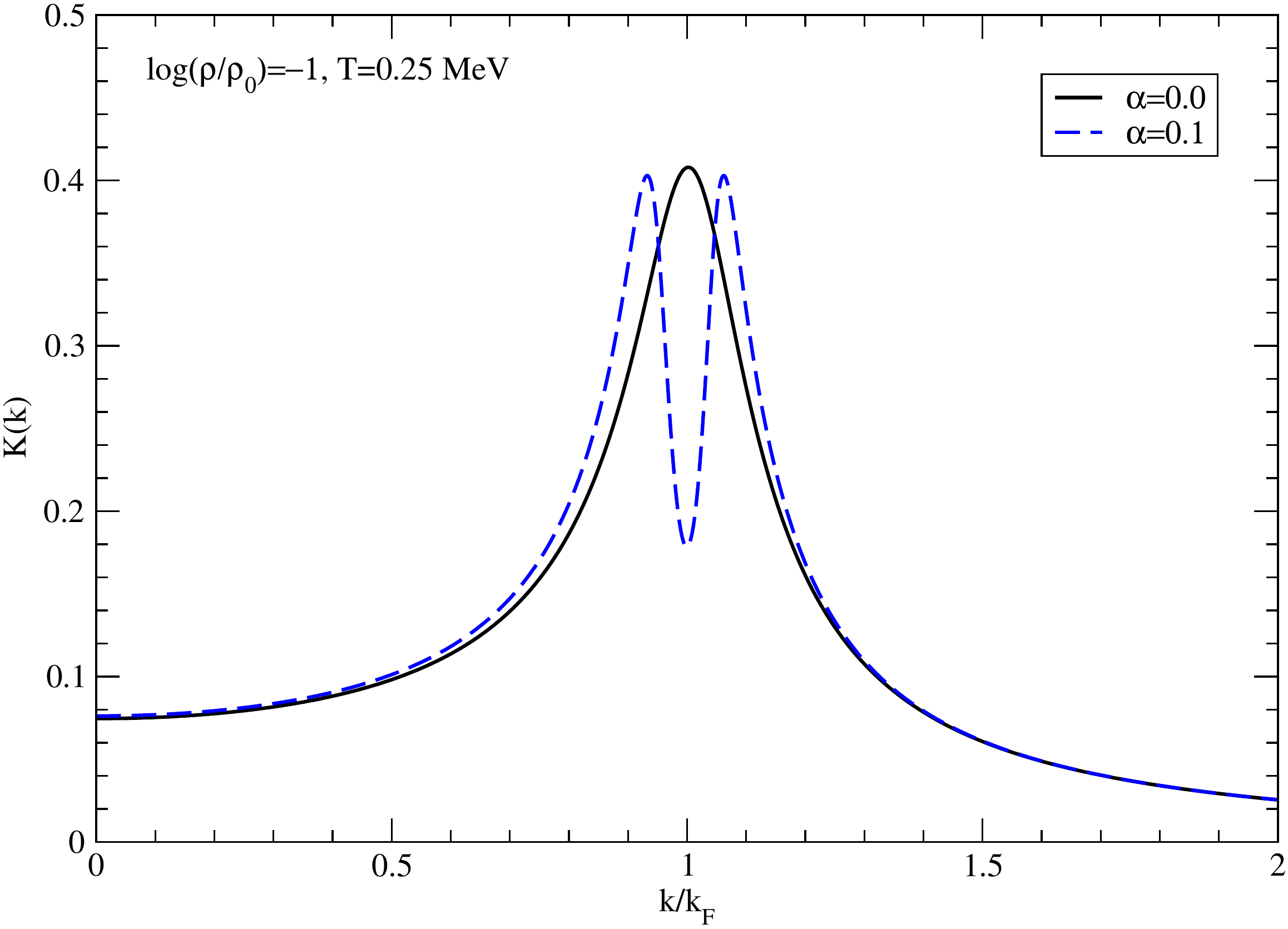}
    \caption[Dependence of the kernel on momentum for various
    polarizations at high density.] {Dependence of the kernel $K(k)$
      on momentum in units of Fermi momentum for fixed
      $\log(\rho/\rho_0)=-1$, $T=0.25$ MeV, and various values of
      polarization indicated in the plot. This figure is analog to
      Fig.~\ref{fig_1_15}.}
    \label{fig_3_04}
  \end{center}
\end{figure}

\begin{figure}[!]
  \begin{center}
    \includegraphics[width=0.8\textwidth]{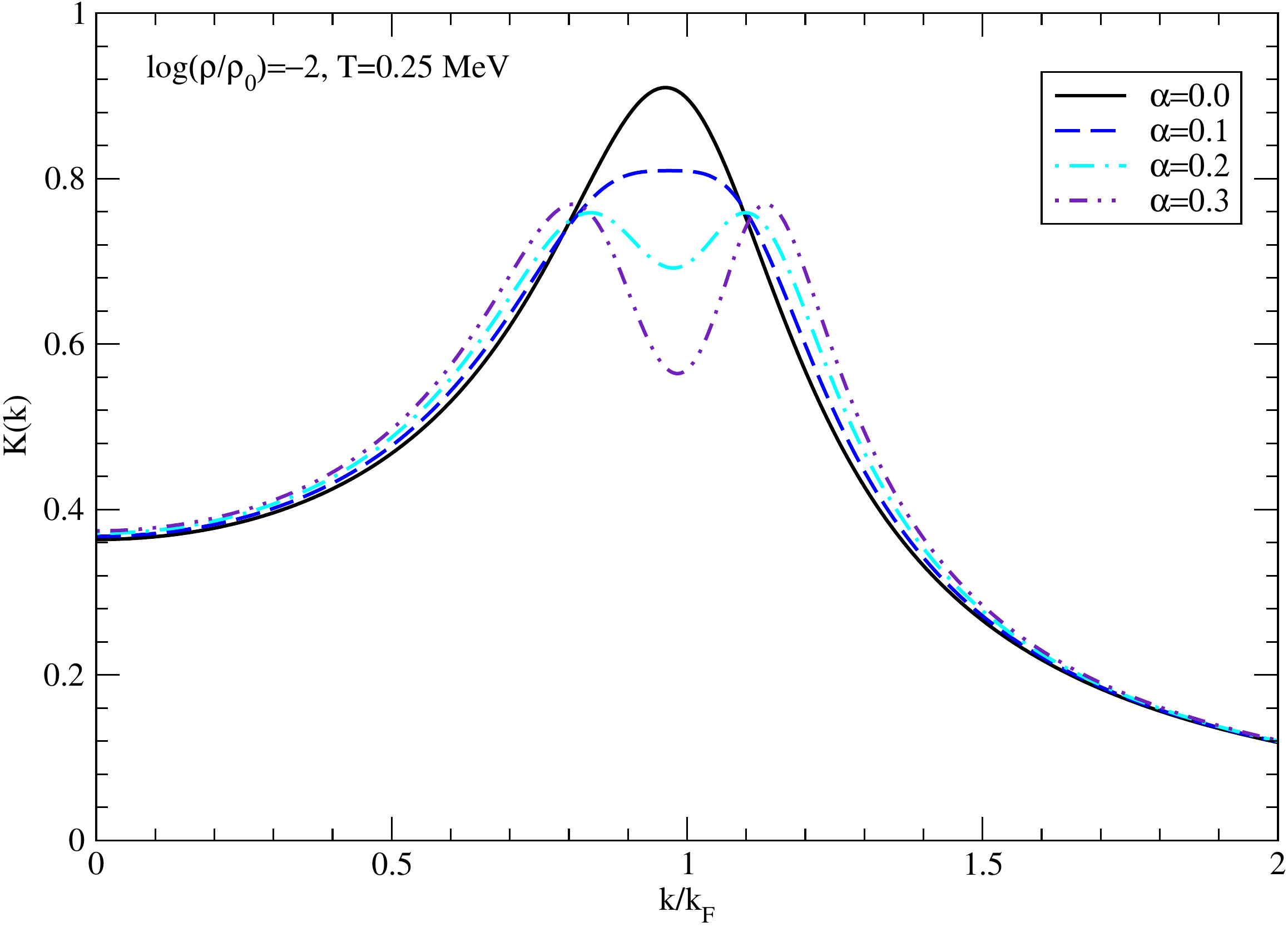}
    \caption[Dependence of the kernel on momentum for various
    polarizations at low density.] {Same as Fig.~\ref{fig_3_04} but
      for $\log(\rho/\rho_0)=-2$ and more values for the
      polarization.}
    \label{fig_3_06}
  \end{center}
\end{figure}

\begin{figure}[!]
  \begin{center}
    \includegraphics[width=0.8\textwidth]{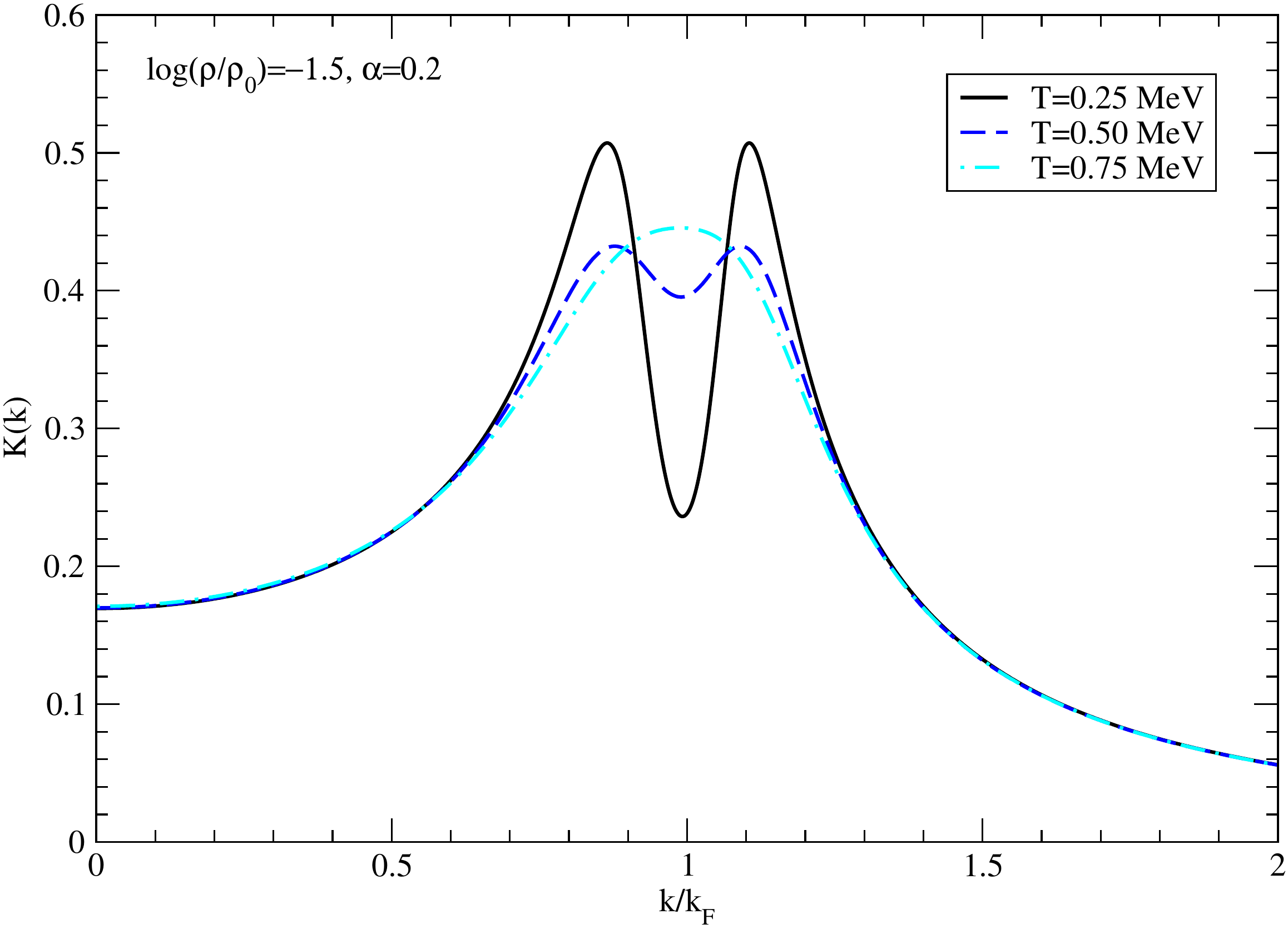}
    \caption[Dependence of the kernel on momentum for various
    temperatures.] {Dependence of the kernel $K(k)$ on momentum in
      units of Fermi momentum for fixed $\log(\rho/\rho_0)=-1$,
      $\alpha=0.2$, and various temperatures indicated in the
      plot. This figure is analog to Fig.~\ref{fig_1_14}. }
    \label{fig_3_07}
  \end{center}
\end{figure}

In Figs.~\ref{fig_3_04}-\ref{fig_3_07} we present the kernel of the
gap equation for various values of density, temperature and
polarization in the BCS phase. Because we do not find a LOFF phase or
a BEC, we do not discuss these regimes here. In the BCS case the
formula for the kernel Eq.~\eqref{eq_1_53} simplifies to
\begin{eqnarray}
  K(k)
  &=&\sum_r\frac{P_r}{2\sqrt{E_{S}^2(k)+\Delta^2(k)}}\,,
\end{eqnarray}
with $P_r^a\rightarrow P_r$ for BCS pairing with $\vecQ=0$ (i.e.,
$E_A = 0$). Figs.~\ref{fig_3_04}-\ref{fig_3_06} show the kernel at
$T=0.25$ MeV for various values of the polarization, the density being
fixed for each figure. As expected in the case of $\alpha=0$ we find a
single peak at the Fermi level. This peak separates into two for
nonvanishing polarizations, which reflects the fact that there are now
two Fermi surfaces for spin-up and spin-down particles. A further
feature seen in Figs.~\ref{fig_3_04}-\ref{fig_3_06} is that at high
densities the peak of the kernel for $\alpha=0$ is located at $k=k_F$
exactly, whereas for low densities this peak is shifted to momenta
below the corresponding $k_F$. In addition at lower densities the
polarization induced two-peak structure is smeared, which is
understood as due to the weakening of the degeneracy of the system.

In Fig.~\ref{fig_3_07} we present the kernel for constant density and
polarization for different temperatures. We clearly see a temperature
induced smearing of the polarization induced two peak structure
resulting, which eventually results in a one peak structure at high
temperatures. Further details to the behavior of the kernel under
asymmetrical conditions can be found in subsection~\ref{subsec_1_3_5}.

\subsubsection{The Cooper-pair wave function}
Next we want to discuss the Cooper-pair wave function $\Psi(r)$ and
the quantity $r^2\vert\Psi(r)\vert^2$, which is the second moment of
the density distribution of Cooper-pairs. Having at our disposal the
wave function we can also access the coherence length
$\xi_\mathrm{rms}$ of the condensate numerically. This then will be
compared with the analytical BCS expression for the coherence length
$\xi_a$ and the interparticle distance $d$. The definitions of these
quantities are analogous to those defined in
Eqs.~\eqref{eq_1_54}-\eqref{eq_1_59}; nevertheless we list them for
convenience. The wave function is given by
\begin{eqnarray}
  \Psi(r)&=& \frac{\sqrt{N}}{2\pi^2r} \int_0^\infty
             dp\,p\,[K(p,\Delta)-K(p,0)]\sin(pr)\,,
\end{eqnarray}
and obeys the normalisation
\begin{eqnarray}
  1&=&N\int d^3r \vert \Psi(r)\vert^2\,.
\end{eqnarray}
The root-mean-square (rms) value for the coherence length is given by
\begin{eqnarray}
  \xi_{\rm rms} &=& \sqrt{\langle r^2\rangle}\,,
\end{eqnarray}
where
\begin{eqnarray}
  \langle r^2\rangle &\equiv& \int d^3r\, r^2
                              \vert\Psi(r)\vert^2\,.
\end{eqnarray}
The analytical BCS for the coherence length is given by
\begin{eqnarray}
  \xi_a &=& \frac{\hbar^2 k_F}{\pi m^* \Delta}\,,
\end{eqnarray}
where now $\Delta$ is the pairing gap in the $^1S_0$ channel and $m^*$
is the effective mass of neutrons. Finally the interparticle distance
is simply related to the net density of the system
\begin{eqnarray}
  d &=& \left(\frac3{4\pi\rho}\right)^{1/3}\,.
\end{eqnarray}
Further discussion of these quantities can be found in
subsection~\ref{subsec_1_3_6}.

Table~\ref{table_3_1} displays the quantities defined above for fixed
temperature $T=0.25$ MeV and vanishing polarization and for several
values of the density, specifically $\log(\rho/\rho_0)=-1.0$,
$\log(\rho/\rho_0)=-1.5$ and $\log(\rho/\rho_0)=-2.0$. We list for
each density $k_F$, $\Delta$, $m^*/m$, $\mu_n$, $d$,
$\xi_\mathrm{rms}$ and $\xi_a$. It is seen that at high densities
$\xi_\mathrm{rms}\approx \xi_a$, i.e., the BCS analytical expression
provides a good approximation to the numerically computed coherence
length. This is not the case in the low density limit and we should
rely only on the numerical value $\xi_\mathrm{rms}$. The comparison of
the numerically computed coherence length with the interparticle
distance shows a clear signature of BCS-BEC crossover, because for
$\log (\rho/\rho_0) = -1$ we find $\xi_\mathrm{rms}/d \approx 2$,
whereas for $\log (\rho/\rho_0) = -2$ we find that
$\xi_\mathrm{rms}/d \approx 0.45$. We will trace the signatures of
BCS-BEC crossover in other variables for spin-polarized neutron matter
below.

\begin{figure}[!]
  \begin{center}
    \includegraphics[width=0.9\textwidth]{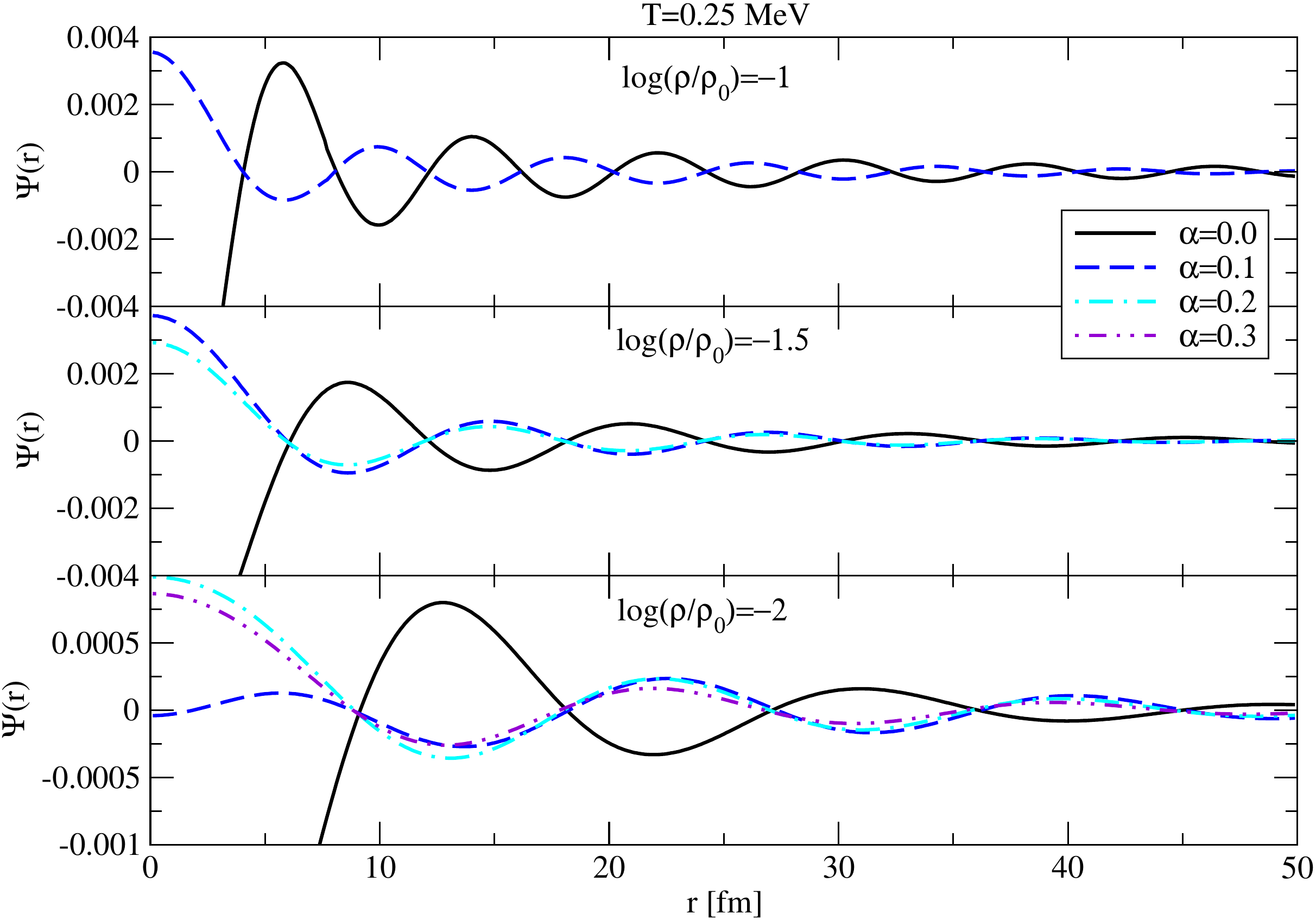}
    \caption[Dependence of $\Psi(r)$ on $r$ for three densities and
    four polarizations.] {Dependence of $\Psi(r)$ on $r$ at fixed
      temperature $T=0.25$ MeV. Different panels show different
      densities and different colors show different values of the
      polarization as indicated in the plot. This figure is analog to
      Fig.~\ref{fig_1_18}.}
    \label{fig_3_08}
  \end{center}
\end{figure}

\begin{figure}[!]
  \begin{center}
    \includegraphics[width=0.9\textwidth]{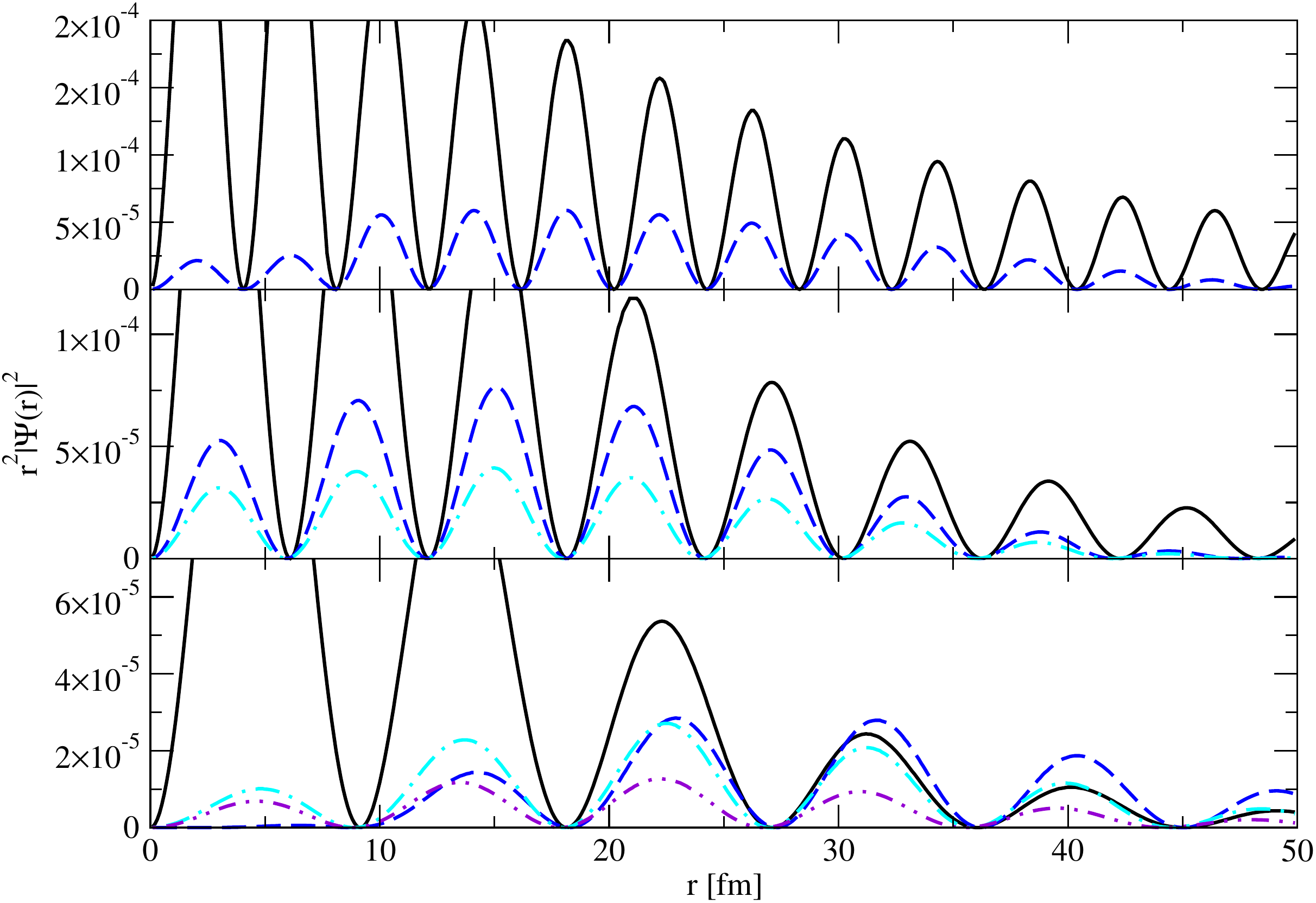}
    \caption[Dependence of $r^2|\Psi(r)|^2$ on $r$ for three densities
    and four polarizations.] {Dependence of $r^2|\Psi(r)|^2$ on
      $r$. The color code, the arrangement of the figures and the
      values for density, temperature and polarization are the same as
      in Fig.~\ref{fig_3_08}. This figure is analog to
      Fig.~\ref{fig_1_19}.}
    \label{fig_3_09}
  \end{center}
\end{figure}

In Fig.~\ref{fig_3_08} we show the Cooper-pair wave function $\Psi(r)$
as a function of radial distance for various densities and
polarization at fixed temperature. In all cases we find strongly
oscillating wave functions. For nonvanishing polarization, the wave
function experiences a sign change (or, in other words, the
oscillations are in counter-phase to the unpolarized case). An
increasing polarization decreases the amplitude $\Psi(r)$, which is
consistent with the fact that the pairing gap is reduced. Furthermore,
the oscillation periods are given roughly by $2\pi/k_F$, therefore
decreasing the densities and $k_F$ leads to the increase of the period
of the oscillations. The degree of polarization does not effect the
period, which is given by $k_F$. In Fig.~\ref{fig_3_09} we show
$r^2\vert\Psi(r)\vert^2$ as a function of radial distance under
conditions fully analogous to Fig.~\ref{fig_3_08}. The oscillatory
behavior discussed for the previous figure is reflected quite
naturally in this quantity as well. A feature that is better visible
here is that for the lowest density the maxima of the function for
different polarizations are shifted with respect to each other; also
the overall maximum reached for each polarization is not at the same
value of $r$. We conclude that general features of the wave functions
of $^1S_0$ pairing neutron matter discussed in this subsubsection are
the same as those of $\SD$ pairing nuclear matter, described in detail
in subsection~\ref{subsec_1_3_6}.

\subsubsection{Occupation numbers}
In this subsubsection we describe the occupation numbers of spin-up
and spin-down neutrons in spin-polarized pure neutron matter. The
occupation numbers are given by the integrand of
Eq.~\eqref{eq_3_20}. It simplifies in $^1S_0$ paired pure neutron
matter in the BCS phase to the following
\begin{eqnarray}
  n_{\uparrow/\downarrow}(k)
  &=&\frac12\left(1+\frac{E_S}{\sqrt{E_S^2+\Delta^2}}\right)f(E_\mp)\nonumber\\
  &&+\frac12\left(1-\frac{E_S}{\sqrt{E_S^2+\Delta^2}}\right)
     [1-f(E_\pm)]\,,
\end{eqnarray}
with $E_r^a\rightarrow E_r$ for BCS pairing with $\vecQ=0$ (i.e.,
$E_A = 0$). We note in passing that because of different degeneracy
factor in neutron matter (no summation over spin) the maximum of
functions $ n_{\uparrow/\downarrow}(k)$ is $1$, instead of $2$ as in
nuclear matter.

\begin{figure}[!]
  \begin{center}
    \includegraphics[width=0.8\textwidth]{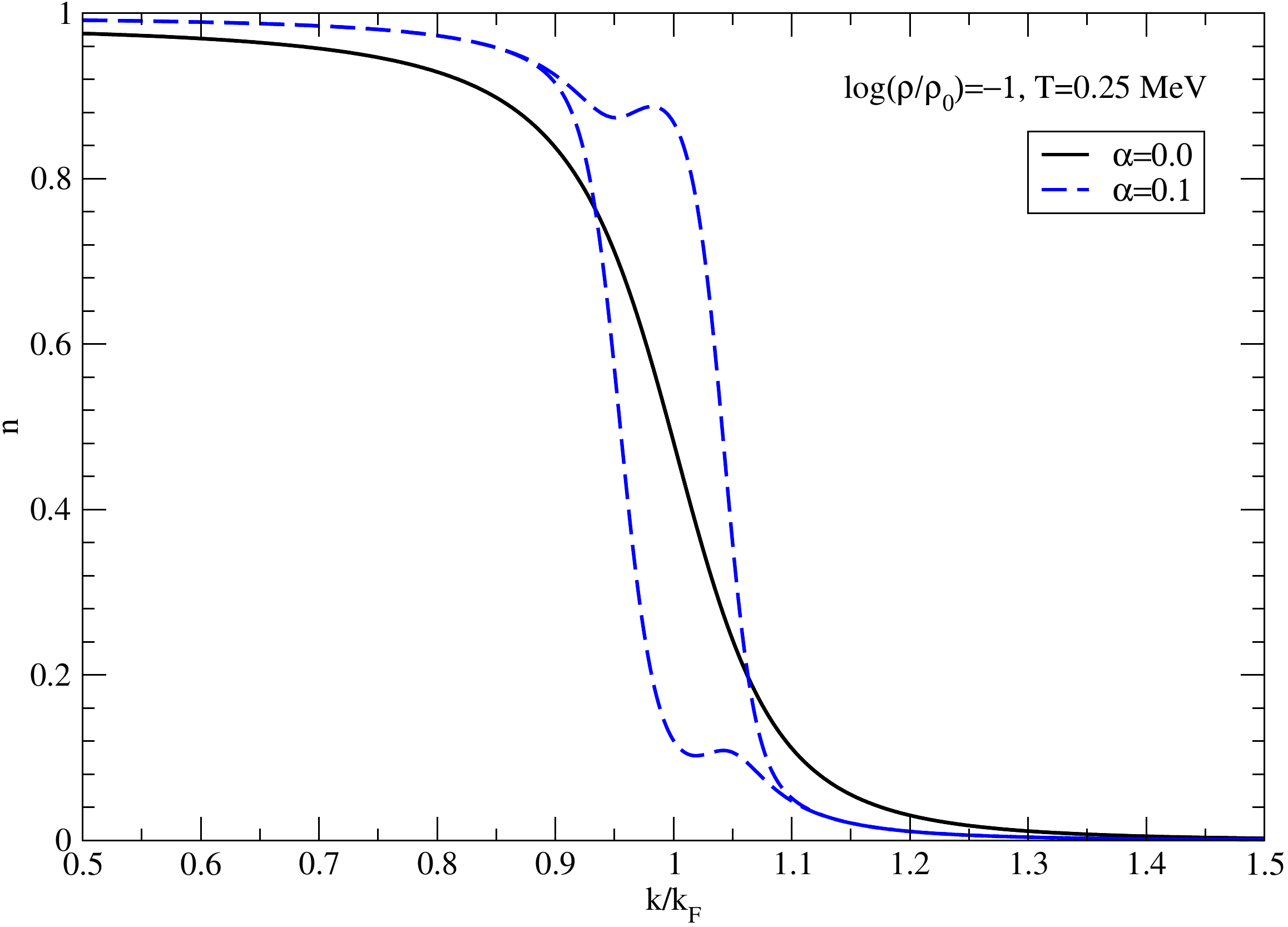}
    \caption[Dependence of the spin-up and spin-down neutron
    occupation numbers on momentum for various polarizations at high
    density.] {Dependence of the spin-up and spin-down neutron
      occupation numbers on momentum $k$ (in units of Fermi momentum)
      for fixed $\log(\rho/\rho_0)=-1$, $T=0.25$ MeV, and various
      values of polarization indicated in the plot. This figure is
      analog to Fig.~\ref{fig_1_24}.}
    \label{fig_3_10}
  \end{center}
\end{figure}

\begin{figure}[!]
  \begin{center}
    \includegraphics[width=0.8\textwidth]{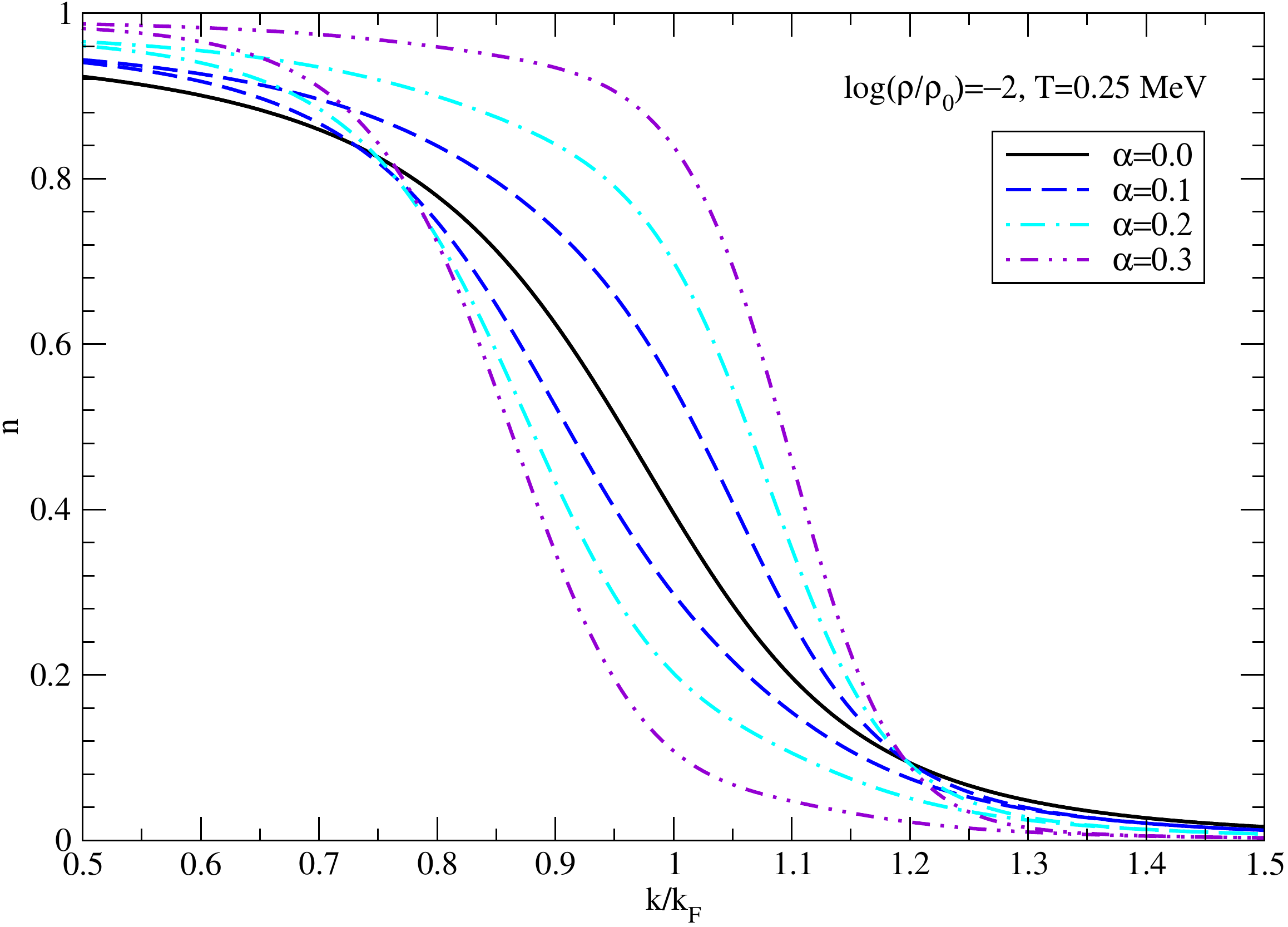}
    \caption[Dependence of the spin-up and spin-down neutron
    occupation numbers on momentum for various polarizations at low
    density.] {Same as Fig.~\ref{fig_3_10} but for
      $\log(\rho/\rho_0)=-2$ and more values for the polarization.}
    \label{fig_3_12}
  \end{center}
\end{figure}

In Figs.~\ref{fig_3_10}-\ref{fig_3_12} we display the occupation
numbers of spin-up and down neutrons at fixed temperature $T=0.25$
MeV, fixed densities $\log (\rho/\rho_0)$ $= -1$, $-1.5$ and $-2$,
respectively. The polarizations are shown in the figures. In the case
of vanishing polarization (solid lines) the Fermi step-function-like
occupation in the high-density limit changes its slope into a
``flatter'' distribution at lower densities; this corresponds to a
more diffuse Fermi surface at low densities. At finite polarizations
the occupation numbers of spin-up and down particles split in the
region around $k_F$. In fact, the locations of the drop in the
occupations of these populations agree well with their corresponding
Fermi wave-vectors. At high density the polarization induced splitting
results in a ``breach'' for large polarizations with $n_{n\uparrow}=1$
and $n_{n\downarrow}=0$ around $k_F$. The breach remains intact at
lower densities, but the slope of the corresponding occupation numbers
changes as already observed for the case of unpolarized matter.

\begin{figure}[!]
  \begin{center}
    \includegraphics[width=0.8\textwidth]{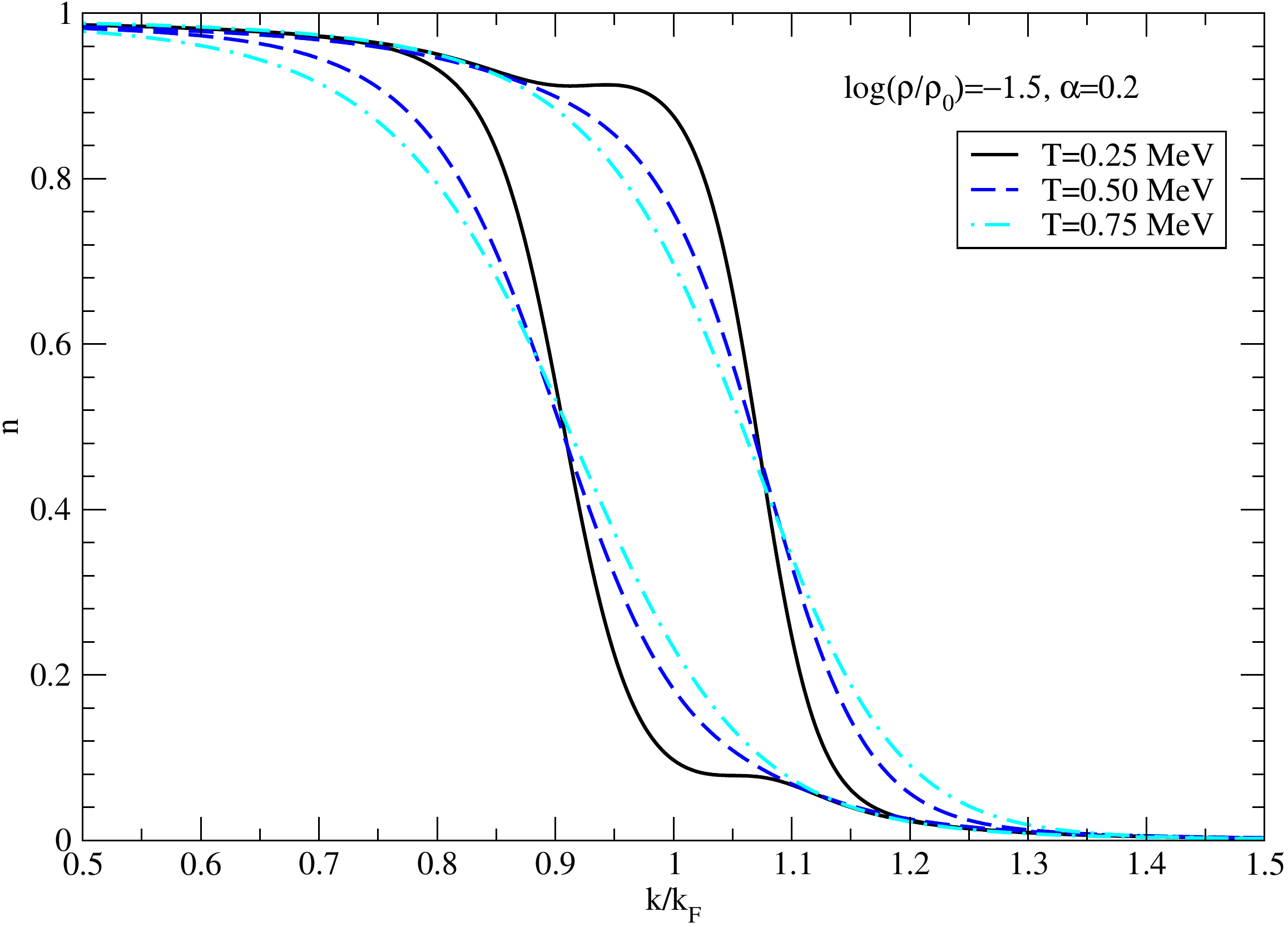}
    \caption[Dependence of the spin-up and spin-down neutron
    occupation numbers on momentum for various temperatures.]
    {Dependence of the spin-up and spin-down neutron occupation
      numbers on momentum $k$ (in units of Fermi momentum) for fixed
      $\log(\rho/\rho_0)=-1.5$, $\alpha=0.2$, and various temperatures
      indicated in the plot. This figure is analog to
      Fig.~\ref{fig_1_09}.}
    \label{fig_3_13}
  \end{center}
\end{figure}

In \ref{fig_3_13} we show the occupation numbers of spin-up and
spin-down neutrons at fixed density $\log(\rho/\rho_0)=-1.5$ and fixed
polarization $\alpha=0.2$ for different temperatures. The effect of
temperature is to induce smearing of the occupation numbers with
increasing temperature. More details can be found in
subsection~\ref{subsec_1_3_3}, where we discuss the temperature
dependence of occupation numbers for $\SD$ condensate and in
subsection~\ref{subsec_1_3_7} where we give a general analysis of the
occupation numbers.

\subsubsection{Quasiparticle spectra}
\begin{figure}[!]
  \begin{center}
    \includegraphics[width=\textwidth]{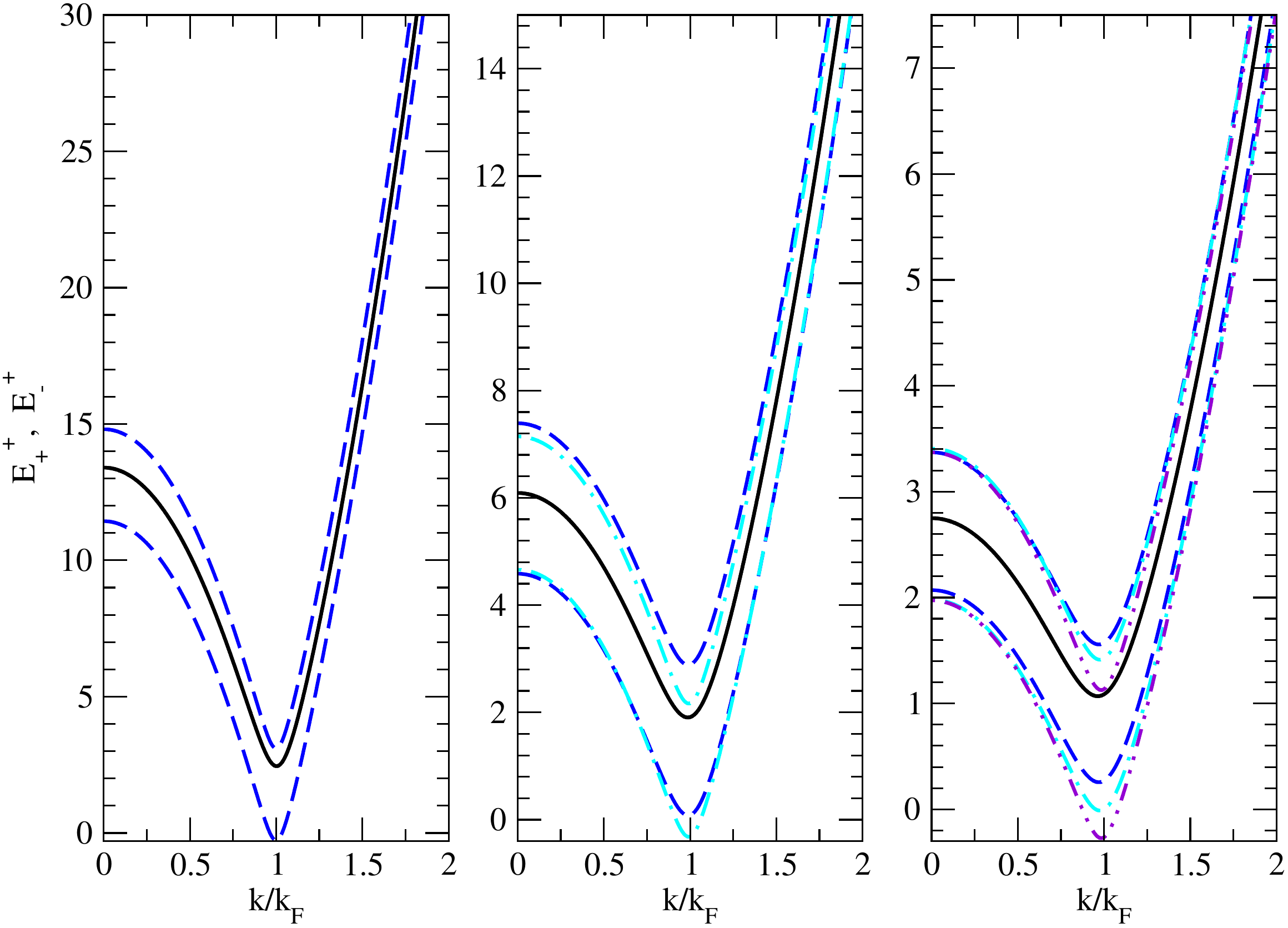}
    \caption[Dispersion relations for quasiparticle spectra in the
    case of the BCS condensate, as functions of momentum.] {Dispersion
      relations for quasiparticle spectra in the case of the BCS
      condensate, as functions of momentum in units of Fermi
      momentum. For each polarization, the upper branch corresponds to
      $E^+_+$, and the lower to the $E^+_-$ solution. The color code,
      the arrangement of the figures and the values for density,
      temperature and polarization are the same as in
      Fig.~\ref{fig_3_08}. This figure is analog to
      Fig.~\ref{fig_1_26}.}
    \label{fig_3_14}
  \end{center}
\end{figure}

As a final intrinsic quantity of interest we consider the dispersion
relations for quasiparticle excitations about the $^1S_0$
condensate. Because we do not find LOFF phase the quasiparticle
branches $E_{\pm}^{-} = E_{\pm}^{+}$ are degenerated and we can drop
the superscript altogether. Thus the quasiparticle spectrum is given
by
\begin{eqnarray}
  E_{\pm}(k) &=&
                 \sqrt{\left(\frac{k^2}{2m^*}-\bar\mu\right)^2+\Delta^2}\pm
                 \delta\mu\,.
\end{eqnarray}
These dispersion relations are shown in Fig.~\ref{fig_3_14} for
various values of density and polarization for fixed temperature
$T=0.25$ MeV. In each case the spectrum has a minimum at $k_F$. At
finite polarization there is a splitting of spectra of spin-up and
spin-down neutrons. Furthermore, for low densities the spectrum of
minority (spin-down neutrons) crosses zero, which means that their
spectrum is gapless. The overall behavior of the dispersion relations
are analogous to that of the $\SD$ condensate which we discussed in
subsection~\ref{subsec_1_3_8}.

\subsection{Magnetic field strength}
As mentioned earlier, the spin-polarization in pure neutron matter can
be induced by a magnetic field. In this subsection we want to discuss
the strength of the magnetic field needed to create a certain
polarization and compare the corresponding energy $\varepsilon_B$
given by Eq.~\eqref{eq_3_25} with the temperature. We note however
that the magnetic field is linearly related to the shift in the
chemical potentials $\delta\mu$ and in essence our study is equivalent
to the study of chemical potentials shift for given spin-polarization.

\begin{figure}[!]
  \begin{center}
    \includegraphics[width=0.8\textwidth]{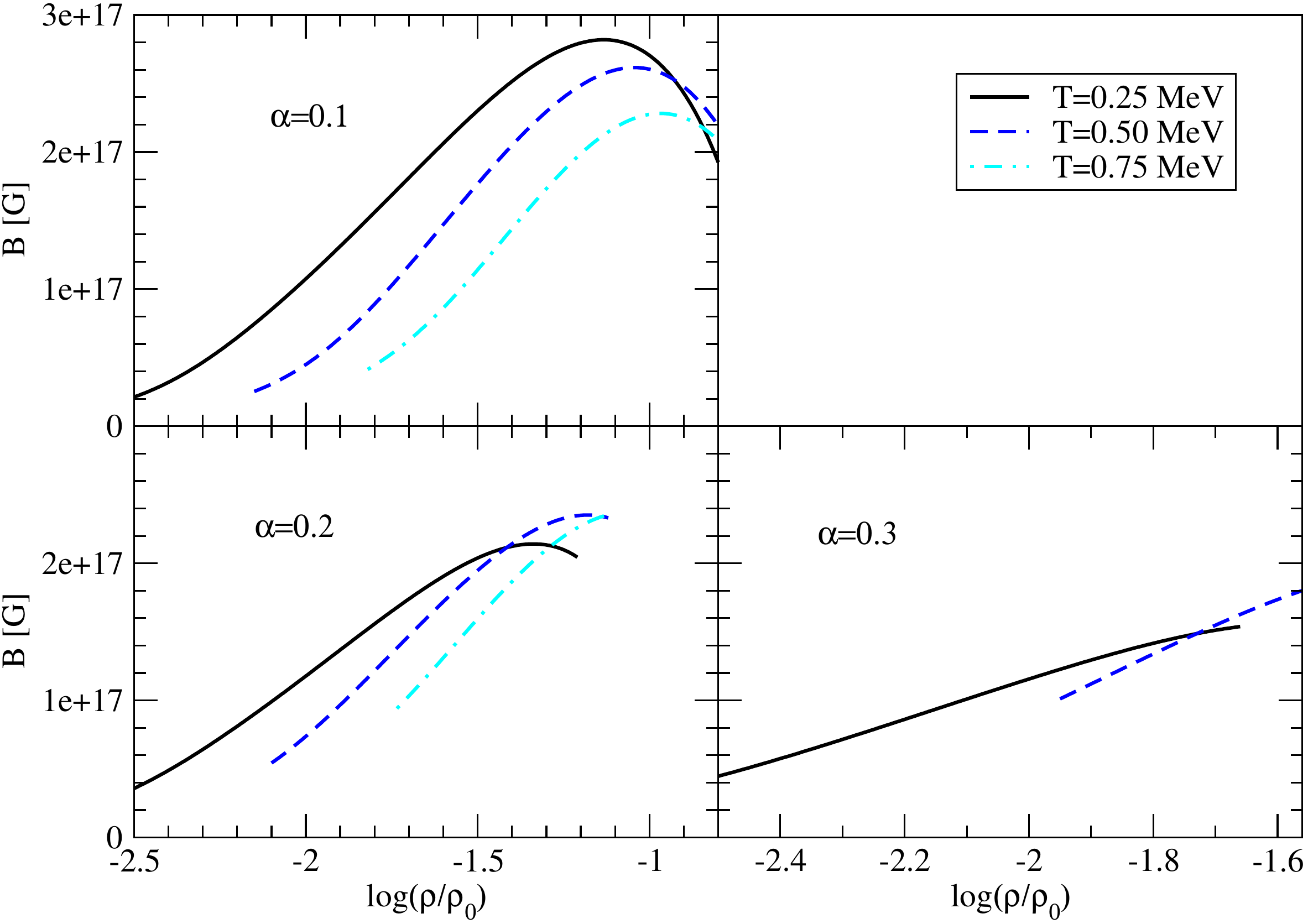}
    \caption[The needed magnetic field to create a certain
    spin-polarization as a function of the density with fixed
    polarisation.] {The needed magnetic field to create a certain
      spin-polarization as a function of the density. In each panel a
      certain polarization is fixed. Different temperatures are
      presented with different colors.}
    \label{fig_3_15}
  \end{center}
\end{figure}

\begin{figure}[!]
  \begin{center}
    \includegraphics[width=0.8\textwidth]{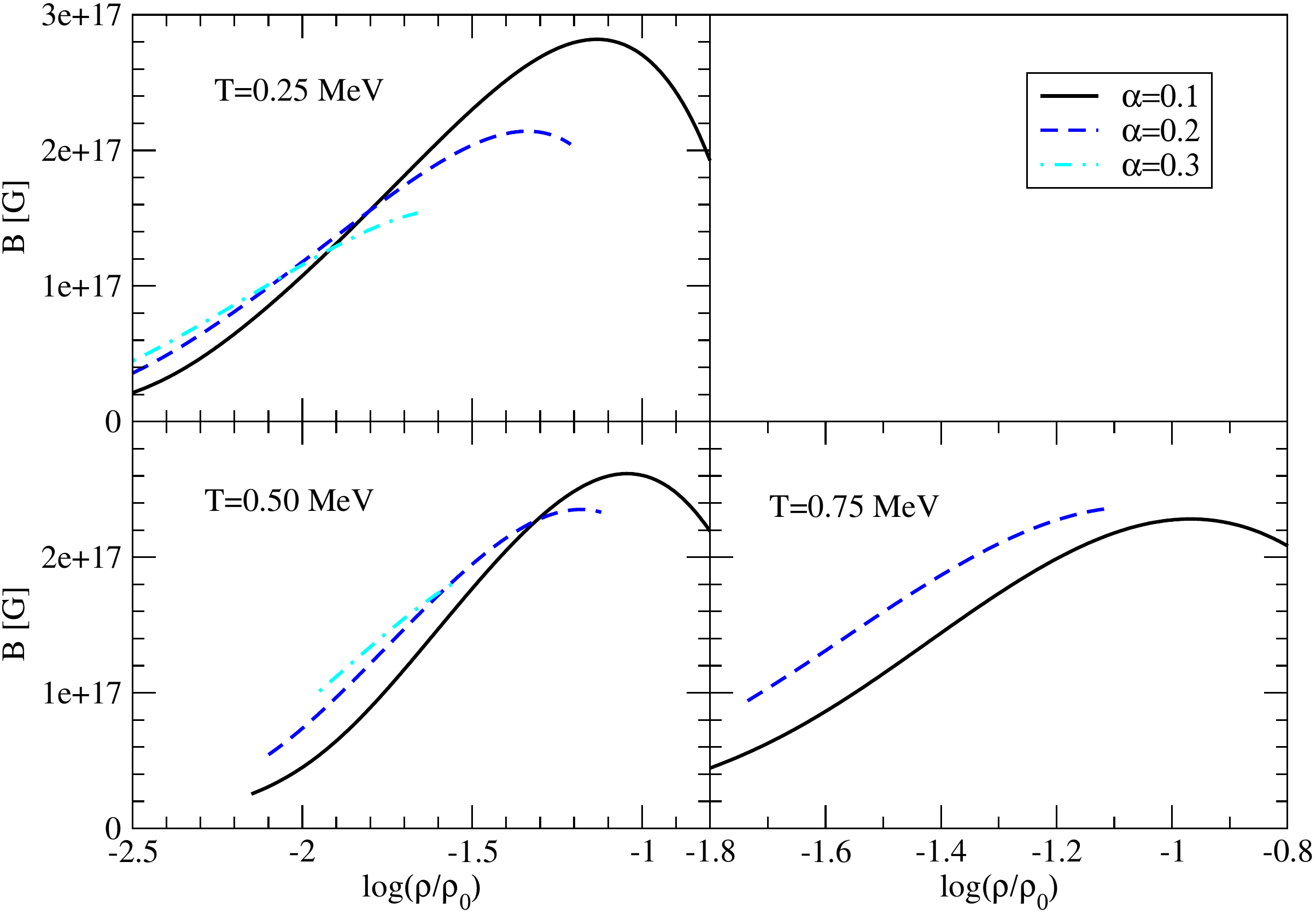}
    \caption[The needed magnetic field to create a certain
    spin-polarization as a function of the density with fixed
    temperature.] {The needed magnetic field to create a certain
      spin-polarization as a function of the density. In each panel a
      certain temperature is fixed. Different values of polarization
      are presented with different colors.}
    \label{fig_3_16}
  \end{center}
\end{figure}

In Fig.~\ref{fig_3_15} we display the magnetic field as a function of
density at constant polarization and temperature. It is seen that to
obtain a given spin-polarization stronger magnetic fields are needed
for higher densities. That is, the dense matter is harder to polarize
than the low density matter. This trend is reversed at and above
approximately a tenth of the saturation density. The physical content
of this observation is difficult to access because the chemical
potential shift is non-trivially related to the polarization and the
pairing gap. It is further seen from Fig.~\ref{fig_3_15} that finite
temperature matter is more easily polarizable at low densities, but
this trend may reverse again at high densities. In Fig.~\ref{fig_3_16}
we fixed the temperature in each panel and present different
polarizations with different colors. We then learn that a strong
magnetic field is needed to achieve a large polarization in the
low-density matter. This trend may again reverse in the high-density
matter.

\begin{figure}[!]
  \begin{center}
    \includegraphics[width=0.8\textwidth]{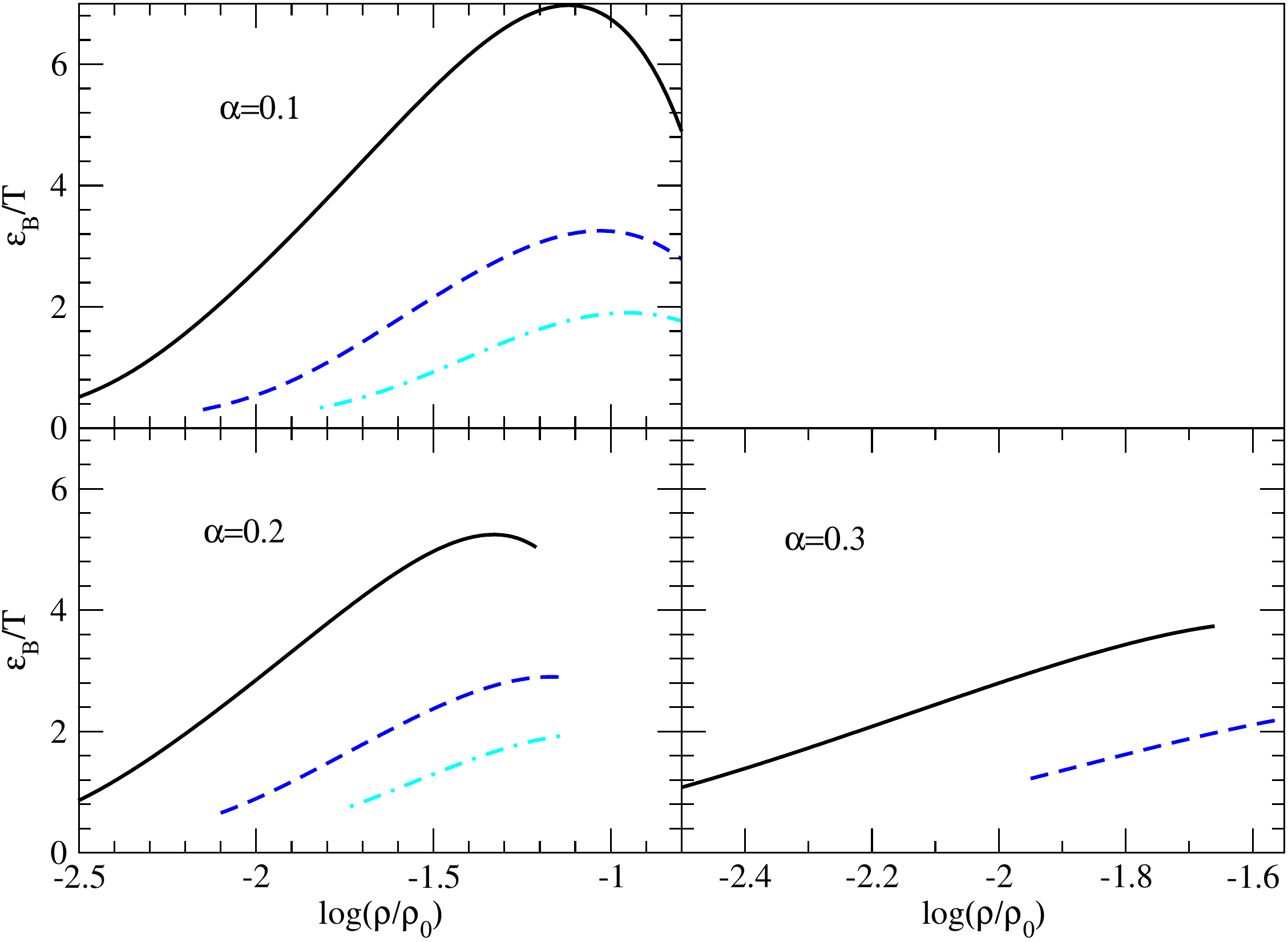}
    \caption[The magnetic energy divided by the temperature as a
    function of the density with fixed polarisation.] {The magnetic
      energy divided by the temperature as a function of the density
      for different temperatures and values of polarization. The color
      code is the same as in Fig.~\ref{fig_3_15}.}
    \label{eq_3_17}
  \end{center}
\end{figure}

\begin{figure}[!]
  \begin{center}
    \includegraphics[width=0.8\textwidth]{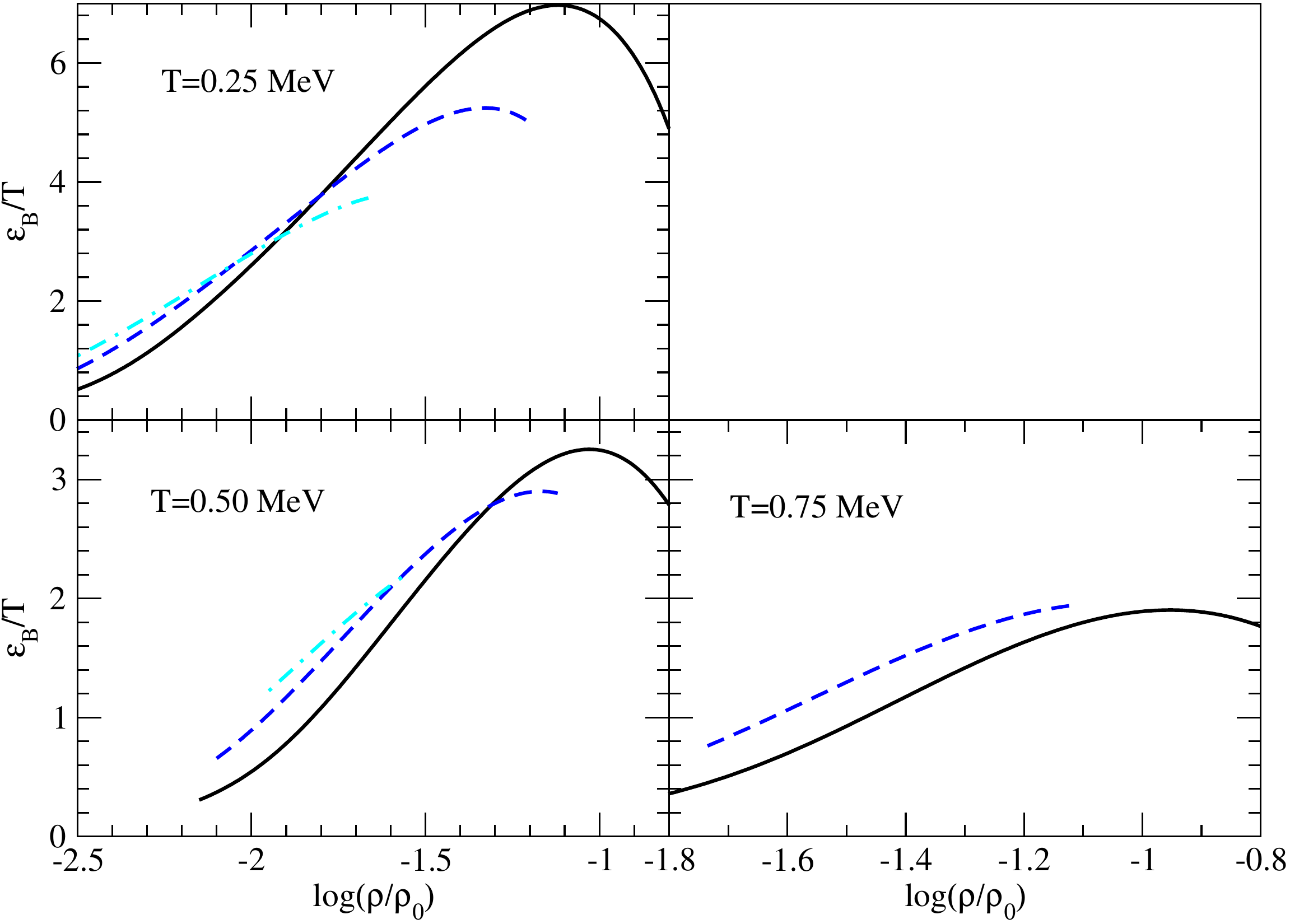}
    \caption[The magnetic energy divided by the temperature as a
    function of the density with fixed temperature.] {The magnetic
      energy divided by the temperature as a function of the density
      for different temperatures and values of polarization. The color
      code is the same as in Fig.~\ref{fig_3_16}.}
    \label{eq_3_18}
  \end{center}
\end{figure}

It is physically interesting to consider the ratio of the magnetic
energy to the temperature in the parameter range discussed above,
i.e., we are interested in
\begin{eqnarray}
  \frac{\varepsilon_B}{T}\approx-\frac{g_n}{2} \frac{e\hbar}{2mc}
  \frac{B}{T} .
\end{eqnarray}
This ratio shown in Figs.~\ref{eq_3_17} and \ref{eq_3_18}, where the
arrangement of the figures and the color code are analogous to
Fig.~\ref{fig_3_15} and Fig.~\ref{fig_3_16}, respectively. We see that
almost in the complete range of the parameter space
$\varepsilon_B/T > 1 $; exceptions are the extreme low density limits
in various plots.


\begin{figure}[!]
  \begin{center}
    \includegraphics[width=0.8\textwidth]{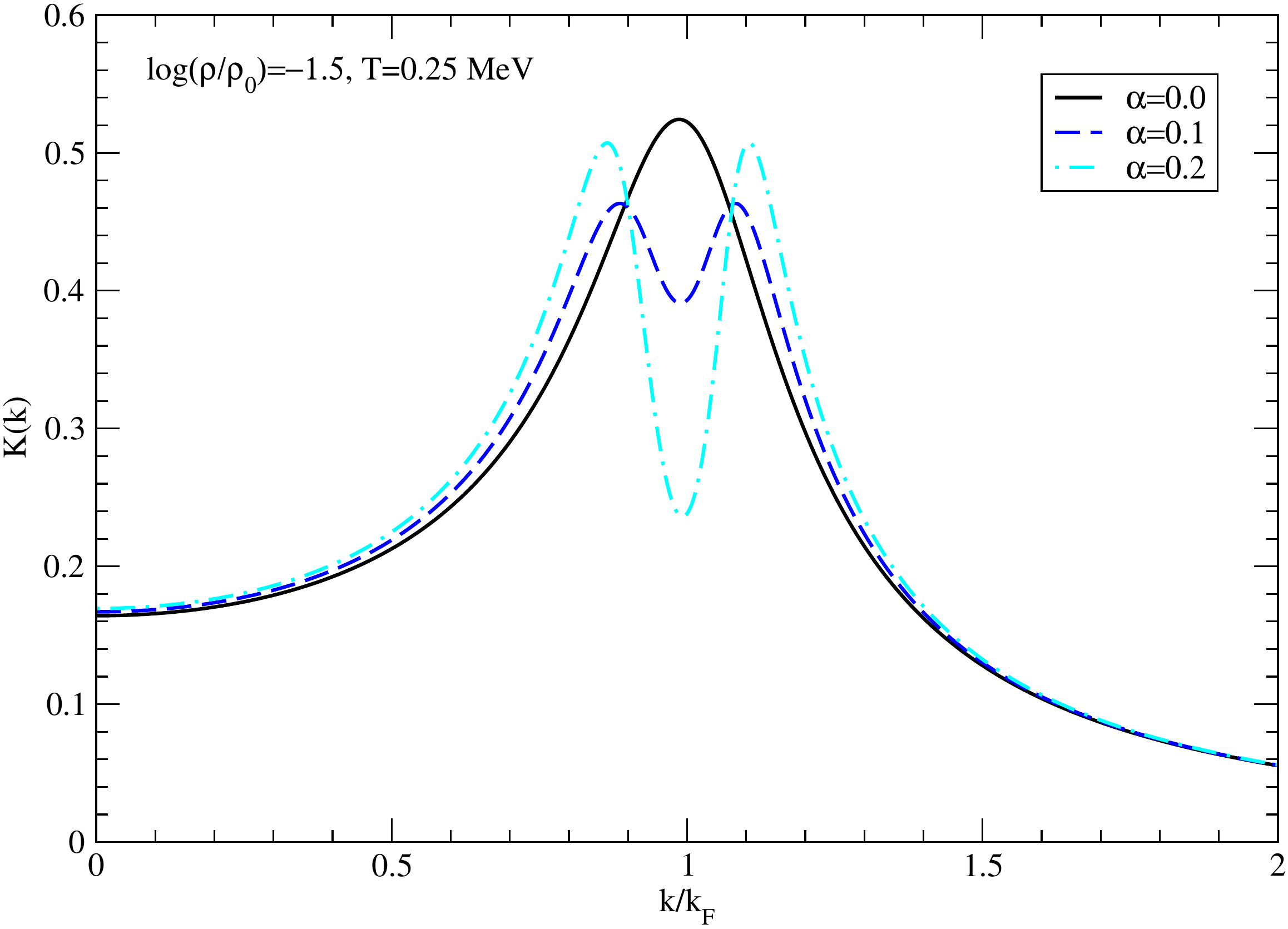}
    \caption[Dependence of the kernel on momentum for various
    polarizations at intermediate density.] {Same as
      Fig.~\ref{fig_3_04} but for $\log(\rho/\rho_0)=-1.5$ and three
      values for the polarization.}
  \end{center}
\end{figure}



\begin{figure}[!]
  \begin{center}
    \includegraphics[width=0.8\textwidth]{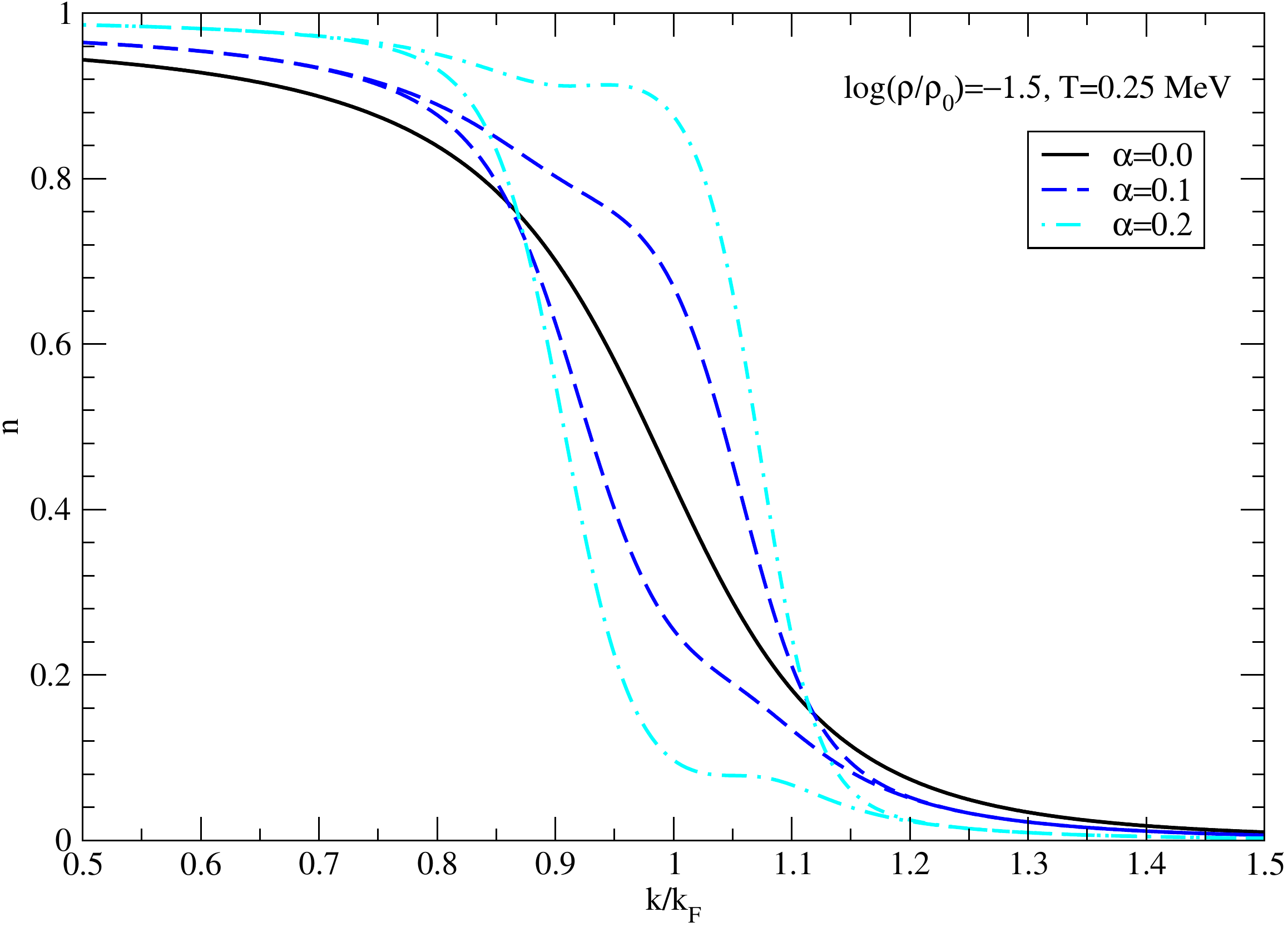}
    \caption[Dependence of the spin-up and spin-down neutron
    occupation numbers on momentum for various polarizations at
    intermediate density.] {Same as Fig.~\ref{fig_3_10} but for
      $\log(\rho/\rho_0)=-1.5$ and more values for the polarization.}
  \end{center}
\end{figure}




\section{Conclusion}
The phase diagram of low-density spin-polarized neutron matter studied
in this chapter has a simpler phase structure compared to the phase
diagram of low-density isospin asymmetric nuclear matter studied in
chapter~\ref{chap_1}. Because two neutrons can not form bound pairs,
there exists no a priori BEC of neutron-neutron pairs. Moreover, there
exists no LOFF phase, because the pairing gap is much too low compared
to the chemical potential. Since the phase diagram consists of only
BCS and unpaired phase, we have two critical temperatures at
non-vanishing polarization.

Our analysis of this chapter can be summarized as follows:
\begin{itemize}
\item At low density, spin-polarization does not affect the pairing
  significantly. For high densities and high polarizations, the
  pairing gap and hence $T_C$ are significantly suppressed. At finite
  polarization and low temperatures we find a lower critical
  temperature due to the combination of the polarization induced
  separation and the temperature induced smearing of the Fermi edges.

\item We studied some intrinsic features of spin-polarized neutron
  condensate, specifically, the gap, the kernel of the gap equation,
  the condensate wave functions, the occupation numbers and the
  quasiparticle spectrum. The similarities and differences to the
  low-density isospin asymmetric nuclear matter studied in
  chapter~\ref{chap_1} have been highlighted. In the following we list
  the main features only.

\item The gap has non-trivial dependence on the polarization and
  temperature. At finite polarizations the gap may be increased by
  increasing the temperature because of the restoration of the
  coherence among the spin-up and down population by the temperature.

\item The kernel of the gap equation has a double peak structure
  compared to the non-polarized case. This structure is most
  pronounced in the high-density and low-temperature limit. Decreasing
  the density (or increasing the temperature) smears out this
  structure.

\item The Cooper-pair wave functions have an oscillating behavior. At
  finite polarization the oscillations are in counter-phase to those
  in the unpolarized case. The period of the oscillations is defined
  by the wave vector as $2\pi/k_F$ and is not affected by the
  polarization.

\item The occupation numbers show a separation of the majority and
  minority components by a ``breach'' around the Fermi momentum. This
  is most pronounced in the high-density and low-temperature limit, in
  which case the spin-down (minority) components is almost
  extinct. For high temperatures or low densities, this ``breach'' is
  smeared out.

\item The study of dispersion relations shows that they have a
  standard BCS form in the unpolarized case and they become split into
  two branches at finite polarization which retains the general BCS
  shape. The minima of these spectra are at $k=k_F$, as it should. At
  larger polarizations the energy spectrum of the spin-down particles
  crosses the zero-energy level, which is a signature of gapless
  superconductivity. In other words, the Fermi surface of spin-down
  particles then features locations where modes can be excited without
  any energy cost.

\item Finally, we studied the influence of the magnetic field on the
  spin-polarization. At low densities a relatively weak magnetic field
  is needed to generate a certain polarization. The magnetic field
  needed to generate a certain polarization in general increases with
  decreasing temperature and with increasing polarization. The energy
  of the magnetic field is in general higher than the temperature,
  except for low densities and high temperatures.
\end{itemize}

\clearpage{\pagestyle{empty}\cleardoublepage}
\begin{appendix}
\phantomsection
\addcontentsline{toc}{chapter}{Appendices}
\chapter[Matsubara summations] {Matsubara summations}
\label{app_1}

First we rewrite the equation for $F_{np}^{\pm}$:
\begin{subequations}
  \begin{eqnarray}
    F_{np}^{\pm} &=& \frac{-i\Delta}{(ik_{\nu}-E^+_{\pm})(ik_{\nu}+E^-_{\mp})}\\
    \Rightarrow F_{np}^{\pm} &=& \frac{-i\Delta}{E^+_{\pm}+E^-_{\mp}}\cdot\left(\frac1{ik_{\nu}-E^+_{\pm}}-\frac1{ik_{\nu}+E^-_{\mp}}\right)\\
    \Rightarrow F_{np}^{\pm} &=& \frac{-i\Delta}{2\sqrt{E_S^2+\Delta^2}}\cdot\left(\frac1{ik_{\nu}-E^+_{\pm}}-\frac1{ik_{\nu}+E^-_{\mp}}\right)\,.
  \end{eqnarray}
\end{subequations}

For the Matsubara summation we first solve
\begin{subequations}
  \begin{eqnarray}
    S(\pm E)&=&\frac1\beta\sum_\nu\frac1{ik_{\nu}\pm E}=\frac1\beta\sum_\nu g(ik_\nu)\,,\\
    I(\pm E)&=&\underset{R\rightarrow\infty}{\lim}\int\frac{dz}{2\pi i}g(z)f(z) =\sum_i\mathrm{Res}(z_i)\,,
  \end{eqnarray}
\end{subequations}
with $z_i$ being the poles of $g(z)f(z)$ and $f(z)$ being the Fermi
function.

\begin{subequations}
  \begin{eqnarray}
    f(z)&=&\frac1{e^{\beta z}+1}\Rightarrow z_n=\frac{(2n+1)\pi i}\beta\\
    \Rightarrow R_n&=&\underset{z\rightarrow (2n+1)\pi i/\beta}{\lim}\,\frac{z-(2n+1)\pi i/\beta}{e^{\beta z}+1}g(z)\nonumber\\
        &=&\underset{z\rightarrow (2n+1)\pi i/\beta}{\lim}\,\frac{z-(2n+1)\pi i/\beta}{-1+\beta z+(2n+1)\pi i+1}g(z)\nonumber\\
        &=&\underset{z\rightarrow (2n+1)\pi i/\beta}{\lim}\,\frac{z-(2n+1)\pi i/\beta}{\beta z+(2n+1)\pi i}g(z_n)\nonumber\\
        &=&-\frac1\beta g(z_n)\,.
  \end{eqnarray}
\end{subequations}

The next residuum we obtain at the pole of $g(z)$
\begin{subequations}
  \begin{eqnarray}
    R_1&=&\underset{z\rightarrow \mp E}{\lim}\,\frac{z\pm E}{z\pm E}f(z)\\
       &=&f(\mp E)\,,\qquad\qquad\qquad f(-E)=1-f(E)\,.
  \end{eqnarray}
\end{subequations}

Thus all together we obtain:
\begin{subequations}
  \begin{eqnarray}
    I(\pm E)&=&-\frac1\beta\sum_\nu g(ik_\nu)+f(\mp E)\\
    \Rightarrow \frac1\beta\sum_\nu\frac1{ik_{\nu}\pm E}&=&f(\mp E)\\
    \Rightarrow \frac1\beta\sum_\nu F_{np}^{\pm} &=& \frac{-i\Delta}{2\sqrt{E_S^2+\Delta^2}}\nonumber\\
            &&\cdot\frac1\beta\sum_\nu\left(\frac1{ik_{\nu}-E^+_{\pm}}-\frac1{ik_{\nu}+E^-_{\mp}}\right)\\
            &=&\frac{-i\Delta}{2\sqrt{E_S^2+\Delta^2}}\left(f(E^+_{\pm})-f(-E^-_{\mp})\right)\\
            &=&\frac{i\Delta}{2\sqrt{E_S^2+\Delta^2}}\left(1-f(E^+_{\pm})-f(E^-_{\mp})\right)\,.
  \end{eqnarray}
\end{subequations}

Analogue we obtain:
\begin{eqnarray}
  \frac1\beta\sum_\nu F_{pn}^{\pm} &=&\frac{-i\Delta}{2\sqrt{E_S^2+\Delta^2}}\left(1-f(E^+_{\mp})-f(E^-_{\pm})\right)\,.
\end{eqnarray}

Next we need the Matsubara summation of $G$:
\begin{eqnarray}
  G_{n/p}^{\pm} &=&
                    \frac{ik_{\nu}\pm\epsilon_{p/n}^{\mp}}{(ik_{\nu}-E^+_{\mp/\pm})(ik_{\nu}+E^-_{\pm/\mp})}\,.
\end{eqnarray}

For this purpose we introduce the following
\begin{subequations}
  \begin{eqnarray}
    S(\pm\epsilon)&=&\frac1\beta\sum_\nu\frac{ik_{\nu}\pm\epsilon}{(ik_{\nu}-E^+)(ik_{\nu}+E^-)} =\frac1\beta\sum_\nu g(ik_\nu)\,,\\
    I(\pm\epsilon)&=&\underset{R\rightarrow\infty}{\lim}\int\frac{dz}{2\pi i}g(z)f(z) =\sum_i\mathrm{Res}(z_i)\,,
  \end{eqnarray}
\end{subequations}

and obtain:
\begin{subequations}
  \begin{align}
    z_n=&(2n+1)\pi i/\beta &R_n=&-\frac1\beta g(z_n)\,,\\
    z_1=&E^+&R_1=&\frac{E^+\pm\epsilon}{E^++E^-}f(E^+)\,,\\
    z_2=&-E^-&R_2=&\frac{E^-\mp\epsilon}{E^++E^-}(1-f(E^-))\,,
  \end{align}
\end{subequations}
which can be solved to
\begin{eqnarray}
  &&\frac1\beta\sum_\nu\frac{ik_{\nu}\pm\epsilon}{(ik_{\nu}-E^+)(ik_{\nu}+E^-)}\nonumber\\
  &=&\frac{E^+\pm\epsilon}{2\sqrt{E_S^2+\Delta^2}}f(E^+)+\frac{E^-\mp\epsilon}{2\sqrt{E_S^2+\Delta^2}}(1-f(E^-))\,.
\end{eqnarray}

Thus we obtain:
\begin{subequations}
  \begin{eqnarray}
    &&\frac1\beta\sum_\nu G_{n/p}^{\pm} =
       \frac1\beta\sum_\nu \frac{ik_{\nu}\pm\epsilon_{p/n}^{\mp}}{(ik_{\nu}-E^+_{\mp/\pm})(ik_{\nu}+E^-_{\pm/\mp})}\\
    &=&\frac{E^+_{\mp/\pm}\pm\epsilon_{p/n}^{\mp}}{2\sqrt{E_S^2+\Delta^2}}f(E^+_{\mp/\pm})+\frac{E^-_{\pm/\mp}\mp\epsilon_{p/n}^{\mp}}{2\sqrt{E_S^2+\Delta^2}}(1-f(E^-_{\pm/\mp}))\\
    &=&\frac12\left(1\pm\frac{E_S}{\sqrt{E_S^2+\Delta^2}}\right)f(E^+_{\mp/\pm})\nonumber\\
    &&+\frac12\left(1\mp\frac{E_S}{\sqrt{E_S^2+\Delta^2}}\right)(1-f(E^-_{\pm/\mp}))\,.
  \end{eqnarray}
\end{subequations}

These summations will be needed for further calculations.

\clearpage{\pagestyle{empty}\cleardoublepage}

\chapter[Description of the TDHF Code] {Description of the TDHF Code}
\label{app_2}

This appendix gives a brief overview of the code Sky3D adopted form
Ref.~\cite{2014CoPhC.185.2195M}; there a more detailed discussion can
be found.

\section{Local densities and currents}
\label{sec_b_1}
The code calculates with a set of single-particle (s.p.) wave
functions $\psi_\alpha$, with $\alpha\leq\Omega$, with $\Omega$
denoting the size of the active single particle (s.p.) space. For
non-occupied states, $\psi_\alpha$ vanishes.

For the description of the Skyrme-energy-density functional only a few
local densities and currents are needed. The time-even fields are
given by
\begin{subequations}
  \label{eq_b_1}
  \begin{align}
    \rho_q(\vecr)=&\sum_{\alpha\in q}\sum_s\left|\psi_\alpha(\vecr,s)\right|^2 &&\text{density}\,,\label{eq_b_1a}\\
    \vecJ_q(\vecr)=&-\mathrm i\sum_{\alpha\in q}\sum_{ss'}\psi^*_\alpha(\vecr,s)\nabla\times\vecsigma_{ss'}\psi_a(\vecr,s') &&\text{spin-orbit density}\,,\\
    \tau_q(\vecr)=&\sum_{\alpha\in q}\sum_s\left|\nabla\psi_\alpha(\vecr,s)\right|^2 \label{eq_b_1c}&&\text{kinetic density}\,,
  \end{align}
  and for the time-odd fields one has
  \begin{align}
    \vecs_q(\vecr)=&\sum_{\alpha\in q}\sum_{ss'}\psi^*_\alpha(\vecr,s)\vecsigma_{ss'}\psi_a(\vecr,s') &&\text{spin density}\,,\\
    \vecj_q(\vecr)=&\Im\left\{\sum_{\alpha\in q}\sum_s\psi^*_\alpha(\vecr,s)\nabla\psi_\alpha(\vecr,s)\right\} &&\text{current density}\,,\label{eq_b_1e}
  \end{align}
\end{subequations}
with $q$ labels the isospin with $q=p$ for protons and $q=n$ for
neutrons. A local density/current without index $q$ is the total
density/current; for example $\rho=\rho_p+\rho_n$. The variable $s$
represents the two spinor components of the wave functions. The terms
are understood as functions of $\vecr$. In addition to these terms,
the pairing density~\cite{2014CoPhC.185.2195M} and the time-odd
kinetic spin-density can be defined; moreover it is possible to define
the spin-orbit density as a tensor, see
Ref.~\cite{2011JPhG...38c3101E}.

\section{The energy-density functional}
\label{sec_b_2}
The terms of Eq.~\eqref{eq_2_01} using the force coefficients of
appendix~\ref{sec_b_3} are given by
\begin{subequations}
  \begin{itemize}
  \item $T$: the total kinetic energy given by
    \begin{eqnarray}
      T=\sum_q\frac{\hbar^2}{2m_q}\int d^3r\,\tau_q\,,
    \end{eqnarray}
    with $\tau_q$ being the kinetic density of Eq.~\ref{eq_b_1c}.
  \item $E_0$: the $b_0$ and $b'_0$-dependent part is given by
    \begin{eqnarray}
      E_0=\int d^3r\left(\frac{b_0}2\rho^2-\frac{b'_0}2\sum_q \rho^2_q\right)\,.\label{eq_b_2b}
    \end{eqnarray}
  \item $E_1$: kinetic terms which contain the coefficients $b_1$ and
    $b'_1$:
    \begin{eqnarray}
      E_1=\int d^3r\left(b_1\left[\rho\tau-\vecj^2\right] -b'_1\sum_q\left[\rho_q\tau_q-\vecj_q^2\right]\right)\,.
    \end{eqnarray}
  \item $E_2$: terms which contain the coefficients $b_2$ and
    $b'_2$. They include the Laplacians of the densities
    \begin{eqnarray}
      E_2&=&\int d^3r\left(-\frac{b_2}2\rho\nabla\rho +\frac{b'_2}2\sum_q\rho_q\nabla\rho_q\right)\,.
    \end{eqnarray}
  \item $E_3$: the many-body contribution
    \begin{eqnarray}
      E_3=\int d^3r\left(\frac{b_3}3\rho^{\alpha+2} -\frac{b'_3}3\rho^\alpha\sum_q\rho_q^2\right)\,.\label{eq_b_2e}
    \end{eqnarray}
  \item $E_{ls}$: the spin-orbit energy is given by
    \begin{eqnarray}
      E_{ls}&=&\int d^3r\left( -b_4\left[\rho\nabla\cdot\vecJ+\vecs\cdot\left(\nabla\times\vecj\right)\right]\right.\nonumber\\
            &&\left. -b'_4\sum_q\left[\rho_q\nabla\cdot\vecJ_q+\vecs_q\cdot\left(\nabla\times\vecj_q\right)\right]\right)\,.\label{eq_b_2f}\\\nonumber
    \end{eqnarray}
  \item $E_C$: the Coulomb energy. It consists of the Hartree term,
    which includes the standard expression of a charge distribution of
    the own field of the nucleus and the exchange term in the Slater
    approximation. It is given by:
    \begin{eqnarray}
      E_C&=&\frac{e^2}2\int d^3r d^3r'\frac{\rho_p(\vecr)\rho_p(\vecr')}{\left|\vecr-\vecr'\right|}\nonumber\\
         &&-\int d^3r\frac{3e^2}4\left(\frac3\pi\right)^{1/3}\rho_p^{4/3}\,,
    \end{eqnarray}
    with $e^2 = 1.43989$ MeV$\cdot$fm being the elementary charge
    unit.
  \item $E_\mathrm{corr}$: this term is for all additional corrections
    beyond mean field. Most calculations consider at least the
    center-of-mass correction $E_{cm}$.
  \end{itemize}
\end{subequations}
This set-up ignores tensor spin-orbit and spin-spin coupling, these
may be important for magnetic excitations and odd nuclei. They could
have a significant influence of the studies of chapter~\ref{chap_2},
because the magnetic field breaks the time-reversal
invariance. However, these applications are not the main objective of
the TDHF approach. A detailed description of all conceivable bilinear
forms in the densities and currents -- \eqref{eq_b_1} and additional
ones -- up to second order derivatives and a discussion of the
importance of the single terms can be found in
Ref.~\cite{2011JPhG...38c3101E}. Only time reversal invariant
combinations are allowed. In particular time-odd currents and
densities need to appear in bilinear form to render the functional
time-reversal invariant~\cite{2011JPhG...38c3101E}. Only the term
$\propto\rho^2$ is of zeroth order derivative, all other terms are of
second order. Taking all terms into account leads to a second term of
zeroth order derivative, being proportional to $\vecs^2$. All coupling
constants in Eqs.~\eqref{eq_b_2b} to~\eqref{eq_b_2f} might depend on
the density $\rho$. This dependence is approximated by the terms
proportional to $\rho^\alpha$ in
Eq.~\eqref{eq_b_2e}~\cite{2011JPhG...38c3101E}.

The most general form of the energy-density functional contains 23
free parameters. Galilean invariance reduces the free parameters to
17. Further assumptions and introducing a Skyrme force
from~\cite{1959NucPh...9..615S} reduces the number of independent
parameters to 10~\cite{2011JPhG...38c3101E}. The Skyrme forces work
best for closed shells. For even states it works better than for
uneven states~\cite{1959NucPh...9..615S}. In chapter~\ref{chap_2} we
analyse the double magic \isotope[16]{O} core and the even
\isotope[12]{C} and \isotope[20]{Ne} cores.

\section{Force coefficients}
\label{sec_b_3}
In the formulations above the parameters $b_0,\,b'_0,\,\dots\,b'_4$
are used, which are related to the constants appearing in the Skyrme
force, which is a density-dependent force with zero-range of
interaction. These relations are given by
\begin{eqnarray}
  \begin{split}
    b_0&=t_0\left(1+\tfrac12x_0\right)\,,\\
    b'_0&=t_0\left(\tfrac12+x_0\right)\,,\\
    b_1&=\tfrac14\left[t_1\left(1+\tfrac12x_1\right)+t_2\left(1+\tfrac12x_2\right)\right]\,,\\
    b'_1&=\tfrac14\left[t_1\left(\tfrac12+x_1\right)-t_2\left(\tfrac12+x_2\right)\right]\,,\\
    b_2&=\tfrac18\left[3t_1\left(1+\tfrac12x_1\right)-t_2\left(1+\tfrac12x_2\right)\right]\,,\\
    b'_2&=\tfrac18\left[3t_1\left(\tfrac12+x_1\right)+t_2\left(\tfrac12+x_2\right)\right]\,,\\
    b_3&=\tfrac14t_3\left(1+\tfrac12x_3\right)\,,\\
    b'_3&=\tfrac14t_3\left(\tfrac12+x_3\right)\,,\\
    b_4&=\tfrac12t_4\,.\\
  \end{split}
\end{eqnarray}
The coefficient $b'_4$ is usually fixed to $b'_4=\frac12t_4$ for most
traditional Skyrme forces, but it can be handled separately as a free
parameter. In addition to the $b$ and $b'$ parameters, the power
coefficient $\alpha$ is included, it is needed e.g. in
Eq.~\eqref{eq_b_2e}. For the input of the force one thus needs to
specify the values of the $t_i,\,x_i$ coefficients.

\section{The single-particle Hamiltonian}
\label{sec_b_4}
The first term in Eq.~\eqref{eq_2_02} is the local part of the mean
field. It acts on the wave functions like a local potential. Its
definition is as follows
\begin{subequations}
  \begin{eqnarray}
    \label{eq_b_4a} U_q&=&b_0\rho-b'_0\rho_q+b_1\tau-b'_1\tau_q-b_2\nabla\rho+b'_2\nabla\rho_q\nonumber\\
                       &&+b_3\frac{\alpha+2}3\rho^{\alpha+1}-b'_3\frac23\rho^\alpha\rho_q-b'_3\frac\alpha3\rho^{\alpha-1}\sum_{q'}\rho^2_{q'}\nonumber\\
                       &&-b_4\nabla\cdot\vecJ-b'_4\nabla\cdot\vecJ_q\,.
  \end{eqnarray}
  The second term in Eq.~\eqref{eq_2_02} referes to the kinetic energy
  of nucleons, where the ``effective mass'' is introduced by replacing
  the free-space factor $\hbar^2/(2m)$ by the isospin and space
  dependent factor
  \begin{eqnarray}
    B_q&=&\frac{\hbar^2}{2m_q}+b_1\rho-b'_1\rho_q\,.
  \end{eqnarray}
  It is clear that this factor depends on a particular
  parameterization of the Skyrme force as well as on the isospin. The
  third term in Eq.~\eqref{eq_2_02} is the spin-orbit potential which
  is given by
  \begin{eqnarray}
    \label{eq_b_4c}
    \vecW_q&=&b_4\nabla\rho+b'_4\nabla\rho_q\,.
  \end{eqnarray}
  Eqs. \eqref{eq_b_4a}-\eqref{eq_b_4c} above are time-even
  contribution to the Hamiltonian operator $\hat h_q$. Dynamical
  effects may enter due to the fourth and fifth terms in
  Eq.~\eqref{eq_2_02} which are time-odd and involve contributions
  from current and spin-density,
  \begin{eqnarray}
    \vecA_q&=&-2b_1\vecj+2b'_1\vecj_q-b_4\nabla\times\vecs-b'_4\nabla\times\vecs_q\,,\\
    \vecS_q&=&-b_4\nabla\times\vecj-b'_4\nabla\times\vecj_q\,.
  \end{eqnarray}
\end{subequations}

\section{Static Hartree-Fock}
\label{sec_b_5}
The code Sky3D uses an iterative method for the solution of the
problem at hand. The wave function in the step $n+1$ is related to the
wave-function at step $n$ by the relation
\begin{eqnarray}
  \psi_\alpha^{(n+1)}=\mathcal{O}\left\{\psi_\alpha^{(n)}-\frac\delta{\hat T+E_0}\left(\hat h^{(n)}-\left<\psi_\alpha^{(n)}\middle|\hat h^{(n)}\middle|\psi_\alpha^{(n)}\right>\right)\psi_\alpha^{(n)}\right\}\,,
  \label{eq_b_5}
\end{eqnarray}
with $\hat T=\hat p^2/(2m)$ being the operator of the kinetic energy,
$\mathcal O$ means orthonormalization of the whole set of new wave
functions, the upper index $n$ is the iteration number. To accelerate
the iteration a damping of the kinetic term is performed. This
kinetic-energy damping is particularly suited if one employs the FFT
(fast Fourier transformation) method. The damped gradient step has two
numerical parameters: the step size $\delta$ and the damping regulator
$E_0$. The latter should be chosen of order of the depth of the local
potential $U_q$, hereby $E_0=100$ MeV is found to be a save choice,
the step size should be in the range $\delta=0.1\dots0.8$. Larger
values lead to a faster iteration, but they are more likely to result
in wrong values. The optimal value depends among other things on the
choice of the Skyrme parametrization. The use of $m^*/m\approx1$
allows larger values of $\delta$ and analogous lower values are needed
for low $m^*/m$.

After performing one of these wave function iteration steps, the
densities of Eqs.~\eqref{eq_b_1a} to \eqref{eq_b_1e} are updated and
new mean fields are computed according to Eq.~\eqref{eq_2_02}. This
provides the starting point for the next iteration. The iterations are
continued until sufficient convergence is achieved. As convergence
criterion the average energy variance or the fluctuation of single
particle states is used:
\begin{subequations}
  \begin{eqnarray}
    \overline{\Delta\varepsilon}&=&\sqrt\frac{\sum_\alpha\Delta\varepsilon_\alpha^2}{\sum_\alpha1}\,,\\
    \Delta\varepsilon_\alpha^2&=&\left<\psi_\alpha\middle|\hat h^2\middle|\psi_\alpha\right>-\varepsilon_\alpha^2\,,\\
    \varepsilon_\alpha&=&\left<\psi_\alpha\middle|\hat h\middle|\psi_\alpha\right>\,,
  \end{eqnarray}
\end{subequations}
with the single particle energy $\varepsilon_\alpha$ being the
expectation value of the Hamiltonian. It becomes the eigenvalue of
Eq.~\eqref{eq_2_03} once the convergence is achieved, i.e.,
$\Delta\varepsilon_\alpha \approx 0$. When the total variance
$\overline{\Delta\varepsilon}$ vanishes, a minimum energy of
mean-field is reached. However, the minimum found with this method may
be a metastable local minimum.

The initialization in the code Sky3D is realized by implementing the
wave functions of a deformed harmonic oscillator. Hereby the states
with lowest oscillator energy are implemented first. This
initialization influences the initial state and the resulting final
state, see Ref.~\cite{2014CoPhC.185.2195M} for further details.

\section{Observables}
\label{sec_b_6}
The output of the code are the observables, which in the case of
Hartree-Fock theory are the energy of the nucleus and the densities of
nucleons. The density distribution can be described in terms of
multipole moments. To treat the center of mass motion the Cartesian
center-of-mass vector for neutrons and protons is introduced
\begin{subequations}
  \begin{eqnarray}
    \vecR_q=\frac{\int d^3r\vecr\rho_q(\vecr)}{A}\,,
  \end{eqnarray}
  where $A=\int d^3r\rho(\vecr)$ is the total mass number. Furthermore
  one can introduce the iso-scalar or total center-of-mass vector as
  \begin{eqnarray}
    \vecR_{T=0}=\frac{\int d^3r\vecr(\rho_p+\rho_n)(\vecr)}{A}\,,
  \end{eqnarray}
  as well as an iso-vector center-of-mass vector
  \begin{eqnarray}
    \vecR_{T=1}=\frac{\int d^3r\vecr\left(\frac{N}{A}\rho_p-\frac{Z}{A}\rho_n\right)(\vecr)}{A}\,.
  \end{eqnarray}
  The same relations can be written for the corresponding quadrupole
  moments, for example,
  \begin{eqnarray}
    \mathcal{Q}_{kl}^{q}&=&\int d^3r\left(3\left(r_k-R_k\right)\left(r_l-R_l\right)\right.\nonumber\\
                        &&\left.-\delta_{kl}\sum_i\left(r_i-R_i\right)^2\right)\rho^{q}\left(\vecr\right)\,,
  \end{eqnarray}
  and other defined in a simular way. The matrix $\mathcal{Q}_{kl}$ is
  not invariant under rotations of the coordinate frame. The preferred
  coordinate system is the system of principle axes, where there are
  only three non-vanishing components $Q_{xx}$, $Q_{yy}$ and $Q_{zz}$
  with the trace $Q_{xx}+Q_{yy}+Q_{zz}=0$. The general matrix
  $\mathcal{Q}_{kl}$ can be diagonalized by appropriate rotations.

  It is also useful to introduce the spherical moments for the
  quadrupole case
  \begin{eqnarray}
    Q_{2m}^{q}=\int d^3rr^2Y_{2m}\rho^{q}\left(\vecr-\vecR\right)\,,
  \end{eqnarray}
  with $r=\left|\vecr\right|$ being the absolute of $\vecr$ and
  $Y_{2m}$ being the spherical harmonics. The latter are often
  expressed as dimensionless quadrupole moments:
  \begin{eqnarray}
    a_m=\frac{4\pi}5\frac{Q_{2m}}{AR^2}\,,
  \end{eqnarray}
  with $R=r_0A^{1/3}$ being a fixed radius of the nucleus which
  depends only on the total mass number $A$. This expression again
  could be calculated for any type of moments, but in practice it is
  mainly used for isoscalar moments.

  Having dimensionless moments is an advantage, because they are free
  of an overall scale which is removed by the division by the factor
  $AR^2$. This description allows a characterization of the shape of
  the nucleus. The general $a_m$ are not invariant under rotations of
  the coordinate frame. A unique characterization is obtained by a
  transformation into the system of principle-axes. Here we have the
  following conditions: $a_{\pm1}=0$ and $a_2=a_{-2}$, thus there are
  two remaining shape parameters: $a_0$ and $a_2$. They can be
  re-expressed by the total deformation $\beta$ and the triaxiality
  $\gamma$, which are often called the Bohr-Mottelson parameters and
  are defined in the following way
  \begin{eqnarray}
    \label{eq_b_7g}
    \beta=\sqrt{a_0^2+2a_2^2}\,,\quad\gamma=\arctan\left(\frac{\sqrt2a_2}{a_0}\right)\,.
  \end{eqnarray}
  The triaxiality $\gamma$ is handled like an angle. In general it can
  take values between $0\degree$ and $360\degree$. However the
  physical relevant region is from $0\degree$ to $60\degree$. The
  other sectors lead to equivalent configurations \cite{1995GM}.

  The r.m.s. radii of neutrons and protons are defined as follows
  \begin{eqnarray}
    \label{eq_b_7h} r_\mathrm{rms}^{q}=\sqrt{\frac{\int d^3r\left(\vecr-\vecR\right)^2\rho^{q}\left(\vecr\right) }{\int d^3r\rho^{q}\left(\vecr\right)}}\,.
  \end{eqnarray}
\end{subequations}
Similarly one can define the total r.m.s. radius of a nucleus by
replacing $\rho^{q}$ by the total density $\rho$.

The total energy $E_\mathrm{tot}$ can be computed in two alternative
ways. One is to use the formula~\eqref{eq_2_01}. The alternative
starts with the equation for the energy~\cite{1995GM}
\begin{eqnarray}
  E_\mathrm{tot,\,HF}=\frac12\sum_\alpha\left(t_\alpha+\varepsilon_\alpha\right)\,,\label{eq_b_8}
\end{eqnarray}
with
$t_\alpha=\left<\psi_\alpha\middle|\hat T\middle|\psi_\alpha\right>$
being the s.p. kinetic energy and the quantity $\varepsilon_\alpha$ is
defined as
\begin{eqnarray}
  \varepsilon_\alpha&=&t_\alpha+u_\alpha\,,\quad u_\alpha=\sum_\beta\left[v_{\alpha\beta\alpha\beta}-v_{\alpha\beta\beta\alpha}\right]=\varepsilon_\alpha-t_\alpha\,,\nonumber
\end{eqnarray}
with $u_\alpha$ being the s.p. mean-field potential energy and $v$ the
two body interaction; leading to:
\begin{eqnarray}
  E_\mathrm{tot,\,HF}=\sum_\alpha t_\alpha+\frac12\sum_{\alpha\beta}\left[v_{\alpha\beta\alpha\beta}-v_{\alpha\beta\beta\alpha}\right]\,.\nonumber
\end{eqnarray}
In the case of Skyrme forces, additional rearrangement energies should
be added~\cite{2014CoPhC.185.2195M}. In the code Sky3D the total
energy is computed in both ways: from the straightforward Skyrme
energy of Eq.~\eqref{eq_2_01} and using a modified version of
Eq.~\ref{eq_b_8}.

\clearpage{\pagestyle{empty}\cleardoublepage}

\end{appendix}

\phantomsection
\addcontentsline{toc}{chapter}{References}
\bibliographystyle{unsrt} {\renewcommand{\MakeUppercase}{}
\bibliography{martin_stein_dissertation.bib}{}}

\clearpage{\pagestyle{empty}\cleardoublepage}
\phantomsection
\addcontentsline{toc}{chapter}{Acknowledgements}
\thispagestyle{empty}
\chapter*{Acknowledgements} I would like to thank all who have helped
me during the time, I did my PhD. A special thank goes to my
supervisors PD\ Dr.\ Armen Sedrakian and Prof.\ Dr.\ Joachim A.\
Maruhn, who have always supported and helped me. Moreover, I thank
Prof.\ Dr.\ Dr.\ h.c.\ Paul-Gerhard Reinhard for fruitful
discussions. I also would like to thank Prof.\ Dr.\ Xu-Guang Huang and
Prof.\ Dr.\ John W. Clark, with whom I published several
papers. Besides, I thank the computer administrations of the Service
Center and the CSC cluster FUCHS for helpful answers and keeping the
computers and computer systems in good conditions. I also want to
thank HGS-HIRe for the possibility to participate in the program of
the graduate school with many interesting and helpful events and
answers to many questions. Moreover, I thank F\&E GSI/GU and HIC for
FAIR Frankfurt for financial support. Furthermore, I thank my parents
for supporting me in many ways, in particular by spending much time in
proofreading my thesis.

\vspace*{3\baselineskip}

\noindent Frankfurt am Main, August 2015 \hfill Martin Stein

\end{document}